\pdfoutput=1
\newcommand*{\ATLASLATEXPATH}{latex/}
\documentclass[cernpreprint,txfonts,UKenglish,texlive=2011]{\ATLASLATEXPATH atlasdoc}

\usepackage{\ATLASLATEXPATH atlaspackage}
\usepackage{\ATLASLATEXPATH atlasbiblatex}

\usepackage{\ATLASLATEXPATH atlascontribute}

\usepackage{\ATLASLATEXPATH atlasphysics}
\usepackage{\ATLASLATEXPATH atlasprocess}

\graphicspath{{logos/}{figures/}}
\usepackage{rotating}
 
\usepackage{ttbarZpaper-defs}

\AtlasTitle{\boldmath Measurements of top-quark pair to \Zboson-boson cross-section ratios at $\sqs=13, 8, 7$~TeV with the ATLAS detector }

\author{The ATLAS Collaboration}

\AtlasRefCode{STDM-2016-02}

\PreprintIdNumber{CERN-EP-2016-271}

\arXivId{1612.03636}

\HepDataRecord{75536}

\AtlasJournalRef{JHEP 02 (2017) 117}
\AtlasDOI{10.1007/JHEP02(2017)117}

\AtlasAbstract{%

Ratios of top-quark pair to \Zboson-boson cross sections measured from proton--proton collisions at the LHC centre-of-mass energies of \sqs~=~13~TeV, 8~TeV, and 7~TeV are presented by the ATLAS Collaboration. Single ratios, at a given \sqs for the two processes and at different \sqs for each process, 
as well as double ratios of the two processes at different \sqs, are evaluated.  
The ratios are constructed using previously published ATLAS measurements of the \ttbar and \Zboson-boson production 
cross sections, corrected to a common phase space where required, and a new analysis
of \Zllpm where $\ell=e,\mu$ at $\sqs=13$~TeV performed with data collected in 2015 with an integrated luminosity of $3.2$~\ifb.
Correlations of systematic uncertainties are taken into account when evaluating the uncertainties in the ratios. 
The correlation model is also used to evaluate the  combined cross section of the $Z\rightarrow  e^+e^-$ and the $Z\rightarrow \mu^+ \mu^-$ channels for each \sqs value.
The results are compared to calculations performed at next-to-next-to-leading-order accuracy using recent
sets of parton distribution functions. The data demonstrate significant power to constrain the gluon distribution function for the Bjorken-$x$ values near 0.1 and the light-quark sea for $x<0.02$. 
}

\renewcommand*{\TeV}{\ifmmode {\mathrm{\ Te\kern -0.1em V}}\else
                   \textrm{Te\kern -0.1em V}\fi}%
\renewcommand*{\GeV}{\ifmmode {\mathrm{\ Ge\kern -0.1em V}}\else
                   \textrm{Ge\kern -0.1em V}\fi}%

\newcommand{\sigtttot}{\ensuremath{\sigma^{\textrm{tot}}_{\ttbar}}}
\newcommand{\sigtt}{\ensuremath{\sigma_{\ttbar}}}
\newcommand{\sigttfid}{\ensuremath{\sigma^{\textrm{fid}}_{\ttbar \to e\mu + X}}}

\newcommand{\sigzfidee}{\ensuremath{\sigma^{\textrm{fid}}_{\Zee}} }
\newcommand{\sigzfidmm}{\ensuremath{\sigma^{\textrm{fid}}_{\Zmm}} }

\newcommand{\ratztt}{\ensuremath{R_{\ttbar/\Zboson}}}
\newcommand{\ratttztt}{\ensuremath{R^{\textrm{tot/tot}}_{\ttbar/\Zboson}}}
\newcommand{\ratttztf}{\ensuremath{R^{\textrm{tot/fid}}_{\ttbar/\Zboson}}}

\newcommand{\ratzf}[2]{\ensuremath{R^{\textrm{fid}}_{\Zboson_{#1}/\Zboson_{#2}}}}
\newcommand{\ratzt}[2]{\ensuremath{R^{\textrm{tot}}_{\Zboson_{#1}/\Zboson_{#2}}}}
\newcommand{\rattt}[2]{\ensuremath{R^{\textrm{tot}}_{\ttbar_{#1}/\ttbar_{#2}}}}

\hypersetup{pdftitle={ATLAS document},pdfauthor={The ATLAS Collaboration}}

\begin{document}

\maketitle

\tableofcontents

\newpage
\section{Introduction}
\label{sec:intro}
Precision measurements of top-quark-pair~\cite{TOPADDENDUM2,Aaboud:2016pbd,TOPQ-2011-02,CMS-TOP-13-004,CMS-TOP-12-006,CMS-TOP-11-003,Khachatryan:2015uqb} and \Zboson-boson~\cite{STDM-2011-06,STDM-2012-20,STDM-2014-12,Aad:2016naf,CMS-EWK-10-005,CMS-SMP-12-011,Chatrchyan:2013tia,CMS:2014jea} 
 production by the ATLAS~\cite{PERF-2007-01} and CMS~\cite{CMS-TDR-08-001}
collaborations at the CERN Large Hadron Collider (LHC)~\cite{1748-0221-3-08-S08001} provide important tests of the  Standard Model (SM). 
The experimental precision of such measurements has reached the few-percent level in the case of the total \ttbar production cross section, \stottt, and the sub-percent level for the \Zboson-boson production cross section with subsequent \Zllpm decay within the fiducial region defined by the detector acceptance, \sfidZ. This experimental precision is complemented by an accurate determination 
of the proton--proton, $pp$, collision luminosity, which has reached a precision of approximately 2\%~\cite{DAPR-2011-01,Aaboud:2016hhf}.
These measurements are compared with theoretical predictions performed at next-to-next-to-leading-order (NNLO) and next-to-next-to-leading-log (NNLL) accuracy in quantum chromodynamics (QCD) for  \stottt~\cite{Czakon:2013goa,Czakon:2012pz,Czakon:2012zr,Baernreuther:2012ws,Cacciari:2011hy,Beneke:2011mq,Czakon:2011xx} and at NNLO QCD plus next-to-leading-order (NLO) electroweak (EW) accuracy for \sfidZ~\cite{Catani:2007vq,Catani:2009sm,Anastasiou:2003ds,FEWZ2,FEWZ3,FEWZ4,Bardin:2012jk,Arbuzov:2012dx}. Quantitative comparisons to the predictions can be used to impose constraints on a 
number of Standard Model parameters such as the parton distribution functions (PDF), the strong coupling constant ($\alpha_{\textrm{S}}$) 
and the top-quark mass (\mtop). 

Further tests may be performed by examining the centre-of-mass-energy (\sqs) dependence of the cross sections. Top-quark-pair and \Zboson-boson production at various \sqs values sample different Bjorken-$x$ regions, with higher energies sampling smaller average $x$. This dependence leads to a strong increase of the gluon-fusion-dominated \ttbar production cross section with \sqs while the increase of the \qqbar-dominated \Zboson-boson production cross section is more moderate. However, the luminosity uncertainties associated with such measurements are dominated by effects uncorrelated between different centre-of-mass energies and data-taking periods, thereby limiting the precision with which cross sections measured at different $\sqrt{s}$ values can be directly compared.

The luminosity uncertainties as well as some of the experimental uncertainties can cancel when ratios of cross sections are evaluated. The predictions of the ratios of \Zboson-boson production cross sections at different centre-of-mass energies are only moderately affected by PDF uncertainties, opening the possibility to use such measurements to cross-normalise other measurements made at different \sqs values or in different running periods, as well as providing cross-checks on the corresponding integrated-luminosity ratios and their uncertainties.

Given that the \ttbar and \Zboson-boson production dynamics are driven to a large extent by different PDFs, the ratio of these cross sections at a given centre-of-mass energy has a significant sensitivity to the  gluon-to-quark PDF ratio~\cite{Rojo:2015acz,Mangano:2012mh}.  Double ratios of  \ttbar to \Zboson-boson cross sections, i.e.\ the ratio of the ratio of the two processes at two energies, provide sensitive tests of the Standard Model predictions which do not depend significantly on the determination of the luminosity.

This paper reports an evaluation of single ratios and double ratios of the \ttbar and \Zboson-boson\footnote{Throughout this paper, $\Zboson/\gamma^*$-boson production is denoted simply by \Zboson-boson production.} production cross sections at $\sqrt s = 13, 8, 7$~TeV.
Previously published ATLAS results  for \ttbar and \Zboson-boson production at $\sqs=7$~and~8~TeV~\cite{STDM-2012-20,STDM-2014-12,TOPADDENDUM2} 
as well as for \ttbar production at $\sqs=13$~TeV~\cite{Aaboud:2016pbd} are used in the evaluation.
For the ratios involving $13$~TeV data, a new analysis of \Zllpm, where $\ell=e, \mu$,  is performed using the data collected in 2015 with an integrated luminosity of $3.2$~\ifb. This measurement uses the same methodology as a previous measurement performed at 13~TeV~\cite{Aad:2016naf} but is specifically designed to be fully synchronised with the corresponding \ttbar selection at the same energy. A detailed evaluation of the correlations of the systematic uncertainties for the different ATLAS results for the two processes and three centre-of-mass energies is performed, resulting in significant cancellations of some of the uncertainties in the ratios.  The correlation model is also used to evaluate the  combined cross section times branching ratios of the $Z\rightarrow  e^+e^-$ and the $Z\rightarrow \mu^+ \mu^-$ channels for each \sqs value, and the resulting measurements are reported together with the corresponding correlation matrix.
The data are compared to the state-of-the-art calculations performed at the highest-available order in perturbative theory, using several of the modern PDF sets. A quantitative study of projected PDF uncertainties with the inclusion of these results shows that the ATLAS measurements presented in this paper can have a significant impact in constraining  the gluon and light-quark sea distributions.

The paper is organised as follows.  The ATLAS detector is described in Section~\ref{sec:detector} and the theoretical predictions for the cross sections and their ratios are summarised in Section~\ref{sec:theory}. Section~\ref{sec:zll} describes the new measurement of the \Zboson-boson production cross section times the branching ratio for \Zllpm at $\sqs=13$~TeV.  The cross-section single and double ratios as well as the combined cross sections, including full correlation information, are evaluated and compared to theoretical predictions in Section~\ref{sec:rat}, and the ability  of these data to further constrain the PDF distributions is discussed. Section~\ref{sec:conclusion} summarises the results obtained in the paper. An Appendix contains additional predictions that use the total rather than fiducial \Zboson-boson cross sections and also presents all experimental results in tabular form.
\section{ATLAS detector}
\label{sec:detector}

\newcommand{\AtlasCoordFootnote}{%
ATLAS uses a right-handed coordinate system with its origin at the nominal interaction point (IP)
in the centre of the detector and the $z$-axis along the beam pipe.
The $x$-axis points from the IP to the centre of the LHC ring,
and the $y$-axis points upwards.
Cylindrical coordinates $(r,\phi)$ are used in the transverse plane, 
$\phi$ being the azimuthal angle around the $z$-axis. 
The pseudorapidity is defined in terms of the polar angle $\theta$ as $\eta = -\ln \tan(\theta/2)$ and
the rapidity is given by $y = \frac{1}{2} \ln \left( \frac{E+p_z}{E-p_z} \right)$, where $E$ is the jet/particle energy and $p_z$ is the $z$-component of
the jet/particle momentum. }

The ATLAS detector~\cite{PERF-2007-01} at the LHC is a multi-purpose particle detector with a forward-backward symmetric cylindrical geometry and a near $4\pi$ coverage in  solid angle.\footnote{\AtlasCoordFootnote} It consists of an inner tracking detector, electromagnetic and hadronic calorimeters, and a muon spectrometer. 

The inner detector is surrounded by a thin superconducting solenoid magnet and includes silicon detectors, which provide precision tracking in the pseudorapidity range $|\eta|<2.5$, and a transition-radiation tracker providing additional tracking and electron identification information for $|\eta|<2.0$. 
For the $\sqs=13$~TeV data-taking period, the inner detector also includes a silicon-pixel insertable B-layer~\cite{ibl}, providing an additional layer of tracking information close to the interaction point.
A lead/liquid-argon (LAr) electromagnetic calorimeter covers the region $|\eta|<3.2$. Hadronic calorimetry is provided by a steel/scintillator-tile calorimeter for $|\eta|<1.7$ and two copper/LAr hadronic endcap calorimeters for $1.5< |\eta|<3.2$. The forward region is covered by additional coarser-granularity LAr calorimeters up to $|\eta| = 4.9$. 
The muon spectrometer consists of three large superconducting toroids each containing eight coils, precision tracking chambers covering the region $|\eta|<2.7$, and separate trigger chambers up to $|\eta|=2.4$. 

For the data taken at 7~and 8~TeV, a three-level trigger system was used. The first-level trigger is implemented in hardware and uses a subset of the detector information. This is followed by two software-based trigger levels that together reduce the accepted event rate to approximately \SI{400}{\hertz}.
For the data taken at 13~TeV, the trigger was changed~\cite{ATL-DAQ-PUB-2016-001} to a two-level system, using custom hardware followed by a software-based level which runs offline reconstruction software, reducing the event rate to approximately 1~kHz.

The data used in this paper were collected by the ATLAS detector in 2011, 2012, and 2015 and correspond to total integrated luminosities of $4.6, 20.2$, and $3.2$~fb$^{-1}$ at $\sqrt s = 7, 8,$ and $13$~TeV, respectively.

\section{Theoretical predictions}
\label{sec:theory}
In this section, predictions are presented at  NNLO+NNLL accuracy for the production cross section of a top-quark pair and at NNLO accuracy for the production cross section of a \Zboson boson times the branching ratio of the decay into a lepton pair of flavour $\ell^{+}\ell^{-}=e^{+}e^{-}\mathrm{~or~} \mu^{+}\mu^{-}$ within the dilepton invariant mass range   $ 66 < m_{\ell\ell} < 116{\GeV}$. The total cross sections for these processes, denoted respectively by \stottt and \stotZ, are calculated for the centre-of-mass energies $\sqrt s = 13, 8, 7$~\TeV. Also presented are predictions at NNLO accuracy for the \Zboson-boson production cross section times the same branching ratio within a fiducial region defined by the detector acceptance, $\sfidZ = \stotZ \cdot A$, where the acceptance factor $A$ is  expressed as the fraction of decays satisfying the matching fiducial acceptance (geometric and kinematic requirements) at the Monte Carlo generator level. The \Zboson-boson fiducial phase space is defined by the lepton transverse momentum $\pT^{\ell}>25$~GeV, the lepton pseudorapidity $|\eta_{\ell}|<$~2.5, and   $66 < m_{\ell\ell} < 116{\GeV}$. Predictions of top-quark-pair fiducial cross sections are not yet available at NNLO accuracy.

\subsection{\Zboson-boson cross-section predictions}
\label{sec:ztheory}

Theoretical predictions of the fiducial and total \Zboson-boson production cross sections times the branching ratio of the decay into a lepton pair \Zllpm at $\sqrt s = 13, 8, 7$~TeV are computed using a version of \textsc{DYNNLO}~1.5~\cite{Catani:2007vq,Catani:2009sm} optimised for speed of computation, for both the central values and all variations reflecting systematic uncertainties, thereby providing NNLO QCD calculations. Electroweak corrections at NLO, calculated with \textsc{Fewz}~3.1~\cite{Anastasiou:2003ds,FEWZ2,FEWZ3,FEWZ4}, are calculated in the $G_\mu$ EW scheme~\cite{Hollik:1988ii}. The cross sections are calculated for \Zboson-boson decays into leptons at Born level, i.e.\ before the decay leptons emit photons via final-state radiation, to match the definition of the cross sections measured in data. Thus, the following components are included: virtual QED and weak corrections, initial-state radiation (ISR), and interference between ISR and FSR~\cite{Dittmaier:2009cr}. 
The NNLO PDFs  \CT~\cite{Dulat:2015mca}, \NNPDF~\cite{Ball:2014uwa}, \MMHT~\cite{Harland-Lang:2014zoa}, \ABM~\cite{Alekhin:2013nda}, \HERAPDF~\cite{Abramowicz:2015mha}, 
and \ATLASepWZ~\cite{Aad:2012sb} are used in the comparisons to data. The \CT PDF set is used as the baseline for the predictions. 

The systematic uncertainties in the predictions are dominated by the knowledge of proton PDFs. 
These uncertainties are obtained from the sum in quadrature of the differences between predictions obtained with the central PDF values and those obtained using the variations (eigenvectors) of the respective PDF sets.
  Where appropriate, asymmetric uncertainties are determined using separate sums of negative and positive variations. The \CT uncertainties are rescaled from 90\%  to 68\% confidence level (CL).
The uncertainties due to the strong coupling constant are estimated following the prescription given with the \CT PDF, varying $\alpha_{\textrm{S}}$ by $\pm 0.001$ to correspond to $68\%$ CL. 
The QCD scale uncertainties are defined by the envelope of variations in which the renormalisation ($\mu_{\textrm{R}}$) and factorisation ($\mu_{\textrm{F}}$) scales are changed by factors of two with an additional constraint of $0.5\leq \mu_{\textrm{R}} / \mu_{\textrm{F}} \leq2$. The dynamic scale $m_{\ell\ell}$ is used as the central value for the \Zboson-boson predictions. 
The limitations in the NNLO calculations, referred to as the ``intrinsic'' uncertainties, are estimated by comparing the predictions calculated with the optimised version of \textsc{DYNNLO}~1.5 to the ones obtained with \textsc{Fewz}~3.1. For the total cross-section predictions, these differences are found to be $<0.2$\% and hence are negligible. For the fiducial cross-section predictions, these differences are larger due to a feature of the calculations involving leptons with symmetric \pT requirements, resulting in consistently larger values from \textsc{Fewz}. The differences are calculated using the \CT PDF to obtain the central value in both cases, and are approximately 0.7\% at all three $\sqrt s$ values. 

The predictions of the fiducial cross sections, together with their uncertainties, are given in Table~\ref{tab:xsec_predictions_Zfidtttot} while the predictions of the total cross sections are given in Table~\ref{tab:xsec_predictions_ZtotZZtt3} of Appendix~\ref{AppA}.

\begin{table}[t]
\centering
\begin{tabular}{|l|l|l|l||l|l|l|} 
\hline  
					& \multicolumn{3}{c||}{\sfidZ}	& \multicolumn{3}{c|}{\stottt} \T \B \\
\hline
\multicolumn{1}{|c|}{\sqs\ [\TeV]}	& \multicolumn{1}{c|}{13} &  \multicolumn{1}{c|}{8} &  \multicolumn{1}{c||}{7}		&  \multicolumn{1}{c|}{13}	&  \multicolumn{1}{c|}{8}	&  \multicolumn{1}{c|}{7} \T	\\ \hline 
\multicolumn{1}{|l|}{Central value [pb]}     &\numRP{743.9}{0}&\numRP{486.0}{0} &\numRP{432.0}{0}& \numRP{841.8}{0}&\numRP{258.9}{0}&\numRP{181.7}{0} \T	\\ \hline
\multicolumn{1}{|l|}{Uncertainties [\%]}& & & & & & \\
\multicolumn{1}{|l|}{\hspace{0.2cm}PDF}              &$^{+\numRP{2.73}{1}}_{\numRP{-3.36}{1}}$&$^{+\numRP{2.49}{1}}_{\numRP{-3.11}{1}}$&$^{+\numRP{2.46}{1}}_{\numRP{-3.03}{1}}$&$^{+\numRP{2.60}{1}}_{\numRP{-2.65}{1}}$&$^{+\numRP{3.91}{1}}_{\numRP{-3.44}{1}}$&$^{+\numRP{4.40}{1}}_{\numRP{-3.70}{1}}$ \T \B \\
\multicolumn{1}{|l|}{\hspace{0.2cm}$\alpha_{\textrm{S}}$} &$^{+\numRP{0.94}{1}}_{\numRP{-1.08}{1}}$&$^{+\numRP{0.97}{1}}_{\numRP{-0.80}{1}}$&$^{+\numRP{1.04}{1}}_{\numRP{-0.67}{1}}$&$^{+\numRP{1.86}{1}}_{\numRP{-1.84}{1}}$&$^{+\numRP{2.11}{1}}_{\numRP{-2.06}{1}}$&$^{+\numRP{2.19}{1}}_{\numRP{-2.12}{1}}$ \T \B \\
\multicolumn{1}{|l|}{\hspace{0.2cm}Scale}            &$^{+\numRP{0.46}{1}}_{\numRP{-0.78}{1}}$&$^{+\numRP{0.51}{1}}_{\numRP{-0.49}{1}}$&$^{+\numRP{0.74}{1}}_{\numRP{-0.32}{1}}$&$^{+\numRP{2.44}{1}}_{\numRP{-3.61}{1}}$&$^{+\numRP{2.58}{1}}_{\numRP{-3.50}{1}}$&$^{+\numRP{2.63}{1}}_{\numRP{-3.47}{1}}$ \T \B \\
\multicolumn{1}{|l|}{\hspace{0.2cm}Intrinsic \Zboson}        &$^{+\numRP{0.70}{1}}_{\numRP{-0.70}{1}}$&$^{+\numRP{0.70}{1}}_{\numRP{-0.70}{1}}$&$^{+\numRP{0.70}{1}}_{\numRP{-0.70}{1}}$& N/A & N/A & N/A \T \B \\
\multicolumn{1}{|l|}{\hspace{0.2cm}\mtop}            & N/A & N/A & N/A & $^{+\numRP{2.79}{1}}_{\numRP{-2.70}{1}}$ & $^{+\numRP{3.00}{1}}_{\numRP{-2.90}{1}}$ & $^{+\numRP{3.07}{1}}_{\numRP{-2.97}{1}}$ \T \B \\ \cline{2-7}
\multicolumn{1}{|l|}{\hspace{0.2cm}Total} &$^{+\numRP{3.0063}{1}}_{-\numRP{3.6816}{1}}$&$^{+\numRP{2.8091}{1}}_{-\numRP{3.3230}{1}}$&$^{+\numRP{2.8585}{1}}_{-\numRP{3.1972}{1}}$&$^{+\numRP{4.89592}{0}}_{-\numRP{5.54797}{0}}$&$^{+\numRP{5.98415}{0}}_{-\numRP{6.01831}{0}}$&$^{+\numRP{6.40469}{0}}_{-\numRP{6.24019}{0}}$ \T \B \\ \hline
\end{tabular}  
\caption{Predictions of the fiducial cross section, \sfidZ, and the total cross section, \stottt, at $\sqrt s = 13, 8, 7$~TeV using the \CT PDF.  The uncertainties, given in \%, correspond to variations of: \CT eigenvector set at 68\% CL, $\alpha_{\textrm{S}}$, QCD scale, intrinsic \Zboson-boson prediction, and top-quark mass, as described in the text.  The statistical uncertainties in the predictions are $\le 1$~pb for the \Zboson boson and $\le 0.1$~pb for \ttbar and are not given in the table. The notation N/A means ``not applicable''.} 
\label{tab:xsec_predictions_Zfidtttot}
\end{table}

\subsection{\ttb cross-section predictions}
\label{sec:ttbartheory}

Theoretical predictions~\cite{Czakon:2013goa,Czakon:2012pz,Czakon:2012zr,Baernreuther:2012ws,Cacciari:2011hy,Beneke:2011mq} of the total \ttb production cross sections at $\sqrt s = 13, 8, 7$~TeV are computed using Top++v2.0~\cite{Czakon:2011xx} for the central values and for all variations reflecting systematic uncertainties, thereby providing NNLO+NNLL resummed QCD calculations. The systematic uncertainties in the predictions are performed as for those of the \Zboson boson, with the following exceptions. Since there is no alternative calculation of the NNLO \ttb cross section available, no intrinsic uncertainty is assigned to its cross-section prediction. It was verified that the code Hathor v1.5~\cite{Aliev:2010zk}, which implements the exact NNLO $t\overline{t}$ cross sections, matches the results obtained with Top++v2.0. 
The \ttb production cross section also has a significant dependence on the value of the top-quark mass, \mtop. A systematic uncertainty is assessed by varying the  mass of the top quark by $\pm 1$~GeV from the baseline value of 172.5~GeV used to obtain the central value of the predictions, resulting in an uncertainty in the cross section of approximately 3\%. The predictions of the total cross sections, together with their uncertainties, are given in Table~\ref{tab:xsec_predictions_Zfidtttot}.

\subsection{Predictions of ratios of cross sections}
\label{sec:ratiotheory} 
 
The \Zboson-boson cross-section measurements made in a fiducial phase space require only a small extrapolation from the experimental phase space and hence benefit from significantly reduced theoretical uncertainties in comparison to the measurements extrapolated to the total phase space. For this reason, the \Zboson-boson fiducial cross sections are primarily used in the measurements of the cross-section ratios. The predictions given in Table~\ref{tab:xsec_predictions_Zfidtttot} are used to build cross-section ratios for 
\begin{itemize}
\item a given process at the different $\sqrt s$: $\ratzf{i}{j}= \sfidZs{i} / \sfidZs{j}$ and $\rattt{i}{j}= \stottts{i} / \stottts{j}$,
\item different processes at the same  $\sqrt s$: $\ratttztf(i~\mathrm{TeV})= \stottts{i}/\sfidZs{i}$, 
\item different processes at the different $\sqrt s$: $\ratttztf(i/j)=\left[\stottts{i}/\sfidZs{i}\right]/\left[\stottts{j}/\sfidZs{j}\right]$ denoted in this paper as double ratios, 
\end{itemize}
where $i,j=13, 8, 7$.
The first set of predictions is presented in Table~\ref{tab:xsec_ratio_pred_ZZtt} while the latter two are presented in Table~\ref{tab:xsec_ratio_pred_ZZtt2}. The corresponding ratios using the total \Zboson-boson production cross sections rather than the fiducial ones are given in Tables~\ref{tab:xsec_predictions_ZtotZZtt3} and~\ref{tab:xsec_ratio_pred_ZZtt4} of Appendix~\ref{AppA}.

The choice of correlation model when combining the theoretical uncertainties in the ratios is not unique. For this paper, the treatment of the systematic uncertainties is taken as follows. 
The  PDF uncertainties are considered as correlated, eigenvector by eigenvector, between predictions. The 
QCD scale uncertainties are treated as uncorrelated between processes but correlated, variation by variation, at the different $\sqrt s$ values for a given process.   
The $\alpha_{\textrm{S}}$  uncertainties are  correlated between predictions. The \Zboson-boson intrinsic and \mtop uncertainties are both considered as correlated at the different $\sqrt s$ values within their respective processes. In the few cases where the coherent variation of a source of systematic uncertainty in the numerator and in the denominator of a ratio results in variations of the same sign, only the largest variation is added in the total uncertainty of the corresponding sign.

\begin{table}[t]
\centering
\begin{tabular}{|l|l|l|l||l|l|l|} 
\hline  
					& \multicolumn{3}{c||}{\ratzf{i}{j}}	& \multicolumn{3}{c|}{\rattt{i}{j}} \T \B \\
\hline
\multicolumn{1}{|c|}{$i/j$}	& \multicolumn{1}{c|}{13/7} &  \multicolumn{1}{c|}{13/8} &  \multicolumn{1}{c||}{8/7}		&  \multicolumn{1}{c|}{13/7}	&  \multicolumn{1}{c|}{13/8}	&  \multicolumn{1}{c|}{8/7} \T	\\ \hline 
\multicolumn{1}{|l|}{Central value}     &\numRP{1.7221}{3}&\numRP{1.5308}{3} &\numRP{1.1250}{3}& \numRP{4.6335}{3}&\numRP{3.2512}{3}&\numRP{1.4252}{3}	\\ \hline
\multicolumn{1}{|l|}{Uncertainties [\%]}& & & & & & \\
\multicolumn{1}{|l|}{\hspace{0.2cm}PDF}               &$^{+\numRP{1.02201}{1}}_{\numRP{-0.929098}{1}}$&$^{+\numRP{0.803501}{1}}_{\numRP{-0.718579}{1}}$&$^{+\numRP{0.222222}{2}}_{\numRP{-0.213333}{2}}$&$^{+\numRP{1.86036}{1}}_{\numRP{-2.33301}{1}}$&$^{+\numRP{1.40256}{1}}_{\numRP{-1.75935}{1}}$&$^{+\numRP{0.456076}{1}}_{\numRP{-0.582374}{1}}$ \T \B \\
\multicolumn{1}{|l|}{\hspace{0.2cm}$\alpha_{\textrm{S}}$}     &$^{-\numRP{0.10}{1}}_{\numRP{-0.41}{1}}$&
$^{-0.1}_{\numRP{-0.28}{1}}$
&$^{-\numRP{0.08}{1}}_{\numRP{-0.13}{1}}$&$^{-\numRP{0.32}{2}}_{+\numRP{0.29}{2}}$&$^{-\numRP{0.25}{2}}_{+\numRP{0.22}{2}}$&$^{-\numRP{0.08}{2}}_{+\numRP{0.07}{2}}$ \T \B \\
\multicolumn{1}{|l|}{\hspace{0.2cm}Scale}              &$^{+\numRP{0.03}{2}}_{-\numRP{0.60}{2}}$&$^{+\numRP{0.02}{2}}_{-\numRP{0.29}{2}}$&$^{+\numRP{0.02}{2}}_{-\numRP{0.31}{2}}$&$^{+\numRP{0.19}{2}}_{-\numRP{0.26}{2}}$&$^{+\numRP{0.13}{2}}_{-\numRP{0.19}{2}}$&$^{+\numRP{0.05}{2}}_{-\numRP{0.07}{2}}$ \T \B \\ 
\multicolumn{1}{|l|}{\hspace{0.2cm}\mtop}  & N/A & N/A & N/A           &$^{+\numRP{0.29}{2}}_{-\numRP{0.29}{2}}$ & $^{+\numRP{0.22}{2}}_{-\numRP{0.22}{2}}$ & $^{+\numRP{0.07}{2}}_{-\numRP{0.07}{2}}$ \T \B \\  \cline{2-7}
\cline{2-7}
\multicolumn{1}{|l|}{\hspace{0.2cm}Total} &$^{+\numRP{1.02245}{1}}_{-\numRP{1.18025}{1}}$&$^{+\numRP{0.804061}{1}}_{-\numRP{0.825167}{1}}$&$^{+\numRP{0.223120}{2}}_{-\numRP{0.398134}{2}}$&$^{+\numRP{1.915}{1}}_{-\numRP{2.388}{1}}$&$^{+\numRP{1.444}{1}}_{-\numRP{1.801}{1}}$&$^{+\numRP{0.470}{1}}_{-\numRP{0.596}{1}}$ \T \B \\ \hline
\end{tabular}   
\caption{Predictions of the cross-section ratios \ratzf{i}{j} and \rattt{i}{j} at the different $\sqrt s$ values where $i/j=13/7, 13/8, $ and $8/7$ using the \CT PDF.  The uncertainties, given in \%, correspond to variations of: \CT eigenvector set at 68\% CL, $\alpha_{\textrm{S}}$, and QCD scale, as described in the text.  The statistical uncertainties in the predictions are $\le 0.002$ for the \Zboson process and $\le 0.001$ for the \ttbar process and are not given in the table. The notation N/A means ``not applicable''.}
\label{tab:xsec_ratio_pred_ZZtt} 
\end{table}

\begin{table}[t]
\centering
\begin{tabular}{|l|l|l|l||l|l|l|} 
\hline  
					& \multicolumn{3}{c||}{$\ratttztf(i~\mathrm{TeV})$}	& \multicolumn{3}{c|}{$\ratttztf(i/j)$} \T \B \\
\hline
\multicolumn{1}{|c|}{$i$ or $i/j$}	& \multicolumn{1}{c|}{13} &  \multicolumn{1}{c|}{8} &  \multicolumn{1}{c||}{7}		&  \multicolumn{1}{c|}{13/7}	&  \multicolumn{1}{c|}{13/8}	&  \multicolumn{1}{c|}{8/7} \T	\\ \hline 
\multicolumn{1}{|l|}{Central value}     &\numRP{1.1315}{3}&\numRP{0.5328}{3} &\numRP{0.4206}{3}& \numRP{2.6906}{3}&\numRP{2.1238}{3}&\numRP{1.2669}{3}	\\ \hline
\multicolumn{1}{|l|}{Uncertainties [\%]}& & & & & & \\
\multicolumn{1}{|l|}{\hspace{0.2cm}PDF}               &$^{+\numRP{5.65371}{0}}_{\numRP{-4.68198}{0}}$&$^{+\numRP{6.56660}{0}}_{\numRP{-5.06567}{0}}$&$^{+\numRP{6.88836}{0}}_{\numRP{-5.22565}{0}}$&$^{+\numRP{1.52360}{1}}_{\numRP{-2.04385}{1}}$&$^{+\numRP{1.12994}{1}}_{\numRP{-1.55367}{1}}$&$^{+\numRP{0.394633}{1}}_{\numRP{-0.473560}{1}}$ \T \B \\
\multicolumn{1}{|l|}{\hspace{0.2cm}$\alpha_{\textrm{S}}$}     &$^{+\numRP{0.91}{1}}_{\numRP{-0.77}{1}}$&$^{+\numRP{1.14}{1}}_{\numRP{-1.26}{1}}$&$^{+\numRP{1.14}{1}}_{\numRP{-1.46}{1}}$&$^{-\numRP{0.22}{2}}_{+\numRP{0.70}{2}}$&$^{-\numRP{0.22}{2}}_{+\numRP{0.50}{2}}$&$^{-\numRP{0.00}{2}}_{+\numRP{0.20}{2}}$ \T \B \\
\multicolumn{1}{|l|}{\hspace{0.2cm}Scale}              &$^{+\numRP{2.56}{1}}_{-\numRP{3.64}{1}}$&$^{+\numRP{2.63}{1}}_{-\numRP{3.54}{1}}$&$^{+\numRP{2.65}{1}}_{-\numRP{3.55}{1}}$&$^{+\numRP{0.62}{2}}_{-\numRP{0.27}{2}}$&$^{+\numRP{0.32}{2}}_{-\numRP{0.20}{2}}$&$^{+\numRP{0.31}{2}}_{-\numRP{0.07}{2}}$ \T \B \\ 
\multicolumn{1}{|l|}{\hspace{0.2cm}Intrinsic  \Zboson}        &$^{+\numRP{0.70}{1}}_{-\numRP{0.70}{1}}$&$^{+\numRP{0.70}{1}}_{-\numRP{0.70}{1}}$&$^{+\numRP{0.70}{1}}_{-\numRP{0.70}{1}}$& $^{+0.00}_{-0.00}$ & $^{+0.00}_{-0.00}$ & $^{+0.00}_{-0.00}$ \T \B \\
\multicolumn{1}{|l|}{\hspace{0.2cm}\mtop}            &$^{+\numRP{2.79}{1}}_{\numRP{-2.70}{1}}$ & $^{+\numRP{3.00}{1}}_{\numRP{-2.90}{1}}$ & $^{+\numRP{3.07}{1}}_{\numRP{-2.97}{1}}$ & $^{+\numRP{0.29}{2}}_{\numRP{-0.29}{2}}$ & $^{+\numRP{0.22}{2}}_{\numRP{-0.22}{2}}$ & $^{+\numRP{0.07}{2}}_{\numRP{-0.07}{2}}$ \T \B \\  \cline{2-7}
\multicolumn{1}{|l|}{\hspace{0.2cm}Total} &$^{+\numRP{6.91841}{0}}_{-\numRP{6.58034}{0}}$&$^{+\numRP{7.82242}{0}}_{-\numRP{6.96427}{0}}$&$^{+\numRP{8.12708}{0}}_{-\numRP{7.21162}{0}}$&$^{+\numRP{1.814}{1}}_{-\numRP{2.093}{1}}$&$^{+\numRP{1.295}{1}}_{-\numRP{1.597}{1}}$&$^{+\numRP{0.548}{1}}_{-\numRP{0.484}{1}}$ \T \B \\ \hline

\end{tabular}  
\caption{
Predictions of the cross-section ratios $\ratttztf(i~\mathrm{TeV})$ and $\ratttztf(i/j)$ at the different $\sqrt s$ values where $i,j=13, 8, 7$ using the \CT PDF.  The uncertainties, given in \%, correspond to variations of: \CT eigenvector set at 68\% CL, $\alpha_{\textrm{S}}$, QCD scale,  intrinsic \Zboson-boson prediction, and top-quark mass, as described in the text.  The statistical uncertainties in the predictions are $\le 0.001$ for $\ratttztf(i~\mathrm{TeV})$  and $\le 0.003$ for $\ratttztf(i/j)$ and are not given in the table. 
}
\label{tab:xsec_ratio_pred_ZZtt2} 
\end{table}

\section{Analysis of $\Zllpm$ at $\sqs = 13$~TeV}
\label{sec:zll}
\subsection{Data set and simulated event samples}
\label{sec:zllsim}
The data sets used in this analysis  of $\Zllpm$ at $\sqs = 13$~TeV were collected  by the ATLAS detector during the period of August to November 2015. During this period, the LHC circulated \SI{6.5}{\TeV} proton beams with a \SI{25}{ns} bunch spacing.  The peak delivered instantaneous luminosity was $\mathcal{L}=\SI{5e33}{\per \cm \squared \per \s}$ and the mean number of $pp$ interactions per bunch crossing (hard scattering and pile-up events) was $\langle \mu \rangle = 13$.  The data set corresponds to a total integrated luminosity of 3.2~fb$^{-1}$.

Monte Carlo simulations are used to evaluate the selection efficiency for signal events and the contribution of several background processes to the analysed data set. All of the samples are processed with the {\textsc{Geant}4}-based simulation~\cite{Agostinelli:2002hh} of the ATLAS detector~\cite{SOFT-2010-01}.

Events containing a \Zboson boson decaying to a lepton pair, \Zllpm where $\ell=e, \mu, \tau$, and events from the leptonic decay of \Wboson bosons are generated with the \POWHEG-Box~v2 Monte Carlo program~\cite{Nason:2004rx,Frixione:2007vw,Alioli:2010xd,Alioli:2008gx,Frixione:2007nw} interfaced to the \PYTHIA~v.8.186~\cite{,Sjostrand:2007gs} parton shower model.
The CT10 PDF set~\cite{Lai:2010vv} is used in the matrix element and the AZNLO~\cite{STDM-2012-23} set of generator-parameter values (tune) is used, with the CTEQ6L1~\cite{Pumplin:2002vw} PDF set, for the modelling of non-perturbative effects.
The EvtGen~v.1.2.0 program~\cite{EvtGen} is used for properties of the bottom and charm hadron decays, and
\textsc{Photos}++ version 3.52~\cite{Golonka:2005pn,Davidson:2010ew} is used for QED emissions from electroweak vertices and charged leptons.
Samples of top-quark pairs are generated with the \POWHEG-Box v2 generator, which uses the four-flavour scheme for the NLO matrix element calculations together with the fixed four-flavour PDF set CT10f4. The top-quark-spin correlations are preserved in these samples and the top-quark mass is set to 172.5~GeV.
The parton shower, fragmentation, and underlying event are simulated using \PYTHIA~v.6.428~\cite{Sjostrand:2006za} with the CTEQ6L1 PDF set and the corresponding Perugia 2012 tune (P2012)~\cite{Skands:2010ak}. The EvtGen~v1.2.0 program is used for properties of the bottom and charm hadron decays.
Diboson processes are simulated using the \SHERPA~v2.1.1 generator~\cite{Gleisberg:2008ta}.
Multiple overlaid $pp$ collisions are simulated with the soft QCD processes of \PYTHIA~v.8.186 using the A2 tune~\cite{ATLAS-PHYS-PUB-2012-003} and the \textsc{MSTW2008LO} PDF~\cite{Martin:2009iq}.

The Monte Carlo events are reweighted so that the $\mu$ distribution matches the one observed in the data. Correction factors are applied to the simulated events to account for the differences observed between the data and MC simulation in the trigger, identification, reconstruction, and isolation efficiencies for the selected electron and muon candidates. Electron-energy- and muon-momentum-calibration corrections are applied as well.

For the comparison to the data distributions, the signal MC simulations are normalised to the cross sections measured by this analysis. 
The remaining simulations are  normalised to the predictions of the highest-order available QCD calculations, with uncertainties of 5\% for the single-boson processes and 6\% for the diboson and top-quark processes.

\subsection{Event selection}
\label{sec:zllevent}

The selections of electron and muon candidates from the decay of the \Zboson boson are designed to be fully synchronised to the $t\bar{t}$ selection at 13~TeV~\cite{Aaboud:2016pbd} e.g.\ using the same lepton trigger, identification, and kinematical requirements on the same data set.

Candidate events are selected using triggers which require at least one electron or muon to exceed 
transverse momentum thresholds of $\pT=\SI{24}{\GeV}$ or \SI{20}{\GeV}, respectively, with some
isolation requirements for the muon trigger. To recover possible efficiency losses at high momenta, additional electron triggers  with thresholds of $\pT \ge \SI{60}{\GeV}$ 
and a muon trigger with a threshold of $\pT=\SI{50}{\GeV}$ are included. Candidate events are required to have a primary vertex, defined as the vertex with the highest sum of track $\pT^2$, with at least two associated tracks with $\pT > \SI{400}{\MeV}$.

Electron candidates are required to have $\pT>\SI{25}{\GeV}$ and to pass ``medium'' likelihood-based identification requirements~\cite{ATLAS-CONF-2016-024} optimised for the 2015 operating conditions, within the fiducial region $|\eta|<2.47$, excluding candidates in the transition region between the barrel and endcap electromagnetic calorimeters, $1.37<|\eta|<1.52$. Muon candidates are considered for $|\eta|<2.4$ with $\pT>\SI{25}{\GeV}$ and must pass ``medium'' identification requirements~\cite{Aad:2016jkr} also optimised for the 2015 operating conditions. At least one of the lepton candidates is required to match the lepton that triggered the event.
The electron and muon candidates must also satisfy \pT-dependent cone-based isolation requirements,
using tracking detector and calorimeter information described in Refs.~\cite{Aad:2015yja,Aad:2014lwa}, respectively. The isolation requirements are
tuned so that the lepton-isolation efficiency is at least 90\% for $\pT>\SI{25}{\GeV}$, increasing to 99\% at 60~GeV. 
Both the electron and muon tracks are required to be associated with the primary vertex, using constraints on the transverse impact-parameter significance, $|d_0|/\delta d_0$, where $d_0$ is the transverse impact parameter and $\delta d_0$ is its uncertainty, and on the longitudinal impact parameter, $z_0$, corrected for the reconstructed position of the primary vertex. The transverse impact-parameter significance is required to be less than five for electrons and three for muons, while the absolute value of the corrected $z_0$ multiplied by the sine of the track polar angle is required to be less than 0.5~mm.

Events containing a \Zboson-boson candidate are chosen by requiring exactly two selected leptons
of the same flavour but of opposite charge with an invariant mass of  $66 < m_{\ell\ell} < \SI{116}{GeV}$.
A total of 1,367,026~candidates and 1,735,197~candidates pass all requirements in the electron and muon channels, respectively.

\subsection{Background processes }
\label{sec:zllbg}

Contributions from the single-boson (\Wln and \Ztautaupm), diboson, and top-quark-pair components of the background are estimated from the Monte Carlo samples described in Section~\ref{sec:zllsim}.  The \Ztautaupm process with the subsequent leptonic decay of the $\tau$  is treated as a background.

Events involving semileptonic decays of heavy quarks, hadrons misidentified as leptons, and, in the case of the electron channel, electrons from photon conversions (all referred to collectively as ``multijet'' events) are a minor background in this analysis. The multijet background is estimated in both channels using data-driven methods. The transverse impact-parameter distribution $d_0$ times the value of the charge of the lepton, to take into account the direction of photon radiation, is used in a template fit in the region where the transverse impact-parameter requirement is inverted. The contribution of multijet events to the event selection in both channels is found to be $<0.1$\% and therefore is neglected in the calculation of the central value of the cross section but contributes  0.05\%  to the cross-section uncertainty.
 
The total background event rate contributing to the \Zllpm selection in both channels is approximately 0.5\%, dominated by $t\overline{t}$ production  while the sum of all electroweak backgrounds is 0.2\%.

\subsection{Cross-section measurement and estimation of the systematic uncertainties}
\label{sec:method}

The methodology for the evaluation of the inclusive fiducial cross section is the same as in previous ATLAS publications~\cite{STDM-2014-12,Aad:2016naf}. The fiducial production cross section of a \Zboson boson times the branching ratio of the decay of the \Zboson boson into a lepton pair of flavour $\ell^{+}\ell^{-}=e^{+}e^{-}\mathrm{~or~} \mu^{+}\mu^{-}$  can be expressed as a ratio of the numbers of background-subtracted data events $N$ to the product of the integrated luminosity of the data $\mathcal{L}$  and a correction factor $C$:
\begin{equation}
\sfidZ = \frac{ N}{\mathcal{L} \cdot C}.
\label{EQ::ZFid}
\end{equation}
The correction factor $C$ is the ratio of the total number of simulated events which pass the final \Zboson-boson selection requirements after reconstruction to the total number of simulated events within the fiducial acceptance defined in  Section~\ref{sec:theory}. This factor,  defined at Born level,   includes the efficiencies for triggering on, reconstructing, and identifying the \Zboson-boson decay products within the acceptance, and also accounts for the slight difference between the fiducial and reconstructed phase spaces.  The contribution from the \Ztautaupm process with the subsequent leptonic decay of the $\tau$ is considered as a background and is not part of the fiducial definition. The total cross section, evaluated by extrapolating to the full phase space by use of the acceptance factor $A$ ($\stotZ = \sfidZ / A$), is further elaborated in Appendix~\ref{AppB}.

The experimental systematic uncertainties in the measurements of the cross section enter via the evaluation of the correction factor  and the luminosity in the denominator of Eq.~(\ref{EQ::ZFid}), as well as through the estimation of the background subtracted from the candidate events in its numerator.

The sources of systematic uncertainties in the correction factors $C$, summarised in Table~\ref{tab:systotZ}, are as follows. 
\begin{itemize}
\item \textit{Trigger:} The lepton trigger efficiency is estimated in simulation, with a dedicated data-driven analysis
  performed to obtain the simulation-to-data trigger correction factors and the corresponding uncertainties. 
\item \textit{Reconstruction, identification, and isolation:} The lepton selection efficiencies as determined from simulation are corrected with simulation-to-data correction factors and their associated uncertainties~\cite{Aad:2016jkr, ATLAS-CONF-2016-024}. 
\item \textit{Energy, momentum scale/resolution:} Uncertainties in the lepton calibrations~\cite{Aad:2016jkr} are assessed as they can cause a change of acceptance because of migration of events across the \pT\ threshold and $m_{\ell\ell}$ boundaries. 
\item \textit{Charge identification:} Electron charge misidentification may occur when electrons radiate in the inner regions in the detector  and the resulting photons
subsequently convert and are reconstructed as high-\pT\ tracks. A particle with reconstructed charge opposite to the parent electron may then be accidentally associated with the energy deposit in the calorimeter. The effect of electrons having their charge misidentified is studied~\cite{Aad:2016naf} using a control sample of $Z\rightarrow e^+e^-$ events in which both electrons are reconstructed with the same charge and is found to be well described by the Monte Carlo simulation, within the statistical uncertainty of the control sample. An uncertainty is assessed to cover any small residual differences between data and simulation. The probability of charge misidentification is negligible in the muon channel. 
\item \textit{Pile-up:} Incorrect modelling of pile-up effects can lead to acceptance changes and is accounted for with dedicated studies.  
\item \textit{PDF:} The impact of the PDF uncertainty is estimated by propagating \NNPDF PDF variations to the correction factor. 
\item \textit{$p_{\text{T}}^{\ell\ell}$ mismodelling:} Mismodelling in the simulation at high dilepton transverse momentum, $p_{\text{T}}^{\ell\ell}$, has been studied in detail in the context of a $\sqrt s=8$~TeV \Zboson-boson analysis~\cite{STDM-2014-12}. The effect is estimated here by reweighting the simulated $p_{\text{T}}^{\ell\ell}$  distribution to a fourth-order polynomial derived from a fit to the corresponding data distribution. It has a small impact on the measured fiducial cross section, as established in a previous \Zboson-boson cross-section analysis at $13$~TeV~\cite{Aad:2016naf} and confirmed for this paper. 
\end{itemize}

The systematic uncertainties from the background estimation contribute negligibly to the experimental cross-section uncertainty. The cross sections have a 2.1\% uncertainty in the measurement of the integrated luminosity, which is derived, following a methodology similar to that detailed in Refs.~\cite{DAPR-2011-01,Aaboud:2016hhf}, from a preliminary calibration of the luminosity scale using a pair of $x$--$y$ beam separation scans  performed in August 2015.  Finally, there exists an uncertainty related to knowledge of the beam energy, taken as $0.66\%$ of the beam-energy value~\cite{Wenninger:1546734}, and propagated to the cross section with the VRAP~0.9 program~\cite{vrap}. Apart from the determination of the luminosity, the dominant experimental systematic uncertainties in the cross-section evaluations are the lepton reconstruction and identification efficiencies. 

\renewcommand{\tabcolsep}{1.5pt}

\begin{table}[t]
\begin{center} 
\small
\begin{tabular}{|l|rr|}
\hline
$\delta C/C$ [\%]  &   $Z \rightarrow e^+e^-$ &  $Z \rightarrow \mu^+ \mu^-$  \\
\hline
Lepton trigger                         &    $< 0.1$ &    0.1 \\ 
Lepton reconstruction, identification  &    0.4 &    0.7 \\ 
Lepton isolation                       &    0.1 &    0.4 \\ 
Lepton scale and resolution            &    0.2 &    0.1 \\ 
Charge identification                  &    0.1 &  --  \\ 
Pile-up modelling                      &    $< 0.1$ &    $< 0.1$ \\ 
PDF                                    &    0.1 &   $< 0.1$ \\ 
$p_{\text{T}}^{\ell\ell}$ mismodelling      &    0.1 &    $< 0.1$ \\ 
 \hline 
 Total                                 &    0.5 &    0.8 \\ 
\hline 
\end{tabular}
\end{center}
\caption{Relative systematic uncertainties, in \%, in the correction factors $C$ in the electron and muon channels. 
\label{tab:systotZ}}
\end{table}
\renewcommand{\tabcolsep}{6pt}

\subsection{Cross-section results}
\label{sec:zllres}
Distributions of the lepton $\eta$ and \pT, and of the dilepton \pT and invariant mass after applying all selection criteria  are shown in Figures~\ref{fig:ZllLeptonEtaPt} and~\ref{fig:ZllBosonPtMass}. Good agreement between data and simulation is observed in the lepton $\eta$ and in the dilepton invariant-mass distributions. As can be seen from the figure, agreement is also achieved in the lepton \pT  distribution after reweighting the simulated dilepton transverse momentum, $p_{\text{T}}^{\ell\ell}$,  to the data, as explained in Section~\ref{sec:method}.  

\begin{figure}[t]
  \centering
  \includegraphics[width=0.48\textwidth]{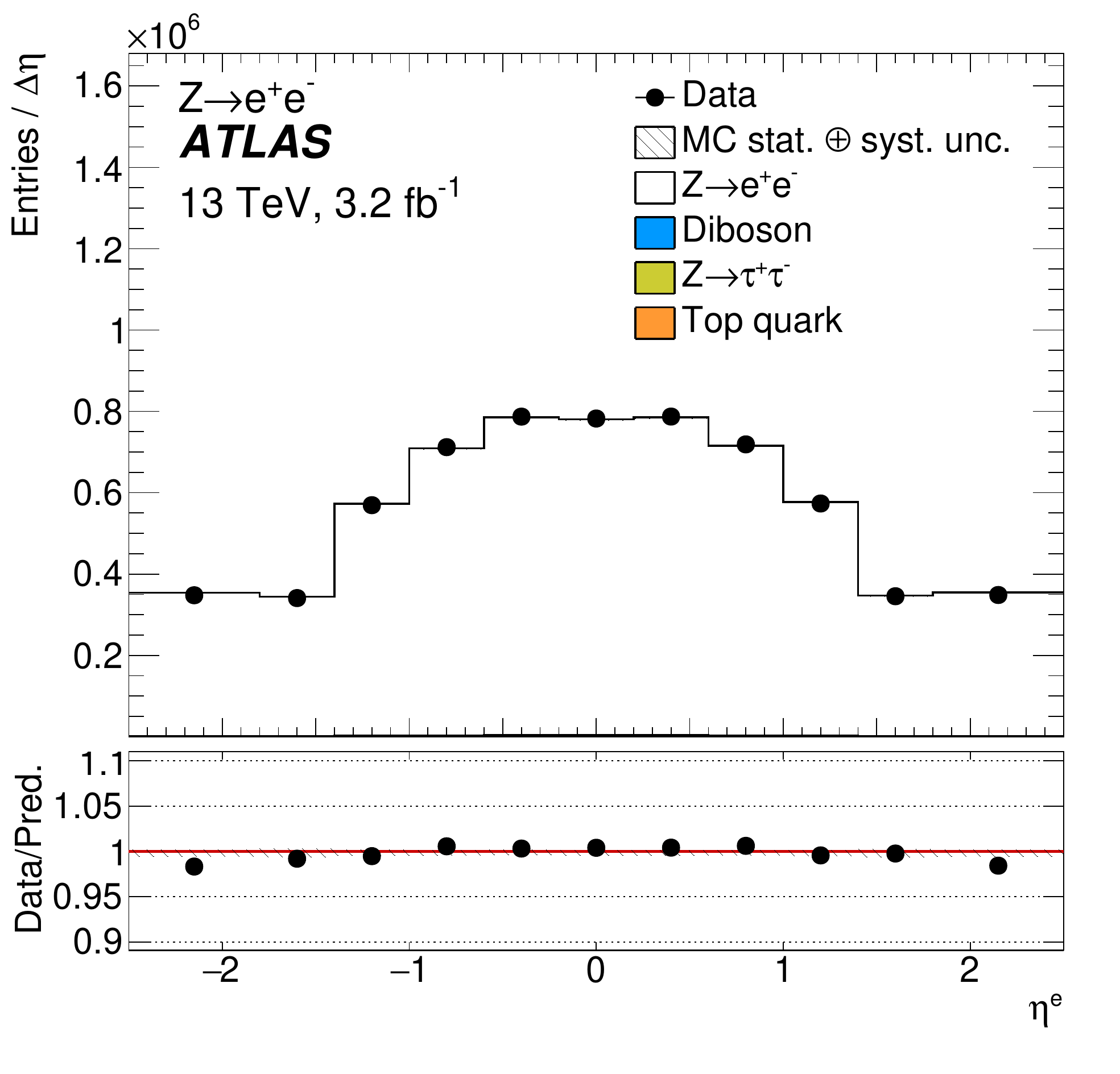}
  \includegraphics[width=0.48\textwidth]{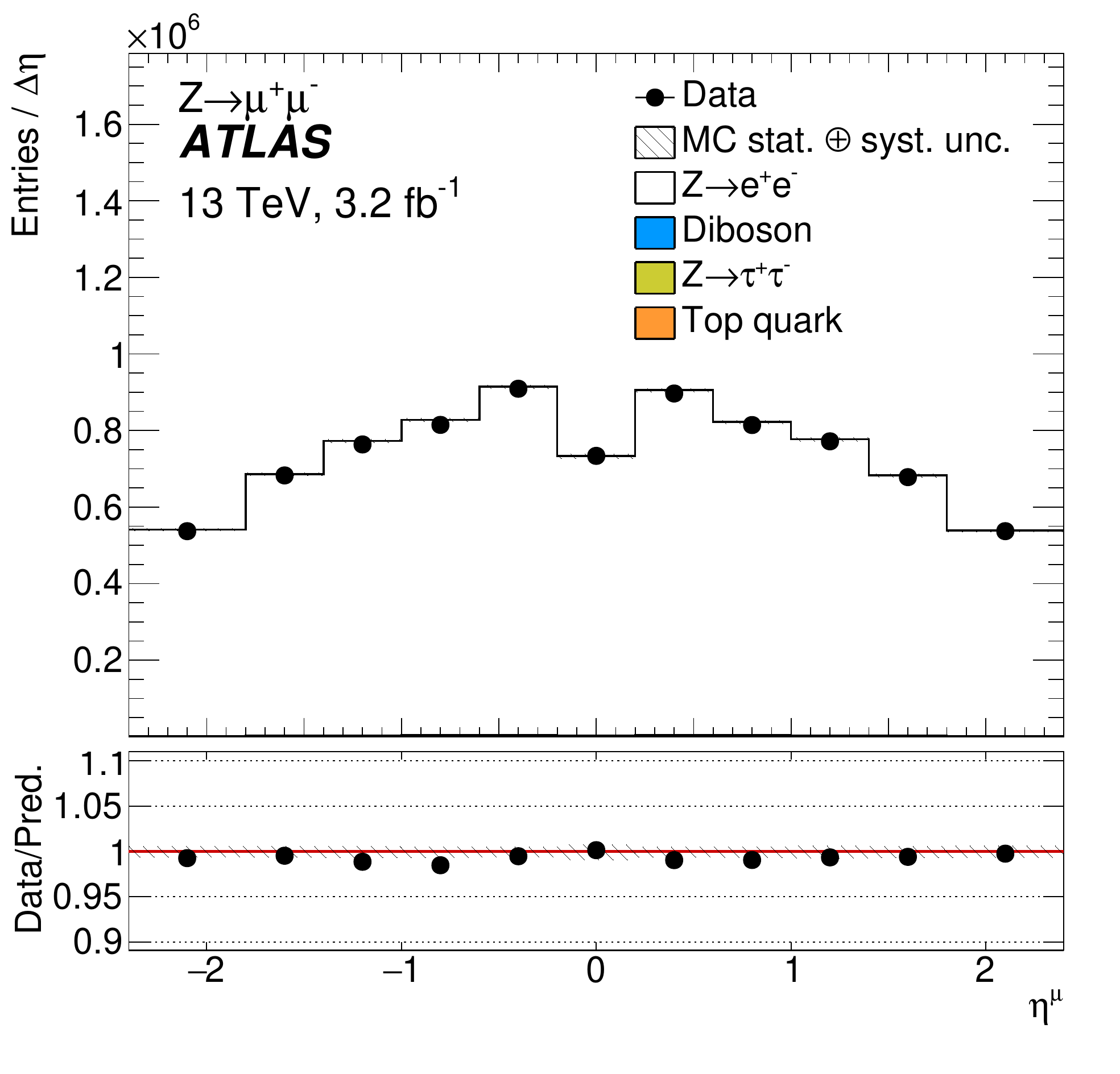}
  \includegraphics[width=0.48\textwidth]{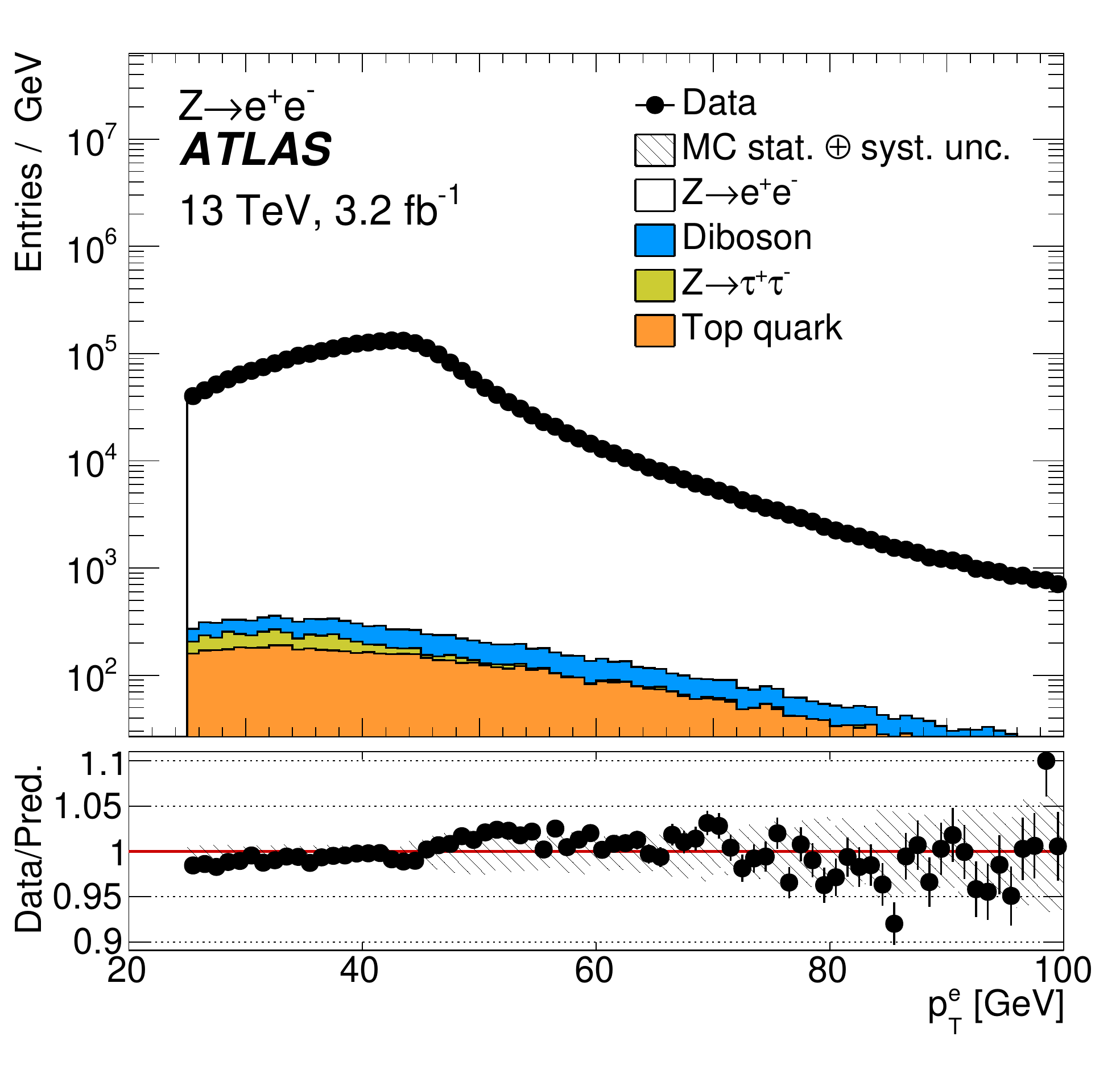}
  \includegraphics[width=0.48\textwidth]{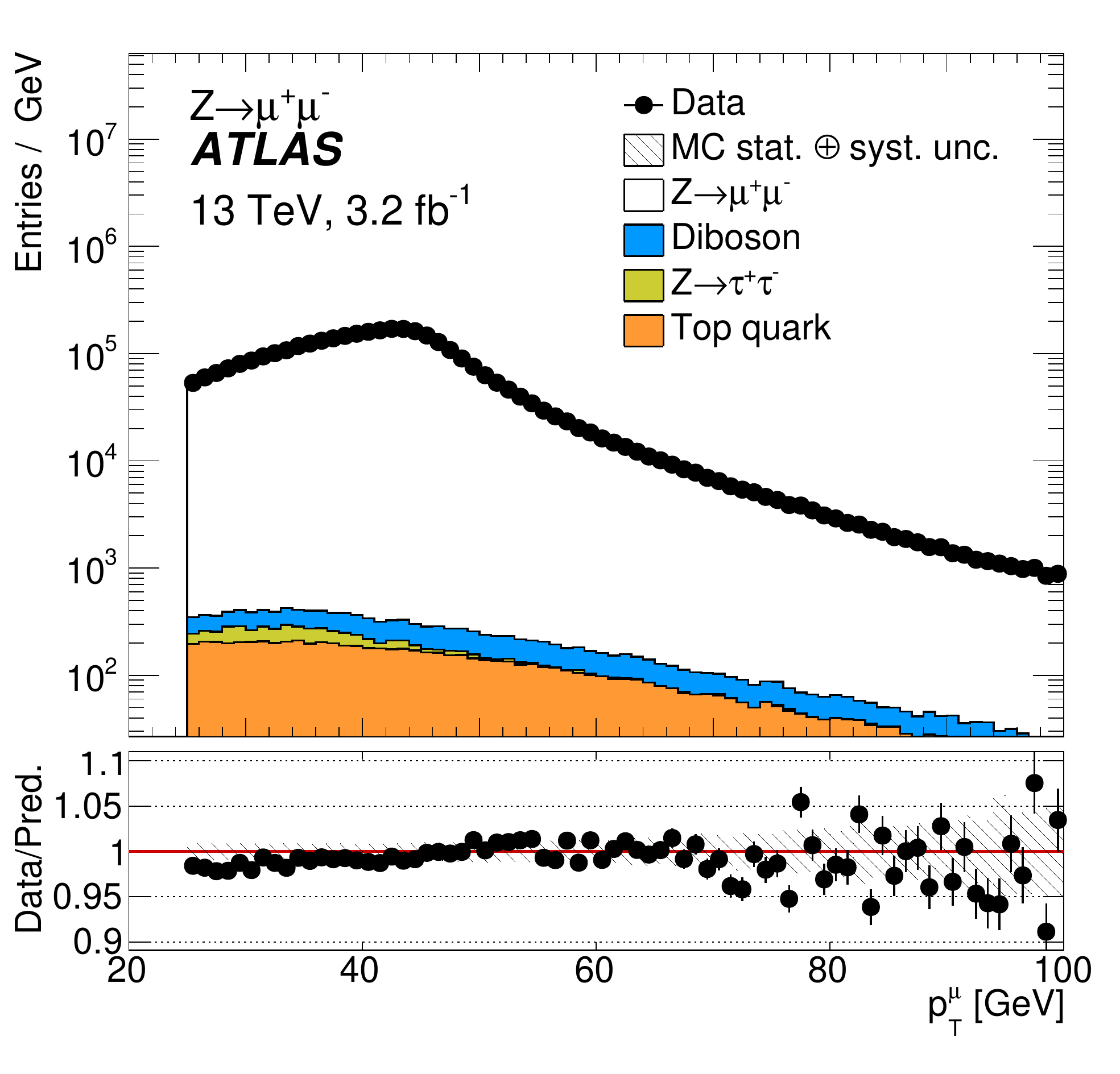}
  \caption{Lepton pseudorapidity (top) and transverse momentum (bottom) distributions from the $Z\rightarrow  e^+e^-$ selection (left) and the $Z\rightarrow \mu^+ \mu^-$  selection (right). Due to the unequal bin widths used in the lepton pseudorapidity distributions, these distributions are plotted divided by the bin width. The background processes are heavily suppressed and not visible on the linear scale.
The systematic uncertainties for the signal and background distributions are combined in the shaded band, while the statistical uncertainty is shown on the data points. The luminosity  uncertainties are not included. There are two lepton entries in  the histogram for each candidate event.
}
  \label{fig:ZllLeptonEtaPt}
\end{figure}

\begin{figure}[t]
  \centering
  \includegraphics[width=0.48\textwidth]{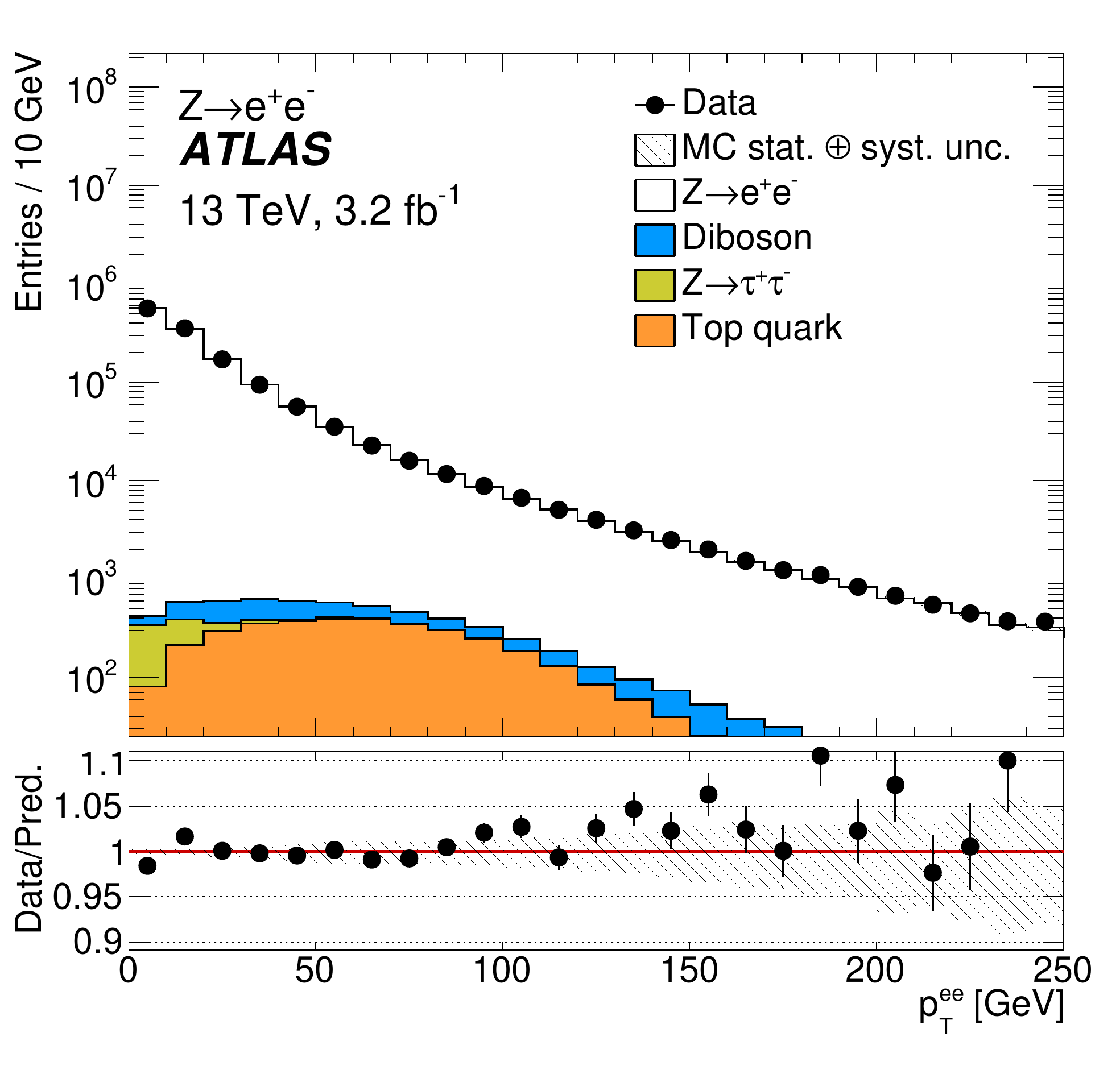}
  \includegraphics[width=0.48\textwidth]{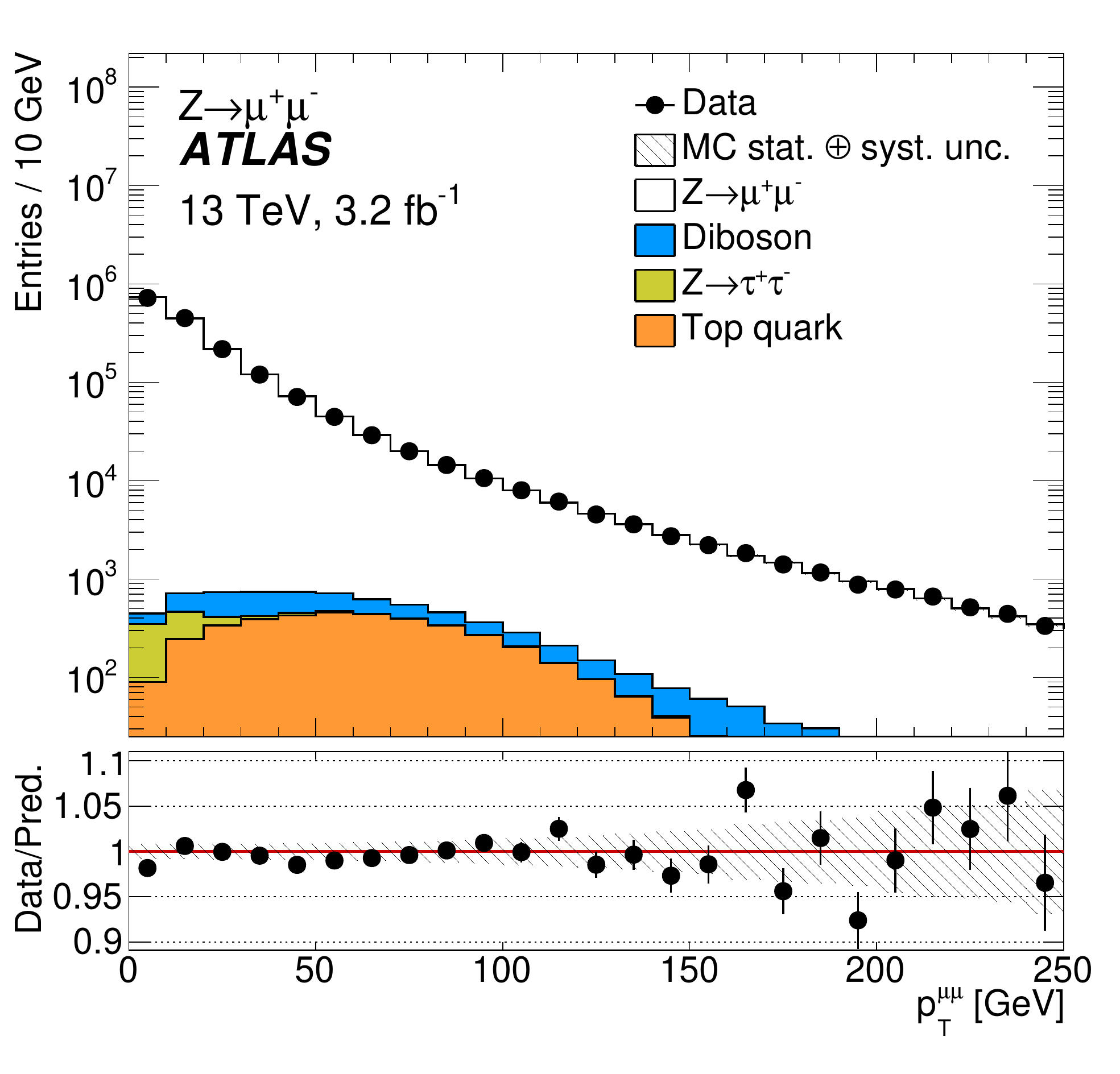}
  \includegraphics[width=0.48\textwidth]{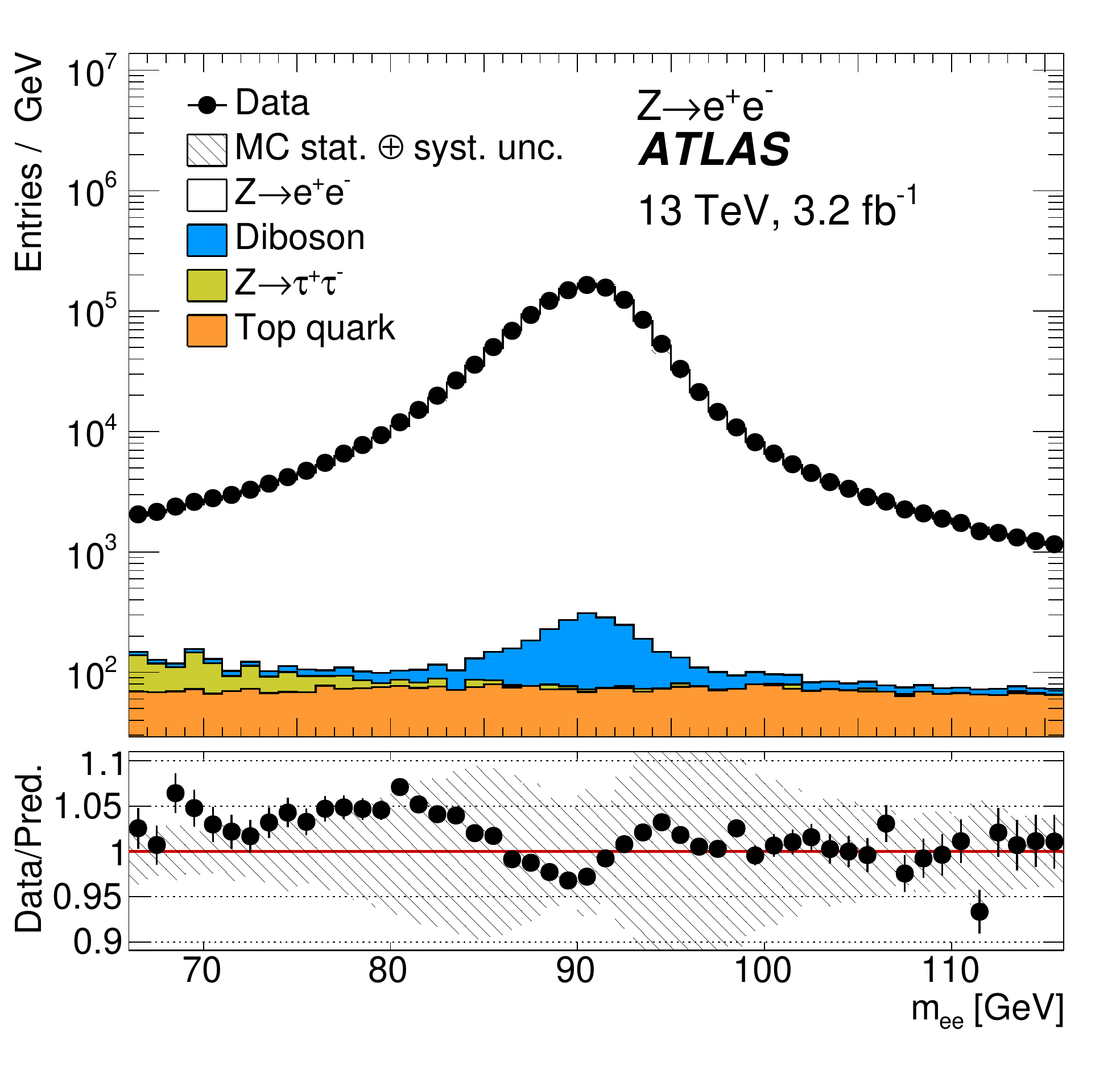}
  \includegraphics[width=0.48\textwidth]{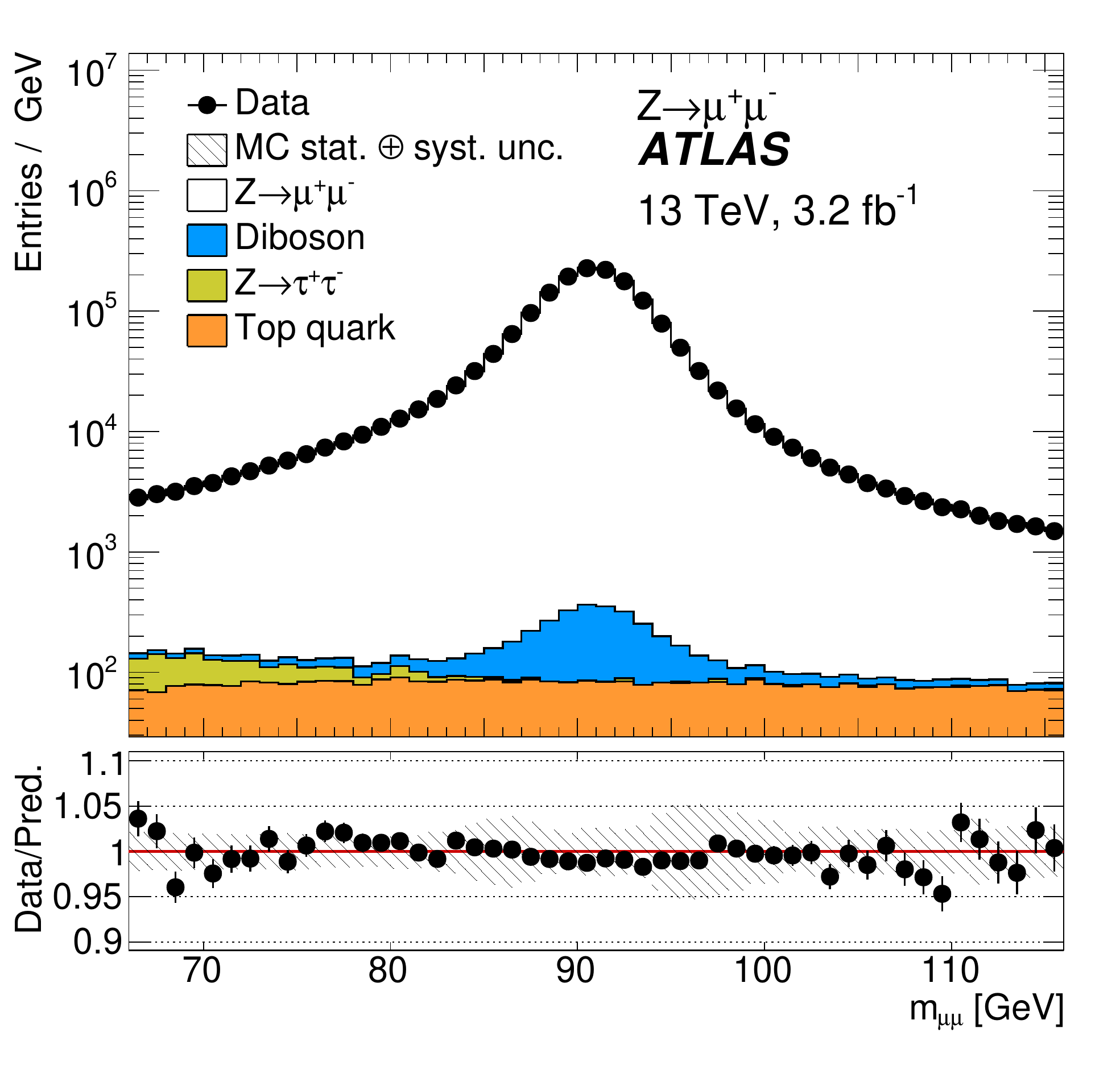}  
  \caption{Dilepton  transverse momentum (top) and invariant mass (bottom) distributions 
from the $Z\rightarrow  e^+e^-$ selection (left) and the $Z\rightarrow \mu^+ \mu^-$  selection (right). 
The systematic uncertainties for the signal and background distributions are combined in the shaded band, while the statistical uncertainty is shown on the data points. The luminosity  uncertainties are not included. 
}
  \label{fig:ZllBosonPtMass}
\end{figure}

All elements necessary to calculate the cross sections for \Zboson-boson production and decay in the electron and muon channels with 3.2~fb$^{-1}$ of data are summarised in Table~\ref{FiducialZXSections}. The measured fiducial cross sections are also presented in this table, along with their statistical, experimental systematic, luminosity, and beam-energy uncertainties, except for the $C$ factor where the total uncertainty is quoted. The fiducial phase space for this measurement is presented in Section~\ref{sec:theory}. These numbers are in agreement, within experimental systematic uncertainties, with the previous ATLAS measurement~\cite{Aad:2016naf} of \Zboson-boson production in the combined electron and muon channels in the same fiducial phase space and which uses an independent data set at $\sqrt s = 13$~TeV: $779 \pm 3~\text{(stat)} \pm 6~\text{(syst)} \pm 16~\text{(lumi)}$~pb. The results are also compatible with the NNLO prediction shown in Section~\ref{sec:ztheory} of $744^{+22}_{-28}$~(tot)~pb.

\begin{table}[t]
\begin{center}
\footnotesize
\begin{tabular}{|l|c|c|}
\hline 
& \multicolumn{1}{c|}{ $Z\rightarrow  e^+e^-$} &  \multicolumn{1}{c|}{ $Z\rightarrow  \mu^+\mu^-$ } \\ \hline
    Events         & $1,360,680 \pm 1170 \text{~(stat)} \pm 760 \text{~(syst)} \pm 130 \text{~(lumi)}$ & $1,727,700 \pm 1320 \text{~(stat)} \pm 950 \text{~(syst)}\pm 160 \text{~(lumi)}$ \T \\
    $C$        & $0.554 \pm 0.003 \text{~(tot)}$ & $0.706 \pm 0.006 \text{~(tot)}$ \\
    \sfidZ [pb]           & $\numRP{778.3}{0} \pm \numRP{0.7}{0} \text{~(stat)} \pm \numRP{4.0}{0} \text{~(syst)}  \pm \numRP{5.4}{0} \text{~(beam)} \pm \numRP{16.3}{0} \text{~(lumi)}$  &  $ \numRP{774.4}{0} \pm \numRP{0.6}{0} \text{~(stat)} \pm \numRP{6.2}{0} \text{~(syst)} \pm \numRP{5.3}{0} \text{~(beam)}  \pm \numRP{16.3}{0} \text{~(lumi)}$ \B \\ 
\hline
\end{tabular}
\caption{The observed numbers of signal events after background subtraction are shown for the electron and muon channels along with the correction factors~$C$ and the  \Zboson-boson fiducial cross sections. The statistical, systematic, beam-energy, and luminosity uncertainties are quoted in that order except for the $C$ factor where the total uncertainty is quoted. 
\label{FiducialZXSections}}
\end{center}

\end{table}

\section{Analysis of ratios}
\label{sec:rat}

\newcommand{\TTZratioVal}{   1.051}
\newcommand{\TTZratioErrSt}{   0.010}
\newcommand{\TTZratioErrSy}{   0.035}
\newcommand{\TTZratioErrTot}{   0.037}

\newcommand{\TTtotalThirtZfiducialThirtRatioVal}{   1.053}
\newcommand{\TTtotalThirtZfiducialThirtRatioErrSt}{   0.010}
\newcommand{\TTtotalThirtZfiducialThirtRatioErrSyWithLumiUnc}{   0.036}
\newcommand{\TTtotalThirtZfiducialThirtRatioErrSyNoLumiUnc}{   0.036}
\newcommand{\TTtotalThirtZfiducialThirtRatioErrTot}{   0.037}
\newcommand{\TTtotalThirtZfiducialThirtRatioErrStPercent}{    0.92}
\newcommand{\TTtotalThirtZfiducialThirtRatioErrSyPercentWithLumiUnc}{    3.40}
\newcommand{\TTtotalThirtZfiducialThirtRatioErrSyPercentNoLumiUnc}{    3.39}
\newcommand{\TTtotalThirtZfiducialThirtRatioErrTotPercent}{    3.52}
\newcommand{\TTtotalThirtZfiducialThirtRatioLumiErrPercent}{    0.21}
\newcommand{\TTtotalThirtZfiducialThirtRatioLumiErrAbs}{   0.002}

\newcommand{\TTtotalThirtZtotalThirtRatioVal}{   0.416}
\newcommand{\TTtotalThirtZtotalThirtRatioErrSt}{   0.004}
\newcommand{\TTtotalThirtZtotalThirtRatioErrSyWithLumiUnc}{   0.016}
\newcommand{\TTtotalThirtZtotalThirtRatioErrSyNoLumiUnc}{   0.016}
\newcommand{\TTtotalThirtZtotalThirtRatioErrTot}{   0.016}
\newcommand{\TTtotalThirtZtotalThirtRatioErrStPercent}{    0.92}
\newcommand{\TTtotalThirtZtotalThirtRatioErrSyPercentWithLumiUnc}{    3.83}
\newcommand{\TTtotalThirtZtotalThirtRatioErrSyPercentNoLumiUnc}{    3.83}
\newcommand{\TTtotalThirtZtotalThirtRatioErrTotPercent}{    3.94}
\newcommand{\TTtotalThirtZtotalThirtRatioLumiErrPercent}{    0.21}
\newcommand{\TTtotalThirtZtotalThirtRatioLumiErrAbs}{   0.001}

\newcommand{\TTfiducialThirtZfiducialThirtRatioVal}{ 0.01280}
\newcommand{\TTfiducialThirtZfiducialThirtRatioErrSt}{ 0.00012}
\newcommand{\TTfiducialThirtZfiducialThirtRatioErrSyWithLumiUnc}{ 0.00034}
\newcommand{\TTfiducialThirtZfiducialThirtRatioErrSyNoLumiUnc}{ 0.00033}
\newcommand{\TTfiducialThirtZfiducialThirtRatioErrTot}{ 0.00036}
\newcommand{\TTfiducialThirtZfiducialThirtRatioErrStPercent}{    0.91}
\newcommand{\TTfiducialThirtZfiducialThirtRatioErrSyPercentWithLumiUnc}{    2.62}
\newcommand{\TTfiducialThirtZfiducialThirtRatioErrSyPercentNoLumiUnc}{    2.61}
\newcommand{\TTfiducialThirtZfiducialThirtRatioErrTotPercent}{    2.78}
\newcommand{\TTfiducialThirtZfiducialThirtRatioLumiErrPercent}{    0.21}
\newcommand{\TTfiducialThirtZfiducialThirtRatioLumiErrAbs}{ 0.00003}

\newcommand{\TTbarEighttotalToZeightfiducialRatioVal}{   0.480}
\newcommand{\TTbarEighttotalToZeightfiducialratioErrSt}{   0.003}
\newcommand{\TTbarEighttotalToZeightfiducialratioErrSyWithLumiUnc}{   0.012}
\newcommand{\TTbarEighttotalToZeightfiducialratioErrSyNoLumiUnc}{   0.012}
\newcommand{\TTbarEighttotalToZeightfiducialratioErrTot}{   0.013}
\newcommand{\TTbarEighttotalToZeightfiducialratioErrStPercent}{    0.71}
\newcommand{\TTbarEighttotalToZeightfiducialratioErrSyPercentWithLumiUnc}{    2.57}
\newcommand{\TTbarEighttotalToZeightfiducialratioErrSyPercentNoLumiUnc}{    2.57}
\newcommand{\TTbarEighttotalToZeightfiducialratioErrTotPercent}{    2.67}
\newcommand{\TTbarEighttotalToZeightfiducialratioLumiErrPercent}{    0.20}
\newcommand{\TTbarEighttotalToZeightfiducialratioLumiErrAbs}{   0.001}

\newcommand{\TTbarEighttotalToZeighttotalRatioVal}{   0.211}
\newcommand{\TTbarEighttotalToZeighttotalratioErrSt}{   0.001}
\newcommand{\TTbarEighttotalToZeighttotalratioErrSyWithLumiUnc}{   0.007}
\newcommand{\TTbarEighttotalToZeighttotalratioErrSyNoLumiUnc}{   0.007}
\newcommand{\TTbarEighttotalToZeighttotalratioErrTot}{   0.007}
\newcommand{\TTbarEighttotalToZeighttotalratioErrStPercent}{    0.71}
\newcommand{\TTbarEighttotalToZeighttotalratioErrSyPercentWithLumiUnc}{    3.10}
\newcommand{\TTbarEighttotalToZeighttotalratioErrSyPercentNoLumiUnc}{    3.09}
\newcommand{\TTbarEighttotalToZeighttotalratioErrTotPercent}{    3.18}
\newcommand{\TTbarEighttotalToZeighttotalratioLumiErrPercent}{    0.20}
\newcommand{\TTbarEighttotalToZeighttotalratioLumiErrAbs}{   0.000}

\newcommand{\TTbarEightfiducialToZeightfiducialRatioVal}{ 0.00602}
\newcommand{\TTbarEightfiducialToZeightfiducialratioErrSt}{ 0.00004}
\newcommand{\TTbarEightfiducialToZeightfiducialratioErrSyWithLumiUnc}{ 0.00014}
\newcommand{\TTbarEightfiducialToZeightfiducialratioErrSyNoLumiUnc}{ 0.00014}
\newcommand{\TTbarEightfiducialToZeightfiducialratioErrTot}{ 0.00015}
\newcommand{\TTbarEightfiducialToZeightfiducialratioErrStPercent}{    0.72}
\newcommand{\TTbarEightfiducialToZeightfiducialratioErrSyPercentWithLumiUnc}{    2.36}
\newcommand{\TTbarEightfiducialToZeightfiducialratioErrSyPercentNoLumiUnc}{    2.35}
\newcommand{\TTbarEightfiducialToZeightfiducialratioErrTotPercent}{    2.47}
\newcommand{\TTbarEightfiducialToZeightfiducialratioLumiErrPercent}{    0.20}
\newcommand{\TTbarEightfiducialToZeightfiducialratioLumiErrAbs}{ 0.00001}

\newcommand{\TTbarSeventotalToZsevenfiducialRatioVal}{   0.406}
\newcommand{\TTbarSeventotalToZsevenfiducialratioErrSt}{   0.007}
\newcommand{\TTbarSeventotalToZsevenfiducialratioErrSyWithLumiUnc}{   0.011}
\newcommand{\TTbarSeventotalToZsevenfiducialratioErrSyNoLumiUnc}{   0.011}
\newcommand{\TTbarSeventotalToZsevenfiducialratioErrTot}{   0.013}
\newcommand{\TTbarSeventotalToZsevenfiducialratioErrStPercent}{    1.69}
\newcommand{\TTbarSeventotalToZsevenfiducialratioErrSyPercentWithLumiUnc}{    2.59}
\newcommand{\TTbarSeventotalToZsevenfiducialratioErrSyPercentNoLumiUnc}{    2.59}
\newcommand{\TTbarSeventotalToZsevenfiducialratioErrTotPercent}{    3.10}
\newcommand{\TTbarSeventotalToZsevenfiducialratioLumiErrPercent}{    0.18}
\newcommand{\TTbarSeventotalToZsevenfiducialratioLumiErrAbs}{   0.001}

\newcommand{\TTbarSeventotalToZseventotalRatioVal}{   0.184}
\newcommand{\TTbarSeventotalToZseventotalratioErrSt}{   0.003}
\newcommand{\TTbarSeventotalToZseventotalratioErrSyWithLumiUnc}{   0.006}
\newcommand{\TTbarSeventotalToZseventotalratioErrSyNoLumiUnc}{   0.006}
\newcommand{\TTbarSeventotalToZseventotalratioErrTot}{   0.007}
\newcommand{\TTbarSeventotalToZseventotalratioErrStPercent}{    1.69}
\newcommand{\TTbarSeventotalToZseventotalratioErrSyPercentWithLumiUnc}{    3.15}
\newcommand{\TTbarSeventotalToZseventotalratioErrSyPercentNoLumiUnc}{    3.14}
\newcommand{\TTbarSeventotalToZseventotalratioErrTotPercent}{    3.57}
\newcommand{\TTbarSeventotalToZseventotalratioLumiErrPercent}{    0.18}
\newcommand{\TTbarSeventotalToZseventotalratioLumiErrAbs}{   0.000}

\newcommand{\TTbarSevenfiducialToZsevenfiducialRatioVal}{ 0.00511}
\newcommand{\TTbarSevenfiducialToZsevenfiducialratioErrSt}{ 0.00009}
\newcommand{\TTbarSevenfiducialToZsevenfiducialratioErrSyWithLumiUnc}{ 0.00013}
\newcommand{\TTbarSevenfiducialToZsevenfiducialratioErrSyNoLumiUnc}{ 0.00013}
\newcommand{\TTbarSevenfiducialToZsevenfiducialratioErrTot}{ 0.00015}
\newcommand{\TTbarSevenfiducialToZsevenfiducialratioErrStPercent}{    1.68}
\newcommand{\TTbarSevenfiducialToZsevenfiducialratioErrSyPercentWithLumiUnc}{    2.46}
\newcommand{\TTbarSevenfiducialToZsevenfiducialratioErrSyPercentNoLumiUnc}{    2.46}
\newcommand{\TTbarSevenfiducialToZsevenfiducialratioErrTotPercent}{    2.98}
\newcommand{\TTbarSevenfiducialToZsevenfiducialratioLumiErrPercent}{    0.18}
\newcommand{\TTbarSevenfiducialToZsevenfiducialratioLumiErrAbs}{ 0.00001}

\newcommand{\AveCentralZEightTevabsvalFid}{ 505.8}
\newcommand{\AveStatZEightTevabsvalFid}{   0.1}
\newcommand{\AveSystZEightTevabsvalFid}{   4.2}
\newcommand{\AveTotUncZEightTevabsvalFid}{   4.2}
\newcommand{\AveStatZEightTevpercentsFid}{   0.0}
\newcommand{\AveSystZEightTevpercentsFid}{   0.8}
\newcommand{\AveTotUncZEightTevpercentsFid}{   0.8}
\newcommand{\AveLumiZEightTevFid}{   9.6}
\newcommand{\AveCentralZThirteenTevabsvalFid}{ 777.1}
\newcommand{\AveStatZThirteenTevabsvalFid}{   0.5}
\newcommand{\AveSystZThirteenTevabsvalFid}{   6.4}
\newcommand{\AveTotUncZThirteenTevabsvalFid}{   6.4}
\newcommand{\AveStatZThirteenTevpercentsFid}{   0.1}
\newcommand{\AveSystZThirteenTevpercentsFid}{   0.8}
\newcommand{\AveTotUncZThirteenTevpercentsFid}{   0.8}
\newcommand{\AveLumiZThirteenTevFid}{  16.3}
\newcommand{\AveChisqZThirteenTevtoZEightTevFid}{0.4}
\newcommand{\CorrCombZEightTevcorZEightTevFid}{ 1.000}
\newcommand{\CorrCombZThirteenTevcorZEightTevFid}{ 0.665}
\newcommand{\CorrCombZThirteenTevcorZThirteenTevFid}{ 1.000}
\newcommand{\RatCombCentZthirteenTeVoverZeightTeVabsvalFid}{   1.537}
\newcommand{\RatCombStatZthirteenTeVoverZeightTeVabsvalFid}{   0.001}
\newcommand{\RatCombSystNoLumiUncZthirteenTeVoverZeightTeVabsvalFid}{   0.010}
\newcommand{\RatCombSystWithLumiUncZthirteenTeVoverZeightTeVabsvalFid}{   0.045}
\newcommand{\RatCombLumiUncZthirteenTeVoverZeightTeVabsvalFid}{   0.044}
\newcommand{\RatCombTotUncZthirteenTeVoverZeightTeVabsvalFid}{ 0.04474}
\newcommand{\RatCombTotUncZthirteenTeVoverZeightTeVpercentsFid}{    2.91}
\newcommand{\RatCombStatZthirteenTeVoverZeightTeVpercentsFid}{    0.06}
\newcommand{\RatCombSystWithLumiZthirteenTeVoverZeightTeVpercentsFid}{    2.91}
\newcommand{\RatCombSystNoLumiUncZthirteenTeVoverZeightTeVpercentsFid}{    0.67}
\newcommand{\RatCombLumiUncZthirteenTeVoverZeightTeVpercentsFid}{    2.83}

\newcommand{\AveCentralZEightTevabsvalTot}{1153.8}
\newcommand{\AveStatZEightTevabsvalTot}{   0.3}
\newcommand{\AveSystZEightTevabsvalTot}{   9.6}
\newcommand{\AveTotUncZEightTevabsvalTot}{   9.6}
\newcommand{\AveStatZEightTevpercentsTot}{   0.0}
\newcommand{\AveSystZEightTevpercentsTot}{   0.8}
\newcommand{\AveTotUncZEightTevpercentsTot}{   0.8}
\newcommand{\AveLumiZEightTevTot}{  21.9}
\newcommand{\AveCentralZThirteenTevabsvalTot}{1969.1}
\newcommand{\AveStatZThirteenTevabsvalTot}{   1.2}
\newcommand{\AveSystZThirteenTevabsvalTot}{  16.1}
\newcommand{\AveTotUncZThirteenTevabsvalTot}{  16.1}
\newcommand{\AveStatZThirteenTevpercentsTot}{   0.1}
\newcommand{\AveSystZThirteenTevpercentsTot}{   0.8}
\newcommand{\AveTotUncZThirteenTevpercentsTot}{   0.8}
\newcommand{\AveLumiZThirteenTevTot}{  41.4}
\newcommand{\AveChisqZThirteenTevtoZEightTevTot}{0.4}
\newcommand{\CorrCombZEightTevcorZEightTevTot}{ 1.000}
\newcommand{\CorrCombZThirteenTevcorZEightTevTot}{ 0.665}
\newcommand{\CorrCombZThirteenTevcorZThirteenTevTot}{ 1.000}
\newcommand{\RatCombCentZthirteenTeVoverZeightTeVabsvalTot}{   1.707}
\newcommand{\RatCombStatZthirteenTeVoverZeightTeVabsvalTot}{   0.001}
\newcommand{\RatCombSystNoLumiUncZthirteenTeVoverZeightTeVabsvalTot}{   0.013}
\newcommand{\RatCombSystWithLumiUncZthirteenTeVoverZeightTeVabsvalTot}{   0.050}
\newcommand{\RatCombLumiUncZthirteenTeVoverZeightTeVabsvalTot}{   0.048}
\newcommand{\RatCombTotUncZthirteenTeVoverZeightTeVabsvalTot}{ 0.05011}
\newcommand{\RatCombTotUncZthirteenTeVoverZeightTeVpercentsTot}{    2.94}
\newcommand{\RatCombStatZthirteenTeVoverZeightTeVpercentsTot}{    0.06}
\newcommand{\RatCombSystWithLumiZthirteenTeVoverZeightTeVpercentsTot}{    2.94}
\newcommand{\RatCombSystNoLumiUncZthirteenTeVoverZeightTeVpercentsTot}{    0.77}
\newcommand{\RatCombLumiUncZthirteenTeVoverZeightTeVpercentsTot}{    2.83}

\newcommand{\AveTTBARCentralTTbarEightTevabsvalFid}{   3.0}
\newcommand{\AveTTBARStatTTbarEightTevabsvalFid}{   0.0}
\newcommand{\AveTTBARSystTTbarEightTevabsvalFid}{   0.1}
\newcommand{\AveTTBARTotUncTTbarEightTevabsvalFid}{   0.1}
\newcommand{\AveTTBARStatTTbarEightTevpercentsFid}{   0.7}
\newcommand{\AveTTBARSystTTbarEightTevpercentsFid}{   2.6}
\newcommand{\AveTTBARTotUncTTbarEightTevpercentsFid}{   2.7}
\newcommand{\AveTTBARLumiTTbarEightTevFid}{   0.1}
\newcommand{\AveTTBARCentralTTbarThirteenTevabsvalFid}{   9.9}
\newcommand{\AveTTBARStatTTbarThirteenTevabsvalFid}{   0.1}
\newcommand{\AveTTBARSystTTbarThirteenTevabsvalFid}{   0.3}
\newcommand{\AveTTBARTotUncTTbarThirteenTevabsvalFid}{   0.3}
\newcommand{\AveTTBARStatTTbarThirteenTevpercentsFid}{   0.9}
\newcommand{\AveTTBARSystTTbarThirteenTevpercentsFid}{   2.9}
\newcommand{\AveTTBARTotUncTTbarThirteenTevpercentsFid}{   3.1}
\newcommand{\AveTTBARLumiTTbarThirteenTevFid}{   0.2}
\newcommand{\AveTTBARChisqTTbarThirteenTevtoTTBARTTbarEightTevFid}{0.0}
\newcommand{\CorrCombTTbarEightTevcorTTbarEightTevFid}{ 1.000}
\newcommand{\CorrCombTTbarThirteenTevcorTTbarEightTevFid}{ 0.518}
\newcommand{\CorrCombTTbarThirteenTevcorTTbarThirteenTevFid}{ 1.000}
\newcommand{\RatCombCentTTthirteenTeVoverTTeightTeVabsvalFid}{   3.270}
\newcommand{\RatCombStatTTthirteenTeVoverTTeightTeVabsvalFid}{   0.038}
\newcommand{\RatCombSystNoLumiUncTTthirteenTeVoverTTeightTeVabsvalFid}{   0.086}
\newcommand{\RatCombSystWithLumiUncTTthirteenTeVoverTTeightTeVabsvalFid}{   0.133}
\newcommand{\RatCombLumiUncTTthirteenTeVoverTTeightTeVabsvalFid}{   0.102}
\newcommand{\RatCombTotUncTTthirteenTeVoverTTeightTeVabsvalFid}{   0.139}
\newcommand{\RatCombTotUncTTthirteenTeVoverTTeightTeVpercentsFid}{    4.24}
\newcommand{\RatCombStatTTthirteenTeVoverTTeightTeVpercentsFid}{    1.16}
\newcommand{\RatCombSystWithLumiTTthirteenTeVoverTTeightTeVpercentsFid}{    4.07}
\newcommand{\RatCombSystNoLumiUncTTthirteenTeVoverTTeightTeVpercentsFid}{    2.62}
\newcommand{\RatCombLumiUncTTthirteenTeVoverTTeightTeVpercentsFid}{    3.12}

\newcommand{\AveTTBARCentralTTbarEightTevabsvalTot}{ 242.9}
\newcommand{\AveTTBARStatTTbarEightTevabsvalTot}{   1.7}
\newcommand{\AveTTBARSystTTbarEightTevabsvalTot}{   6.9}
\newcommand{\AveTTBARTotUncTTbarEightTevabsvalTot}{   7.1}
\newcommand{\AveTTBARStatTTbarEightTevpercentsTot}{   0.7}
\newcommand{\AveTTBARSystTTbarEightTevpercentsTot}{   2.8}
\newcommand{\AveTTBARTotUncTTbarEightTevpercentsTot}{   2.9}
\newcommand{\AveTTBARLumiTTbarEightTevTot}{   5.1}
\newcommand{\AveTTBARCentralTTbarThirteenTevabsvalTot}{ 817.5}
\newcommand{\AveTTBARStatTTbarThirteenTevabsvalTot}{   7.5}
\newcommand{\AveTTBARSystTTbarThirteenTevabsvalTot}{  29.8}
\newcommand{\AveTTBARTotUncTTbarThirteenTevabsvalTot}{  30.8}
\newcommand{\AveTTBARStatTTbarThirteenTevpercentsTot}{   0.9}
\newcommand{\AveTTBARSystTTbarThirteenTevpercentsTot}{   3.6}
\newcommand{\AveTTBARTotUncTTbarThirteenTevpercentsTot}{   3.8}
\newcommand{\AveTTBARLumiTTbarThirteenTevTot}{  18.9}
\newcommand{\AveTTBARChisqTTbarThirteenTevtoTTBARTTbarEightTevTot}{0.0}
\newcommand{\CorrCombTTbarEightTevcorTTbarEightTevTot}{ 1.000}
\newcommand{\CorrCombTTbarThirteenTevcorTTbarEightTevTot}{ 0.465}
\newcommand{\CorrCombTTbarThirteenTevcorTTbarThirteenTevTot}{ 1.000}
\newcommand{\RatCombCentTTthirteenTeVoverTTeightTeVabsvalTot}{   3.365}
\newcommand{\RatCombStatTTthirteenTeVoverTTeightTeVabsvalTot}{   0.039}
\newcommand{\RatCombSystNoLumiUncTTthirteenTeVoverTTeightTeVabsvalTot}{   0.112}
\newcommand{\RatCombSystWithLumiUncTTthirteenTeVoverTTeightTeVabsvalTot}{   0.154}
\newcommand{\RatCombLumiUncTTthirteenTeVoverTTeightTeVabsvalTot}{   0.105}
\newcommand{\RatCombTotUncTTthirteenTeVoverTTeightTeVabsvalTot}{   0.159}
\newcommand{\RatCombTotUncTTthirteenTeVoverTTeightTeVpercentsTot}{    4.72}
\newcommand{\RatCombStatTTthirteenTeVoverTTeightTeVpercentsTot}{    1.16}
\newcommand{\RatCombSystWithLumiTTthirteenTeVoverTTeightTeVpercentsTot}{    4.57}
\newcommand{\RatCombSystNoLumiUncTTthirteenTeVoverTTeightTeVpercentsTot}{    3.34}
\newcommand{\RatCombLumiUncTTthirteenTeVoverTTeightTeVpercentsTot}{    3.12}

\newcommand{\AveCentralZSevenTevabsvalTot}{ 994.9}
\newcommand{\AveStatZSevenTevabsvalTot}{   0.6}
\newcommand{\AveSystZSevenTevabsvalTot}{   6.8}
\newcommand{\AveTotUncZSevenTevabsvalTot}{   6.8}
\newcommand{\AveStatZSevenTevpercentsTot}{   0.1}
\newcommand{\AveSystZSevenTevpercentsTot}{   0.7}
\newcommand{\AveTotUncZSevenTevpercentsTot}{   0.7}
\newcommand{\AveLumiZSevenTevTot}{  17.9}
\newcommand{\AveCentralZThirteenTeVabsvalTot}{1969.4}
\newcommand{\AveStatZThirteenTeVabsvalTot}{   1.2}
\newcommand{\AveSystZThirteenTeVabsvalTot}{  16.1}
\newcommand{\AveTotUncZThirteenTeVabsvalTot}{  16.1}
\newcommand{\AveStatZThirteenTeVpercentsTot}{   0.1}
\newcommand{\AveSystZThirteenTeVpercentsTot}{   0.8}
\newcommand{\AveTotUncZThirteenTeVpercentsTot}{   0.8}
\newcommand{\AveLumiZThirteenTeVTot}{  41.4}
\newcommand{\AveChisqZThirteenTeVtoZSevenTevTot}{0.5}
\newcommand{\CorrCombZSevenTevcorZSevenTevTot}{ 1.000}
\newcommand{\CorrCombZThirteenTeVcorZSevenTevTot}{ 0.772}
\newcommand{\CorrCombZThirteenTeVcorZThirteenTeVTot}{ 1.000}
\newcommand{\RatCombCentZthirteenTeVoverZsevenTeVabsvalTot}{   1.979}
\newcommand{\RatCombStatZthirteenTeVoverZsevenTeVabsvalTot}{   0.002}
\newcommand{\RatCombSystNoLumiUncZthirteenTeVoverZsevenTeVabsvalTot}{   0.014}
\newcommand{\RatCombSystWithLumiUncZthirteenTeVoverZsevenTeVabsvalTot}{   0.057}
\newcommand{\RatCombLumiUncZthirteenTeVoverZsevenTeVabsvalTot}{   0.055}
\newcommand{\RatCombTotUncZthirteenTeVoverZsevenTeVabsvalTot}{ 0.05657}
\newcommand{\RatCombTotUncZthirteenTeVoverZsevenTeVpercentsTot}{    2.86}
\newcommand{\RatCombStatZthirteenTeVoverZsevenTeVpercentsTot}{    0.09}
\newcommand{\RatCombSystWithLumiZthirteenTeVoverZsevenTeVpercentsTot}{    2.86}
\newcommand{\RatCombSystNoLumiUncZthirteenTeVoverZsevenTeVpercentsTot}{    0.71}
\newcommand{\RatCombLumiUncZthirteenTeVoverZsevenTeVpercentsTot}{    2.77}

\newcommand{\AveCentralZSevenTevabsvalFid}{ 450.8}
\newcommand{\AveStatZSevenTevabsvalFid}{   0.3}
\newcommand{\AveSystZSevenTevabsvalFid}{   3.1}
\newcommand{\AveTotUncZSevenTevabsvalFid}{   3.1}
\newcommand{\AveStatZSevenTevpercentsFid}{   0.1}
\newcommand{\AveSystZSevenTevpercentsFid}{   0.7}
\newcommand{\AveTotUncZSevenTevpercentsFid}{   0.7}
\newcommand{\AveLumiZSevenTevFid}{   8.1}
\newcommand{\AveCentralZThirteenTeVabsvalFid}{ 777.3}
\newcommand{\AveStatZThirteenTeVabsvalFid}{   0.5}
\newcommand{\AveSystZThirteenTeVabsvalFid}{   6.3}
\newcommand{\AveTotUncZThirteenTeVabsvalFid}{   6.4}
\newcommand{\AveStatZThirteenTeVpercentsFid}{   0.1}
\newcommand{\AveSystZThirteenTeVpercentsFid}{   0.8}
\newcommand{\AveTotUncZThirteenTeVpercentsFid}{   0.8}
\newcommand{\AveLumiZThirteenTeVFid}{  16.3}
\newcommand{\AveChisqZThirteenTeVtoZSevenTevFid}{0.5}
\newcommand{\CorrCombZSevenTevcorZSevenTevFid}{ 1.000}
\newcommand{\CorrCombZThirteenTeVcorZSevenTevFid}{ 0.772}
\newcommand{\CorrCombZThirteenTeVcorZThirteenTeVFid}{ 1.000}
\newcommand{\RatCombCentZthirteenTeVoverZsevenTeVabsvalFid}{   1.724}
\newcommand{\RatCombStatZthirteenTeVoverZsevenTeVabsvalFid}{   0.001}
\newcommand{\RatCombSystNoLumiUncZthirteenTeVoverZsevenTeVabsvalFid}{   0.009}
\newcommand{\RatCombSystWithLumiUncZthirteenTeVoverZsevenTeVabsvalFid}{   0.049}
\newcommand{\RatCombLumiUncZthirteenTeVoverZsevenTeVabsvalFid}{   0.048}
\newcommand{\RatCombTotUncZthirteenTeVoverZsevenTeVabsvalFid}{ 0.04854}
\newcommand{\RatCombTotUncZthirteenTeVoverZsevenTeVpercentsFid}{    2.81}
\newcommand{\RatCombStatZthirteenTeVoverZsevenTeVpercentsFid}{    0.09}
\newcommand{\RatCombSystWithLumiZthirteenTeVoverZsevenTeVpercentsFid}{    2.81}
\newcommand{\RatCombSystNoLumiUncZthirteenTeVoverZsevenTeVpercentsFid}{    0.52}
\newcommand{\RatCombLumiUncZthirteenTeVoverZsevenTeVpercentsFid}{    2.77}

\newcommand{\AveTTBARCentralTTbarSevenTevabsvalTot}{ 182.9}
\newcommand{\AveTTBARStatTTbarSevenTevabsvalTot}{   3.1}
\newcommand{\AveTTBARSystTTbarSevenTevabsvalTot}{   5.3}
\newcommand{\AveTTBARTotUncTTbarSevenTevabsvalTot}{   6.1}
\newcommand{\AveTTBARStatTTbarSevenTevpercentsTot}{   1.7}
\newcommand{\AveTTBARSystTTbarSevenTevpercentsTot}{   2.9}
\newcommand{\AveTTBARTotUncTTbarSevenTevpercentsTot}{   3.3}
\newcommand{\AveTTBARLumiTTbarSevenTevTot}{   3.6}
\newcommand{\AveTTBARCentralTTbarThirteenTeVabsvalTot}{ 817.5}
\newcommand{\AveTTBARStatTTbarThirteenTeVabsvalTot}{   7.5}
\newcommand{\AveTTBARSystTTbarThirteenTeVabsvalTot}{  29.8}
\newcommand{\AveTTBARTotUncTTbarThirteenTeVabsvalTot}{  30.8}
\newcommand{\AveTTBARStatTTbarThirteenTeVpercentsTot}{   0.9}
\newcommand{\AveTTBARSystTTbarThirteenTeVpercentsTot}{   3.6}
\newcommand{\AveTTBARTotUncTTbarThirteenTeVpercentsTot}{   3.8}
\newcommand{\AveTTBARLumiTTbarThirteenTeVTot}{  18.9}
\newcommand{\AveTTBARChisqTTbarThirteenTeVtoTTBARTTbarSevenTevTot}{0.0}
\newcommand{\CorrCombTTbarSevenTevcorTTbarSevenTevTot}{ 1.000}
\newcommand{\CorrCombTTbarThirteenTeVcorTTbarSevenTevTot}{ 0.425}
\newcommand{\CorrCombTTbarThirteenTeVcorTTbarThirteenTeVTot}{ 1.000}
\newcommand{\RatCombCentTTthirteenTeVoverTTsevenTeVabsvalTot}{   4.470}
\newcommand{\RatCombStatTTthirteenTeVoverTTsevenTeVabsvalTot}{   0.086}
\newcommand{\RatCombSystNoLumiUncTTthirteenTeVoverTTsevenTeVabsvalTot}{   0.149}
\newcommand{\RatCombSystWithLumiUncTTthirteenTeVoverTTsevenTeVabsvalTot}{   0.202}
\newcommand{\RatCombLumiUncTTthirteenTeVoverTTsevenTeVabsvalTot}{   0.136}
\newcommand{\RatCombTotUncTTthirteenTeVoverTTsevenTeVabsvalTot}{   0.219}
\newcommand{\RatCombTotUncTTthirteenTeVoverTTsevenTeVpercentsTot}{    4.91}
\newcommand{\RatCombStatTTthirteenTeVoverTTsevenTeVpercentsTot}{    1.92}
\newcommand{\RatCombSystWithLumiTTthirteenTeVoverTTsevenTeVpercentsTot}{    4.51}
\newcommand{\RatCombSystNoLumiUncTTthirteenTeVoverTTsevenTeVpercentsTot}{    3.33}
\newcommand{\RatCombLumiUncTTthirteenTeVoverTTsevenTeVpercentsTot}{    3.04}

\newcommand{\AveTTBARCentralTTbarSevenTevabsvalFid}{   2.3}
\newcommand{\AveTTBARStatTTbarSevenTevabsvalFid}{   0.0}
\newcommand{\AveTTBARSystTTbarSevenTevabsvalFid}{   0.1}
\newcommand{\AveTTBARTotUncTTbarSevenTevabsvalFid}{   0.1}
\newcommand{\AveTTBARStatTTbarSevenTevpercentsFid}{   1.7}
\newcommand{\AveTTBARSystTTbarSevenTevpercentsFid}{   2.8}
\newcommand{\AveTTBARTotUncTTbarSevenTevpercentsFid}{   3.2}
\newcommand{\AveTTBARLumiTTbarSevenTevFid}{   0.0}
\newcommand{\AveTTBARCentralTTbarThirteenTeVabsvalFid}{   9.9}
\newcommand{\AveTTBARStatTTbarThirteenTeVabsvalFid}{   0.1}
\newcommand{\AveTTBARSystTTbarThirteenTeVabsvalFid}{   0.3}
\newcommand{\AveTTBARTotUncTTbarThirteenTeVabsvalFid}{   0.3}
\newcommand{\AveTTBARStatTTbarThirteenTeVpercentsFid}{   0.9}
\newcommand{\AveTTBARSystTTbarThirteenTeVpercentsFid}{   2.9}
\newcommand{\AveTTBARTotUncTTbarThirteenTeVpercentsFid}{   3.1}
\newcommand{\AveTTBARLumiTTbarThirteenTeVFid}{   0.2}
\newcommand{\AveTTBARChisqTTbarThirteenTeVtoTTBARTTbarSevenTevFid}{0.0}
\newcommand{\CorrCombTTbarSevenTevcorTTbarSevenTevFid}{ 1.000}
\newcommand{\CorrCombTTbarThirteenTeVcorTTbarSevenTevFid}{ 0.463}
\newcommand{\CorrCombTTbarThirteenTeVcorTTbarThirteenTeVFid}{ 1.000}
\newcommand{\RatCombCentTTthirteenTeVoverTTsevenTeVabsvalFid}{   4.322}
\newcommand{\RatCombStatTTthirteenTeVoverTTsevenTeVabsvalFid}{   0.083}
\newcommand{\RatCombSystNoLumiUncTTthirteenTeVoverTTsevenTeVabsvalFid}{   0.116}
\newcommand{\RatCombSystWithLumiUncTTthirteenTeVoverTTsevenTeVabsvalFid}{   0.176}
\newcommand{\RatCombLumiUncTTthirteenTeVoverTTsevenTeVabsvalFid}{   0.131}
\newcommand{\RatCombTotUncTTthirteenTeVoverTTsevenTeVabsvalFid}{   0.194}
\newcommand{\RatCombTotUncTTthirteenTeVoverTTsevenTeVpercentsFid}{    4.49}
\newcommand{\RatCombStatTTthirteenTeVoverTTsevenTeVpercentsFid}{    1.91}
\newcommand{\RatCombSystWithLumiTTthirteenTeVoverTTsevenTeVpercentsFid}{    4.06}
\newcommand{\RatCombSystNoLumiUncTTthirteenTeVoverTTsevenTeVpercentsFid}{    2.69}
\newcommand{\RatCombLumiUncTTthirteenTeVoverTTsevenTeVpercentsFid}{    3.04}

\newcommand{\AveCentralZSevenTeVabsvalTot}{ 994.8}
\newcommand{\AveStatZSevenTeVabsvalTot}{   0.6}
\newcommand{\AveSystZSevenTeVabsvalTot}{   6.8}
\newcommand{\AveTotUncZSevenTeVabsvalTot}{   6.8}
\newcommand{\AveStatZSevenTeVpercentsTot}{   0.1}
\newcommand{\AveSystZSevenTeVpercentsTot}{   0.7}
\newcommand{\AveTotUncZSevenTeVpercentsTot}{   0.7}
\newcommand{\AveLumiZSevenTeVTot}{  17.9}
\newcommand{\AveCentralZEightTeVabsvalTot}{1153.7}
\newcommand{\AveStatZEightTeVabsvalTot}{   0.3}
\newcommand{\AveSystZEightTeVabsvalTot}{   9.6}
\newcommand{\AveTotUncZEightTeVabsvalTot}{   9.6}
\newcommand{\AveStatZEightTeVpercentsTot}{   0.0}
\newcommand{\AveSystZEightTeVpercentsTot}{   0.8}
\newcommand{\AveTotUncZEightTeVpercentsTot}{   0.8}
\newcommand{\AveLumiZEightTeVTot}{  21.9}
\newcommand{\AveChisqZEightTeVtoZSevenTeVTot}{0.4}
\newcommand{\CorrCombZSevenTeVcorZSevenTeVTot}{ 1.000}
\newcommand{\CorrCombZEightTeVcorZSevenTeVTot}{ 0.687}
\newcommand{\CorrCombZEightTeVcorZEightTeVTot}{ 1.000}
\newcommand{\RatCombCentZeightTeVoverZsevenTeVabsvalTot}{   1.160}
\newcommand{\RatCombStatZeightTeVoverZsevenTeVabsvalTot}{   0.001}
\newcommand{\RatCombSystNoLumiUncZeightTeVoverZsevenTeVabsvalTot}{   0.007}
\newcommand{\RatCombSystWithLumiUncZeightTeVoverZsevenTeVabsvalTot}{   0.031}
\newcommand{\RatCombLumiUncZeightTeVoverZsevenTeVabsvalTot}{   0.030}
\newcommand{\RatCombTotUncZeightTeVoverZsevenTeVabsvalTot}{ 0.03123}
\newcommand{\RatCombTotUncZeightTeVoverZsevenTeVpercentsTot}{    2.69}
\newcommand{\RatCombStatZeightTeVoverZsevenTeVpercentsTot}{    0.07}
\newcommand{\RatCombSystWithLumiZeightTeVoverZsevenTeVpercentsTot}{    2.69}
\newcommand{\RatCombSystNoLumiUncZeightTeVoverZsevenTeVpercentsTot}{    0.63}
\newcommand{\RatCombLumiUncZeightTeVoverZsevenTeVpercentsTot}{    2.62}

\newcommand{\AveCentralZSevenTeVabsvalFid}{ 450.8}
\newcommand{\AveStatZSevenTeVabsvalFid}{   0.3}
\newcommand{\AveSystZSevenTeVabsvalFid}{   3.1}
\newcommand{\AveTotUncZSevenTeVabsvalFid}{   3.1}
\newcommand{\AveStatZSevenTeVpercentsFid}{   0.1}
\newcommand{\AveSystZSevenTeVpercentsFid}{   0.7}
\newcommand{\AveTotUncZSevenTeVpercentsFid}{   0.7}
\newcommand{\AveLumiZSevenTeVFid}{   8.1}
\newcommand{\AveCentralZEightTeVabsvalFid}{ 505.8}
\newcommand{\AveStatZEightTeVabsvalFid}{   0.1}
\newcommand{\AveSystZEightTeVabsvalFid}{   4.2}
\newcommand{\AveTotUncZEightTeVabsvalFid}{   4.2}
\newcommand{\AveStatZEightTeVpercentsFid}{   0.0}
\newcommand{\AveSystZEightTeVpercentsFid}{   0.8}
\newcommand{\AveTotUncZEightTeVpercentsFid}{   0.8}
\newcommand{\AveLumiZEightTeVFid}{   9.6}
\newcommand{\AveChisqZEightTeVtoZSevenTeVFid}{0.4}
\newcommand{\CorrCombZSevenTeVcorZSevenTeVFid}{ 1.000}
\newcommand{\CorrCombZEightTeVcorZSevenTeVFid}{ 0.687}
\newcommand{\CorrCombZEightTeVcorZEightTeVFid}{ 1.000}
\newcommand{\RatCombCentZeightTeVoverZsevenTeVabsvalFid}{   1.122}
\newcommand{\RatCombStatZeightTeVoverZsevenTeVabsvalFid}{   0.001}
\newcommand{\RatCombSystNoLumiUncZeightTeVoverZsevenTeVabsvalFid}{   0.007}
\newcommand{\RatCombSystWithLumiUncZeightTeVoverZsevenTeVabsvalFid}{   0.030}
\newcommand{\RatCombLumiUncZeightTeVoverZsevenTeVabsvalFid}{   0.029}
\newcommand{\RatCombTotUncZeightTeVoverZsevenTeVabsvalFid}{ 0.03016}
\newcommand{\RatCombTotUncZeightTeVoverZsevenTeVpercentsFid}{    2.69}
\newcommand{\RatCombStatZeightTeVoverZsevenTeVpercentsFid}{    0.07}
\newcommand{\RatCombSystWithLumiZeightTeVoverZsevenTeVpercentsFid}{    2.69}
\newcommand{\RatCombSystNoLumiUncZeightTeVoverZsevenTeVpercentsFid}{    0.61}
\newcommand{\RatCombLumiUncZeightTeVoverZsevenTeVpercentsFid}{    2.62}

\newcommand{\AveTTBARCentralTTbarSevenTeVabsvalTot}{ 182.9}
\newcommand{\AveTTBARStatTTbarSevenTeVabsvalTot}{   3.1}
\newcommand{\AveTTBARSystTTbarSevenTeVabsvalTot}{   5.3}
\newcommand{\AveTTBARTotUncTTbarSevenTeVabsvalTot}{   6.1}
\newcommand{\AveTTBARStatTTbarSevenTeVpercentsTot}{   1.7}
\newcommand{\AveTTBARSystTTbarSevenTeVpercentsTot}{   2.9}
\newcommand{\AveTTBARTotUncTTbarSevenTeVpercentsTot}{   3.3}
\newcommand{\AveTTBARLumiTTbarSevenTeVTot}{   3.6}
\newcommand{\AveTTBARCentralTTbarEightTeVabsvalTot}{ 242.9}
\newcommand{\AveTTBARStatTTbarEightTeVabsvalTot}{   1.7}
\newcommand{\AveTTBARSystTTbarEightTeVabsvalTot}{   6.9}
\newcommand{\AveTTBARTotUncTTbarEightTeVabsvalTot}{   7.1}
\newcommand{\AveTTBARStatTTbarEightTeVpercentsTot}{   0.7}
\newcommand{\AveTTBARSystTTbarEightTeVpercentsTot}{   2.8}
\newcommand{\AveTTBARTotUncTTbarEightTeVpercentsTot}{   2.9}
\newcommand{\AveTTBARLumiTTbarEightTeVTot}{   5.1}
\newcommand{\AveTTBARChisqTTbarEightTeVtoTTBARTTbarSevenTeVTot}{0.0}
\newcommand{\CorrCombTTbarSevenTeVcorTTbarSevenTeVTot}{ 1.000}
\newcommand{\CorrCombTTbarEightTeVcorTTbarSevenTeVTot}{ 0.775}
\newcommand{\CorrCombTTbarEightTeVcorTTbarEightTeVTot}{ 1.000}
\newcommand{\RatCombCentTTeightTeVoverTTsevenTeVabsvalTot}{   1.328}
\newcommand{\RatCombStatTTeightTeVoverTTsevenTeVabsvalTot}{   0.024}
\newcommand{\RatCombSystNoLumiUncTTeightTeVoverTTsevenTeVabsvalTot}{   0.015}
\newcommand{\RatCombSystWithLumiUncTTeightTeVoverTTsevenTeVabsvalTot}{   0.041}
\newcommand{\RatCombLumiUncTTeightTeVoverTTsevenTeVabsvalTot}{   0.038}
\newcommand{\RatCombTotUncTTeightTeVoverTTsevenTeVabsvalTot}{   0.048}
\newcommand{\RatCombTotUncTTeightTeVoverTTsevenTeVpercentsTot}{    3.60}
\newcommand{\RatCombStatTTeightTeVoverTTsevenTeVpercentsTot}{    1.83}
\newcommand{\RatCombSystWithLumiTTeightTeVoverTTsevenTeVpercentsTot}{    3.09}
\newcommand{\RatCombSystNoLumiUncTTeightTeVoverTTsevenTeVpercentsTot}{    1.11}
\newcommand{\RatCombLumiUncTTeightTeVoverTTsevenTeVpercentsTot}{    2.89}

\newcommand{\AveTTBARCentralTTbarSevenTeVabsvalFid}{   2.3}
\newcommand{\AveTTBARStatTTbarSevenTeVabsvalFid}{   0.0}
\newcommand{\AveTTBARSystTTbarSevenTeVabsvalFid}{   0.1}
\newcommand{\AveTTBARTotUncTTbarSevenTeVabsvalFid}{   0.1}
\newcommand{\AveTTBARStatTTbarSevenTeVpercentsFid}{   1.7}
\newcommand{\AveTTBARSystTTbarSevenTeVpercentsFid}{   2.8}
\newcommand{\AveTTBARTotUncTTbarSevenTeVpercentsFid}{   3.2}
\newcommand{\AveTTBARLumiTTbarSevenTeVFid}{   0.0}
\newcommand{\AveTTBARCentralTTbarEightTeVabsvalFid}{   3.0}
\newcommand{\AveTTBARStatTTbarEightTeVabsvalFid}{   0.0}
\newcommand{\AveTTBARSystTTbarEightTeVabsvalFid}{   0.1}
\newcommand{\AveTTBARTotUncTTbarEightTeVabsvalFid}{   0.1}
\newcommand{\AveTTBARStatTTbarEightTeVpercentsFid}{   0.7}
\newcommand{\AveTTBARSystTTbarEightTeVpercentsFid}{   2.6}
\newcommand{\AveTTBARTotUncTTbarEightTeVpercentsFid}{   2.7}
\newcommand{\AveTTBARLumiTTbarEightTeVFid}{   0.1}
\newcommand{\AveTTBARChisqTTbarEightTeVtoTTBARTTbarSevenTeVFid}{0.0}
\newcommand{\CorrCombTTbarSevenTeVcorTTbarSevenTeVFid}{ 1.000}
\newcommand{\CorrCombTTbarEightTeVcorTTbarSevenTeVFid}{ 0.756}
\newcommand{\CorrCombTTbarEightTeVcorTTbarEightTeVFid}{ 1.000}
\newcommand{\RatCombCentTTeightTeVoverTTsevenTeVabsvalFid}{   1.322}
\newcommand{\RatCombStatTTeightTeVoverTTsevenTeVabsvalFid}{   0.024}
\newcommand{\RatCombSystNoLumiUncTTeightTeVoverTTsevenTeVabsvalFid}{   0.015}
\newcommand{\RatCombSystWithLumiUncTTeightTeVoverTTsevenTeVabsvalFid}{   0.041}
\newcommand{\RatCombLumiUncTTeightTeVoverTTsevenTeVabsvalFid}{   0.038}
\newcommand{\RatCombTotUncTTeightTeVoverTTsevenTeVabsvalFid}{   0.048}
\newcommand{\RatCombTotUncTTeightTeVoverTTsevenTeVpercentsFid}{    3.60}
\newcommand{\RatCombStatTTeightTeVoverTTsevenTeVpercentsFid}{    1.83}
\newcommand{\RatCombSystWithLumiTTeightTeVoverTTsevenTeVpercentsFid}{    3.10}
\newcommand{\RatCombSystNoLumiUncTTeightTeVoverTTsevenTeVpercentsFid}{    1.12}
\newcommand{\RatCombLumiUncTTeightTeVoverTTsevenTeVpercentsFid}{    2.89}

\newcommand{\AveStatZThirteenTevForDouRthirtToeightpercentsttTotZTot}{   0.1}
\newcommand{\AveSystZThirteenTevForDouRthirtToeightpercentsttTotZTot}{   2.3}
\newcommand{\AveTotUncZThirteenTevForDouRthirtToeightpercentsttTotZTot}{   2.3}
\newcommand{\AveLumiZThirteenTevForDouRthirtToeightabsolutettTotZTot}{  41.3}
\newcommand{\AveCentralZThirteenTevForDouRthirtToeightabsvalttTotZTot}{1968.9}
\newcommand{\AveStatZThirteenTevForDouRthirtToeightabsvalttTotZTot}{   1.2}
\newcommand{\AveSystZThirteenTevForDouRthirtToeightabsvalttTotZTot}{  44.4}
\newcommand{\AveTotUncZThirteenTevForDouRthirtToeightabsvalttTotZTot}{  44.4}
\newcommand{\AveStatTTEightTevForDouRthirtToeightpercentsttTotZTot}{   0.7}
\newcommand{\AveSystTTEightTevForDouRthirtToeightpercentsttTotZTot}{   3.5}
\newcommand{\AveTotUncTTEightTevForDouRthirtToeightpercentsttTotZTot}{   3.6}
\newcommand{\AveLumiTTEightTevForDouRthirtToeightabsolutettTotZTot}{   5.1}
\newcommand{\AveCentralTTEightTevForDouRthirtToeightabsvalttTotZTot}{ 242.8}
\newcommand{\AveStatTTEightTevForDouRthirtToeightabsvalttTotZTot}{   1.7}
\newcommand{\AveSystTTEightTevForDouRthirtToeightabsvalttTotZTot}{   8.6}
\newcommand{\AveTotUncTTEightTevForDouRthirtToeightabsvalttTotZTot}{   8.7}
\newcommand{\AveStatZEightTevForDouRthirtToeightpercentsttTotZTot}{   0.0}
\newcommand{\AveSystZEightTevForDouRthirtToeightpercentsttTotZTot}{   2.1}
\newcommand{\AveTotUncZEightTevForDouRthirtToeightpercentsttTotZTot}{   2.1}
\newcommand{\AveLumiZEightTevForDouRthirtToeightabsolutettTotZTot}{  21.9}
\newcommand{\AveCentralZEightTevForDouRthirtToeightabsvalttTotZTot}{1153.7}
\newcommand{\AveStatZEightTevForDouRthirtToeightabsvalttTotZTot}{   0.3}
\newcommand{\AveSystZEightTevForDouRthirtToeightabsvalttTotZTot}{  23.9}
\newcommand{\AveTotUncZEightTevForDouRthirtToeightabsvalttTotZTot}{  23.9}
\newcommand{\AveStatTTThirteenTevForDouRthirtToeightpercentsttTotZTot}{   0.9}
\newcommand{\AveSystTTThirteenTevForDouRthirtToeightpercentsttTotZTot}{   4.3}
\newcommand{\AveTotUncTTThirteenTevForDouRthirtToeightpercentsttTotZTot}{   4.4}
\newcommand{\AveLumiTTThirteenTevForDouRthirtToeightabsolutettTotZTot}{  18.9}
\newcommand{\AveCentralTTThirteenTevForDouRthirtToeightabsvalttTotZTot}{ 818.1}
\newcommand{\AveStatTTThirteenTevForDouRthirtToeightabsvalttTotZTot}{   7.5}
\newcommand{\AveSystTTThirteenTevForDouRthirtToeightabsvalttTotZTot}{  35.3}
\newcommand{\AveTotUncTTThirteenTevForDouRthirtToeightabsvalttTotZTot}{  36.1}
\newcommand{\AveChisqDoublRatioTTThirteenTevForDouRthirtToeightTotZEightTevForDouRthirtToeightTottoTTEightTevForDouRthirtToeightTotZThirteenTevForDouRthirtToeightTot}{0.4}
\newcommand{\CorrCombZThirteenTevForDouRthirtToeightcorZThirteenTevForDouRthirtToeightttTotZTot}{ 1.000}
\newcommand{\CorrCombTTEightTevForDouRthirtToeightcorZThirteenTevForDouRthirtToeightttTotZTot}{ 0.157}
\newcommand{\CorrCombZEightTevForDouRthirtToeightcorZThirteenTevForDouRthirtToeightttTotZTot}{ 0.097}
\newcommand{\CorrCombTTThirteenTevForDouRthirtToeightcorZThirteenTevForDouRthirtToeightttTotZTot}{ 0.612}
\newcommand{\CorrCombTTEightTevForDouRthirtToeightcorTTEightTevForDouRthirtToeightttTotZTot}{ 1.000}
\newcommand{\CorrCombZEightTevForDouRthirtToeightcorTTEightTevForDouRthirtToeightttTotZTot}{ 0.679}
\newcommand{\CorrCombTTThirteenTevForDouRthirtToeightcorTTEightTevForDouRthirtToeightttTotZTot}{ 0.324}
\newcommand{\CorrCombZEightTevForDouRthirtToeightcorZEightTevForDouRthirtToeightttTotZTot}{ 1.000}
\newcommand{\CorrCombTTThirteenTevForDouRthirtToeightcorZEightTevForDouRthirtToeightttTotZTot}{ 0.104}
\newcommand{\CorrCombTTThirteenTevForDouRthirtToeightcorTTThirteenTevForDouRthirtToeightttTotZTot}{ 1.000}
\newcommand{\RatCombCentDoubRTTthirtOverZthirtTOtteightOverZeightabsvalttTotZTot}{   1.975}
\newcommand{\RatCombStatDoubRTTthirtOverZthirtTOtteightOverZeightabsvalttTotZTot}{   0.023}
\newcommand{\RatCombSystWithLumiUncDoubRTTthirtOverZthirtTOtteightOverZeightabsvalttTotZTot}{   0.067}
\newcommand{\RatCombTotUncDoubRTTthirtOverZthirtTOtteightOverZeightabsvalttTotZTot}{   0.071}
\newcommand{\RatCombTotUncDoubRTTthirtOverZthirtTOtteightOverZeightpercentsttTotZTot}{    3.59}
\newcommand{\RatCombStatDoubRTTthirtOverZthirtTOtteightOverZeightpercentsttTotZTot}{    1.16}
\newcommand{\RatCombSystWithLumiDoubRTTthirtOverZthirtTOtteightOverZeightpercentsttTotZTot}{    3.40}
\newcommand{\RatCombSystNoLumiUncDoubRTTthirtOverZthirtTOtteightOverZeightabsvalttTotZTot}{   0.067}
\newcommand{\RatCombSystNoLumiDoubRTTthirtOverZthirtTOtteightOverZeightpercentsttTotZTot}{    3.38}
\newcommand{\RatCombTotUncNoLumiDoubRTTthirtOverZthirtTOtteightOverZeightabsvalttTotZTot}{   0.071}
\newcommand{\RatCombTotUncNoLumiDoubRTTthirtOverZthirtTOtteightOverZeightpercentsttTotZTot}{    3.58}
\newcommand{\RatCombLumiUncDoubRTTthirtOverZthirtTOtteightOverZeightabsvalttTotZTot}{   0.006}
\newcommand{\RatCombLumiUncDoubRTTthirtOverZthirtTOtteightOverZeightpercentsttTotZTot}{    0.29}

\newcommand{\AveStatZThirteenTevForDouRthirtToeightpercentsttTotZFid}{   0.1}
\newcommand{\AveSystZThirteenTevForDouRthirtToeightpercentsttTotZFid}{   2.3}
\newcommand{\AveTotUncZThirteenTevForDouRthirtToeightpercentsttTotZFid}{   2.3}
\newcommand{\AveLumiZThirteenTevForDouRthirtToeightabsolutettTotZFid}{  16.3}
\newcommand{\AveCentralZThirteenTevForDouRthirtToeightabsvalttTotZFid}{ 777.1}
\newcommand{\AveStatZThirteenTevForDouRthirtToeightabsvalttTotZFid}{   0.5}
\newcommand{\AveSystZThirteenTevForDouRthirtToeightabsvalttTotZFid}{  17.5}
\newcommand{\AveTotUncZThirteenTevForDouRthirtToeightabsvalttTotZFid}{  17.5}
\newcommand{\AveStatTTEightTevForDouRthirtToeightpercentsttTotZFid}{   0.7}
\newcommand{\AveSystTTEightTevForDouRthirtToeightpercentsttTotZFid}{   3.5}
\newcommand{\AveTotUncTTEightTevForDouRthirtToeightpercentsttTotZFid}{   3.6}
\newcommand{\AveLumiTTEightTevForDouRthirtToeightabsolutettTotZFid}{   5.1}
\newcommand{\AveCentralTTEightTevForDouRthirtToeightabsvalttTotZFid}{ 242.7}
\newcommand{\AveStatTTEightTevForDouRthirtToeightabsvalttTotZFid}{   1.7}
\newcommand{\AveSystTTEightTevForDouRthirtToeightabsvalttTotZFid}{   8.6}
\newcommand{\AveTotUncTTEightTevForDouRthirtToeightabsvalttTotZFid}{   8.7}
\newcommand{\AveStatZEightTevForDouRthirtToeightpercentsttTotZFid}{   0.0}
\newcommand{\AveSystZEightTevForDouRthirtToeightpercentsttTotZFid}{   2.1}
\newcommand{\AveTotUncZEightTevForDouRthirtToeightpercentsttTotZFid}{   2.1}
\newcommand{\AveLumiZEightTevForDouRthirtToeightabsolutettTotZFid}{   9.6}
\newcommand{\AveCentralZEightTevForDouRthirtToeightabsvalttTotZFid}{ 505.6}
\newcommand{\AveStatZEightTevForDouRthirtToeightabsvalttTotZFid}{   0.1}
\newcommand{\AveSystZEightTevForDouRthirtToeightabsvalttTotZFid}{  10.5}
\newcommand{\AveTotUncZEightTevForDouRthirtToeightabsvalttTotZFid}{  10.5}
\newcommand{\AveStatTTThirteenTevForDouRthirtToeightpercentsttTotZFid}{   0.9}
\newcommand{\AveSystTTThirteenTevForDouRthirtToeightpercentsttTotZFid}{   4.3}
\newcommand{\AveTotUncTTThirteenTevForDouRthirtToeightpercentsttTotZFid}{   4.4}
\newcommand{\AveLumiTTThirteenTevForDouRthirtToeightabsolutettTotZFid}{  18.9}
\newcommand{\AveCentralTTThirteenTevForDouRthirtToeightabsvalttTotZFid}{ 818.0}
\newcommand{\AveStatTTThirteenTevForDouRthirtToeightabsvalttTotZFid}{   7.5}
\newcommand{\AveSystTTThirteenTevForDouRthirtToeightabsvalttTotZFid}{  35.3}
\newcommand{\AveTotUncTTThirteenTevForDouRthirtToeightabsvalttTotZFid}{  36.1}
\newcommand{\AveChisqDoublRatioTTThirteenTevForDouRthirtToeightTotZEightTevForDouRthirtToeightFidtoTTEightTevForDouRthirtToeightTotZThirteenTevForDouRthirtToeightFid}{0.4}
\newcommand{\CorrCombZThirteenTevForDouRthirtToeightcorZThirteenTevForDouRthirtToeightttTotZFid}{ 1.000}
\newcommand{\CorrCombTTEightTevForDouRthirtToeightcorZThirteenTevForDouRthirtToeightttTotZFid}{ 0.157}
\newcommand{\CorrCombZEightTevForDouRthirtToeightcorZThirteenTevForDouRthirtToeightttTotZFid}{ 0.097}
\newcommand{\CorrCombTTThirteenTevForDouRthirtToeightcorZThirteenTevForDouRthirtToeightttTotZFid}{ 0.612}
\newcommand{\CorrCombTTEightTevForDouRthirtToeightcorTTEightTevForDouRthirtToeightttTotZFid}{ 1.000}
\newcommand{\CorrCombZEightTevForDouRthirtToeightcorTTEightTevForDouRthirtToeightttTotZFid}{ 0.679}
\newcommand{\CorrCombTTThirteenTevForDouRthirtToeightcorTTEightTevForDouRthirtToeightttTotZFid}{ 0.324}
\newcommand{\CorrCombZEightTevForDouRthirtToeightcorZEightTevForDouRthirtToeightttTotZFid}{ 1.000}
\newcommand{\CorrCombTTThirteenTevForDouRthirtToeightcorZEightTevForDouRthirtToeightttTotZFid}{ 0.104}
\newcommand{\CorrCombTTThirteenTevForDouRthirtToeightcorTTThirteenTevForDouRthirtToeightttTotZFid}{ 1.000}
\newcommand{\RatCombCentDoubRTTthirtOverZthirtTOtteightOverZeightabsvalttTotZFid}{   2.193}
\newcommand{\RatCombStatDoubRTTthirtOverZthirtTOtteightOverZeightabsvalttTotZFid}{   0.026}
\newcommand{\RatCombSystWithLumiUncDoubRTTthirtOverZthirtTOtteightOverZeightabsvalttTotZFid}{   0.074}
\newcommand{\RatCombTotUncDoubRTTthirtOverZthirtTOtteightOverZeightabsvalttTotZFid}{   0.078}
\newcommand{\RatCombTotUncDoubRTTthirtOverZthirtTOtteightOverZeightpercentsttTotZFid}{    3.57}
\newcommand{\RatCombStatDoubRTTthirtOverZthirtTOtteightOverZeightpercentsttTotZFid}{    1.16}
\newcommand{\RatCombSystWithLumiDoubRTTthirtOverZthirtTOtteightOverZeightpercentsttTotZFid}{    3.38}
\newcommand{\RatCombSystNoLumiUncDoubRTTthirtOverZthirtTOtteightOverZeightabsvalttTotZFid}{   0.074}
\newcommand{\RatCombSystNoLumiDoubRTTthirtOverZthirtTOtteightOverZeightpercentsttTotZFid}{    3.36}
\newcommand{\RatCombTotUncNoLumiDoubRTTthirtOverZthirtTOtteightOverZeightabsvalttTotZFid}{   0.078}
\newcommand{\RatCombTotUncNoLumiDoubRTTthirtOverZthirtTOtteightOverZeightpercentsttTotZFid}{    3.55}
\newcommand{\RatCombLumiUncDoubRTTthirtOverZthirtTOtteightOverZeightabsvalttTotZFid}{   0.008}
\newcommand{\RatCombLumiUncDoubRTTthirtOverZthirtTOtteightOverZeightpercentsttTotZFid}{    0.37}

\newcommand{\AveStatZThirteenTevForDouRthirtToeightpercentsttFidZFid}{   0.1}
\newcommand{\AveSystZThirteenTevForDouRthirtToeightpercentsttFidZFid}{   2.3}
\newcommand{\AveTotUncZThirteenTevForDouRthirtToeightpercentsttFidZFid}{   2.3}
\newcommand{\AveLumiZThirteenTevForDouRthirtToeightabsolutettFidZFid}{  16.3}
\newcommand{\AveCentralZThirteenTevForDouRthirtToeightabsvalttFidZFid}{ 777.1}
\newcommand{\AveStatZThirteenTevForDouRthirtToeightabsvalttFidZFid}{   0.5}
\newcommand{\AveSystZThirteenTevForDouRthirtToeightabsvalttFidZFid}{  17.5}
\newcommand{\AveTotUncZThirteenTevForDouRthirtToeightabsvalttFidZFid}{  17.5}
\newcommand{\AveStatTTEightTevForDouRthirtToeightpercentsttFidZFid}{   0.7}
\newcommand{\AveSystTTEightTevForDouRthirtToeightpercentsttFidZFid}{   3.4}
\newcommand{\AveTotUncTTEightTevForDouRthirtToeightpercentsttFidZFid}{   3.5}
\newcommand{\AveLumiTTEightTevForDouRthirtToeightabsolutettFidZFid}{   0.1}
\newcommand{\AveCentralTTEightTevForDouRthirtToeightabsvalttFidZFid}{   3.0}
\newcommand{\AveStatTTEightTevForDouRthirtToeightabsvalttFidZFid}{   0.0}
\newcommand{\AveSystTTEightTevForDouRthirtToeightabsvalttFidZFid}{   0.1}
\newcommand{\AveTotUncTTEightTevForDouRthirtToeightabsvalttFidZFid}{   0.1}
\newcommand{\AveStatZEightTevForDouRthirtToeightpercentsttFidZFid}{   0.0}
\newcommand{\AveSystZEightTevForDouRthirtToeightpercentsttFidZFid}{   2.1}
\newcommand{\AveTotUncZEightTevForDouRthirtToeightpercentsttFidZFid}{   2.1}
\newcommand{\AveLumiZEightTevForDouRthirtToeightabsolutettFidZFid}{   9.6}
\newcommand{\AveCentralZEightTevForDouRthirtToeightabsvalttFidZFid}{ 505.6}
\newcommand{\AveStatZEightTevForDouRthirtToeightabsvalttFidZFid}{   0.1}
\newcommand{\AveSystZEightTevForDouRthirtToeightabsvalttFidZFid}{  10.5}
\newcommand{\AveTotUncZEightTevForDouRthirtToeightabsvalttFidZFid}{  10.5}
\newcommand{\AveStatTTThirteenTevForDouRthirtToeightpercentsttFidZFid}{   0.9}
\newcommand{\AveSystTTThirteenTevForDouRthirtToeightpercentsttFidZFid}{   3.7}
\newcommand{\AveTotUncTTThirteenTevForDouRthirtToeightpercentsttFidZFid}{   3.8}
\newcommand{\AveLumiTTThirteenTevForDouRthirtToeightabsolutettFidZFid}{   0.2}
\newcommand{\AveCentralTTThirteenTevForDouRthirtToeightabsvalttFidZFid}{   9.9}
\newcommand{\AveStatTTThirteenTevForDouRthirtToeightabsvalttFidZFid}{   0.1}
\newcommand{\AveSystTTThirteenTevForDouRthirtToeightabsvalttFidZFid}{   0.4}
\newcommand{\AveTotUncTTThirteenTevForDouRthirtToeightabsvalttFidZFid}{   0.4}
\newcommand{\AveChisqDoublRatioTTThirteenTevForDouRthirtToeightFidZEightTevForDouRthirtToeightFidtoTTEightTevForDouRthirtToeightFidZThirteenTevForDouRthirtToeightFid}{0.4}
\newcommand{\CorrCombZThirteenTevForDouRthirtToeightcorZThirteenTevForDouRthirtToeightttFidZFid}{ 1.000}
\newcommand{\CorrCombTTEightTevForDouRthirtToeightcorZThirteenTevForDouRthirtToeightttFidZFid}{ 0.163}
\newcommand{\CorrCombZEightTevForDouRthirtToeightcorZThirteenTevForDouRthirtToeightttFidZFid}{ 0.097}
\newcommand{\CorrCombTTThirteenTevForDouRthirtToeightcorZThirteenTevForDouRthirtToeightttFidZFid}{ 0.702}
\newcommand{\CorrCombTTEightTevForDouRthirtToeightcorTTEightTevForDouRthirtToeightttFidZFid}{ 1.000}
\newcommand{\CorrCombZEightTevForDouRthirtToeightcorTTEightTevForDouRthirtToeightttFidZFid}{ 0.709}
\newcommand{\CorrCombTTThirteenTevForDouRthirtToeightcorTTEightTevForDouRthirtToeightttFidZFid}{ 0.330}
\newcommand{\CorrCombZEightTevForDouRthirtToeightcorZEightTevForDouRthirtToeightttFidZFid}{ 1.000}
\newcommand{\CorrCombTTThirteenTevForDouRthirtToeightcorZEightTevForDouRthirtToeightttFidZFid}{ 0.120}
\newcommand{\CorrCombTTThirteenTevForDouRthirtToeightcorTTThirteenTevForDouRthirtToeightttFidZFid}{ 1.000}
\newcommand{\RatCombCentDoubRTTthirtOverZthirtTOtteightOverZeightabsvalttFidZFid}{   2.131}
\newcommand{\RatCombStatDoubRTTthirtOverZthirtTOtteightOverZeightabsvalttFidZFid}{   0.025}
\newcommand{\RatCombSystWithLumiUncDoubRTTthirtOverZthirtTOtteightOverZeightabsvalttFidZFid}{   0.057}
\newcommand{\RatCombTotUncDoubRTTthirtOverZthirtTOtteightOverZeightabsvalttFidZFid}{   0.062}
\newcommand{\RatCombTotUncDoubRTTthirtOverZthirtTOtteightOverZeightpercentsttFidZFid}{    2.93}
\newcommand{\RatCombStatDoubRTTthirtOverZthirtTOtteightOverZeightpercentsttFidZFid}{    1.16}
\newcommand{\RatCombSystWithLumiDoubRTTthirtOverZthirtTOtteightOverZeightpercentsttFidZFid}{    2.69}
\newcommand{\RatCombSystNoLumiUncDoubRTTthirtOverZthirtTOtteightOverZeightabsvalttFidZFid}{   0.057}
\newcommand{\RatCombSystNoLumiDoubRTTthirtOverZthirtTOtteightOverZeightpercentsttFidZFid}{    2.67}
\newcommand{\RatCombTotUncNoLumiDoubRTTthirtOverZthirtTOtteightOverZeightabsvalttFidZFid}{   0.062}
\newcommand{\RatCombTotUncNoLumiDoubRTTthirtOverZthirtTOtteightOverZeightpercentsttFidZFid}{    2.91}
\newcommand{\RatCombLumiUncDoubRTTthirtOverZthirtTOtteightOverZeightabsvalttFidZFid}{   0.006}
\newcommand{\RatCombLumiUncDoubRTTthirtOverZthirtTOtteightOverZeightpercentsttFidZFid}{    0.29}

\newcommand{\AveStatZThirteenTevForDouRthirtTosevenpercentsttTotZTot}{   0.1}
\newcommand{\AveSystZThirteenTevForDouRthirtTosevenpercentsttTotZTot}{   2.3}
\newcommand{\AveTotUncZThirteenTevForDouRthirtTosevenpercentsttTotZTot}{   2.3}
\newcommand{\AveLumiZThirteenTevForDouRthirtTosevenabsolutettTotZTot}{  41.4}
\newcommand{\AveCentralZThirteenTevForDouRthirtTosevenabsvalttTotZTot}{1969.1}
\newcommand{\AveStatZThirteenTevForDouRthirtTosevenabsvalttTotZTot}{   1.2}
\newcommand{\AveSystZThirteenTevForDouRthirtTosevenabsvalttTotZTot}{  44.4}
\newcommand{\AveTotUncZThirteenTevForDouRthirtTosevenabsvalttTotZTot}{  44.4}
\newcommand{\AveStatTTSevenTevForDouRthirtTosevenpercentsttTotZTot}{   1.7}
\newcommand{\AveSystTTSevenTevForDouRthirtTosevenpercentsttTotZTot}{   3.5}
\newcommand{\AveTotUncTTSevenTevForDouRthirtTosevenpercentsttTotZTot}{   3.9}
\newcommand{\AveLumiTTSevenTevForDouRthirtTosevenabsolutettTotZTot}{   3.6}
\newcommand{\AveCentralTTSevenTevForDouRthirtTosevenabsvalttTotZTot}{ 182.8}
\newcommand{\AveStatTTSevenTevForDouRthirtTosevenabsvalttTotZTot}{   3.1}
\newcommand{\AveSystTTSevenTevForDouRthirtTosevenabsvalttTotZTot}{   6.4}
\newcommand{\AveTotUncTTSevenTevForDouRthirtTosevenabsvalttTotZTot}{   7.1}
\newcommand{\AveStatZSevenTevForDouRthirtTosevenpercentsttTotZTot}{   0.1}
\newcommand{\AveSystZSevenTevForDouRthirtTosevenpercentsttTotZTot}{   1.9}
\newcommand{\AveTotUncZSevenTevForDouRthirtTosevenpercentsttTotZTot}{   1.9}
\newcommand{\AveLumiZSevenTevForDouRthirtTosevenabsolutettTotZTot}{  17.9}
\newcommand{\AveCentralZSevenTevForDouRthirtTosevenabsvalttTotZTot}{ 994.8}
\newcommand{\AveStatZSevenTevForDouRthirtTosevenabsvalttTotZTot}{   0.6}
\newcommand{\AveSystZSevenTevForDouRthirtTosevenabsvalttTotZTot}{  19.1}
\newcommand{\AveTotUncZSevenTevForDouRthirtTosevenabsvalttTotZTot}{  19.2}
\newcommand{\AveStatTTThirteenTevForDouRthirtTosevenpercentsttTotZTot}{   0.9}
\newcommand{\AveSystTTThirteenTevForDouRthirtTosevenpercentsttTotZTot}{   4.3}
\newcommand{\AveTotUncTTThirteenTevForDouRthirtTosevenpercentsttTotZTot}{   4.4}
\newcommand{\AveLumiTTThirteenTevForDouRthirtTosevenabsolutettTotZTot}{  18.9}
\newcommand{\AveCentralTTThirteenTevForDouRthirtTosevenabsvalttTotZTot}{ 818.0}
\newcommand{\AveStatTTThirteenTevForDouRthirtTosevenabsvalttTotZTot}{   7.5}
\newcommand{\AveSystTTThirteenTevForDouRthirtTosevenabsvalttTotZTot}{  35.3}
\newcommand{\AveTotUncTTThirteenTevForDouRthirtTosevenabsvalttTotZTot}{  36.1}
\newcommand{\AveChisqDoublRatioTTThirteenTevForDouRthirtTosevenTotZSevenTevForDouRthirtTosevenTottoTTSevenTevForDouRthirtTosevenTotZThirteenTevForDouRthirtTosevenTot}{0.5}
\newcommand{\CorrCombZThirteenTevForDouRthirtTosevencorZThirteenTevForDouRthirtToseventtTotZTot}{ 1.000}
\newcommand{\CorrCombTTSevenTevForDouRthirtTosevencorZThirteenTevForDouRthirtToseventtTotZTot}{ 0.145}
\newcommand{\CorrCombZSevenTevForDouRthirtTosevencorZThirteenTevForDouRthirtToseventtTotZTot}{ 0.100}
\newcommand{\CorrCombTTThirteenTevForDouRthirtTosevencorZThirteenTevForDouRthirtToseventtTotZTot}{ 0.612}
\newcommand{\CorrCombTTSevenTevForDouRthirtTosevencorTTSevenTevForDouRthirtToseventtTotZTot}{ 1.000}
\newcommand{\CorrCombZSevenTevForDouRthirtTosevencorTTSevenTevForDouRthirtToseventtTotZTot}{ 0.619}
\newcommand{\CorrCombTTThirteenTevForDouRthirtTosevencorTTSevenTevForDouRthirtToseventtTotZTot}{ 0.312}
\newcommand{\CorrCombZSevenTevForDouRthirtTosevencorZSevenTevForDouRthirtToseventtTotZTot}{ 1.000}
\newcommand{\CorrCombTTThirteenTevForDouRthirtTosevencorZSevenTevForDouRthirtToseventtTotZTot}{ 0.106}
\newcommand{\CorrCombTTThirteenTevForDouRthirtTosevencorTTThirteenTevForDouRthirtToseventtTotZTot}{ 1.000}
\newcommand{\RatCombCentDoubRTTthirtOverZthirtTOttsevenOverZsevenabsvalttTotZTot}{   2.260}
\newcommand{\RatCombStatDoubRTTthirtOverZthirtTOttsevenOverZsevenabsvalttTotZTot}{   0.044}
\newcommand{\RatCombSystWithLumiUncDoubRTTthirtOverZthirtTOttsevenOverZsevenabsvalttTotZTot}{   0.076}
\newcommand{\RatCombTotUncDoubRTTthirtOverZthirtTOttsevenOverZsevenabsvalttTotZTot}{ 0.08735}
\newcommand{\RatCombTotUncDoubRTTthirtOverZthirtTOttsevenOverZsevenpercentsttTotZTot}{    3.86}
\newcommand{\RatCombStatDoubRTTthirtOverZthirtTOttsevenOverZsevenpercentsttTotZTot}{    1.93}
\newcommand{\RatCombSystWithLumiDoubRTTthirtOverZthirtTOttsevenOverZsevenpercentsttTotZTot}{    3.35}
\newcommand{\RatCombSystNoLumiUncDoubRTTthirtOverZthirtTOttsevenOverZsevenabsvalttTotZTot}{   0.075}
\newcommand{\RatCombSystNoLumiDoubRTTthirtOverZthirtTOttsevenOverZsevenpercentsttTotZTot}{    3.33}
\newcommand{\RatCombTotUncNoLumiDoubRTTthirtOverZthirtTOttsevenOverZsevenabsvalttTotZTot}{   0.087}
\newcommand{\RatCombTotUncNoLumiDoubRTTthirtOverZthirtTOttsevenOverZsevenpercentsttTotZTot}{    3.85}
\newcommand{\RatCombLumiUncDoubRTTthirtOverZthirtTOttsevenOverZsevenabsvalttTotZTot}{   0.007}
\newcommand{\RatCombLumiUncDoubRTTthirtOverZthirtTOttsevenOverZsevenpercentsttTotZTot}{    0.32}

\newcommand{\AveStatZThirteenTevForDouRthirtTosevenpercentsttTotZFid}{   0.1}
\newcommand{\AveSystZThirteenTevForDouRthirtTosevenpercentsttTotZFid}{   2.3}
\newcommand{\AveTotUncZThirteenTevForDouRthirtTosevenpercentsttTotZFid}{   2.3}
\newcommand{\AveLumiZThirteenTevForDouRthirtTosevenabsolutettTotZFid}{  16.3}
\newcommand{\AveCentralZThirteenTevForDouRthirtTosevenabsvalttTotZFid}{ 777.2}
\newcommand{\AveStatZThirteenTevForDouRthirtTosevenabsvalttTotZFid}{   0.5}
\newcommand{\AveSystZThirteenTevForDouRthirtTosevenabsvalttTotZFid}{  17.5}
\newcommand{\AveTotUncZThirteenTevForDouRthirtTosevenabsvalttTotZFid}{  17.5}
\newcommand{\AveStatTTSevenTevForDouRthirtTosevenpercentsttTotZFid}{   1.7}
\newcommand{\AveSystTTSevenTevForDouRthirtTosevenpercentsttTotZFid}{   3.5}
\newcommand{\AveTotUncTTSevenTevForDouRthirtTosevenpercentsttTotZFid}{   3.9}
\newcommand{\AveLumiTTSevenTevForDouRthirtTosevenabsolutettTotZFid}{   3.6}
\newcommand{\AveCentralTTSevenTevForDouRthirtTosevenabsvalttTotZFid}{ 182.9}
\newcommand{\AveStatTTSevenTevForDouRthirtTosevenabsvalttTotZFid}{   3.1}
\newcommand{\AveSystTTSevenTevForDouRthirtTosevenabsvalttTotZFid}{   6.4}
\newcommand{\AveTotUncTTSevenTevForDouRthirtTosevenabsvalttTotZFid}{   7.1}
\newcommand{\AveStatZSevenTevForDouRthirtTosevenpercentsttTotZFid}{   0.1}
\newcommand{\AveSystZSevenTevForDouRthirtTosevenpercentsttTotZFid}{   1.9}
\newcommand{\AveTotUncZSevenTevForDouRthirtTosevenpercentsttTotZFid}{   1.9}
\newcommand{\AveLumiZSevenTevForDouRthirtTosevenabsolutettTotZFid}{   8.1}
\newcommand{\AveCentralZSevenTevForDouRthirtTosevenabsvalttTotZFid}{ 450.8}
\newcommand{\AveStatZSevenTevForDouRthirtTosevenabsvalttTotZFid}{   0.3}
\newcommand{\AveSystZSevenTevForDouRthirtTosevenabsvalttTotZFid}{   8.7}
\newcommand{\AveTotUncZSevenTevForDouRthirtTosevenabsvalttTotZFid}{   8.7}
\newcommand{\AveStatTTThirteenTevForDouRthirtTosevenpercentsttTotZFid}{   0.9}
\newcommand{\AveSystTTThirteenTevForDouRthirtTosevenpercentsttTotZFid}{   4.3}
\newcommand{\AveTotUncTTThirteenTevForDouRthirtTosevenpercentsttTotZFid}{   4.4}
\newcommand{\AveLumiTTThirteenTevForDouRthirtTosevenabsolutettTotZFid}{  18.9}
\newcommand{\AveCentralTTThirteenTevForDouRthirtTosevenabsvalttTotZFid}{ 817.9}
\newcommand{\AveStatTTThirteenTevForDouRthirtTosevenabsvalttTotZFid}{   7.5}
\newcommand{\AveSystTTThirteenTevForDouRthirtTosevenabsvalttTotZFid}{  35.3}
\newcommand{\AveTotUncTTThirteenTevForDouRthirtTosevenabsvalttTotZFid}{  36.1}
\newcommand{\AveChisqDoublRatioTTThirteenTevForDouRthirtTosevenTotZSevenTevForDouRthirtTosevenFidtoTTSevenTevForDouRthirtTosevenTotZThirteenTevForDouRthirtTosevenFid}{0.5}
\newcommand{\CorrCombZThirteenTevForDouRthirtTosevencorZThirteenTevForDouRthirtToseventtTotZFid}{ 1.000}
\newcommand{\CorrCombTTSevenTevForDouRthirtTosevencorZThirteenTevForDouRthirtToseventtTotZFid}{ 0.145}
\newcommand{\CorrCombZSevenTevForDouRthirtTosevencorZThirteenTevForDouRthirtToseventtTotZFid}{ 0.100}
\newcommand{\CorrCombTTThirteenTevForDouRthirtTosevencorZThirteenTevForDouRthirtToseventtTotZFid}{ 0.612}
\newcommand{\CorrCombTTSevenTevForDouRthirtTosevencorTTSevenTevForDouRthirtToseventtTotZFid}{ 1.000}
\newcommand{\CorrCombZSevenTevForDouRthirtTosevencorTTSevenTevForDouRthirtToseventtTotZFid}{ 0.619}
\newcommand{\CorrCombTTThirteenTevForDouRthirtTosevencorTTSevenTevForDouRthirtToseventtTotZFid}{ 0.312}
\newcommand{\CorrCombZSevenTevForDouRthirtTosevencorZSevenTevForDouRthirtToseventtTotZFid}{ 1.000}
\newcommand{\CorrCombTTThirteenTevForDouRthirtTosevencorZSevenTevForDouRthirtToseventtTotZFid}{ 0.106}
\newcommand{\CorrCombTTThirteenTevForDouRthirtTosevencorTTThirteenTevForDouRthirtToseventtTotZFid}{ 1.000}
\newcommand{\RatCombCentDoubRTTthirtOverZthirtTOttsevenOverZsevenabsvalttTotZFid}{   2.594}
\newcommand{\RatCombStatDoubRTTthirtOverZthirtTOttsevenOverZsevenabsvalttTotZFid}{   0.050}
\newcommand{\RatCombSystWithLumiUncDoubRTTthirtOverZthirtTOttsevenOverZsevenabsvalttTotZFid}{   0.086}
\newcommand{\RatCombTotUncDoubRTTthirtOverZthirtTOttsevenOverZsevenabsvalttTotZFid}{ 0.09941}
\newcommand{\RatCombTotUncDoubRTTthirtOverZthirtTOttsevenOverZsevenpercentsttTotZFid}{    3.83}
\newcommand{\RatCombStatDoubRTTthirtOverZthirtTOttsevenOverZsevenpercentsttTotZFid}{    1.93}
\newcommand{\RatCombSystWithLumiDoubRTTthirtOverZthirtTOttsevenOverZsevenpercentsttTotZFid}{    3.31}
\newcommand{\RatCombSystNoLumiUncDoubRTTthirtOverZthirtTOttsevenOverZsevenabsvalttTotZFid}{   0.086}
\newcommand{\RatCombSystNoLumiDoubRTTthirtOverZthirtTOttsevenOverZsevenpercentsttTotZFid}{    3.30}
\newcommand{\RatCombTotUncNoLumiDoubRTTthirtOverZthirtTOttsevenOverZsevenabsvalttTotZFid}{   0.099}
\newcommand{\RatCombTotUncNoLumiDoubRTTthirtOverZthirtTOttsevenOverZsevenpercentsttTotZFid}{    3.82}
\newcommand{\RatCombLumiUncDoubRTTthirtOverZthirtTOttsevenOverZsevenabsvalttTotZFid}{   0.008}
\newcommand{\RatCombLumiUncDoubRTTthirtOverZthirtTOttsevenOverZsevenpercentsttTotZFid}{    0.30}

\newcommand{\AveStatZThirteenTevForDouRthirtTosevenpercentsttFidZFid}{   0.1}
\newcommand{\AveSystZThirteenTevForDouRthirtTosevenpercentsttFidZFid}{   2.3}
\newcommand{\AveTotUncZThirteenTevForDouRthirtTosevenpercentsttFidZFid}{   2.3}
\newcommand{\AveLumiZThirteenTevForDouRthirtTosevenabsolutettFidZFid}{  16.3}
\newcommand{\AveCentralZThirteenTevForDouRthirtTosevenabsvalttFidZFid}{ 777.2}
\newcommand{\AveStatZThirteenTevForDouRthirtTosevenabsvalttFidZFid}{   0.5}
\newcommand{\AveSystZThirteenTevForDouRthirtTosevenabsvalttFidZFid}{  17.5}
\newcommand{\AveTotUncZThirteenTevForDouRthirtTosevenabsvalttFidZFid}{  17.5}
\newcommand{\AveStatTTSevenTevForDouRthirtTosevenpercentsttFidZFid}{   1.7}
\newcommand{\AveSystTTSevenTevForDouRthirtTosevenpercentsttFidZFid}{   3.4}
\newcommand{\AveTotUncTTSevenTevForDouRthirtTosevenpercentsttFidZFid}{   3.8}
\newcommand{\AveLumiTTSevenTevForDouRthirtTosevenabsolutettFidZFid}{   0.0}
\newcommand{\AveCentralTTSevenTevForDouRthirtTosevenabsvalttFidZFid}{   2.3}
\newcommand{\AveStatTTSevenTevForDouRthirtTosevenabsvalttFidZFid}{   0.0}
\newcommand{\AveSystTTSevenTevForDouRthirtTosevenabsvalttFidZFid}{   0.1}
\newcommand{\AveTotUncTTSevenTevForDouRthirtTosevenabsvalttFidZFid}{   0.1}
\newcommand{\AveStatZSevenTevForDouRthirtTosevenpercentsttFidZFid}{   0.1}
\newcommand{\AveSystZSevenTevForDouRthirtTosevenpercentsttFidZFid}{   1.9}
\newcommand{\AveTotUncZSevenTevForDouRthirtTosevenpercentsttFidZFid}{   1.9}
\newcommand{\AveLumiZSevenTevForDouRthirtTosevenabsolutettFidZFid}{   8.1}
\newcommand{\AveCentralZSevenTevForDouRthirtTosevenabsvalttFidZFid}{ 450.8}
\newcommand{\AveStatZSevenTevForDouRthirtTosevenabsvalttFidZFid}{   0.3}
\newcommand{\AveSystZSevenTevForDouRthirtTosevenabsvalttFidZFid}{   8.7}
\newcommand{\AveTotUncZSevenTevForDouRthirtTosevenabsvalttFidZFid}{   8.7}
\newcommand{\AveStatTTThirteenTevForDouRthirtTosevenpercentsttFidZFid}{   0.9}
\newcommand{\AveSystTTThirteenTevForDouRthirtTosevenpercentsttFidZFid}{   3.7}
\newcommand{\AveTotUncTTThirteenTevForDouRthirtTosevenpercentsttFidZFid}{   3.8}
\newcommand{\AveLumiTTThirteenTevForDouRthirtTosevenabsolutettFidZFid}{   0.2}
\newcommand{\AveCentralTTThirteenTevForDouRthirtTosevenabsvalttFidZFid}{   9.9}
\newcommand{\AveStatTTThirteenTevForDouRthirtTosevenabsvalttFidZFid}{   0.1}
\newcommand{\AveSystTTThirteenTevForDouRthirtTosevenabsvalttFidZFid}{   0.4}
\newcommand{\AveTotUncTTThirteenTevForDouRthirtTosevenabsvalttFidZFid}{   0.4}
\newcommand{\AveChisqDoublRatioTTThirteenTevForDouRthirtTosevenFidZSevenTevForDouRthirtTosevenFidtoTTSevenTevForDouRthirtTosevenFidZThirteenTevForDouRthirtTosevenFid}{0.5}
\newcommand{\CorrCombZThirteenTevForDouRthirtTosevencorZThirteenTevForDouRthirtToseventtFidZFid}{ 1.000}
\newcommand{\CorrCombTTSevenTevForDouRthirtTosevencorZThirteenTevForDouRthirtToseventtFidZFid}{ 0.148}
\newcommand{\CorrCombZSevenTevForDouRthirtTosevencorZThirteenTevForDouRthirtToseventtFidZFid}{ 0.100}
\newcommand{\CorrCombTTThirteenTevForDouRthirtTosevencorZThirteenTevForDouRthirtToseventtFidZFid}{ 0.702}
\newcommand{\CorrCombTTSevenTevForDouRthirtTosevencorTTSevenTevForDouRthirtToseventtFidZFid}{ 1.000}
\newcommand{\CorrCombZSevenTevForDouRthirtTosevencorTTSevenTevForDouRthirtToseventtFidZFid}{ 0.634}
\newcommand{\CorrCombTTThirteenTevForDouRthirtTosevencorTTSevenTevForDouRthirtToseventtFidZFid}{ 0.317}
\newcommand{\CorrCombZSevenTevForDouRthirtTosevencorZSevenTevForDouRthirtToseventtFidZFid}{ 1.000}
\newcommand{\CorrCombTTThirteenTevForDouRthirtTosevencorZSevenTevForDouRthirtToseventtFidZFid}{ 0.122}
\newcommand{\CorrCombTTThirteenTevForDouRthirtTosevencorTTThirteenTevForDouRthirtToseventtFidZFid}{ 1.000}
\newcommand{\RatCombCentDoubRTTthirtOverZthirtTOttsevenOverZsevenabsvalttFidZFid}{   2.508}
\newcommand{\RatCombStatDoubRTTthirtOverZthirtTOttsevenOverZsevenabsvalttFidZFid}{   0.048}
\newcommand{\RatCombSystWithLumiUncDoubRTTthirtOverZthirtTOttsevenOverZsevenabsvalttFidZFid}{   0.067}
\newcommand{\RatCombTotUncDoubRTTthirtOverZthirtTOttsevenOverZsevenabsvalttFidZFid}{ 0.08254}
\newcommand{\RatCombTotUncDoubRTTthirtOverZthirtTOttsevenOverZsevenpercentsttFidZFid}{    3.29}
\newcommand{\RatCombStatDoubRTTthirtOverZthirtTOttsevenOverZsevenpercentsttFidZFid}{    1.91}
\newcommand{\RatCombSystWithLumiDoubRTTthirtOverZthirtTOttsevenOverZsevenpercentsttFidZFid}{    2.68}
\newcommand{\RatCombSystNoLumiUncDoubRTTthirtOverZthirtTOttsevenOverZsevenabsvalttFidZFid}{   0.067}
\newcommand{\RatCombSystNoLumiDoubRTTthirtOverZthirtTOttsevenOverZsevenpercentsttFidZFid}{    2.66}
\newcommand{\RatCombTotUncNoLumiDoubRTTthirtOverZthirtTOttsevenOverZsevenabsvalttFidZFid}{   0.082}
\newcommand{\RatCombTotUncNoLumiDoubRTTthirtOverZthirtTOttsevenOverZsevenpercentsttFidZFid}{    3.27}
\newcommand{\RatCombLumiUncDoubRTTthirtOverZthirtTOttsevenOverZsevenabsvalttFidZFid}{   0.008}
\newcommand{\RatCombLumiUncDoubRTTthirtOverZthirtTOttsevenOverZsevenpercentsttFidZFid}{    0.32}

\newcommand{\AveStatZEightTevForDouReightTosevenpercentsttTotZTot}{   0.0}
\newcommand{\AveSystZEightTevForDouReightTosevenpercentsttTotZTot}{   2.1}
\newcommand{\AveTotUncZEightTevForDouReightTosevenpercentsttTotZTot}{   2.1}
\newcommand{\AveLumiZEightTevForDouReightTosevenabsolutettTotZTot}{  21.9}
\newcommand{\AveCentralZEightTevForDouReightTosevenabsvalttTotZTot}{1153.5}
\newcommand{\AveStatZEightTevForDouReightTosevenabsvalttTotZTot}{   0.3}
\newcommand{\AveSystZEightTevForDouReightTosevenabsvalttTotZTot}{  23.9}
\newcommand{\AveTotUncZEightTevForDouReightTosevenabsvalttTotZTot}{  23.9}
\newcommand{\AveStatTTSevenTevForDouReightTosevenpercentsttTotZTot}{   1.7}
\newcommand{\AveSystTTSevenTevForDouReightTosevenpercentsttTotZTot}{   3.5}
\newcommand{\AveTotUncTTSevenTevForDouReightTosevenpercentsttTotZTot}{   3.9}
\newcommand{\AveLumiTTSevenTevForDouReightTosevenabsolutettTotZTot}{   3.6}
\newcommand{\AveCentralTTSevenTevForDouReightTosevenabsvalttTotZTot}{ 182.9}
\newcommand{\AveStatTTSevenTevForDouReightTosevenabsvalttTotZTot}{   3.1}
\newcommand{\AveSystTTSevenTevForDouReightTosevenabsvalttTotZTot}{   6.4}
\newcommand{\AveTotUncTTSevenTevForDouReightTosevenabsvalttTotZTot}{   7.1}
\newcommand{\AveStatZSevenTevForDouReightTosevenpercentsttTotZTot}{   0.1}
\newcommand{\AveSystZSevenTevForDouReightTosevenpercentsttTotZTot}{   1.9}
\newcommand{\AveTotUncZSevenTevForDouReightTosevenpercentsttTotZTot}{   1.9}
\newcommand{\AveLumiZSevenTevForDouReightTosevenabsolutettTotZTot}{  17.9}
\newcommand{\AveCentralZSevenTevForDouReightTosevenabsvalttTotZTot}{ 994.7}
\newcommand{\AveStatZSevenTevForDouReightTosevenabsvalttTotZTot}{   0.6}
\newcommand{\AveSystZSevenTevForDouReightTosevenabsvalttTotZTot}{  19.1}
\newcommand{\AveTotUncZSevenTevForDouReightTosevenabsvalttTotZTot}{  19.2}
\newcommand{\AveStatTTEightTevForDouReightTosevenpercentsttTotZTot}{   0.7}
\newcommand{\AveSystTTEightTevForDouReightTosevenpercentsttTotZTot}{   3.5}
\newcommand{\AveTotUncTTEightTevForDouReightTosevenpercentsttTotZTot}{   3.6}
\newcommand{\AveLumiTTEightTevForDouReightTosevenabsolutettTotZTot}{   5.1}
\newcommand{\AveCentralTTEightTevForDouReightTosevenabsvalttTotZTot}{ 242.9}
\newcommand{\AveStatTTEightTevForDouReightTosevenabsvalttTotZTot}{   1.7}
\newcommand{\AveSystTTEightTevForDouReightTosevenabsvalttTotZTot}{   8.6}
\newcommand{\AveTotUncTTEightTevForDouReightTosevenabsvalttTotZTot}{   8.8}
\newcommand{\AveChisqDoublRatioTTEightTevForDouReightTosevenTotZSevenTevForDouReightTosevenTottoTTSevenTevForDouReightTosevenTotZEightTevForDouReightTosevenTot}{0.4}
\newcommand{\CorrCombZEightTevForDouReightTosevencorZEightTevForDouReightToseventtTotZTot}{ 1.000}
\newcommand{\CorrCombTTSevenTevForDouReightTosevencorZEightTevForDouReightToseventtTotZTot}{ 0.138}
\newcommand{\CorrCombZSevenTevForDouReightTosevencorZEightTevForDouReightToseventtTotZTot}{ 0.097}
\newcommand{\CorrCombTTEightTevForDouReightTosevencorZEightTevForDouReightToseventtTotZTot}{ 0.678}
\newcommand{\CorrCombTTSevenTevForDouReightTosevencorTTSevenTevForDouReightToseventtTotZTot}{ 1.000}
\newcommand{\CorrCombZSevenTevForDouReightTosevencorTTSevenTevForDouReightToseventtTotZTot}{ 0.619}
\newcommand{\CorrCombTTEightTevForDouReightTosevencorTTSevenTevForDouReightToseventtTotZTot}{ 0.542}
\newcommand{\CorrCombZSevenTevForDouReightTosevencorZSevenTevForDouReightToseventtTotZTot}{ 1.000}
\newcommand{\CorrCombTTEightTevForDouReightTosevencorZSevenTevForDouReightToseventtTotZTot}{ 0.149}
\newcommand{\CorrCombTTEightTevForDouReightTosevencorTTEightTevForDouReightToseventtTotZTot}{ 1.000}
\newcommand{\RatCombCentDoubRTTeightOverZeightTOttsevenOverZsevenabsvalttTotZTot}{   1.145}
\newcommand{\RatCombStatDoubRTTeightOverZeightTOttsevenOverZsevenabsvalttTotZTot}{   0.021}
\newcommand{\RatCombSystWithLumiUncDoubRTTeightOverZeightTOttsevenOverZsevenabsvalttTotZTot}{   0.015}
\newcommand{\RatCombTotUncDoubRTTeightOverZeightTOttsevenOverZsevenabsvalttTotZTot}{ 0.02580}
\newcommand{\RatCombTotUncDoubRTTeightOverZeightTOttsevenOverZsevenpercentsttTotZTot}{    2.25}
\newcommand{\RatCombStatDoubRTTeightOverZeightTOttsevenOverZsevenpercentsttTotZTot}{    1.83}
\newcommand{\RatCombSystWithLumiDoubRTTeightOverZeightTOttsevenOverZsevenpercentsttTotZTot}{    1.31}
\newcommand{\RatCombSystNoLumiUncDoubRTTeightOverZeightTOttsevenOverZsevenabsvalttTotZTot}{   0.015}
\newcommand{\RatCombSystNoLumiDoubRTTeightOverZeightTOttsevenOverZsevenpercentsttTotZTot}{    1.28}
\newcommand{\RatCombTotUncNoLumiDoubRTTeightOverZeightTOttsevenOverZsevenabsvalttTotZTot}{   0.026}
\newcommand{\RatCombTotUncNoLumiDoubRTTeightOverZeightTOttsevenOverZsevenpercentsttTotZTot}{    2.24}
\newcommand{\RatCombLumiUncDoubRTTeightOverZeightTOttsevenOverZsevenabsvalttTotZTot}{   0.003}
\newcommand{\RatCombLumiUncDoubRTTeightOverZeightTOttsevenOverZsevenpercentsttTotZTot}{    0.28}

\newcommand{\AveStatZEightTevForDouReightTosevenpercentsttTotZFid}{   0.0}
\newcommand{\AveSystZEightTevForDouReightTosevenpercentsttTotZFid}{   2.1}
\newcommand{\AveTotUncZEightTevForDouReightTosevenpercentsttTotZFid}{   2.1}
\newcommand{\AveLumiZEightTevForDouReightTosevenabsolutettTotZFid}{   9.6}
\newcommand{\AveCentralZEightTevForDouReightTosevenabsvalttTotZFid}{ 505.7}
\newcommand{\AveStatZEightTevForDouReightTosevenabsvalttTotZFid}{   0.1}
\newcommand{\AveSystZEightTevForDouReightTosevenabsvalttTotZFid}{  10.5}
\newcommand{\AveTotUncZEightTevForDouReightTosevenabsvalttTotZFid}{  10.5}
\newcommand{\AveStatTTSevenTevForDouReightTosevenpercentsttTotZFid}{   1.7}
\newcommand{\AveSystTTSevenTevForDouReightTosevenpercentsttTotZFid}{   3.5}
\newcommand{\AveTotUncTTSevenTevForDouReightTosevenpercentsttTotZFid}{   3.9}
\newcommand{\AveLumiTTSevenTevForDouReightTosevenabsolutettTotZFid}{   3.6}
\newcommand{\AveCentralTTSevenTevForDouReightTosevenabsvalttTotZFid}{ 182.9}
\newcommand{\AveStatTTSevenTevForDouReightTosevenabsvalttTotZFid}{   3.1}
\newcommand{\AveSystTTSevenTevForDouReightTosevenabsvalttTotZFid}{   6.4}
\newcommand{\AveTotUncTTSevenTevForDouReightTosevenabsvalttTotZFid}{   7.1}
\newcommand{\AveStatZSevenTevForDouReightTosevenpercentsttTotZFid}{   0.1}
\newcommand{\AveSystZSevenTevForDouReightTosevenpercentsttTotZFid}{   1.9}
\newcommand{\AveTotUncZSevenTevForDouReightTosevenpercentsttTotZFid}{   1.9}
\newcommand{\AveLumiZSevenTevForDouReightTosevenabsolutettTotZFid}{   8.1}
\newcommand{\AveCentralZSevenTevForDouReightTosevenabsvalttTotZFid}{ 450.7}
\newcommand{\AveStatZSevenTevForDouReightTosevenabsvalttTotZFid}{   0.3}
\newcommand{\AveSystZSevenTevForDouReightTosevenabsvalttTotZFid}{   8.7}
\newcommand{\AveTotUncZSevenTevForDouReightTosevenabsvalttTotZFid}{   8.7}
\newcommand{\AveStatTTEightTevForDouReightTosevenpercentsttTotZFid}{   0.7}
\newcommand{\AveSystTTEightTevForDouReightTosevenpercentsttTotZFid}{   3.5}
\newcommand{\AveTotUncTTEightTevForDouReightTosevenpercentsttTotZFid}{   3.6}
\newcommand{\AveLumiTTEightTevForDouReightTosevenabsolutettTotZFid}{   5.1}
\newcommand{\AveCentralTTEightTevForDouReightTosevenabsvalttTotZFid}{ 242.9}
\newcommand{\AveStatTTEightTevForDouReightTosevenabsvalttTotZFid}{   1.7}
\newcommand{\AveSystTTEightTevForDouReightTosevenabsvalttTotZFid}{   8.6}
\newcommand{\AveTotUncTTEightTevForDouReightTosevenabsvalttTotZFid}{   8.7}
\newcommand{\AveChisqDoublRatioTTEightTevForDouReightTosevenTotZSevenTevForDouReightTosevenFidtoTTSevenTevForDouReightTosevenTotZEightTevForDouReightTosevenFid}{0.4}
\newcommand{\CorrCombZEightTevForDouReightTosevencorZEightTevForDouReightToseventtTotZFid}{ 1.000}
\newcommand{\CorrCombTTSevenTevForDouReightTosevencorZEightTevForDouReightToseventtTotZFid}{ 0.138}
\newcommand{\CorrCombZSevenTevForDouReightTosevencorZEightTevForDouReightToseventtTotZFid}{ 0.097}
\newcommand{\CorrCombTTEightTevForDouReightTosevencorZEightTevForDouReightToseventtTotZFid}{ 0.678}
\newcommand{\CorrCombTTSevenTevForDouReightTosevencorTTSevenTevForDouReightToseventtTotZFid}{ 1.000}
\newcommand{\CorrCombZSevenTevForDouReightTosevencorTTSevenTevForDouReightToseventtTotZFid}{ 0.619}
\newcommand{\CorrCombTTEightTevForDouReightTosevencorTTSevenTevForDouReightToseventtTotZFid}{ 0.542}
\newcommand{\CorrCombZSevenTevForDouReightTosevencorZSevenTevForDouReightToseventtTotZFid}{ 1.000}
\newcommand{\CorrCombTTEightTevForDouReightTosevencorZSevenTevForDouReightToseventtTotZFid}{ 0.149}
\newcommand{\CorrCombTTEightTevForDouReightTosevencorTTEightTevForDouReightToseventtTotZFid}{ 1.000}
\newcommand{\RatCombCentDoubRTTeightOverZeightTOttsevenOverZsevenabsvalttTotZFid}{   1.184}
\newcommand{\RatCombStatDoubRTTeightOverZeightTOttsevenOverZsevenabsvalttTotZFid}{   0.022}
\newcommand{\RatCombSystWithLumiUncDoubRTTeightOverZeightTOttsevenOverZsevenabsvalttTotZFid}{   0.015}
\newcommand{\RatCombTotUncDoubRTTeightOverZeightTOttsevenOverZsevenabsvalttTotZFid}{ 0.02657}
\newcommand{\RatCombTotUncDoubRTTeightOverZeightTOttsevenOverZsevenpercentsttTotZFid}{    2.24}
\newcommand{\RatCombStatDoubRTTeightOverZeightTOttsevenOverZsevenpercentsttTotZFid}{    1.83}
\newcommand{\RatCombSystWithLumiDoubRTTeightOverZeightTOttsevenOverZsevenpercentsttTotZFid}{    1.29}
\newcommand{\RatCombSystNoLumiUncDoubRTTeightOverZeightTOttsevenOverZsevenabsvalttTotZFid}{   0.015}
\newcommand{\RatCombSystNoLumiDoubRTTeightOverZeightTOttsevenOverZsevenpercentsttTotZFid}{    1.27}
\newcommand{\RatCombTotUncNoLumiDoubRTTeightOverZeightTOttsevenOverZsevenabsvalttTotZFid}{   0.026}
\newcommand{\RatCombTotUncNoLumiDoubRTTeightOverZeightTOttsevenOverZsevenpercentsttTotZFid}{    2.23}
\newcommand{\RatCombLumiUncDoubRTTeightOverZeightTOttsevenOverZsevenabsvalttTotZFid}{   0.003}
\newcommand{\RatCombLumiUncDoubRTTeightOverZeightTOttsevenOverZsevenpercentsttTotZFid}{    0.27}

\newcommand{\AveStatZEightTevForDouReightTosevenpercentsttFidZFid}{   0.0}
\newcommand{\AveSystZEightTevForDouReightTosevenpercentsttFidZFid}{   2.1}
\newcommand{\AveTotUncZEightTevForDouReightTosevenpercentsttFidZFid}{   2.1}
\newcommand{\AveLumiZEightTevForDouReightTosevenabsolutettFidZFid}{   9.6}
\newcommand{\AveCentralZEightTevForDouReightTosevenabsvalttFidZFid}{ 505.7}
\newcommand{\AveStatZEightTevForDouReightTosevenabsvalttFidZFid}{   0.1}
\newcommand{\AveSystZEightTevForDouReightTosevenabsvalttFidZFid}{  10.5}
\newcommand{\AveTotUncZEightTevForDouReightTosevenabsvalttFidZFid}{  10.5}
\newcommand{\AveStatTTSevenTevForDouReightTosevenpercentsttFidZFid}{   1.7}
\newcommand{\AveSystTTSevenTevForDouReightTosevenpercentsttFidZFid}{   3.4}
\newcommand{\AveTotUncTTSevenTevForDouReightTosevenpercentsttFidZFid}{   3.8}
\newcommand{\AveLumiTTSevenTevForDouReightTosevenabsolutettFidZFid}{   0.0}
\newcommand{\AveCentralTTSevenTevForDouReightTosevenabsvalttFidZFid}{   2.3}
\newcommand{\AveStatTTSevenTevForDouReightTosevenabsvalttFidZFid}{   0.0}
\newcommand{\AveSystTTSevenTevForDouReightTosevenabsvalttFidZFid}{   0.1}
\newcommand{\AveTotUncTTSevenTevForDouReightTosevenabsvalttFidZFid}{   0.1}
\newcommand{\AveStatZSevenTevForDouReightTosevenpercentsttFidZFid}{   0.1}
\newcommand{\AveSystZSevenTevForDouReightTosevenpercentsttFidZFid}{   1.9}
\newcommand{\AveTotUncZSevenTevForDouReightTosevenpercentsttFidZFid}{   1.9}
\newcommand{\AveLumiZSevenTevForDouReightTosevenabsolutettFidZFid}{   8.1}
\newcommand{\AveCentralZSevenTevForDouReightTosevenabsvalttFidZFid}{ 450.7}
\newcommand{\AveStatZSevenTevForDouReightTosevenabsvalttFidZFid}{   0.3}
\newcommand{\AveSystZSevenTevForDouReightTosevenabsvalttFidZFid}{   8.7}
\newcommand{\AveTotUncZSevenTevForDouReightTosevenabsvalttFidZFid}{   8.7}
\newcommand{\AveStatTTEightTevForDouReightTosevenpercentsttFidZFid}{   0.7}
\newcommand{\AveSystTTEightTevForDouReightTosevenpercentsttFidZFid}{   3.4}
\newcommand{\AveTotUncTTEightTevForDouReightTosevenpercentsttFidZFid}{   3.5}
\newcommand{\AveLumiTTEightTevForDouReightTosevenabsolutettFidZFid}{   0.1}
\newcommand{\AveCentralTTEightTevForDouReightTosevenabsvalttFidZFid}{   3.0}
\newcommand{\AveStatTTEightTevForDouReightTosevenabsvalttFidZFid}{   0.0}
\newcommand{\AveSystTTEightTevForDouReightTosevenabsvalttFidZFid}{   0.1}
\newcommand{\AveTotUncTTEightTevForDouReightTosevenabsvalttFidZFid}{   0.1}
\newcommand{\AveChisqDoublRatioTTEightTevForDouReightTosevenFidZSevenTevForDouReightTosevenFidtoTTSevenTevForDouReightTosevenFidZEightTevForDouReightTosevenFid}{0.4}
\newcommand{\CorrCombZEightTevForDouReightTosevencorZEightTevForDouReightToseventtFidZFid}{ 1.000}
\newcommand{\CorrCombTTSevenTevForDouReightTosevencorZEightTevForDouReightToseventtFidZFid}{ 0.141}
\newcommand{\CorrCombZSevenTevForDouReightTosevencorZEightTevForDouReightToseventtFidZFid}{ 0.098}
\newcommand{\CorrCombTTEightTevForDouReightTosevencorZEightTevForDouReightToseventtFidZFid}{ 0.707}
\newcommand{\CorrCombTTSevenTevForDouReightTosevencorTTSevenTevForDouReightToseventtFidZFid}{ 1.000}
\newcommand{\CorrCombZSevenTevForDouReightTosevencorTTSevenTevForDouReightToseventtFidZFid}{ 0.634}
\newcommand{\CorrCombTTEightTevForDouReightTosevencorTTSevenTevForDouReightToseventtFidZFid}{ 0.513}
\newcommand{\CorrCombZSevenTevForDouReightTosevencorZSevenTevForDouReightToseventtFidZFid}{ 1.000}
\newcommand{\CorrCombTTEightTevForDouReightTosevencorZSevenTevForDouReightToseventtFidZFid}{ 0.155}
\newcommand{\CorrCombTTEightTevForDouReightTosevencorTTEightTevForDouReightToseventtFidZFid}{ 1.000}
\newcommand{\RatCombCentDoubRTTeightOverZeightTOttsevenOverZsevenabsvalttFidZFid}{   1.178}
\newcommand{\RatCombStatDoubRTTeightOverZeightTOttsevenOverZsevenabsvalttFidZFid}{   0.022}
\newcommand{\RatCombSystWithLumiUncDoubRTTeightOverZeightTOttsevenOverZsevenabsvalttFidZFid}{   0.015}
\newcommand{\RatCombTotUncDoubRTTeightOverZeightTOttsevenOverZsevenabsvalttFidZFid}{ 0.02649}
\newcommand{\RatCombTotUncDoubRTTeightOverZeightTOttsevenOverZsevenpercentsttFidZFid}{    2.25}
\newcommand{\RatCombStatDoubRTTeightOverZeightTOttsevenOverZsevenpercentsttFidZFid}{    1.83}
\newcommand{\RatCombSystWithLumiDoubRTTeightOverZeightTOttsevenOverZsevenpercentsttFidZFid}{    1.31}
\newcommand{\RatCombSystNoLumiUncDoubRTTeightOverZeightTOttsevenOverZsevenabsvalttFidZFid}{   0.015}
\newcommand{\RatCombSystNoLumiDoubRTTeightOverZeightTOttsevenOverZsevenpercentsttFidZFid}{    1.28}
\newcommand{\RatCombTotUncNoLumiDoubRTTeightOverZeightTOttsevenOverZsevenabsvalttFidZFid}{   0.026}
\newcommand{\RatCombTotUncNoLumiDoubRTTeightOverZeightTOttsevenOverZsevenpercentsttFidZFid}{    2.23}
\newcommand{\RatCombLumiUncDoubRTTeightOverZeightTOttsevenOverZsevenabsvalttFidZFid}{   0.003}
\newcommand{\RatCombLumiUncDoubRTTeightOverZeightTOttsevenOverZsevenpercentsttFidZFid}{    0.29}

\newcommand{\AveStatZThirteenTevpercentsttFidZFidGlobalCorr}{  0.06}
\newcommand{\AveSystZThirteenTevpercentsttFidZFidGlobalCorr}{  2.25 }
\newcommand{\AveTotUncZThirteenTevpercentsttFidZFidGlobalCorr}{  2.25}
\newcommand{\AveLumiZThirteenTevabsolutettFidZFidGlobalCorr}{ 16.32}
\newcommand{\AveCentralZThirteenTevabsvalttFidZFidGlobalCorr}{777.29}
\newcommand{\AveStatZThirteenTevabsvalttFidZFidGlobalCorr}{  0.46}
\newcommand{\AveSystZThirteenTevabsvalttFidZFidGlobalCorr}{ 17.51 }
\newcommand{\AveTotUncZThirteenTevabsvalttFidZFidGlobalCorr}{ 17.52}
\newcommand{\AveStatTTThirteenTevpercentsttFidZFidGlobalCorr}{  0.91}
\newcommand{\AveSystTTThirteenTevpercentsttFidZFidGlobalCorr}{  3.74 }
\newcommand{\AveTotUncTTThirteenTevpercentsttFidZFidGlobalCorr}{  3.84}
\newcommand{\AveLumiTTThirteenTevabsolutettFidZFidGlobalCorr}{  0.23}
\newcommand{\AveCentralTTThirteenTevabsvalttFidZFidGlobalCorr}{  9.95}
\newcommand{\AveStatTTThirteenTevabsvalttFidZFidGlobalCorr}{  0.09}
\newcommand{\AveSystTTThirteenTevabsvalttFidZFidGlobalCorr}{  0.37 }
\newcommand{\AveTotUncTTThirteenTevabsvalttFidZFidGlobalCorr}{  0.38}
\newcommand{\AveStatZEightTevpercentsttFidZFidGlobalCorr}{  0.02}
\newcommand{\AveSystZEightTevpercentsttFidZFidGlobalCorr}{  2.07 }
\newcommand{\AveTotUncZEightTevpercentsttFidZFidGlobalCorr}{  2.07}
\newcommand{\AveLumiZEightTevabsolutettFidZFidGlobalCorr}{  9.61}
\newcommand{\AveCentralZEightTevabsvalttFidZFidGlobalCorr}{505.77}
\newcommand{\AveStatZEightTevabsvalttFidZFidGlobalCorr}{  0.12}
\newcommand{\AveSystZEightTevabsvalttFidZFidGlobalCorr}{ 10.48 }
\newcommand{\AveTotUncZEightTevabsvalttFidZFidGlobalCorr}{ 10.48}
\newcommand{\AveStatTTEightTevpercentsttFidZFidGlobalCorr}{  0.72}
\newcommand{\AveSystTTEightTevpercentsttFidZFidGlobalCorr}{  3.38 }
\newcommand{\AveTotUncTTEightTevpercentsttFidZFidGlobalCorr}{  3.45}
\newcommand{\AveLumiTTEightTevabsolutettFidZFidGlobalCorr}{  0.06}
\newcommand{\AveCentralTTEightTevabsvalttFidZFidGlobalCorr}{  3.04}
\newcommand{\AveStatTTEightTevabsvalttFidZFidGlobalCorr}{  0.02}
\newcommand{\AveSystTTEightTevabsvalttFidZFidGlobalCorr}{  0.10 }
\newcommand{\AveTotUncTTEightTevabsvalttFidZFidGlobalCorr}{  0.10}
\newcommand{\AveStatZSevenTevpercentsttFidZFidGlobalCorr}{  0.06}
\newcommand{\AveSystZSevenTevpercentsttFidZFidGlobalCorr}{  1.92 }
\newcommand{\AveTotUncZSevenTevpercentsttFidZFidGlobalCorr}{  1.93}
\newcommand{\AveLumiZSevenTevabsolutettFidZFidGlobalCorr}{  8.11}
\newcommand{\AveCentralZSevenTevabsvalttFidZFidGlobalCorr}{450.74}
\newcommand{\AveStatZSevenTevabsvalttFidZFidGlobalCorr}{  0.28}
\newcommand{\AveSystZSevenTevabsvalttFidZFidGlobalCorr}{  8.67 }
\newcommand{\AveTotUncZSevenTevabsvalttFidZFidGlobalCorr}{  8.68}
\newcommand{\AveStatTTSevenTevpercentsttFidZFidGlobalCorr}{  1.68}
\newcommand{\AveSystTTSevenTevpercentsttFidZFidGlobalCorr}{  3.41 }
\newcommand{\AveTotUncTTSevenTevpercentsttFidZFidGlobalCorr}{  3.80}
\newcommand{\AveLumiTTSevenTevabsolutettFidZFidGlobalCorr}{  0.05}
\newcommand{\AveCentralTTSevenTevabsvalttFidZFidGlobalCorr}{  2.30}
\newcommand{\AveStatTTSevenTevabsvalttFidZFidGlobalCorr}{  0.04}
\newcommand{\AveSystTTSevenTevabsvalttFidZFidGlobalCorr}{  0.08 }
\newcommand{\AveTotUncTTSevenTevabsvalttFidZFidGlobalCorr}{  0.09}
\newcommand{\AveChisqAllChanEnergZThirteenTevTTThirteenTevZEightTevTTEightTevZSevenTevTTSevenTevFidFidGlobalCorr}{0.6}
\newcommand{\CorrCombZThirteenTevcorZThirteenTevttFidZFidGlobalCorr}{ 1.000}
\newcommand{\CorrCombTTThirteenTevcorZThirteenTevttFidZFidGlobalCorr}{ 0.702}
\newcommand{\CorrCombZEightTevcorZThirteenTevttFidZFidGlobalCorr}{ 0.097}
\newcommand{\CorrCombTTEightTevcorZThirteenTevttFidZFidGlobalCorr}{ 0.163}
\newcommand{\CorrCombZSevenTevcorZThirteenTevttFidZFidGlobalCorr}{ 0.100}
\newcommand{\CorrCombTTSevenTevcorZThirteenTevttFidZFidGlobalCorr}{ 0.148}
\newcommand{\CorrCombTTThirteenTevcorTTThirteenTevttFidZFidGlobalCorr}{ 1.000}
\newcommand{\CorrCombZEightTevcorTTThirteenTevttFidZFidGlobalCorr}{ 0.120}
\newcommand{\CorrCombTTEightTevcorTTThirteenTevttFidZFidGlobalCorr}{ 0.331}
\newcommand{\CorrCombZSevenTevcorTTThirteenTevttFidZFidGlobalCorr}{ 0.122}
\newcommand{\CorrCombTTSevenTevcorTTThirteenTevttFidZFidGlobalCorr}{ 0.317}
\newcommand{\CorrCombZEightTevcorZEightTevttFidZFidGlobalCorr}{ 1.000}
\newcommand{\CorrCombTTEightTevcorZEightTevttFidZFidGlobalCorr}{ 0.709}
\newcommand{\CorrCombZSevenTevcorZEightTevttFidZFidGlobalCorr}{ 0.098}
\newcommand{\CorrCombTTSevenTevcorZEightTevttFidZFidGlobalCorr}{ 0.142}
\newcommand{\CorrCombTTEightTevcorTTEightTevttFidZFidGlobalCorr}{ 1.000}
\newcommand{\CorrCombZSevenTevcorTTEightTevttFidZFidGlobalCorr}{ 0.156}
\newcommand{\CorrCombTTSevenTevcorTTEightTevttFidZFidGlobalCorr}{ 0.513}
\newcommand{\CorrCombZSevenTevcorZSevenTevttFidZFidGlobalCorr}{ 1.000}
\newcommand{\CorrCombTTSevenTevcorZSevenTevttFidZFidGlobalCorr}{ 0.634}
\newcommand{\CorrCombTTSevenTevcorTTSevenTevttFidZFidGlobalCorr}{ 1.000}
\newcommand{\AveSystZThirteenTevpercentsttFidZFidGlobalCorrNoLumi}{    0.82}
\newcommand{\AveSystZThirteenTevabsvalttFidZFidGlobalCorrNoLumi}{    6.35}
\newcommand{\AveSystTTThirteenTevpercentsttFidZFidGlobalCorrNoLumi}{    2.93}
\newcommand{\AveSystTTThirteenTevabsvalttFidZFidGlobalCorrNoLumi}{    0.29}
\newcommand{\AveSystZEightTevpercentsttFidZFidGlobalCorrNoLumi}{    0.83}
\newcommand{\AveSystZEightTevabsvalttFidZFidGlobalCorrNoLumi}{    4.18}
\newcommand{\AveSystTTEightTevpercentsttFidZFidGlobalCorrNoLumi}{    2.64}
\newcommand{\AveSystTTEightTevabsvalttFidZFidGlobalCorrNoLumi}{    0.08}
\newcommand{\AveSystZSevenTevpercentsttFidZFidGlobalCorrNoLumi}{    0.68}
\newcommand{\AveSystZSevenTevabsvalttFidZFidGlobalCorrNoLumi}{    3.07}
\newcommand{\AveSystTTSevenTevpercentsttFidZFidGlobalCorrNoLumi}{    2.78}
\newcommand{\AveSystTTSevenTevabsvalttFidZFidGlobalCorrNoLumi}{    0.06}
\newcommand{\AveSystZThirteenTevpercentsttFidZFidGlobalCorrLumiOnly}{    2.10}
\newcommand{\AveSystZThirteenTevabsvalttFidZFidGlobalCorrLumiOnly}{   16.32}
\newcommand{\AveSystTTThirteenTevpercentsttFidZFidGlobalCorrLumiOnly}{    2.32}
\newcommand{\AveSystTTThirteenTevabsvalttFidZFidGlobalCorrLumiOnly}{    0.23}
\newcommand{\AveSystZEightTevpercentsttFidZFidGlobalCorrLumiOnly}{    1.90}
\newcommand{\AveSystZEightTevabsvalttFidZFidGlobalCorrLumiOnly}{    9.61}
\newcommand{\AveSystTTEightTevpercentsttFidZFidGlobalCorrLumiOnly}{    2.11}
\newcommand{\AveSystTTEightTevabsvalttFidZFidGlobalCorrLumiOnly}{    0.06}
\newcommand{\AveSystZSevenTevpercentsttFidZFidGlobalCorrLumiOnly}{    1.80}
\newcommand{\AveSystZSevenTevabsvalttFidZFidGlobalCorrLumiOnly}{    8.11}
\newcommand{\AveSystTTSevenTevpercentsttFidZFidGlobalCorrLumiOnly}{    1.97}
\newcommand{\AveSystTTSevenTevabsvalttFidZFidGlobalCorrLumiOnly}{    0.05}
\newcommand{\AveSystZThirteenTevpercentsttFidZFidGlobalCorrNoLumiNoBeam}{    0.44}
\newcommand{\AveSystZThirteenTevabsvalttFidZFidGlobalCorrNoLumiNoBeam}{    3.39}
\newcommand{\AveSystTTThirteenTevpercentsttFidZFidGlobalCorrNoLumiNoBeam}{    2.52}
\newcommand{\AveSystTTThirteenTevabsvalttFidZFidGlobalCorrNoLumiNoBeam}{    0.25}
\newcommand{\AveSystZEightTevpercentsttFidZFidGlobalCorrNoLumiNoBeam}{    0.55}
\newcommand{\AveSystZEightTevabsvalttFidZFidGlobalCorrNoLumiNoBeam}{    2.77}
\newcommand{\AveSystTTEightTevpercentsttFidZFidGlobalCorrNoLumiNoBeam}{    2.01}
\newcommand{\AveSystTTEightTevabsvalttFidZFidGlobalCorrNoLumiNoBeam}{    0.06}
\newcommand{\AveSystZSevenTevpercentsttFidZFidGlobalCorrNoLumiNoBeam}{    0.32}
\newcommand{\AveSystZSevenTevabsvalttFidZFidGlobalCorrNoLumiNoBeam}{    1.44}
\newcommand{\AveSystTTSevenTevpercentsttFidZFidGlobalCorrNoLumiNoBeam}{    2.13}
\newcommand{\AveSystTTSevenTevabsvalttFidZFidGlobalCorrNoLumiNoBeam}{    0.05}
\newcommand{\AveSystZThirteenTevpercentsttFidZFidGlobalCorrBeamOnly}{    0.69}
\newcommand{\AveSystZThirteenTevabsvalttFidZFidGlobalCorrBeamOnly}{    5.37}
\newcommand{\AveSystTTThirteenTevpercentsttFidZFidGlobalCorrBeamOnly}{    1.49}
\newcommand{\AveSystTTThirteenTevabsvalttFidZFidGlobalCorrBeamOnly}{    0.15}
\newcommand{\AveSystZEightTevpercentsttFidZFidGlobalCorrBeamOnly}{    0.62}
\newcommand{\AveSystZEightTevabsvalttFidZFidGlobalCorrBeamOnly}{    3.13}
\newcommand{\AveSystTTEightTevpercentsttFidZFidGlobalCorrBeamOnly}{    1.71}
\newcommand{\AveSystTTEightTevabsvalttFidZFidGlobalCorrBeamOnly}{    0.05}
\newcommand{\AveSystZSevenTevpercentsttFidZFidGlobalCorrBeamOnly}{    0.60}
\newcommand{\AveSystZSevenTevabsvalttFidZFidGlobalCorrBeamOnly}{    2.71}
\newcommand{\AveSystTTSevenTevpercentsttFidZFidGlobalCorrBeamOnly}{    1.79}
\newcommand{\AveSystTTSevenTevabsvalttFidZFidGlobalCorrBeamOnly}{    0.03}

\newcommand{\AveStatZThirteenTevpercentsttTotZTotGlobalCorr}{  0.06}
\newcommand{\AveSystZThirteenTevpercentsttTotZTotGlobalCorr}{  2.86 }
\newcommand{\AveTotUncZThirteenTevpercentsttTotZTotGlobalCorr}{  2.87}
\newcommand{\AveLumiZThirteenTevabsolutettTotZTotGlobalCorr}{ 41.35}
\newcommand{\AveCentralZThirteenTevabsvalttTotZTotGlobalCorr}{1969.00}
\newcommand{\AveStatZThirteenTevabsvalttTotZTotGlobalCorr}{  1.18}
\newcommand{\AveSystZThirteenTevabsvalttTotZTotGlobalCorr}{ 56.41 }
\newcommand{\AveTotUncZThirteenTevabsvalttTotZTotGlobalCorr}{ 56.42}
\newcommand{\AveStatTTThirteenTevpercentsttTotZTotGlobalCorr}{  0.92}
\newcommand{\AveSystTTThirteenTevpercentsttTotZTotGlobalCorr}{  4.32 }
\newcommand{\AveTotUncTTThirteenTevpercentsttTotZTotGlobalCorr}{  4.41}
\newcommand{\AveLumiTTThirteenTevabsolutettTotZTotGlobalCorr}{ 18.89}
\newcommand{\AveCentralTTThirteenTevabsvalttTotZTotGlobalCorr}{817.82}
\newcommand{\AveStatTTThirteenTevabsvalttTotZTotGlobalCorr}{  7.52}
\newcommand{\AveSystTTThirteenTevabsvalttTotZTotGlobalCorr}{ 35.29 }
\newcommand{\AveTotUncTTThirteenTevabsvalttTotZTotGlobalCorr}{ 36.08}
\newcommand{\AveStatZEightTevpercentsttTotZTotGlobalCorr}{  0.02}
\newcommand{\AveSystZEightTevpercentsttTotZTotGlobalCorr}{  2.69 }
\newcommand{\AveTotUncZEightTevpercentsttTotZTotGlobalCorr}{  2.69}
\newcommand{\AveLumiZEightTevabsolutettTotZTotGlobalCorr}{ 21.92}
\newcommand{\AveCentralZEightTevabsvalttTotZTotGlobalCorr}{1153.50}
\newcommand{\AveStatZEightTevabsvalttTotZTotGlobalCorr}{  0.28}
\newcommand{\AveSystZEightTevabsvalttTotZTotGlobalCorr}{ 31.07 }
\newcommand{\AveTotUncZEightTevabsvalttTotZTotGlobalCorr}{ 31.07}
\newcommand{\AveStatTTEightTevpercentsttTotZTotGlobalCorr}{  0.71}
\newcommand{\AveSystTTEightTevpercentsttTotZTotGlobalCorr}{  3.53 }
\newcommand{\AveTotUncTTEightTevpercentsttTotZTotGlobalCorr}{  3.60}
\newcommand{\AveLumiTTEightTevabsolutettTotZTotGlobalCorr}{  5.10}
\newcommand{\AveCentralTTEightTevabsvalttTotZTotGlobalCorr}{242.70}
\newcommand{\AveStatTTEightTevabsvalttTotZTotGlobalCorr}{  1.72}
\newcommand{\AveSystTTEightTevabsvalttTotZTotGlobalCorr}{  8.57 }
\newcommand{\AveTotUncTTEightTevabsvalttTotZTotGlobalCorr}{  8.74}
\newcommand{\AveStatZSevenTevpercentsttTotZTotGlobalCorr}{  0.06}
\newcommand{\AveSystZSevenTevpercentsttTotZTotGlobalCorr}{  2.62 }
\newcommand{\AveTotUncZSevenTevpercentsttTotZTotGlobalCorr}{  2.62}
\newcommand{\AveLumiZSevenTevabsolutettTotZTotGlobalCorr}{ 17.91}
\newcommand{\AveCentralZSevenTevabsvalttTotZTotGlobalCorr}{994.77}
\newcommand{\AveStatZSevenTevabsvalttTotZTotGlobalCorr}{  0.62}
\newcommand{\AveSystZSevenTevabsvalttTotZTotGlobalCorr}{ 26.07 }
\newcommand{\AveTotUncZSevenTevabsvalttTotZTotGlobalCorr}{ 26.08}
\newcommand{\AveStatTTSevenTevpercentsttTotZTotGlobalCorr}{  1.69}
\newcommand{\AveSystTTSevenTevpercentsttTotZTotGlobalCorr}{  3.51 }
\newcommand{\AveTotUncTTSevenTevpercentsttTotZTotGlobalCorr}{  3.89}
\newcommand{\AveLumiTTSevenTevabsolutettTotZTotGlobalCorr}{  3.62}
\newcommand{\AveCentralTTSevenTevabsvalttTotZTotGlobalCorr}{182.84}
\newcommand{\AveStatTTSevenTevabsvalttTotZTotGlobalCorr}{  3.09}
\newcommand{\AveSystTTSevenTevabsvalttTotZTotGlobalCorr}{  6.41 }
\newcommand{\AveTotUncTTSevenTevabsvalttTotZTotGlobalCorr}{  7.12}
\newcommand{\AveChisqAllChanEnergZThirteenTevTTThirteenTevZEightTevTTEightTevZSevenTevTTSevenTevTotTotGlobalCorr}{0.6}
\newcommand{\CorrCombZThirteenTevcorZThirteenTevttTotZTotGlobalCorr}{ 1.000}
\newcommand{\CorrCombTTThirteenTevcorZThirteenTevttTotZTotGlobalCorr}{ 0.482}
\newcommand{\CorrCombZEightTevcorZThirteenTevttTotZTotGlobalCorr}{ 0.453}
\newcommand{\CorrCombTTEightTevcorZThirteenTevttTotZTotGlobalCorr}{ 0.123}
\newcommand{\CorrCombZSevenTevcorZThirteenTevttTotZTotGlobalCorr}{ 0.477}
\newcommand{\CorrCombTTSevenTevcorZThirteenTevttTotZTotGlobalCorr}{ 0.114}
\newcommand{\CorrCombTTThirteenTevcorTTThirteenTevttTotZTotGlobalCorr}{ 1.000}
\newcommand{\CorrCombZEightTevcorTTThirteenTevttTotZTotGlobalCorr}{ 0.080}
\newcommand{\CorrCombTTEightTevcorTTThirteenTevttTotZTotGlobalCorr}{ 0.324}
\newcommand{\CorrCombZSevenTevcorTTThirteenTevttTotZTotGlobalCorr}{ 0.078}
\newcommand{\CorrCombTTSevenTevcorTTThirteenTevttTotZTotGlobalCorr}{ 0.312}
\newcommand{\CorrCombZEightTevcorZEightTevttTotZTotGlobalCorr}{ 1.000}
\newcommand{\CorrCombTTEightTevcorZEightTevttTotZTotGlobalCorr}{ 0.522}
\newcommand{\CorrCombZSevenTevcorZEightTevttTotZTotGlobalCorr}{ 0.489}
\newcommand{\CorrCombTTSevenTevcorZEightTevttTotZTotGlobalCorr}{ 0.106}
\newcommand{\CorrCombTTEightTevcorTTEightTevttTotZTotGlobalCorr}{ 1.000}
\newcommand{\CorrCombZSevenTevcorTTEightTevttTotZTotGlobalCorr}{ 0.110}
\newcommand{\CorrCombTTSevenTevcorTTEightTevttTotZTotGlobalCorr}{ 0.542}
\newcommand{\CorrCombZSevenTevcorZSevenTevttTotZTotGlobalCorr}{ 1.000}
\newcommand{\CorrCombTTSevenTevcorZSevenTevttTotZTotGlobalCorr}{ 0.455}
\newcommand{\CorrCombTTSevenTevcorTTSevenTevttTotZTotGlobalCorr}{ 1.000}
\newcommand{\AveSystZThirteenTevpercentsttTotZTotGlobalCorrNoLumi}{    1.95}
\newcommand{\AveSystZThirteenTevabsvalttTotZTotGlobalCorrNoLumi}{   38.38}
\newcommand{\AveSystTTThirteenTevpercentsttTotZTotGlobalCorrNoLumi}{    3.64}
\newcommand{\AveSystTTThirteenTevabsvalttTotZTotGlobalCorrNoLumi}{   29.81}
\newcommand{\AveSystZEightTevpercentsttTotZTotGlobalCorrNoLumi}{    1.91}
\newcommand{\AveSystZEightTevabsvalttTotZTotGlobalCorrNoLumi}{   22.02}
\newcommand{\AveSystTTEightTevpercentsttTotZTotGlobalCorrNoLumi}{    2.84}
\newcommand{\AveSystTTEightTevabsvalttTotZTotGlobalCorrNoLumi}{    6.89}
\newcommand{\AveSystZSevenTevpercentsttTotZTotGlobalCorrNoLumi}{    1.91}
\newcommand{\AveSystZSevenTevabsvalttTotZTotGlobalCorrNoLumi}{   18.96}
\newcommand{\AveSystTTSevenTevpercentsttTotZTotGlobalCorrNoLumi}{    2.89}
\newcommand{\AveSystTTSevenTevabsvalttTotZTotGlobalCorrNoLumi}{    5.29}
\newcommand{\AveSystZThirteenTevpercentsttTotZTotGlobalCorrLumiOnly}{    2.09}
\newcommand{\AveSystZThirteenTevabsvalttTotZTotGlobalCorrLumiOnly}{   41.34}
\newcommand{\AveSystTTThirteenTevpercentsttTotZTotGlobalCorrLumiOnly}{    2.33}
\newcommand{\AveSystTTThirteenTevabsvalttTotZTotGlobalCorrLumiOnly}{   18.89}
\newcommand{\AveSystZEightTevpercentsttTotZTotGlobalCorrLumiOnly}{    1.89}
\newcommand{\AveSystZEightTevabsvalttTotZTotGlobalCorrLumiOnly}{   21.92}
\newcommand{\AveSystTTEightTevpercentsttTotZTotGlobalCorrLumiOnly}{    2.10}
\newcommand{\AveSystTTEightTevabsvalttTotZTotGlobalCorrLumiOnly}{    5.10}
\newcommand{\AveSystZSevenTevpercentsttTotZTotGlobalCorrLumiOnly}{    1.79}
\newcommand{\AveSystZSevenTevabsvalttTotZTotGlobalCorrLumiOnly}{   17.89}
\newcommand{\AveSystTTSevenTevpercentsttTotZTotGlobalCorrLumiOnly}{    1.99}
\newcommand{\AveSystTTSevenTevabsvalttTotZTotGlobalCorrLumiOnly}{    3.62}
\newcommand{\AveSystZThirteenTevpercentsttTotZTotGlobalCorrNoLumiNoBeam}{    1.82}
\newcommand{\AveSystZThirteenTevabsvalttTotZTotGlobalCorrNoLumiNoBeam}{   35.90}
\newcommand{\AveSystTTThirteenTevpercentsttTotZTotGlobalCorrNoLumiNoBeam}{    3.32}
\newcommand{\AveSystTTThirteenTevabsvalttTotZTotGlobalCorrNoLumiNoBeam}{   27.16}
\newcommand{\AveSystZEightTevpercentsttTotZTotGlobalCorrNoLumiNoBeam}{    1.81}
\newcommand{\AveSystZEightTevabsvalttTotZTotGlobalCorrNoLumiNoBeam}{   20.83}
\newcommand{\AveSystTTEightTevpercentsttTotZTotGlobalCorrNoLumiNoBeam}{    2.26}
\newcommand{\AveSystTTEightTevabsvalttTotZTotGlobalCorrNoLumiNoBeam}{    5.48}
\newcommand{\AveSystZSevenTevpercentsttTotZTotGlobalCorrNoLumiNoBeam}{    1.81}
\newcommand{\AveSystZSevenTevabsvalttTotZTotGlobalCorrNoLumiNoBeam}{   17.99}
\newcommand{\AveSystTTSevenTevpercentsttTotZTotGlobalCorrNoLumiNoBeam}{    2.27}
\newcommand{\AveSystTTSevenTevabsvalttTotZTotGlobalCorrNoLumiNoBeam}{    4.15}
\newcommand{\AveSystZThirteenTevpercentsttTotZTotGlobalCorrBeamOnly}{    0.70}
\newcommand{\AveSystZThirteenTevabsvalttTotZTotGlobalCorrBeamOnly}{   13.57}
\newcommand{\AveSystTTThirteenTevpercentsttTotZTotGlobalCorrBeamOnly}{    1.49}
\newcommand{\AveSystTTThirteenTevabsvalttTotZTotGlobalCorrBeamOnly}{   12.29}
\newcommand{\AveSystZEightTevpercentsttTotZTotGlobalCorrBeamOnly}{    0.61}
\newcommand{\AveSystZEightTevabsvalttTotZTotGlobalCorrBeamOnly}{    7.14}
\newcommand{\AveSystTTEightTevpercentsttTotZTotGlobalCorrBeamOnly}{    1.72}
\newcommand{\AveSystTTEightTevabsvalttTotZTotGlobalCorrBeamOnly}{    4.18}
\newcommand{\AveSystZSevenTevpercentsttTotZTotGlobalCorrBeamOnly}{    0.61}
\newcommand{\AveSystZSevenTevabsvalttTotZTotGlobalCorrBeamOnly}{    5.99}
\newcommand{\AveSystTTSevenTevpercentsttTotZTotGlobalCorrBeamOnly}{    1.79}
\newcommand{\AveSystTTSevenTevabsvalttTotZTotGlobalCorrBeamOnly}{    3.28}

\newcommand{\AveStatZThirteenTevpercentsttTotZFidGlobalCorr}{  0.06}
\newcommand{\AveSystZThirteenTevpercentsttTotZFidGlobalCorr}{  2.25 }
\newcommand{\AveTotUncZThirteenTevpercentsttTotZFidGlobalCorr}{  2.25}
\newcommand{\AveLumiZThirteenTevabsolutettTotZFidGlobalCorr}{ 16.32}
\newcommand{\AveCentralZThirteenTevabsvalttTotZFidGlobalCorr}{777.28}
\newcommand{\AveStatZThirteenTevabsvalttTotZFidGlobalCorr}{  0.46}
\newcommand{\AveSystZThirteenTevabsvalttTotZFidGlobalCorr}{ 17.51 }
\newcommand{\AveTotUncZThirteenTevabsvalttTotZFidGlobalCorr}{ 17.51}
\newcommand{\AveStatTTThirteenTevpercentsttTotZFidGlobalCorr}{  0.92}
\newcommand{\AveSystTTThirteenTevpercentsttTotZFidGlobalCorr}{  4.31 }
\newcommand{\AveTotUncTTThirteenTevpercentsttTotZFidGlobalCorr}{  4.41}
\newcommand{\AveLumiTTThirteenTevabsolutettTotZFidGlobalCorr}{ 18.90}
\newcommand{\AveCentralTTThirteenTevabsvalttTotZFidGlobalCorr}{817.99}
\newcommand{\AveStatTTThirteenTevabsvalttTotZFidGlobalCorr}{  7.52}
\newcommand{\AveSystTTThirteenTevabsvalttTotZFidGlobalCorr}{ 35.29 }
\newcommand{\AveTotUncTTThirteenTevabsvalttTotZFidGlobalCorr}{ 36.08}
\newcommand{\AveStatZEightTevpercentsttTotZFidGlobalCorr}{  0.02}
\newcommand{\AveSystZEightTevpercentsttTotZFidGlobalCorr}{  2.07 }
\newcommand{\AveTotUncZEightTevpercentsttTotZFidGlobalCorr}{  2.07}
\newcommand{\AveLumiZEightTevabsolutettTotZFidGlobalCorr}{  9.61}
\newcommand{\AveCentralZEightTevabsvalttTotZFidGlobalCorr}{505.80}
\newcommand{\AveStatZEightTevabsvalttTotZFidGlobalCorr}{  0.12}
\newcommand{\AveSystZEightTevabsvalttTotZFidGlobalCorr}{ 10.48 }
\newcommand{\AveTotUncZEightTevabsvalttTotZFidGlobalCorr}{ 10.48}
\newcommand{\AveStatTTEightTevpercentsttTotZFidGlobalCorr}{  0.71}
\newcommand{\AveSystTTEightTevpercentsttTotZFidGlobalCorr}{  3.53 }
\newcommand{\AveTotUncTTEightTevpercentsttTotZFidGlobalCorr}{  3.60}
\newcommand{\AveLumiTTEightTevabsolutettTotZFidGlobalCorr}{  5.11}
\newcommand{\AveCentralTTEightTevabsvalttTotZFidGlobalCorr}{242.78}
\newcommand{\AveStatTTEightTevabsvalttTotZFidGlobalCorr}{  1.72}
\newcommand{\AveSystTTEightTevabsvalttTotZFidGlobalCorr}{  8.57 }
\newcommand{\AveTotUncTTEightTevabsvalttTotZFidGlobalCorr}{  8.74}
\newcommand{\AveStatZSevenTevpercentsttTotZFidGlobalCorr}{  0.06}
\newcommand{\AveSystZSevenTevpercentsttTotZFidGlobalCorr}{  1.92 }
\newcommand{\AveTotUncZSevenTevpercentsttTotZFidGlobalCorr}{  1.93}
\newcommand{\AveLumiZSevenTevabsolutettTotZFidGlobalCorr}{  8.11}
\newcommand{\AveCentralZSevenTevabsvalttTotZFidGlobalCorr}{450.76}
\newcommand{\AveStatZSevenTevabsvalttTotZFidGlobalCorr}{  0.28}
\newcommand{\AveSystZSevenTevabsvalttTotZFidGlobalCorr}{  8.67 }
\newcommand{\AveTotUncZSevenTevabsvalttTotZFidGlobalCorr}{  8.68}
\newcommand{\AveStatTTSevenTevpercentsttTotZFidGlobalCorr}{  1.69}
\newcommand{\AveSystTTSevenTevpercentsttTotZFidGlobalCorr}{  3.51 }
\newcommand{\AveTotUncTTSevenTevpercentsttTotZFidGlobalCorr}{  3.89}
\newcommand{\AveLumiTTSevenTevabsolutettTotZFidGlobalCorr}{  3.62}
\newcommand{\AveCentralTTSevenTevabsvalttTotZFidGlobalCorr}{182.87}
\newcommand{\AveStatTTSevenTevabsvalttTotZFidGlobalCorr}{  3.09}
\newcommand{\AveSystTTSevenTevabsvalttTotZFidGlobalCorr}{  6.41 }
\newcommand{\AveTotUncTTSevenTevabsvalttTotZFidGlobalCorr}{  7.12}
\newcommand{\AveChisqAllChanEnergZThirteenTevTTThirteenTevZEightTevTTEightTevZSevenTevTTSevenTevFidTotGlobalCorr}{0.6}
\newcommand{\CorrCombZThirteenTevcorZThirteenTevttTotZFidGlobalCorr}{ 1.000}
\newcommand{\CorrCombTTThirteenTevcorZThirteenTevttTotZFidGlobalCorr}{ 0.612}
\newcommand{\CorrCombZEightTevcorZThirteenTevttTotZFidGlobalCorr}{ 0.097}
\newcommand{\CorrCombTTEightTevcorZThirteenTevttTotZFidGlobalCorr}{ 0.156}
\newcommand{\CorrCombZSevenTevcorZThirteenTevttTotZFidGlobalCorr}{ 0.100}
\newcommand{\CorrCombTTSevenTevcorZThirteenTevttTotZFidGlobalCorr}{ 0.145}
\newcommand{\CorrCombTTThirteenTevcorTTThirteenTevttTotZFidGlobalCorr}{ 1.000}
\newcommand{\CorrCombZEightTevcorTTThirteenTevttTotZFidGlobalCorr}{ 0.105}
\newcommand{\CorrCombTTEightTevcorTTThirteenTevttTotZFidGlobalCorr}{ 0.324}
\newcommand{\CorrCombZSevenTevcorTTThirteenTevttTotZFidGlobalCorr}{ 0.106}
\newcommand{\CorrCombTTSevenTevcorTTThirteenTevttTotZFidGlobalCorr}{ 0.312}
\newcommand{\CorrCombZEightTevcorZEightTevttTotZFidGlobalCorr}{ 1.000}
\newcommand{\CorrCombTTEightTevcorZEightTevttTotZFidGlobalCorr}{ 0.679}
\newcommand{\CorrCombZSevenTevcorZEightTevttTotZFidGlobalCorr}{ 0.097}
\newcommand{\CorrCombTTSevenTevcorZEightTevttTotZFidGlobalCorr}{ 0.138}
\newcommand{\CorrCombTTEightTevcorTTEightTevttTotZFidGlobalCorr}{ 1.000}
\newcommand{\CorrCombZSevenTevcorTTEightTevttTotZFidGlobalCorr}{ 0.149}
\newcommand{\CorrCombTTSevenTevcorTTEightTevttTotZFidGlobalCorr}{ 0.542}
\newcommand{\CorrCombZSevenTevcorZSevenTevttTotZFidGlobalCorr}{ 1.000}
\newcommand{\CorrCombTTSevenTevcorZSevenTevttTotZFidGlobalCorr}{ 0.620}
\newcommand{\CorrCombTTSevenTevcorTTSevenTevttTotZFidGlobalCorr}{ 1.000}
\newcommand{\AveSystZThirteenTevpercentsttTotZFidGlobalCorrNoLumi}{    0.82}
\newcommand{\AveSystZThirteenTevabsvalttTotZFidGlobalCorrNoLumi}{    6.34}
\newcommand{\AveSystTTThirteenTevpercentsttTotZFidGlobalCorrNoLumi}{    3.64}
\newcommand{\AveSystTTThirteenTevabsvalttTotZFidGlobalCorrNoLumi}{   29.81}
\newcommand{\AveSystZEightTevpercentsttTotZFidGlobalCorrNoLumi}{    0.83}
\newcommand{\AveSystZEightTevabsvalttTotZFidGlobalCorrNoLumi}{    4.18}
\newcommand{\AveSystTTEightTevpercentsttTotZFidGlobalCorrNoLumi}{    2.84}
\newcommand{\AveSystTTEightTevabsvalttTotZFidGlobalCorrNoLumi}{    6.89}
\newcommand{\AveSystZSevenTevpercentsttTotZFidGlobalCorrNoLumi}{    0.68}
\newcommand{\AveSystZSevenTevabsvalttTotZFidGlobalCorrNoLumi}{    3.07}
\newcommand{\AveSystTTSevenTevpercentsttTotZFidGlobalCorrNoLumi}{    2.89}
\newcommand{\AveSystTTSevenTevabsvalttTotZFidGlobalCorrNoLumi}{    5.29}
\newcommand{\AveSystZThirteenTevpercentsttTotZFidGlobalCorrLumiOnly}{    2.10}
\newcommand{\AveSystZThirteenTevabsvalttTotZFidGlobalCorrLumiOnly}{   16.32}
\newcommand{\AveSystTTThirteenTevpercentsttTotZFidGlobalCorrLumiOnly}{    2.31}
\newcommand{\AveSystTTThirteenTevabsvalttTotZFidGlobalCorrLumiOnly}{   18.89}
\newcommand{\AveSystZEightTevpercentsttTotZFidGlobalCorrLumiOnly}{    1.90}
\newcommand{\AveSystZEightTevabsvalttTotZFidGlobalCorrLumiOnly}{    9.61}
\newcommand{\AveSystTTEightTevpercentsttTotZFidGlobalCorrLumiOnly}{    2.10}
\newcommand{\AveSystTTEightTevabsvalttTotZFidGlobalCorrLumiOnly}{    5.10}
\newcommand{\AveSystZSevenTevpercentsttTotZFidGlobalCorrLumiOnly}{    1.80}
\newcommand{\AveSystZSevenTevabsvalttTotZFidGlobalCorrLumiOnly}{    8.11}
\newcommand{\AveSystTTSevenTevpercentsttTotZFidGlobalCorrLumiOnly}{    1.99}
\newcommand{\AveSystTTSevenTevabsvalttTotZFidGlobalCorrLumiOnly}{    3.62}
\newcommand{\AveSystZThirteenTevpercentsttTotZFidGlobalCorrNoLumiNoBeam}{    0.44}
\newcommand{\AveSystZThirteenTevabsvalttTotZFidGlobalCorrNoLumiNoBeam}{    3.38}
\newcommand{\AveSystTTThirteenTevpercentsttTotZFidGlobalCorrNoLumiNoBeam}{    3.32}
\newcommand{\AveSystTTThirteenTevabsvalttTotZFidGlobalCorrNoLumiNoBeam}{   27.16}
\newcommand{\AveSystZEightTevpercentsttTotZFidGlobalCorrNoLumiNoBeam}{    0.55}
\newcommand{\AveSystZEightTevabsvalttTotZFidGlobalCorrNoLumiNoBeam}{    2.77}
\newcommand{\AveSystTTEightTevpercentsttTotZFidGlobalCorrNoLumiNoBeam}{    2.26}
\newcommand{\AveSystTTEightTevabsvalttTotZFidGlobalCorrNoLumiNoBeam}{    5.48}
\newcommand{\AveSystZSevenTevpercentsttTotZFidGlobalCorrNoLumiNoBeam}{    0.32}
\newcommand{\AveSystZSevenTevabsvalttTotZFidGlobalCorrNoLumiNoBeam}{    1.44}
\newcommand{\AveSystTTSevenTevpercentsttTotZFidGlobalCorrNoLumiNoBeam}{    2.27}
\newcommand{\AveSystTTSevenTevabsvalttTotZFidGlobalCorrNoLumiNoBeam}{    4.15}
\newcommand{\AveSystZThirteenTevpercentsttTotZFidGlobalCorrBeamOnly}{    0.69}
\newcommand{\AveSystZThirteenTevabsvalttTotZFidGlobalCorrBeamOnly}{    5.36}
\newcommand{\AveSystTTThirteenTevpercentsttTotZFidGlobalCorrBeamOnly}{    1.49}
\newcommand{\AveSystTTThirteenTevabsvalttTotZFidGlobalCorrBeamOnly}{   12.29}
\newcommand{\AveSystZEightTevpercentsttTotZFidGlobalCorrBeamOnly}{    0.62}
\newcommand{\AveSystZEightTevabsvalttTotZFidGlobalCorrBeamOnly}{    3.13}
\newcommand{\AveSystTTEightTevpercentsttTotZFidGlobalCorrBeamOnly}{    1.72}
\newcommand{\AveSystTTEightTevabsvalttTotZFidGlobalCorrBeamOnly}{    4.18}
\newcommand{\AveSystZSevenTevpercentsttTotZFidGlobalCorrBeamOnly}{    0.60}
\newcommand{\AveSystZSevenTevabsvalttTotZFidGlobalCorrBeamOnly}{    2.71}
\newcommand{\AveSystTTSevenTevpercentsttTotZFidGlobalCorrBeamOnly}{    1.79}
\newcommand{\AveSystTTSevenTevabsvalttTotZFidGlobalCorrBeamOnly}{    3.28}

\newcommand{\GroupedSystPercLumiErrorZThirtTeVttTotZTotGlobalCorr}{  2.10}
\newcommand{\GroupedSystPercLumiErrorZEightTeVttTotZTotGlobalCorr}{  1.90}
\newcommand{\GroupedSystPercLumiErrorZSevenTeVttTotZTotGlobalCorr}{  1.80}
\newcommand{\GroupedSystPercLumiErrorTTThirtTeVttTotZTotGlobalCorr}{  2.31}
\newcommand{\GroupedSystPercLumiErrorTTEightTeVttTotZTotGlobalCorr}{  2.10}
\newcommand{\GroupedSystPercLumiErrorTTSevenTeVttTotZTotGlobalCorr}{  1.98}
\newcommand{\GroupedSystPercEBeamErrorZThirtTeVttTotZTotGlobalCorr}{  0.69}
\newcommand{\GroupedSystPercEBeamErrorZEightTeVttTotZTotGlobalCorr}{  0.62}
\newcommand{\GroupedSystPercEBeamErrorZSevenTeVttTotZTotGlobalCorr}{  0.60}
\newcommand{\GroupedSystPercEBeamErrorTTThirtTeVttTotZTotGlobalCorr}{  1.50}
\newcommand{\GroupedSystPercEBeamErrorTTEightTeVttTotZTotGlobalCorr}{  1.72}
\newcommand{\GroupedSystPercEBeamErrorTTSevenTeVttTotZTotGlobalCorr}{  1.79}
\newcommand{\GroupedSystPercMuTrigErrorZThirtTeVttTotZTotGlobalCorr}{  0.12}
\newcommand{\GroupedSystPercMuTrigErrorZEightTeVttTotZTotGlobalCorr}{  0.55}
\newcommand{\GroupedSystPercMuTrigErrorZSevenTeVttTotZTotGlobalCorr}{  0.05}
\newcommand{\GroupedSystPercMuTrigErrorTTThirtTeVttTotZTotGlobalCorr}{  0.05}
\newcommand{\GroupedSystPercMuTrigErrorTTEightTeVttTotZTotGlobalCorr}{  0.17}
\newcommand{\GroupedSystPercMuTrigErrorTTSevenTeVttTotZTotGlobalCorr}{  0.19}
\newcommand{\GroupedSystPercMuRecIdErrorZThirtTeVttTotZTotGlobalCorr}{  0.68}
\newcommand{\GroupedSystPercMuRecIdErrorZEightTeVttTotZTotGlobalCorr}{  0.45}
\newcommand{\GroupedSystPercMuRecIdErrorZSevenTeVttTotZTotGlobalCorr}{  0.30}
\newcommand{\GroupedSystPercMuRecIdErrorTTThirtTeVttTotZTotGlobalCorr}{  0.44}
\newcommand{\GroupedSystPercMuRecIdErrorTTEightTeVttTotZTotGlobalCorr}{  0.42}
\newcommand{\GroupedSystPercMuRecIdErrorTTSevenTeVttTotZTotGlobalCorr}{  0.31}
\newcommand{\GroupedSystPercMuIsoErrorZThirtTeVttTotZTotGlobalCorr}{  0.41}
\newcommand{\GroupedSystPercMuIsoErrorZEightTeVttTotZTotGlobalCorr}{  0.04}
\newcommand{\GroupedSystPercMuIsoErrorZSevenTeVttTotZTotGlobalCorr}{  0.15}
\newcommand{\GroupedSystPercMuIsoErrorTTThirtTeVttTotZTotGlobalCorr}{  0.27}
\newcommand{\GroupedSystPercMuIsoErrorTTEightTeVttTotZTotGlobalCorr}{  0.22}
\newcommand{\GroupedSystPercMuIsoErrorTTSevenTeVttTotZTotGlobalCorr}{  0.44}
\newcommand{\GroupedSystPercMuScaleErrorZThirtTeVttTotZTotGlobalCorr}{  0.06}
\newcommand{\GroupedSystPercMuScaleErrorZEightTeVttTotZTotGlobalCorr}{  0.03}
\newcommand{\GroupedSystPercMuScaleErrorZSevenTeVttTotZTotGlobalCorr}{  0.03}
\newcommand{\GroupedSystPercMuScaleErrorTTThirtTeVttTotZTotGlobalCorr}{  0.04}
\newcommand{\GroupedSystPercMuScaleErrorTTEightTeVttTotZTotGlobalCorr}{  0.01}
\newcommand{\GroupedSystPercMuScaleErrorTTSevenTeVttTotZTotGlobalCorr}{  0.14}
\newcommand{\GroupedSystPercElTrigErrorZThirtTeVttTotZTotGlobalCorr}{  0.01}
\newcommand{\GroupedSystPercElTrigErrorZEightTeVttTotZTotGlobalCorr}{  0.19}
\newcommand{\GroupedSystPercElTrigErrorZSevenTeVttTotZTotGlobalCorr}{  0.04}
\newcommand{\GroupedSystPercElTrigErrorTTThirtTeVttTotZTotGlobalCorr}{  0.14}
\newcommand{\GroupedSystPercElTrigErrorTTEightTeVttTotZTotGlobalCorr}{  0.00}
\newcommand{\GroupedSystPercElTrigErrorTTSevenTeVttTotZTotGlobalCorr}{  0.00}
\newcommand{\GroupedSystPercElRecIdErrorZThirtTeVttTotZTotGlobalCorr}{  0.41}
\newcommand{\GroupedSystPercElRecIdErrorZEightTeVttTotZTotGlobalCorr}{  0.80}
\newcommand{\GroupedSystPercElRecIdErrorZSevenTeVttTotZTotGlobalCorr}{  0.26}
\newcommand{\GroupedSystPercElRecIdErrorTTThirtTeVttTotZTotGlobalCorr}{  0.34}
\newcommand{\GroupedSystPercElRecIdErrorTTEightTeVttTotZTotGlobalCorr}{  0.41}
\newcommand{\GroupedSystPercElRecIdErrorTTSevenTeVttTotZTotGlobalCorr}{  0.13}
\newcommand{\GroupedSystPercElIsoErrorZThirtTeVttTotZTotGlobalCorr}{  0.14}
\newcommand{\GroupedSystPercElIsoErrorZEightTeVttTotZTotGlobalCorr}{  0.00}
\newcommand{\GroupedSystPercElIsoErrorTTThirtTeVttTotZTotGlobalCorr}{  0.39}
\newcommand{\GroupedSystPercElIsoErrorTTEightTeVttTotZTotGlobalCorr}{  0.30}
\newcommand{\GroupedSystPercElIsoErrorTTSevenTeVttTotZTotGlobalCorr}{  0.59}
\newcommand{\GroupedSystPercElScaleErrorZThirtTeVttTotZTotGlobalCorr}{  0.25}
\newcommand{\GroupedSystPercElScaleErrorZEightTeVttTotZTotGlobalCorr}{  0.07}
\newcommand{\GroupedSystPercElScaleErrorZSevenTeVttTotZTotGlobalCorr}{  0.08}
\newcommand{\GroupedSystPercElScaleErrorTTThirtTeVttTotZTotGlobalCorr}{  0.20}
\newcommand{\GroupedSystPercElScaleErrorTTEightTeVttTotZTotGlobalCorr}{  0.51}
\newcommand{\GroupedSystPercElScaleErrorTTSevenTeVttTotZTotGlobalCorr}{  0.21}
\newcommand{\GroupedSystPercJesErrorTTThirtTeVttTotZTotGlobalCorr}{  0.38}
\newcommand{\GroupedSystPercJesErrorTTEightTeVttTotZTotGlobalCorr}{  0.72}
\newcommand{\GroupedSystPercJesErrorTTSevenTeVttTotZTotGlobalCorr}{  0.40}
\newcommand{\GroupedSystPercFlTagErrorTTThirtTeVttTotZTotGlobalCorr}{  0.53}
\newcommand{\GroupedSystPercFlTagErrorTTEightTeVttTotZTotGlobalCorr}{  0.40}
\newcommand{\GroupedSystPercFlTagErrorTTSevenTeVttTotZTotGlobalCorr}{  0.46}
\newcommand{\GroupedSystPercBKGErrorZThirtTeVttTotZTotGlobalCorr}{  0.08}
\newcommand{\GroupedSystPercBKGErrorZEightTeVttTotZTotGlobalCorr}{  0.15}
\newcommand{\GroupedSystPercBKGErrorZSevenTeVttTotZTotGlobalCorr}{  0.08}
\newcommand{\GroupedSystPercBKGErrorTTThirtTeVttTotZTotGlobalCorr}{  1.09}
\newcommand{\GroupedSystPercBKGErrorTTEightTeVttTotZTotGlobalCorr}{  1.04}
\newcommand{\GroupedSystPercBKGErrorTTSevenTeVttTotZTotGlobalCorr}{  1.04}
\newcommand{\GroupedSystPercSigAndPDFErrorZThirtTeVttTotZTotGlobalCorr}{  0.12}
\newcommand{\GroupedSystPercSigAndPDFErrorZEightTeVttTotZTotGlobalCorr}{  0.08}
\newcommand{\GroupedSystPercSigAndPDFErrorZSevenTeVttTotZTotGlobalCorr}{  0.27}
\newcommand{\GroupedSystPercSigAndPDFErrorTTThirtTeVttTotZTotGlobalCorr}{  2.98}
\newcommand{\GroupedSystPercSigAndPDFErrorTTEightTeVttTotZTotGlobalCorr}{  1.70}
\newcommand{\GroupedSystPercSigAndPDFErrorTTSevenTeVttTotZTotGlobalCorr}{  1.81}

\newcommand{\GroupedSystPercLumiErrorZThirtTeVttTotZFidGlobalCorr}{  2.10}
\newcommand{\GroupedSystPercLumiErrorZEightTeVttTotZFidGlobalCorr}{  1.90}
\newcommand{\GroupedSystPercLumiErrorZSevenTeVttTotZFidGlobalCorr}{  1.80}
\newcommand{\GroupedSystPercLumiErrorTTThirtTeVttTotZFidGlobalCorr}{  2.31}
\newcommand{\GroupedSystPercLumiErrorTTEightTeVttTotZFidGlobalCorr}{  2.10}
\newcommand{\GroupedSystPercLumiErrorTTSevenTeVttTotZFidGlobalCorr}{  1.98}
\newcommand{\GroupedSystPercEBeamErrorZThirtTeVttTotZFidGlobalCorr}{  0.69}
\newcommand{\GroupedSystPercEBeamErrorZEightTeVttTotZFidGlobalCorr}{  0.62}
\newcommand{\GroupedSystPercEBeamErrorZSevenTeVttTotZFidGlobalCorr}{  0.60}
\newcommand{\GroupedSystPercEBeamErrorTTThirtTeVttTotZFidGlobalCorr}{  1.50}
\newcommand{\GroupedSystPercEBeamErrorTTEightTeVttTotZFidGlobalCorr}{  1.72}
\newcommand{\GroupedSystPercEBeamErrorTTSevenTeVttTotZFidGlobalCorr}{  1.79}
\newcommand{\GroupedSystPercMuTrigErrorZThirtTeVttTotZFidGlobalCorr}{  0.12}
\newcommand{\GroupedSystPercMuTrigErrorZEightTeVttTotZFidGlobalCorr}{  0.55}
\newcommand{\GroupedSystPercMuTrigErrorZSevenTeVttTotZFidGlobalCorr}{  0.05}
\newcommand{\GroupedSystPercMuTrigErrorTTThirtTeVttTotZFidGlobalCorr}{  0.05}
\newcommand{\GroupedSystPercMuTrigErrorTTEightTeVttTotZFidGlobalCorr}{  0.17}
\newcommand{\GroupedSystPercMuTrigErrorTTSevenTeVttTotZFidGlobalCorr}{  0.19}
\newcommand{\GroupedSystPercMuRecIdErrorZThirtTeVttTotZFidGlobalCorr}{  0.68}
\newcommand{\GroupedSystPercMuRecIdErrorZEightTeVttTotZFidGlobalCorr}{  0.45}
\newcommand{\GroupedSystPercMuRecIdErrorZSevenTeVttTotZFidGlobalCorr}{  0.30}
\newcommand{\GroupedSystPercMuRecIdErrorTTThirtTeVttTotZFidGlobalCorr}{  0.44}
\newcommand{\GroupedSystPercMuRecIdErrorTTEightTeVttTotZFidGlobalCorr}{  0.42}
\newcommand{\GroupedSystPercMuRecIdErrorTTSevenTeVttTotZFidGlobalCorr}{  0.31}
\newcommand{\GroupedSystPercMuIsoErrorZThirtTeVttTotZFidGlobalCorr}{  0.41}
\newcommand{\GroupedSystPercMuIsoErrorZEightTeVttTotZFidGlobalCorr}{  0.04}
\newcommand{\GroupedSystPercMuIsoErrorZSevenTeVttTotZFidGlobalCorr}{  0.15}
\newcommand{\GroupedSystPercMuIsoErrorTTThirtTeVttTotZFidGlobalCorr}{  0.27}
\newcommand{\GroupedSystPercMuIsoErrorTTEightTeVttTotZFidGlobalCorr}{  0.22}
\newcommand{\GroupedSystPercMuIsoErrorTTSevenTeVttTotZFidGlobalCorr}{  0.44}
\newcommand{\GroupedSystPercMuScaleErrorZThirtTeVttTotZFidGlobalCorr}{  0.06}
\newcommand{\GroupedSystPercMuScaleErrorZEightTeVttTotZFidGlobalCorr}{  0.03}
\newcommand{\GroupedSystPercMuScaleErrorZSevenTeVttTotZFidGlobalCorr}{  0.03}
\newcommand{\GroupedSystPercMuScaleErrorTTThirtTeVttTotZFidGlobalCorr}{  0.04}
\newcommand{\GroupedSystPercMuScaleErrorTTEightTeVttTotZFidGlobalCorr}{  0.01}
\newcommand{\GroupedSystPercMuScaleErrorTTSevenTeVttTotZFidGlobalCorr}{  0.14}
\newcommand{\GroupedSystPercElTrigErrorZThirtTeVttTotZFidGlobalCorr}{  0.01}
\newcommand{\GroupedSystPercElTrigErrorZEightTeVttTotZFidGlobalCorr}{  0.19}
\newcommand{\GroupedSystPercElTrigErrorZSevenTeVttTotZFidGlobalCorr}{  0.04}
\newcommand{\GroupedSystPercElTrigErrorTTThirtTeVttTotZFidGlobalCorr}{  0.14}
\newcommand{\GroupedSystPercElTrigErrorTTEightTeVttTotZFidGlobalCorr}{  0.00}
\newcommand{\GroupedSystPercElTrigErrorTTSevenTeVttTotZFidGlobalCorr}{  0.00}
\newcommand{\GroupedSystPercElRecIdErrorZThirtTeVttTotZFidGlobalCorr}{  0.41}
\newcommand{\GroupedSystPercElRecIdErrorZEightTeVttTotZFidGlobalCorr}{  0.80}
\newcommand{\GroupedSystPercElRecIdErrorZSevenTeVttTotZFidGlobalCorr}{  0.26}
\newcommand{\GroupedSystPercElRecIdErrorTTThirtTeVttTotZFidGlobalCorr}{  0.34}
\newcommand{\GroupedSystPercElRecIdErrorTTEightTeVttTotZFidGlobalCorr}{  0.41}
\newcommand{\GroupedSystPercElRecIdErrorTTSevenTeVttTotZFidGlobalCorr}{  0.13}
\newcommand{\GroupedSystPercElIsoErrorZThirtTeVttTotZFidGlobalCorr}{  0.14}
\newcommand{\GroupedSystPercElIsoErrorZEightTeVttTotZFidGlobalCorr}{  0.00}
\newcommand{\GroupedSystPercElIsoErrorTTThirtTeVttTotZFidGlobalCorr}{  0.39}
\newcommand{\GroupedSystPercElIsoErrorTTEightTeVttTotZFidGlobalCorr}{  0.30}
\newcommand{\GroupedSystPercElIsoErrorTTSevenTeVttTotZFidGlobalCorr}{  0.59}
\newcommand{\GroupedSystPercElScaleErrorZThirtTeVttTotZFidGlobalCorr}{  0.25}
\newcommand{\GroupedSystPercElScaleErrorZEightTeVttTotZFidGlobalCorr}{  0.07}
\newcommand{\GroupedSystPercElScaleErrorZSevenTeVttTotZFidGlobalCorr}{  0.08}
\newcommand{\GroupedSystPercElScaleErrorTTThirtTeVttTotZFidGlobalCorr}{  0.20}
\newcommand{\GroupedSystPercElScaleErrorTTEightTeVttTotZFidGlobalCorr}{  0.51}
\newcommand{\GroupedSystPercElScaleErrorTTSevenTeVttTotZFidGlobalCorr}{  0.21}
\newcommand{\GroupedSystPercJesErrorTTThirtTeVttTotZFidGlobalCorr}{  0.38}
\newcommand{\GroupedSystPercJesErrorTTEightTeVttTotZFidGlobalCorr}{  0.72}
\newcommand{\GroupedSystPercJesErrorTTSevenTeVttTotZFidGlobalCorr}{  0.40}
\newcommand{\GroupedSystPercFlTagErrorTTThirtTeVttTotZFidGlobalCorr}{  0.53}
\newcommand{\GroupedSystPercFlTagErrorTTEightTeVttTotZFidGlobalCorr}{  0.40}
\newcommand{\GroupedSystPercFlTagErrorTTSevenTeVttTotZFidGlobalCorr}{  0.46}
\newcommand{\GroupedSystPercBKGErrorZThirtTeVttTotZFidGlobalCorr}{  0.08}
\newcommand{\GroupedSystPercBKGErrorZEightTeVttTotZFidGlobalCorr}{  0.15}
\newcommand{\GroupedSystPercBKGErrorZSevenTeVttTotZFidGlobalCorr}{  0.08}
\newcommand{\GroupedSystPercBKGErrorTTThirtTeVttTotZFidGlobalCorr}{  1.09}
\newcommand{\GroupedSystPercBKGErrorTTEightTeVttTotZFidGlobalCorr}{  1.04}
\newcommand{\GroupedSystPercBKGErrorTTSevenTeVttTotZFidGlobalCorr}{  1.04}
\newcommand{\GroupedSystPercSigAndPDFErrorZThirtTeVttTotZFidGlobalCorr}{  0.12}
\newcommand{\GroupedSystPercSigAndPDFErrorZEightTeVttTotZFidGlobalCorr}{  0.08}
\newcommand{\GroupedSystPercSigAndPDFErrorZSevenTeVttTotZFidGlobalCorr}{  0.27}
\newcommand{\GroupedSystPercSigAndPDFErrorTTThirtTeVttTotZFidGlobalCorr}{  2.98}
\newcommand{\GroupedSystPercSigAndPDFErrorTTEightTeVttTotZFidGlobalCorr}{  1.70}
\newcommand{\GroupedSystPercSigAndPDFErrorTTSevenTeVttTotZFidGlobalCorr}{  1.81}

\newcommand{\AveStatZThirteenTevpercentsttTotZFidGlobalCorrNoLumiUnc}{  0.06}
\newcommand{\AveSystZThirteenTevpercentsttTotZFidGlobalCorrNoLumiUnc}{  0.44 }
\newcommand{\AveTotUncZThirteenTevpercentsttTotZFidGlobalCorrNoLumiUnc}{  0.44}
\newcommand{\AveLumiZThirteenTevabsolutettTotZFidGlobalCorrNoLumiUnc}{ 16.32}
\newcommand{\AveCentralZThirteenTevabsvalttTotZFidGlobalCorrNoLumiUnc}{777.35}
\newcommand{\AveStatZThirteenTevabsvalttTotZFidGlobalCorrNoLumiUnc}{  0.46}
\newcommand{\AveSystZThirteenTevabsvalttTotZFidGlobalCorrNoLumiUnc}{  3.38 }
\newcommand{\AveTotUncZThirteenTevabsvalttTotZFidGlobalCorrNoLumiUnc}{  3.41}
\newcommand{\AveStatTTThirteenTevpercentsttTotZFidGlobalCorrNoLumiUnc}{  0.92}
\newcommand{\AveSystTTThirteenTevpercentsttTotZFidGlobalCorrNoLumiUnc}{  3.32 }
\newcommand{\AveTotUncTTThirteenTevpercentsttTotZFidGlobalCorrNoLumiUnc}{  3.45}
\newcommand{\AveLumiTTThirteenTevabsolutettTotZFidGlobalCorrNoLumiUnc}{ 18.90}
\newcommand{\AveCentralTTThirteenTevabsvalttTotZFidGlobalCorrNoLumiUnc}{818.05}
\newcommand{\AveStatTTThirteenTevabsvalttTotZFidGlobalCorrNoLumiUnc}{  7.52}
\newcommand{\AveSystTTThirteenTevabsvalttTotZFidGlobalCorrNoLumiUnc}{ 27.16 }
\newcommand{\AveTotUncTTThirteenTevabsvalttTotZFidGlobalCorrNoLumiUnc}{ 28.19}
\newcommand{\AveStatZEightTevpercentsttTotZFidGlobalCorrNoLumiUnc}{  0.02}
\newcommand{\AveSystZEightTevpercentsttTotZFidGlobalCorrNoLumiUnc}{  0.55 }
\newcommand{\AveTotUncZEightTevpercentsttTotZFidGlobalCorrNoLumiUnc}{  0.55}
\newcommand{\AveLumiZEightTevabsolutettTotZFidGlobalCorrNoLumiUnc}{  9.61}
\newcommand{\AveCentralZEightTevabsvalttTotZFidGlobalCorrNoLumiUnc}{505.84}
\newcommand{\AveStatZEightTevabsvalttTotZFidGlobalCorrNoLumiUnc}{  0.12}
\newcommand{\AveSystZEightTevabsvalttTotZFidGlobalCorrNoLumiUnc}{  2.77 }
\newcommand{\AveTotUncZEightTevabsvalttTotZFidGlobalCorrNoLumiUnc}{  2.77}
\newcommand{\AveStatTTEightTevpercentsttTotZFidGlobalCorrNoLumiUnc}{  0.71}
\newcommand{\AveSystTTEightTevpercentsttTotZFidGlobalCorrNoLumiUnc}{  2.26 }
\newcommand{\AveTotUncTTEightTevpercentsttTotZFidGlobalCorrNoLumiUnc}{  2.37}
\newcommand{\AveLumiTTEightTevabsolutettTotZFidGlobalCorrNoLumiUnc}{  5.11}
\newcommand{\AveCentralTTEightTevabsvalttTotZFidGlobalCorrNoLumiUnc}{242.80}
\newcommand{\AveStatTTEightTevabsvalttTotZFidGlobalCorrNoLumiUnc}{  1.72}
\newcommand{\AveSystTTEightTevabsvalttTotZFidGlobalCorrNoLumiUnc}{  5.48 }
\newcommand{\AveTotUncTTEightTevabsvalttTotZFidGlobalCorrNoLumiUnc}{  5.74}
\newcommand{\AveStatZSevenTevpercentsttTotZFidGlobalCorrNoLumiUnc}{  0.06}
\newcommand{\AveSystZSevenTevpercentsttTotZFidGlobalCorrNoLumiUnc}{  0.32 }
\newcommand{\AveTotUncZSevenTevpercentsttTotZFidGlobalCorrNoLumiUnc}{  0.33}
\newcommand{\AveLumiZSevenTevabsolutettTotZFidGlobalCorrNoLumiUnc}{  8.11}
\newcommand{\AveCentralZSevenTevabsvalttTotZFidGlobalCorrNoLumiUnc}{450.79}
\newcommand{\AveStatZSevenTevabsvalttTotZFidGlobalCorrNoLumiUnc}{  0.28}
\newcommand{\AveSystZSevenTevabsvalttTotZFidGlobalCorrNoLumiUnc}{  1.44 }
\newcommand{\AveTotUncZSevenTevabsvalttTotZFidGlobalCorrNoLumiUnc}{  1.47}
\newcommand{\AveStatTTSevenTevpercentsttTotZFidGlobalCorrNoLumiUnc}{  1.69}
\newcommand{\AveSystTTSevenTevpercentsttTotZFidGlobalCorrNoLumiUnc}{  2.27 }
\newcommand{\AveTotUncTTSevenTevpercentsttTotZFidGlobalCorrNoLumiUnc}{  2.83}
\newcommand{\AveLumiTTSevenTevabsolutettTotZFidGlobalCorrNoLumiUnc}{  3.62}
\newcommand{\AveCentralTTSevenTevabsvalttTotZFidGlobalCorrNoLumiUnc}{182.87}
\newcommand{\AveStatTTSevenTevabsvalttTotZFidGlobalCorrNoLumiUnc}{  3.09}
\newcommand{\AveSystTTSevenTevabsvalttTotZFidGlobalCorrNoLumiUnc}{  4.15 }
\newcommand{\AveTotUncTTSevenTevabsvalttTotZFidGlobalCorrNoLumiUnc}{  5.18}
\newcommand{\AveChisqAllChanEnergZThirteenTevTTThirteenTevZEightTevTTEightTevZSevenTevTTSevenTevFidTotGlobalCorrNoLumiUnc}{0.6}
\newcommand{\CorrCombZThirteenTevcorZThirteenTevttTotZFidGlobalCorrNoLumiUnc}{ 1.000}
\newcommand{\CorrCombTTThirteenTevcorZThirteenTevttTotZFidGlobalCorrNoLumiUnc}{ 0.132}
\newcommand{\CorrCombZEightTevcorZThirteenTevttTotZFidGlobalCorrNoLumiUnc}{ 0.091}
\newcommand{\CorrCombTTEightTevcorZThirteenTevttTotZFidGlobalCorrNoLumiUnc}{ 0.084}
\newcommand{\CorrCombZSevenTevcorZThirteenTevttTotZFidGlobalCorrNoLumiUnc}{ 0.123}
\newcommand{\CorrCombTTSevenTevcorZThirteenTevttTotZFidGlobalCorrNoLumiUnc}{ 0.031}
\newcommand{\CorrCombTTThirteenTevcorTTThirteenTevttTotZFidGlobalCorrNoLumiUnc}{ 1.000}
\newcommand{\CorrCombZEightTevcorTTThirteenTevttTotZFidGlobalCorrNoLumiUnc}{ 0.013}
\newcommand{\CorrCombTTEightTevcorTTThirteenTevttTotZFidGlobalCorrNoLumiUnc}{ 0.315}
\newcommand{\CorrCombZSevenTevcorTTThirteenTevttTotZFidGlobalCorrNoLumiUnc}{ 0.002}
\newcommand{\CorrCombTTSevenTevcorTTThirteenTevttTotZFidGlobalCorrNoLumiUnc}{ 0.274}
\newcommand{\CorrCombZEightTevcorZEightTevttTotZFidGlobalCorrNoLumiUnc}{ 1.000}
\newcommand{\CorrCombTTEightTevcorZEightTevttTotZFidGlobalCorrNoLumiUnc}{ 0.009}
\newcommand{\CorrCombZSevenTevcorZEightTevttTotZFidGlobalCorrNoLumiUnc}{ 0.090}
\newcommand{\CorrCombTTSevenTevcorZEightTevttTotZFidGlobalCorrNoLumiUnc}{ 0.004}
\newcommand{\CorrCombTTEightTevcorTTEightTevttTotZFidGlobalCorrNoLumiUnc}{ 1.000}
\newcommand{\CorrCombZSevenTevcorTTEightTevttTotZFidGlobalCorrNoLumiUnc}{ 0.002}
\newcommand{\CorrCombTTSevenTevcorTTEightTevttTotZFidGlobalCorrNoLumiUnc}{ 0.674}
\newcommand{\CorrCombZSevenTevcorZSevenTevttTotZFidGlobalCorrNoLumiUnc}{ 1.000}
\newcommand{\CorrCombTTSevenTevcorZSevenTevttTotZFidGlobalCorrNoLumiUnc}{ 0.002}
\newcommand{\CorrCombTTSevenTevcorTTSevenTevttTotZFidGlobalCorrNoLumiUnc}{ 1.000}

\newcommand{\XsecFidZmmThirteen}{774.43}
\newcommand{\XsecFidZmmThirteenStatUncPerc}{  0.08}
\newcommand{\XsecFidZmmThirteenStatUncAbs}{  0.59}
\newcommand{\XsecFidZmmThirteenSystUncPerc}{  2.35}
\newcommand{\XsecFidZmmThirteenSystUncAbs}{ 18.23}
\newcommand{\XsecFidZeeThirteen}{778.26}
\newcommand{\XsecFidZeeThirteenStatUncPerc}{  0.09}
\newcommand{\XsecFidZeeThirteenStatUncAbs}{  0.67}
\newcommand{\XsecFidZeeThirteenSystUncPerc}{  2.27}
\newcommand{\XsecFidZeeThirteenSystUncAbs}{ 17.67}
\newcommand{\XsecFidttbarThirteenTeV}{  9.94}
\newcommand{\XsecFidttbarThirteenTeVStatUncPerc}{  0.91}
\newcommand{\XsecFidttbarThirteenTeVStatUncAbs}{  0.09}
\newcommand{\XsecFidttbarThirteenTeVSystUncPerc}{  3.74}
\newcommand{\XsecFidttbarThirteenTeVSystUncAbs}{  0.37}
\newcommand{\XsecFidZmmEightTeVOwnFidPhaseSpace}{536.31}
\newcommand{\XsecFidZmmEightTeV}{504.74}
\newcommand{\XsecFidZmmEightTeVStatUncPerc}{  0.03}
\newcommand{\XsecFidZmmEightTeVStatUncAbs}{  0.15}
\newcommand{\XsecFidZmmEightTeVSystUncPerc}{  2.14}
\newcommand{\XsecFidZmmEightTeVSystUncAbs}{ 10.82}
\newcommand{\XsecFidZeeEightTeVOwnFidPhaseSpace}{538.70}
\newcommand{\XsecFidZeeEightTeV}{506.99}
\newcommand{\XsecFidZeeEightTeVStatUncPerc}{  0.04}
\newcommand{\XsecFidZeeEightTeVStatUncAbs}{  0.20}
\newcommand{\XsecFidZeeEightTeVSystUncPerc}{  2.18}
\newcommand{\XsecFidZeeEightTeVSystUncAbs}{ 11.03}
\newcommand{\XsecFidttbarEightTeV}{  3.04}
\newcommand{\XsecFidttbarEightTeVStatUncPerc}{  0.72}
\newcommand{\XsecFidttbarEightTeVStatUncAbs}{  0.02}
\newcommand{\XsecFidttbarEightTeVSystUncPerc}{  3.38}
\newcommand{\XsecFidttbarEightTeVSystUncAbs}{  0.10}
\newcommand{\XsecFidZmmSevenTeVOwnFidPhaseSpace}{501.38}
\newcommand{\XsecFidZmmSevenTeV}{450.02}
\newcommand{\XsecFidZmmSevenTeVStatUncPerc}{  0.08}
\newcommand{\XsecFidZmmSevenTeVStatUncAbs}{  0.34}
\newcommand{\XsecFidZmmSevenTeVSystUncPerc}{  1.95}
\newcommand{\XsecFidZmmSevenTeVSystUncAbs}{  8.76}
\newcommand{\XsecFidZeeSevenTeVOwnFidPhaseSpace}{502.66}
\newcommand{\XsecFidZeeSevenTeV}{451.17}
\newcommand{\XsecFidZeeSevenTeVStatUncPerc}{  0.10}
\newcommand{\XsecFidZeeSevenTeVStatUncAbs}{  0.45}
\newcommand{\XsecFidZeeSevenTeVSystUncPerc}{  1.93}
\newcommand{\XsecFidZeeSevenTeVSystUncAbs}{  8.71}
\newcommand{\XsecFidttbarSevenTeV}{  2.30}
\newcommand{\XsecFidttbarSevenTeVStatUncPerc}{  1.68}
\newcommand{\XsecFidttbarSevenTeVStatUncAbs}{  0.04}
\newcommand{\XsecFidttbarSevenTeVSystUncPerc}{  3.41}
\newcommand{\XsecFidttbarSevenTeVSystUncAbs}{  0.08}

\newcommand{\XsecTotZmmThirteen}{1962.08}
\newcommand{\XsecTotZmmThirteenStatUncPerc}{  0.08}
\newcommand{\XsecTotZmmThirteenStatUncAbs}{  1.49}
\newcommand{\XsecTotZmmThirteenSystUncPerc}{  2.95}
\newcommand{\XsecTotZmmThirteenSystUncAbs}{ 57.82}
\newcommand{\XsecTotZeeThirteen}{1971.79}
\newcommand{\XsecTotZeeThirteenStatUncPerc}{  0.09}
\newcommand{\XsecTotZeeThirteenStatUncAbs}{  1.70}
\newcommand{\XsecTotZeeThirteenSystUncPerc}{  2.88}
\newcommand{\XsecTotZeeThirteenSystUncAbs}{ 56.80}
\newcommand{\XsecTotttbarThirteenTeV}{817.53}
\newcommand{\XsecTotttbarThirteenTeVStatUncPerc}{  0.92}
\newcommand{\XsecTotttbarThirteenTeVStatUncAbs}{  7.52}
\newcommand{\XsecTotttbarThirteenTeVSystUncPerc}{  4.32}
\newcommand{\XsecTotttbarThirteenTeVSystUncAbs}{ 35.31}
\newcommand{\XsecTotZmmEightTeV}{1151.37}
\newcommand{\XsecTotZmmEightTeVStatUncPerc}{  0.03}
\newcommand{\XsecTotZmmEightTeVStatUncAbs}{  0.34}
\newcommand{\XsecTotZmmEightTeVSystUncPerc}{  2.75}
\newcommand{\XsecTotZmmEightTeVSystUncAbs}{ 31.63}
\newcommand{\XsecTotZeeEightTeV}{1156.50}
\newcommand{\XsecTotZeeEightTeVStatUncPerc}{  0.04}
\newcommand{\XsecTotZeeEightTeVStatUncAbs}{  0.46}
\newcommand{\XsecTotZeeEightTeVSystUncPerc}{  2.77}
\newcommand{\XsecTotZeeEightTeVSystUncAbs}{ 32.05}
\newcommand{\XsecTotttbarEightTeV}{242.94}
\newcommand{\XsecTotttbarEightTeVStatUncPerc}{  0.71}
\newcommand{\XsecTotttbarEightTeVStatUncAbs}{  1.73}
\newcommand{\XsecTotttbarEightTeVSystUncPerc}{  3.53}
\newcommand{\XsecTotttbarEightTeVSystUncAbs}{  8.57}
\newcommand{\XsecTotZmmSevenTeV}{993.22}
\newcommand{\XsecTotZmmSevenTeVStatUncPerc}{  0.00}
\newcommand{\XsecTotZmmSevenTeVStatUncAbs}{  0.75}
\newcommand{\XsecTotZmmSevenTeVSystUncPerc}{  2.64}
\newcommand{\XsecTotZmmSevenTeVSystUncAbs}{ 26.22}
\newcommand{\XsecTotZeeSevenTeV}{995.75}
\newcommand{\XsecTotZeeSevenTeVStatUncPerc}{  0.00}
\newcommand{\XsecTotZeeSevenTeVStatUncAbs}{  0.99}
\newcommand{\XsecTotZeeSevenTeVSystUncPerc}{  2.63}
\newcommand{\XsecTotZeeSevenTeVSystUncAbs}{ 26.16}
\newcommand{\XsecTotttbarSevenTeV}{182.90}
\newcommand{\XsecTotttbarSevenTeVStatUncPerc}{  1.69}
\newcommand{\XsecTotttbarSevenTeVStatUncAbs}{  3.09}
\newcommand{\XsecTotttbarSevenTeVSystUncPerc}{  3.51}
\newcommand{\XsecTotttbarSevenTeVSystUncAbs}{  6.41}

\newcommand{\xzee}{\mbox{$\sigma_{\Zee}$}}
\newcommand{\xzmm}{\mbox{$\sigma_{\Zmm}$}}
\newcommand{\xtt}{\mbox{$\sigma_{\ttbar}$}}
\newcommand{\rttz}{\mbox{$R_{\ttbar/Z}$}}

\newcommand{\xzeeThirtTot}{\mbox{$\sigma^{\Zee}_{\textrm{tot}}({\textrm{13TeV}})$}}
\newcommand{\xzmmThirtTot}{\mbox{$\sigma^{\Zmm}^{\textrm{tot}}({\textrm{13TeV}})$}}
\newcommand{\xzThirtTot}{\mbox{$\sigma_{Z}^{\textrm{tot}}({\textrm{13~TeV}})$}}
\newcommand{\xttThirtTot}{\mbox{$\sigma_{\ttbar}^{\textrm{tot}}({\textrm{13~TeV}})$}}

\newcommand{\xzeeThirtFid}{\mbox{$\sigma_{\Zee}^{\textrm{fid}}({\textrm{13~TeV}})$}}
\newcommand{\xzmmThirtFid}{\mbox{$\sigma_{\Zmm}^{\textrm{fid}}({\textrm{13~TeV}})$}}
\newcommand{\xzThirtFid}{\mbox{$\sigma_{Z}^{\textrm{fid}}({\textrm{13~TeV}})$}}
\newcommand{\xttThirtFid}{\mbox{$\sigma_{\ttbar}^{\textrm{fid}}({\textrm{13~TeV}})$}}

\newcommand{\xzeeEightTot}{\mbox{$\sigma_{\Zee}^{\textrm{tot}}({\textrm{8~TeV}})$}}
\newcommand{\xzmmEightTot}{\mbox{$\sigma_{\Zmm}^{\textrm{tot}}({\textrm{8~TeV}})$}}
\newcommand{\xzEightTot}{\mbox{$\sigma_{Z}^{\textrm{tot}}({\textrm{8~TeV}})$}}
\newcommand{\xttEightTot}{\mbox{$\sigma_{\ttbar}^{\textrm{tot}}({\textrm{8~TeV}})$}}

\newcommand{\xzeeEightFid}{\mbox{$\sigma_{\Zee}^{\textrm{fid}}({\textrm{8~TeV}})$}}
\newcommand{\xzmmEightFid}{\mbox{$\sigma_{\Zmm}^{\textrm{fid}}({\textrm{8~TeV}})$}}
\newcommand{\xzEightFid}{\mbox{$\sigma_{Z}^{\textrm{fid}}({\textrm{8~TeV}})$}}
\newcommand{\xttEightFid}{\mbox{$\sigma_{\ttbar}^{\textrm{fid}}({\textrm{8~TeV}})$}}

\newcommand{\xzeeSevenTot}{\mbox{$\sigma_{\Zee}^{\textrm{tot}}({\textrm{7~TeV}})$}}
\newcommand{\xzmmSevenTot}{\mbox{$\sigma_{\Zmm}^{\textrm{tot}}({\textrm{7~TeV}})$}}
\newcommand{\xzSevenTot}{\mbox{$\sigma_{Z}^{\textrm{tot}}({\textrm{7~TeV}})$}}
\newcommand{\xttSevenTot}{\mbox{$\sigma_{\ttbar}^{\textrm{tot}}({\textrm{7~TeV}})$}}

\newcommand{\xzeeSevenFid}{\mbox{$\sigma_{\Zee}^{\textrm{fid}}({\textrm{7~TeV}})$}}
\newcommand{\xzmmSevenFid}{\mbox{$\sigma_{\Zmm}^{\textrm{fid}}({\textrm{7~TeV}})$}}
\newcommand{\xzSevenFid}{\mbox{$\sigma_{Z}^{\textrm{fid}}({\textrm{7~TeV}})$}}
\newcommand{\xttSevenFid}{\mbox{$\sigma_{\ttbar}^{\textrm{fid}}({\textrm{7~TeV}})$}}

\newcommand{\rttTotThirtToZFidThirt}{\mbox{$R_{\ttbar(tot ~ 13TeV) / Z(fid ~ 13TeV)}$}}
\newcommand{\rttTotThirtToZTotThirt}{\mbox{$R_{\ttbar(tot ~ 13TeV) / Z(tot ~ 13TeV)}$}}
\newcommand{\rttFidThirtToZFidThirt}{\mbox{$R_{\ttbar(fid ~ 13TeV) / Z(fid ~ 13TeV)}$}}
\newcommand{\rttThirtToZThirt}{$\sigma^{13~TeV}_{t \bar{t}} / \sigma^{13~TeV}_{Z}$}

\newcommand{\rttTotEightToZFidEight}{\mbox{$R_{\ttbar(tot ~ 8TeV) / Z(fid ~ 8TeV)}$}}
\newcommand{\rttTotEightToZTotEight}{\mbox{$R_{\ttbar(tot ~ 8TeV) / Z(tot ~ 8TeV)}$}}
\newcommand{\rttFidEightToZFidEight}{\mbox{$R_{\ttbar(fid ~ 8TeV) / Z(fid ~ 8TeV)}$}}
\newcommand{\rttEightToZEight}{$\sigma^{8~TeV}_{t \bar{t}} / \sigma^{8~TeV}_{Z}$}

\newcommand{\rttTotSevenToZFidSeven}{\mbox{$R_{\ttbar(tot ~ 7TeV) / Z(fid ~ 7TeV)}$}}
\newcommand{\rttTotSevenToZTotSeven}{\mbox{$R_{\ttbar(tot ~ 7TeV) / Z(tot ~ 7TeV)}$}}
\newcommand{\rttFidSevenToZFidSeven}{\mbox{$R_{\ttbar(fid ~ 7TeV) / Z(fid ~ 7TeV)}$}}
\newcommand{\rttSevenToZSeven}{$\sigma^{7~TeV}_{t \bar{t}} / \sigma^{7~TeV}_{Z}$}

\newcommand{\rttTotThirtTottTotEight}{\mbox{$R_{\ttbar(tot ~ 13TeV) / \ttbar(tot ~ 8TeV)}$}}
\newcommand{\rttFidThirtTottFidEight}{\mbox{$R_{\ttbar(fid ~ 13TeV) / \ttbar(fid ~ 8TeV)}$}}
\newcommand{\rttThirtTottEight}{$\sigma^{13~TeV}_{t \bar{t}} / \sigma^{8~TeV}_{t \bar{t}}$}

\newcommand{\rZTotThirtToZTotEight}{\mbox{$R_{Z(tot ~ 13TeV) / Z(tot ~ 8TeV)}$}}
\newcommand{\rZFidThirtToZFidEight}{\mbox{$R_{Z(fid ~ 13TeV) / Z(fid ~ 8TeV)}$}}
\newcommand{\rZThirtToZEight}{$\sigma^{13~TeV}_{Z} / \sigma^{8~TeV}_{Z}$}

\newcommand{\rttTotThirtTottTotSeven}{\mbox{$R_{\ttbar(tot ~ 13TeV) / \ttbar(tot ~ 7TeV)}$}}
\newcommand{\rttFidThirtTottFidSeven}{\mbox{$R_{\ttbar(fid ~ 13TeV) / \ttbar(fid ~ 7TeV)}$}}
\newcommand{\rttThirtTottSeven}{$\sigma^{13~TeV}_{t \bar{t}} / \sigma^{7~TeV}_{t \bar{t}}$}

\newcommand{\rZTotThirtToZTotSeven}{\mbox{$R_{Z(tot ~ 13TeV) / Z(tot ~ 7TeV)}$}}
\newcommand{\rZFidThirtToZFidSeven}{\mbox{$R_{Z(fid ~ 13TeV) / Z(fid ~ 7TeV)}$}}
\newcommand{\rZThirtToZSeven}{$\sigma^{13~TeV}_{Z} / \sigma^{7~TeV}_{Z}$}

\newcommand{\rttTotEightTottTotSeven}{\mbox{$R_{\ttbar(tot ~ 8TeV) / \ttbar(tot ~ 7TeV)}$}}
\newcommand{\rttFidEightTottFidSeven}{\mbox{$R_{\ttbar(fid ~ 8TeV) / \ttbar(fid ~ 7TeV)}$}}
\newcommand{\rttEightTottSeven}{$\sigma^{8~TeV}_{t \bar{t}} / \sigma^{7~TeV}_{t \bar{t}}$}

\newcommand{\rZTotEightToZTotSeven}{\mbox{$R_{Z(tot ~ 8TeV) / Z(tot ~ 7TeV)}$}}
\newcommand{\rZFidEightToZFidSeven}{\mbox{$R_{Z(fid ~ 8TeV) / Z(fid ~ 7TeV)}$}}
\newcommand{\rZEightToZSeven}{$\sigma^{8~TeV}_{Z} / \sigma^{7~TeV}_{Z}$}

\newcommand{\dRttThirtToZThirtTTeightToZeight}{$ \frac{\sigma^{13~TeV}_{t \bar{t}}}{\sigma^{13~TeV}_{Z}} / \frac{\sigma^{8~TeV}_{t \bar{t}}}{\sigma^{8~TeV}_{Z}} $}

\newcommand{\dRttThirtToZThirtTTsevenToZseven}{$ \frac{\sigma^{13~TeV}_{t \bar{t}}}{\sigma^{13~TeV}_{Z}} / \frac{\sigma^{7~TeV}_{t \bar{t}}}{\sigma^{7~TeV}_{Z}} $}

\newcommand{\dRttEightToZEightTTsevenToZseven}{$ \frac{\sigma^{8~TeV}_{t \bar{t}}}{\sigma^{8~TeV}_{Z}} / \frac{\sigma^{7~TeV}_{t \bar{t}}}{\sigma^{7~TeV}_{Z}} $}

\subsection{Methodology}

The following ratios are considered in this section: $\ratzf{i}{j}, \rattt{i}{j}, \ratttztf(i~\mathrm{TeV})$, and $\ratttztf(i/j)$ where $i,j=13, 8, 7$
and $i\ne j$. The corresponding ratios using the \Zboson-boson total cross section are reported in Appendix~\ref{AppC}. Ratios using fiducial \ttbar cross sections
are also reported in Appendix~\ref{AppC}, although there are no NNLO calculations available yet for \ttbar production cross sections with requirements on the final-state
leptons.

For the evaluation of the \ttbar/\Zboson ratios, $\ratztt$, the \Zboson-boson cross sections from the electron and muon channels are both employed and taken with the same weight in the ratio, i.e.
\begin{equation}
  \ratztt  = \frac{ \sigtt } { 0.5 \left( \sigma_{\Zee} + \sigma_{\Zmm} \right)}
\end{equation}
since the \ttbar production cross section is measured from the electron and muon pair final state topology.
This ensures the best cancellation of important systematic uncertainties related to lepton reconstruction, identification, and trigger.
For other ratios involving \Zboson bosons, the $Z\rightarrow e^+e^-$ and $Z\rightarrow \mu^+\mu^-$ results are combined (Section~\ref{sec:correlsigma} describes the results of the combination) using the code described in Refs.~\cite{Glazov:2005rn,Aaron:2009bp}, taking into account
correlations of  systematic uncertainties across channels and $\sqrt s$.

\subsection{Inputs to the ratios}

The primary inputs to the ratios are the \Zboson-boson and \ttbar\ production cross sections at $13, 8, 7$~TeV~\cite{STDM-2012-20,STDM-2011-06,Aaboud:2016pbd,TOPADDENDUM2}, 
each obtained with its own experimental selection criteria, measured within an experimental phase space, and reported in a corresponding fiducial phase space or in the total phase space. 
The event topologies of the two processes are independent of the centre-of-mass energy. The \Zboson-boson selections target two isolated, same-flavour, opposite-charge reconstructed leptons, identified as electrons or muons, whose dilepton invariant mass is consistent with that of a \Zboson boson. The \ttbar topology specific to this paper is that of an opposite-charge, isolated electron and muon pair, and additional jets tagged as containing $b$-hadrons. Although the \ttbar fiducial phase space has remained unchanged at $13, 8, 7$~TeV (lepton $\pT>25$~GeV and $|\eta|<2.5$), this has not been the case for the \Zboson-boson measurements, in some part due to the evolution of the trigger requirements as the peak luminosity and the degree of pile-up from the LHC have increased with time. Table~\ref{table:fid_def} reports the fiducial phase space used in the  $13, 8, 7$~TeV measurements of the \Zboson-boson fiducial cross sections. In this paper, all ratios involving \Zboson bosons at 7~and~8~TeV are extrapolated to the 13~TeV phase space using the same methodology as reported in Section~\ref{sec:ztheory}, i.e. computed using an optimised version of \textsc{DYNNLO}~1.5 and the NNLO parton distribution functions \CT. These 13-to-7~TeV and 13-to-8~TeV extrapolation factors, $E$, are multiplicative factors to the cross sections, and are also reported in Table~\ref{table:fid_def}.

\begin{table}[t]
\centering
\begin{tabular}{|l|l|l|l|}
\hline
$\sqrt s$~[TeV]              &  \multicolumn{1}{c|}{13}     &  \multicolumn{1}{c|}{8}      &  \multicolumn{1}{c|}{7} \\ \hline
$p_{\text{T}}^{\ell}>$                  &  25~GeV   & 20~GeV     & 20~GeV  \\ 
$|\eta_{\ell}|<$               & 2.5      & 2.4        & 2.5  \\
$|y_{\ell\ell}|<$               & -       & 2.4        & - \\
$m_{\ell\ell}$         & 66--116~GeV &  66--116~GeV &  66--116~GeV \\ \hline
Extrapolation $E$      &  - & $0.941 \pm 0.001$~(PDF) & $0.898 \pm 0.001$~(PDF)  \\ 
\hline
\end{tabular}
\caption{\Zboson-boson fiducial definition at $\sqrt s = 13, 8, 7$~TeV. The ratios measured in this analysis are calculated in the 13~TeV phase space for all $\sqrt s$. The factor $E$ is used to extrapolate the 7 and 8~TeV results to the common phase space defined by the 13~TeV results. The PDF uncertainty is obtained from the \CT eigenvector set.} 
\label{table:fid_def}
\end{table}

Table~\ref{table:primary_input} summarises the primary inputs, in the common 13~TeV phase space for the \Zboson-boson measurements, that enter the cross-section ratios, including the statistical and total systematic uncertainties, the latter  encompassing experimental, luminosity, beam-energy, and some theoretical uncertainties (as explained in Section~\ref{corrmodel}). These results are taken directly from the publications and from Section~\ref{sec:zll}, with one exception: since the publication of the 8~TeV \Zboson-boson fiducial cross section~\cite{STDM-2014-12}, the 8~TeV luminosity values have been finalised~\cite{Aaboud:2016hhf}, resulting in a slight shift of the integrated luminosity value from  the published 20.3~fb$^{-1}$ to 20.2~fb$^{-1}$ and significantly reducing the uncertainty from 2.8\% to 1.9\%. The 8~TeV results presented here have been updated accordingly. The \ttbar fiducial cross-section results in Table~\ref{table:primary_input} are reported in the phase space defined by lepton $\pT>25$~GeV and $|\eta|<2.5$ and for which the contribution from $W \rightarrow \tau \rightarrow \ell$ decay has been subtracted.
The breakdown of the systematic uncertainties is presented in Table~\ref{tab:systerrorsitemized} while the correlation model for the uncertainties is elaborated in the next subsection.

\begin{table}[t]
\centering
\begin{tabular}{|l|r|r|r|}
\hline
            & \multicolumn{3}{c|}{$\sigma \pm$ stat $\pm$  syst [pb]} \\
$\sqrt s$ [TeV]  & \multicolumn{1}{c|}{13}     &  \multicolumn{1}{c|}{8}      &  \multicolumn{1}{c|}{7} \\ \hline 
\sigzfidee  &  
               $\numRP{\XsecFidZeeThirteen}{1} \pm  \numRP{\XsecFidZeeThirteenStatUncAbs}{1} \pm \numRP{\XsecFidZeeThirteenSystUncAbs}{1} $ &
               $\numRP{\XsecFidZeeEightTeV}{1} \pm  \numRP{\XsecFidZeeEightTeVStatUncAbs}{1} \pm \numRP{\XsecFidZeeEightTeVSystUncAbs}{1} $ &
               $\numRP{\XsecFidZeeSevenTeV}{1} \pm  \numRP{\XsecFidZeeSevenTeVStatUncAbs}{1} \pm \numRP{\XsecFidZeeSevenTeVSystUncAbs}{1} $ 
\T \B \\
\sigzfidmm  &  
               $\numRP{\XsecFidZmmThirteen}{1} \pm  \numRP{\XsecFidZmmThirteenStatUncAbs}{1} \pm \numRP{\XsecFidZmmThirteenSystUncAbs}{1} $ &
               $\numRP{\XsecFidZmmEightTeV}{1} \pm  \numRP{\XsecFidZmmEightTeVStatUncAbs}{1} \pm \numRP{\XsecFidZmmEightTeVSystUncAbs}{1} $ &
               $\numRP{\XsecFidZmmSevenTeV}{1} \pm  \numRP{\XsecFidZmmSevenTeVStatUncAbs}{1} \pm \numRP{\XsecFidZmmSevenTeVSystUncAbs}{1} $
\B \\

\sigttfid   &  
               $\numRP{\XsecFidttbarThirteenTeV}{2} \pm  \numRP{\XsecFidttbarThirteenTeVStatUncAbs}{2} \pm \numRP{\XsecFidttbarThirteenTeVSystUncAbs}{2}$&
               $\numRP{\XsecFidttbarEightTeV}{2} \pm  \numRP{\XsecFidttbarEightTeVStatUncAbs}{2} \pm \numRP{\XsecFidttbarEightTeVSystUncAbs}{2} $ &
               $\numRP{\XsecFidttbarSevenTeV}{2} \pm  \numRP{\XsecFidttbarSevenTeVStatUncAbs}{2} \pm \numRP{\XsecFidttbarSevenTeVSystUncAbs}{2} $ 
 \T \\
\sigtttot   &  
               $\numRP{\XsecTotttbarThirteenTeV}{0} \pm  \numRP{\XsecTotttbarThirteenTeVStatUncAbs}{0} \pm \numRP{\XsecTotttbarThirteenTeVSystUncAbs}{0}$ &
               $\numRP{\XsecTotttbarEightTeV}{0} \pm  \numRP{\XsecTotttbarEightTeVStatUncAbs}{0} \pm \numRP{\XsecTotttbarEightTeVSystUncAbs}{0} $ &
               $\numRP{\XsecTotttbarSevenTeV}{0} \pm  \numRP{\XsecTotttbarSevenTeVStatUncAbs}{0} \pm \numRP{\XsecTotttbarSevenTeVSystUncAbs}{0} $
 \B \\
\hline 
\end{tabular}
\caption{Fiducial and total cross sections at $\sqrt s = 13, 8, 7$~TeV that form the primary input to the cross-section ratios. The \Zboson-boson cross sections are provided in the common 13~TeV phase space. The systematic uncertainties include experimental, luminosity, beam-energy, and some theoretical uncertainties (see text).  }
\label{table:primary_input}
\end{table}

\subsection{Correlation model}
\label{corrmodel}
\begin{table}[t]
\centering
\begin{tabular}{|l|ccc|ccc|}
\hline
                       & \multicolumn{3}{c|}{$\delta$ \sfidZ} &     \multicolumn{3}{c|}{$\delta$ \stottt} \T \\
 Systematic [\%] \hspace{0.3cm} / \hspace{0.5cm} \sqs [TeV]              & $13$ &  $8$ &  $7$ &  $13$ &  $8$ &  $7$ \\
\hline
  Luminosity                     & 
   \numRP{\GroupedSystPercLumiErrorZThirtTeVttTotZFidGlobalCorr}{1}   & 
   \numRP{\GroupedSystPercLumiErrorZEightTeVttTotZFidGlobalCorr}{1}   & 
   \numRP{\GroupedSystPercLumiErrorZSevenTeVttTotZFidGlobalCorr}{1}  &
   \numRP{\GroupedSystPercLumiErrorTTThirtTeVttTotZFidGlobalCorr}{1}  & 
   \numRP{\GroupedSystPercLumiErrorTTEightTeVttTotZFidGlobalCorr}{1}  & 
   \numRP{\GroupedSystPercLumiErrorTTSevenTeVttTotZFidGlobalCorr}{1} \\
  Beam energy                    & 
   \numRP{\GroupedSystPercEBeamErrorZThirtTeVttTotZFidGlobalCorr}{1}  & 
   \numRP{\GroupedSystPercEBeamErrorZEightTeVttTotZFidGlobalCorr}{1}  & 
   \numRP{\GroupedSystPercEBeamErrorZSevenTeVttTotZFidGlobalCorr}{1}  & 
   \numRP{\GroupedSystPercEBeamErrorTTThirtTeVttTotZFidGlobalCorr}{1} & 
   \numRP{\GroupedSystPercEBeamErrorTTEightTeVttTotZFidGlobalCorr}{1} & 
   \numRP{\GroupedSystPercEBeamErrorTTSevenTeVttTotZFidGlobalCorr}{1} \\
\hline
  Muon (lepton) trigger                   & 
   \numRP{\GroupedSystPercMuTrigErrorZThirtTeVttTotZFidGlobalCorr}{1}  & 
   \numRP{\GroupedSystPercMuTrigErrorZEightTeVttTotZFidGlobalCorr}{1}  & 
   \numRP{\GroupedSystPercMuTrigErrorZSevenTeVttTotZFidGlobalCorr}{1}  &
   \numRP{\GroupedSystPercMuTrigErrorTTThirtTeVttTotZFidGlobalCorr}{1} & 
   \numRP{\GroupedSystPercMuTrigErrorTTEightTeVttTotZFidGlobalCorr}{1} & 
   \numRP{\GroupedSystPercMuTrigErrorTTSevenTeVttTotZFidGlobalCorr}{1} \\
  Muon reconstruction/ID               & 
   \numRP{\GroupedSystPercMuRecIdErrorZThirtTeVttTotZFidGlobalCorr}{1} & 
   \numRP{\GroupedSystPercMuRecIdErrorZEightTeVttTotZFidGlobalCorr}{1} & 
   \numRP{\GroupedSystPercMuRecIdErrorZSevenTeVttTotZFidGlobalCorr}{1} &
   \numRP{\GroupedSystPercMuRecIdErrorTTThirtTeVttTotZFidGlobalCorr}{1} & 
   \numRP{\GroupedSystPercMuRecIdErrorTTEightTeVttTotZFidGlobalCorr}{1} & 
   \numRP{\GroupedSystPercMuRecIdErrorTTSevenTeVttTotZFidGlobalCorr}{1} \\
  Muon isolation                 & 
   \numRP{\GroupedSystPercMuIsoErrorZThirtTeVttTotZFidGlobalCorr}{1}    & 
   \numRP{\GroupedSystPercMuIsoErrorZEightTeVttTotZFidGlobalCorr}{1}    & 
   \numRP{\GroupedSystPercMuIsoErrorZSevenTeVttTotZFidGlobalCorr}{1} &
   \numRP{\GroupedSystPercMuIsoErrorTTThirtTeVttTotZFidGlobalCorr}{1}   & 
   \numRP{\GroupedSystPercMuIsoErrorTTEightTeVttTotZFidGlobalCorr}{1}   & 
   \numRP{\GroupedSystPercMuIsoErrorTTSevenTeVttTotZFidGlobalCorr}{1}   \\
  Muon momentum scale            & 
   \numRP{\GroupedSystPercMuScaleErrorZThirtTeVttTotZFidGlobalCorr}{1}  & 
   \numRP{\GroupedSystPercMuScaleErrorZEightTeVttTotZFidGlobalCorr}{1}  & 
   \numRP{\GroupedSystPercMuScaleErrorZSevenTeVttTotZFidGlobalCorr}{1} &
   \numRP{\GroupedSystPercMuScaleErrorTTThirtTeVttTotZFidGlobalCorr}{1} & 
   \numRP{\GroupedSystPercMuScaleErrorTTEightTeVttTotZFidGlobalCorr}{1} & 
   \numRP{\GroupedSystPercMuScaleErrorTTSevenTeVttTotZFidGlobalCorr}{1}  \\
\hline
  Electron trigger               & 
   \numRP{\GroupedSystPercElTrigErrorZThirtTeVttTotZFidGlobalCorr}{1}   & 
   \numRP{\GroupedSystPercElTrigErrorZEightTeVttTotZFidGlobalCorr}{1}   & 
   \numRP{\GroupedSystPercElTrigErrorZSevenTeVttTotZFidGlobalCorr}{1}  &
   \numRP{\GroupedSystPercElTrigErrorTTThirtTeVttTotZFidGlobalCorr}{1}  &  ---                                                      & --- \\
  Electron reconstruction/ID           & 
   \numRP{\GroupedSystPercElRecIdErrorZThirtTeVttTotZFidGlobalCorr}{1}  & 
   \numRP{\GroupedSystPercElRecIdErrorZEightTeVttTotZFidGlobalCorr}{1}  & 
   \numRP{\GroupedSystPercElRecIdErrorZSevenTeVttTotZFidGlobalCorr}{1}  &
   \numRP{\GroupedSystPercElRecIdErrorTTThirtTeVttTotZFidGlobalCorr}{1} & 
   \numRP{\GroupedSystPercElRecIdErrorTTEightTeVttTotZFidGlobalCorr}{1} & 
   \numRP{\GroupedSystPercElRecIdErrorTTSevenTeVttTotZFidGlobalCorr}{1} \\
  Electron isolation             & 
   \numRP{\GroupedSystPercElIsoErrorZThirtTeVttTotZFidGlobalCorr}{1}    & 
   \numRP{\GroupedSystPercElIsoErrorZEightTeVttTotZFidGlobalCorr}{1}    &  ---                                                      &
   \numRP{\GroupedSystPercElIsoErrorTTThirtTeVttTotZFidGlobalCorr}{1}   & 
   \numRP{\GroupedSystPercElIsoErrorTTEightTeVttTotZFidGlobalCorr}{1}   & 
   \numRP{\GroupedSystPercElIsoErrorTTSevenTeVttTotZFidGlobalCorr}{1}   \\
  Electron energy scale          & 
   \numRP{\GroupedSystPercElScaleErrorZThirtTeVttTotZFidGlobalCorr}{1}  & 
   \numRP{\GroupedSystPercElScaleErrorZEightTeVttTotZFidGlobalCorr}{1}  & 
   \numRP{\GroupedSystPercElScaleErrorZSevenTeVttTotZFidGlobalCorr}{1}  &
   \numRP{\GroupedSystPercElScaleErrorTTThirtTeVttTotZFidGlobalCorr}{1} & 
   \numRP{\GroupedSystPercElScaleErrorTTEightTeVttTotZFidGlobalCorr}{1} & 
   \numRP{\GroupedSystPercElScaleErrorTTSevenTeVttTotZFidGlobalCorr}{1} \\
\hline 
  Jet energy scale               &  ---   &  ---   &  ---   & 
   \numRP{\GroupedSystPercJesErrorTTThirtTeVttTotZFidGlobalCorr}{1} & 
   \numRP{\GroupedSystPercJesErrorTTEightTeVttTotZFidGlobalCorr}{1} & 
   \numRP{\GroupedSystPercJesErrorTTSevenTeVttTotZFidGlobalCorr}{1} \\
  $b$-tagging                &  ---   &  ---   &  ---   & 
   \numRP{\GroupedSystPercFlTagErrorTTThirtTeVttTotZFidGlobalCorr}{1} & 
   \numRP{\GroupedSystPercFlTagErrorTTEightTeVttTotZFidGlobalCorr}{1} & 
   \numRP{\GroupedSystPercFlTagErrorTTSevenTeVttTotZFidGlobalCorr}{1} \\
\hline 
  Background                     & 
   \numRP{\GroupedSystPercBKGErrorZThirtTeVttTotZFidGlobalCorr}{1}     & 
   \numRP{\GroupedSystPercBKGErrorZEightTeVttTotZFidGlobalCorr}{1}    & 
   \numRP{\GroupedSystPercBKGErrorZSevenTeVttTotZFidGlobalCorr}{1}    &
   \numRP{\GroupedSystPercBKGErrorTTThirtTeVttTotZFidGlobalCorr}{1}    & 
   \numRP{\GroupedSystPercBKGErrorTTEightTeVttTotZFidGlobalCorr}{1}   & 
   \numRP{\GroupedSystPercBKGErrorTTSevenTeVttTotZFidGlobalCorr}{1}    \\
  Signal modelling  (incl. PDF)   &  
   \numRP{\GroupedSystPercSigAndPDFErrorZThirtTeVttTotZFidGlobalCorr}{1}   & 
   \numRP{\GroupedSystPercSigAndPDFErrorZEightTeVttTotZFidGlobalCorr}{1} & 
   \numRP{\GroupedSystPercSigAndPDFErrorZSevenTeVttTotZFidGlobalCorr}{1} &    
   \numRP{\GroupedSystPercSigAndPDFErrorTTThirtTeVttTotZFidGlobalCorr}{1} &  
   \numRP{\GroupedSystPercSigAndPDFErrorTTEightTeVttTotZFidGlobalCorr}{1}   & 
   \numRP{\GroupedSystPercSigAndPDFErrorTTSevenTeVttTotZFidGlobalCorr}{1}\\
\hline 
\end{tabular}

\caption{\label{tab:systerrorsitemized} Systematic uncertainties in \%, $\delta$,  for the measurement of \Zboson-boson and \ttbar production at $\sqrt s = 13, 8, 7$~TeV. Values listed as 0.0 are $<0.05$\%. Values listed as ``--'' have no corresponding uncertainty. The entry ``(lepton)'' in ``Muon (lepton) trigger'' refers to the \ttbar trigger for the 7~and~8~TeV data set which quotes a single uncertainty for the combined effects of the uncertainties in the electron and muon triggers and so there is a corresponding entry ``--'' for the electron trigger for the 7~and~8~TeV \ttbar data set. 
}
\end{table}

\begin{table}[t]
\begin{center}
\begin{tabular}{|l|ccc|ccc|}
\hline
                       & \multicolumn{3}{c|}{$\delta$ \sfidZ} &     \multicolumn{3}{c|}{$\delta$ \stottt} \T \\
  Source  \hspace{0.5cm} / \hspace{0.5cm} \sqs [TeV]             & $13$ &  $8$ &  $7$ &  $13$ &  $8$ &  $7$ \\
\hline
  Luminosity                     &     A         &     B        &      C        &     A          &       B       &      C   \\
  Beam energy                    &     A         &     A        &      A        &      A           &      A       &      A  \\
\hline
  Muon (lepton) trigger          &     A         &     A$^*$        &      A         &    A          &       B        &    B    \\
  Muon reconstruction/ID         &     A         &     B        &      C         &     A          &    D         &   D       \\
  Muon isolation                 &     A        &     A       &     A        &     B         &    C         &   D       \\
  Muon momentum scale            &     A        &     A       &     A         &    A          &    A         &   A       \\
\hline
  Electron trigger               &    A           &       A       &      A         &     A          &  ---             &    ---  \\
  Electron reconstruction/ID           &   A            &      B       &    C           &    A           &  D            &   D   \\
  Electron isolation             &   A           &      A       &      ---        &    B           &  C            &   D \\
  Electron energy scale          &   A          &      A      &  A            &    A          &  A           &   A \\
\hline 
  Jet energy scale               &    ---         &    ---        &  ---   &  A   &  B  & B \\
  $b$-tagging                &    ---         &    ---        &  ---   &  A   &  B  & B \\
\hline 
  Background                     &    A          &   A          &   A            &   B            & B             &  B  \\
  Signal modelling  (incl. PDF)          &    A          &   A          &   A            &   B$^*$            &  B            &  B \\
\hline 
\end{tabular}
\end{center}
\caption{\label{tab:corrmodel} Correlation model for the systematic uncertainties, $\delta$, of the measurements of \Zboson-boson and \ttbar production at $\sqrt s = 13, 8, 7$~TeV. Entries in different rows are  uncorrelated with each other. Entries within a row with the same letter are fully correlated.
Entries within a row with a starred letter are mostly correlated with the entries with the same letter (most of the individual sources of uncertainties within a group are taken as correlated). Entries with different letters within a row are either fully or mostly uncorrelated with each other. This table uses the same categories as Table~\ref{tab:systerrorsitemized}.}
\end{table}

The correlation model used in this analysis is summarised in Table~\ref{tab:corrmodel}. The groups listed in the table may be represented by a single source, or by several individual sources of systematic uncertainties (nuisance parameters). The groups of sources are:
\begin{itemize}
\item \textit{Luminosity} is considered to be correlated for the measurements performed at the same $\sqs$ but uncorrelated for data at different $\sqs$. 
\item \textit{Beam energy} uncertainty is $0.66\%$ of the beam-energy value~\cite{Wenninger:1546734} and is considered to be fully correlated for all data sets. 
\item \textit{Muon trigger} is a small source of uncertainty for most analyses. It is considered to be correlated for all \Zboson-boson measurements and for the \ttbar measurement at 13~TeV, and separately between the two \ttbar  analyses at 7~and~8~TeV, following the prescription of Ref.~\cite{TOPADDENDUM2}.
\item \textit{Muon reconstruction/identification} is described by several nuisance parameters. The treatment is fully synchronised for the 13~TeV measurements. The \Zboson-boson measurements at 7~and~8~TeV are considered uncorrelated with each other and with the $\ttbar$ measurements since different muon reconstruction algorithms were employed for these measurements. However, the measurements of $\ttbar$ at 7~and~8~TeV are assumed to be correlated since they use the same reconstruction algorithm.   
\item \textit{Muon isolation} is a small and similar source of uncertainty for all \Zboson-boson measurements and thus it is considered to be correlated amongst the measurements. For \ttbar analyses, the muon isolation uncertainty is determined in situ, to account for different hadronic environments, and has significant statistical uncertainties. For these reasons,  these uncertainties are considered to be uncorrelated with each other and with the \Zboson-boson uncertainties.  
\item \textit{Muon momentum scale} is a small source of uncertainty for all measurements. It is validated in situ by comparing the  invariant mass distributions of muon pairs in data and simulation. Similar levels of agreement are observed for all data-taking periods, and thus all measurements are considered to be correlated. 
\item \textit{Electron trigger} is a small source of uncertainty for all measurements and is considered to be fully correlated amongst all measurements.
\item \textit{Electron reconstruction/identification} is treated similarly to the muon reconstruction/identification.
\item \textit{Electron isolation} is treated similarly to the muon isolation. 
\item \textit{Electron energy scale} is  treated and validated similarly to the muon momentum scale. 
\item \textit{Jet energy scale} only affects the \ttbar measurements and is described by several nuisance parameters. The uncertainty is correlated for 7~and~8~TeV data, following the prescription of  Ref.~\cite{TOPADDENDUM2}, and mostly uncorrelated with 13~TeV data, in part due to the in-situ corrections. The impact of this source on the \ttbar measurements is small.
\item \textit{$b$-tagging} also only affects the \ttbar measurements. The source is considered to be correlated for 7~and~8~TeV data but uncorrelated with 13~TeV data since  the installation of the new insertable B-layer in the inner detector and re-optimised $b$-tagging algorithms  used at 13~TeV resulted in significantly improved $b$-tagging performance.
\item \textit{Background} is treated as fully correlated for all $\sqrt s$ within a given process.
The main uncertainty for this source is driven by the theoretical uncertainties in the cross sections of the background processes, and the leading background sources are very different for the \Zboson-boson and \ttbar measurements.
\item \textit{Signal modelling} uncertainty is small for  the fiducial \Zboson-boson measurements. 
Signal modelling is the leading source of uncertainty for the \ttbar measurements. Several sources of uncertainty, such as uncertainties related to signal and background MC generators and to PDFs, are considered to be correlated across the different $\sqs$ values.
An additional source of uncertainty is included only for the \ttbar measurement at 13~TeV, due to the level of agreement observed in events with at least three $b$-tagged jets~\cite{Aaboud:2016pbd}.
\end{itemize}

The correlation model described above corresponds to a fully synchronised analysis of \Zboson-boson and \ttbar data at 13~TeV. It also follows the prescription given in Ref.~\cite{TOPADDENDUM2} for the \ttbar measurements at 7~and~8~TeV. The stability of the results relative to the correlation assumptions was verified by altering the model for the sources of uncertainty where the level of correlation is not precisely known, such as lepton reconstruction and identification at 7~and~8~TeV, resulting in only small changes in the uncertainties.

\subsection{Results}
\label{sec:ratiosres}

In this section, a representative set of total \ttbar and fiducial \Zboson-boson cross sections and their ratios are compared to the theory predictions.
The full set of single-ratio and double-ratio results for the various combinations of fiducial and total cross sections is given in Appendix~\ref{AppC}.

\subsubsection{Single ratios at a given \sqs}
\label{sec:ratiosresones}
\begin{figure}[t]
\centering
    \includegraphics[width=0.49\textwidth]{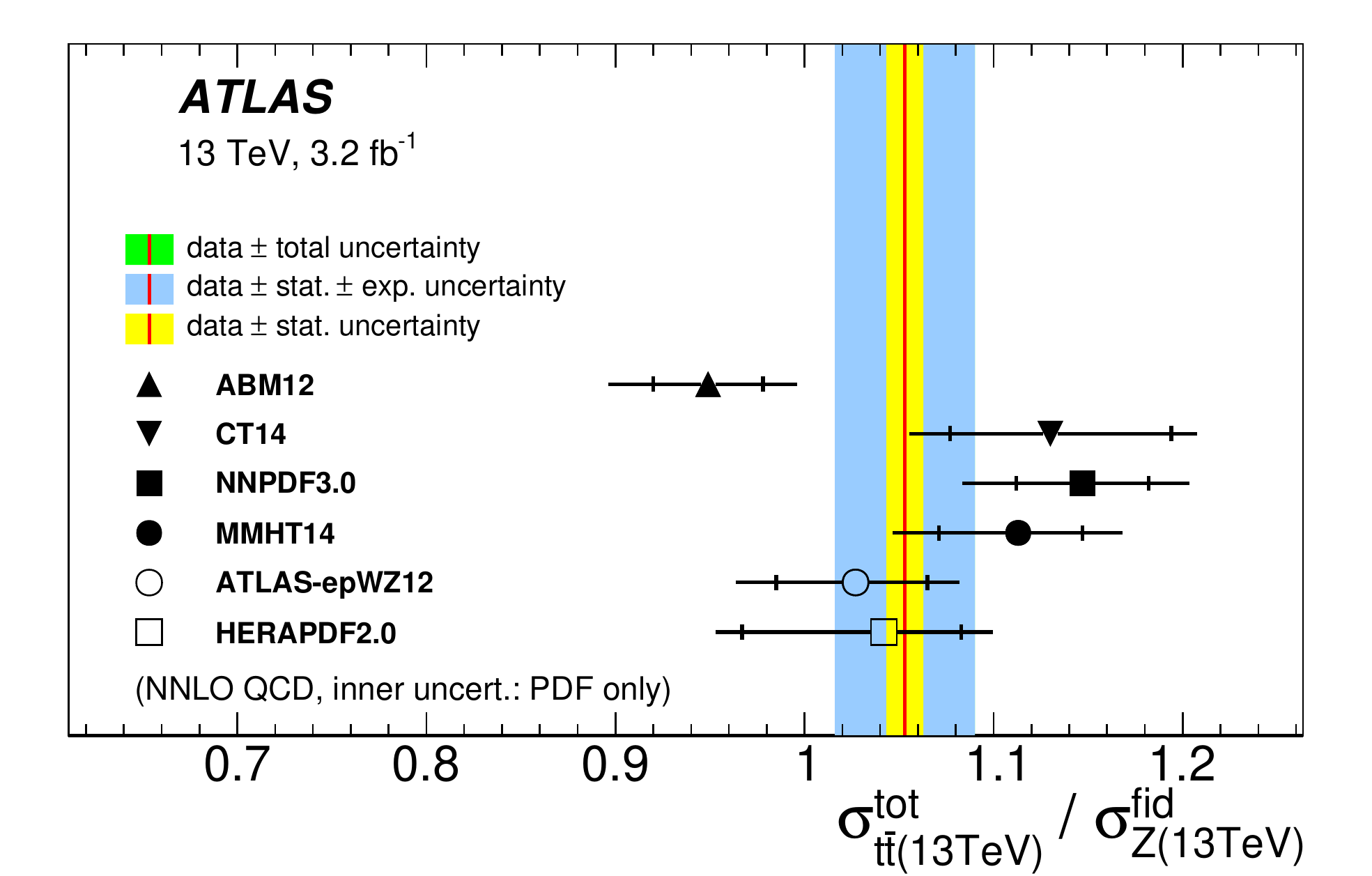}
    \includegraphics[width=0.49\textwidth]{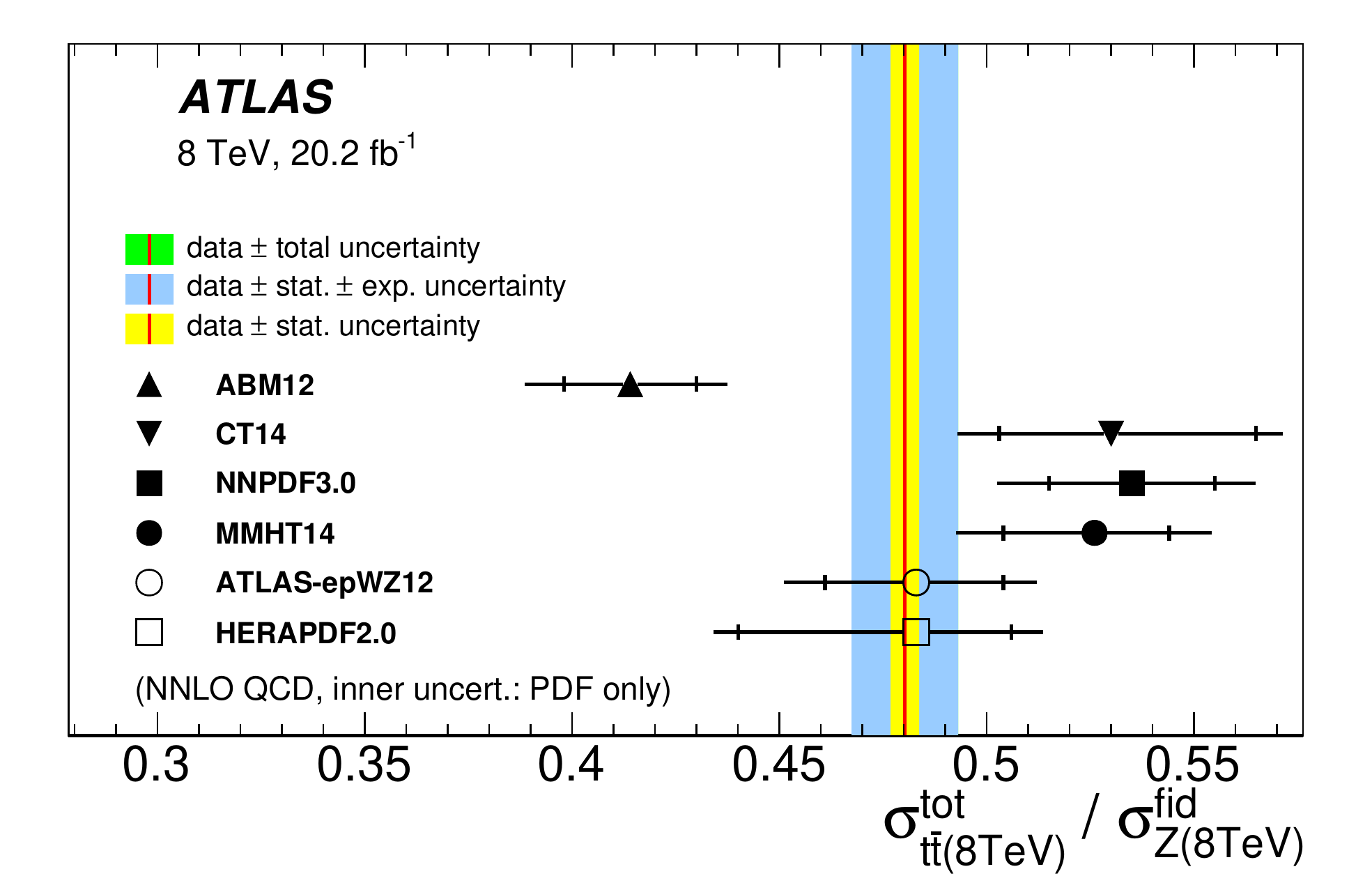}
 \includegraphics[width=0.49\textwidth]{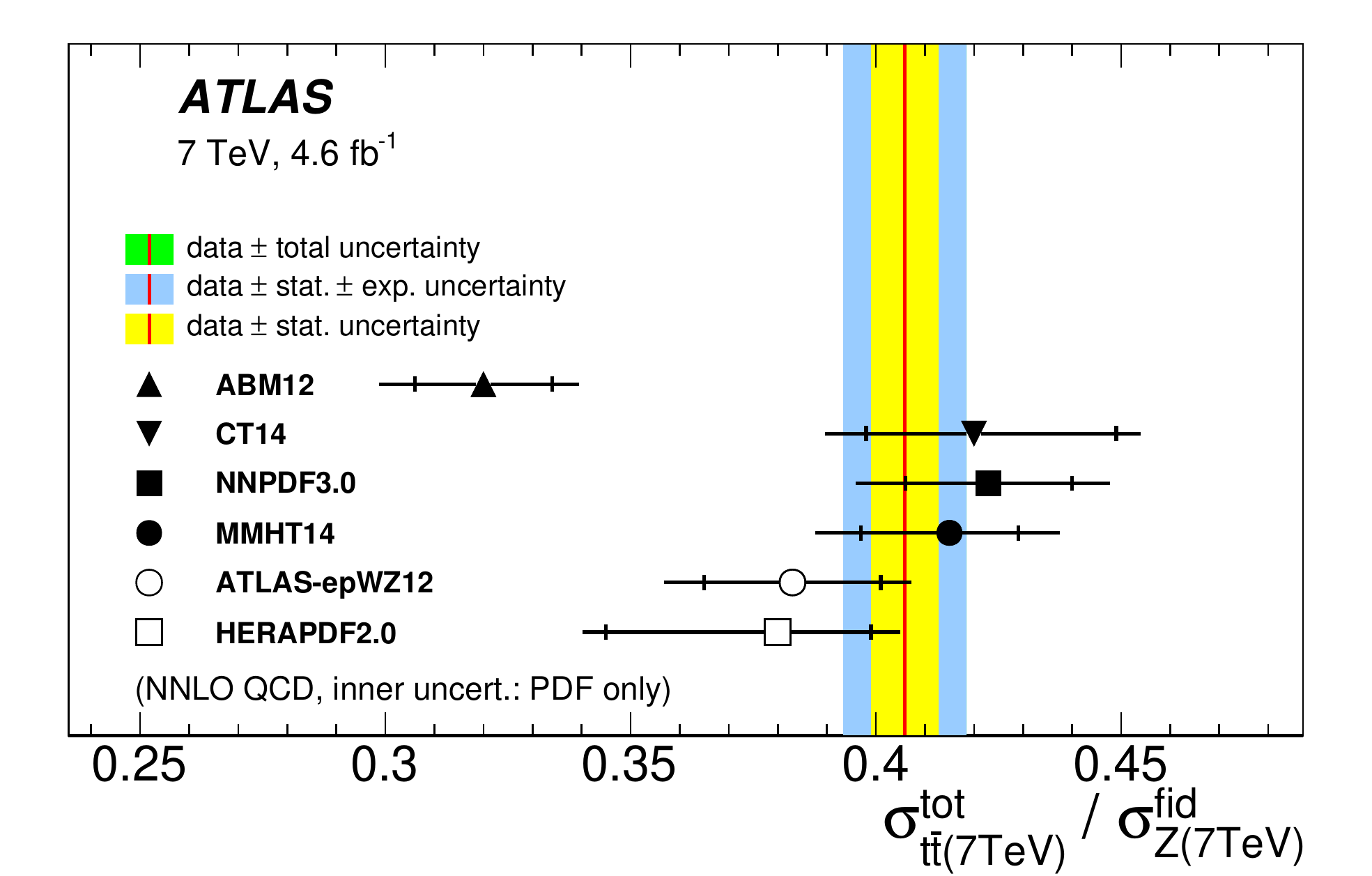}

  \caption{
The ratios $\ratttztf(i~\mathrm{TeV})$, for $i=13, 8, 7$  compared to predictions based on different PDF sets.
The inner shaded band corresponds to the statistical uncertainty, the middle band to the statistical  and experimental systematic uncertainties added in quadrature, while the outer band shows the total uncertainty, including the luminosity uncertainty. The latter is not visible since 
the luminosity uncertainties almost entirely cancel in these ratios. 
The theory predictions are given with the corresponding
PDF uncertainties shown as inner  bars while the outer bars include all other uncertainties added in quadrature.
  \label{fig:ttbarToZfid_at_13_8_7}}
\end{figure}

The single ratios  \ratttztf~are compared in Figure~\ref{fig:ttbarToZfid_at_13_8_7} to the theoretical
predictions based on different PDF sets.
For all centre-of-mass energies, the predictions follow a similar pattern for the following three groups of PDFs. 
The \ABM set yields the lowest
values. The three PDF sets used in the PDF4LHC prescription~\cite{Butterworth:2015oua}, \CT, \NNPDF, and \MMHT, predict 
the largest ratios. The HERA-based \HERAPDF and \ATLASepWZ sets are in the middle. 
The spread of the predictions is beyond the PDF uncertainties for the three groups of PDFs while the quoted
PDF uncertainties are similar in size, with the \HERAPDF errors being the largest and \ABM the smallest.
This pattern could be explained by the differences in the gluon density and the $\alpha_{\textrm{S}}$ value used in the PDF sets.
The \ABM, \HERAPDF and \ATLASepWZ sets do not include collider jet data, which typically lead to a lower gluon density for
the $x$ values where the $t\bar{t}$ data at the LHC are sensitive. In addition, the \ABM set uses a lower value of $\alpha_{\textrm{S}}$. 
The size of the error bars depends on the data sets used in the PDF fits and also on the statistical model
used for the analysis. 

The ATLAS data are more precise than most of the theory predictions, suggesting the data have strong 
constraining power. The experimental uncertainties are the smallest
for the 8~TeV measurement. The 7~TeV result has a sizeable statistical uncertainty, while
the systematic uncertainty at 13~TeV is larger than at both 7~and~8~TeV, mostly due to a larger
\ttbar modelling uncertainty. For the most precise measurement, at 8~TeV, the data agree best with the \HERAPDF and \ATLASepWZ PDF sets while they deviate by 1.6--2.1$\sigma$ from
the PDF4LHC PDFs, where $\sigma$ is the total experimental uncertainty plus the luminosity uncertainty 
(but agree well when including the respective prediction uncertainties), and by $2.6\sigma$ from the \ABM PDF. A similar but less significant pattern is observed for the 
13~TeV data. The 7~TeV data are most consistent with the \MMHT PDF set. 
The data are between the predictions of the  PDF4LHC PDFs and the HERA-based PDFs \HERAPDF and \ATLASepWZ, deviating most from the \ABM prediction.  
The difference between data and predictions for the 7 and 8~TeV results is consistent with the results published by ATLAS for the
ratio of \ttbar cross sections at these two energies~\cite{TOPADDENDUM2}, as is discussed in Section~\ref{sec:ratiosresdiffs}.
 
\subsubsection{Single ratios at different \sqs}
\label{sec:ratiosresdiffs}

\begin{figure}[t]
  \centering

    \includegraphics[width=0.49\textwidth]{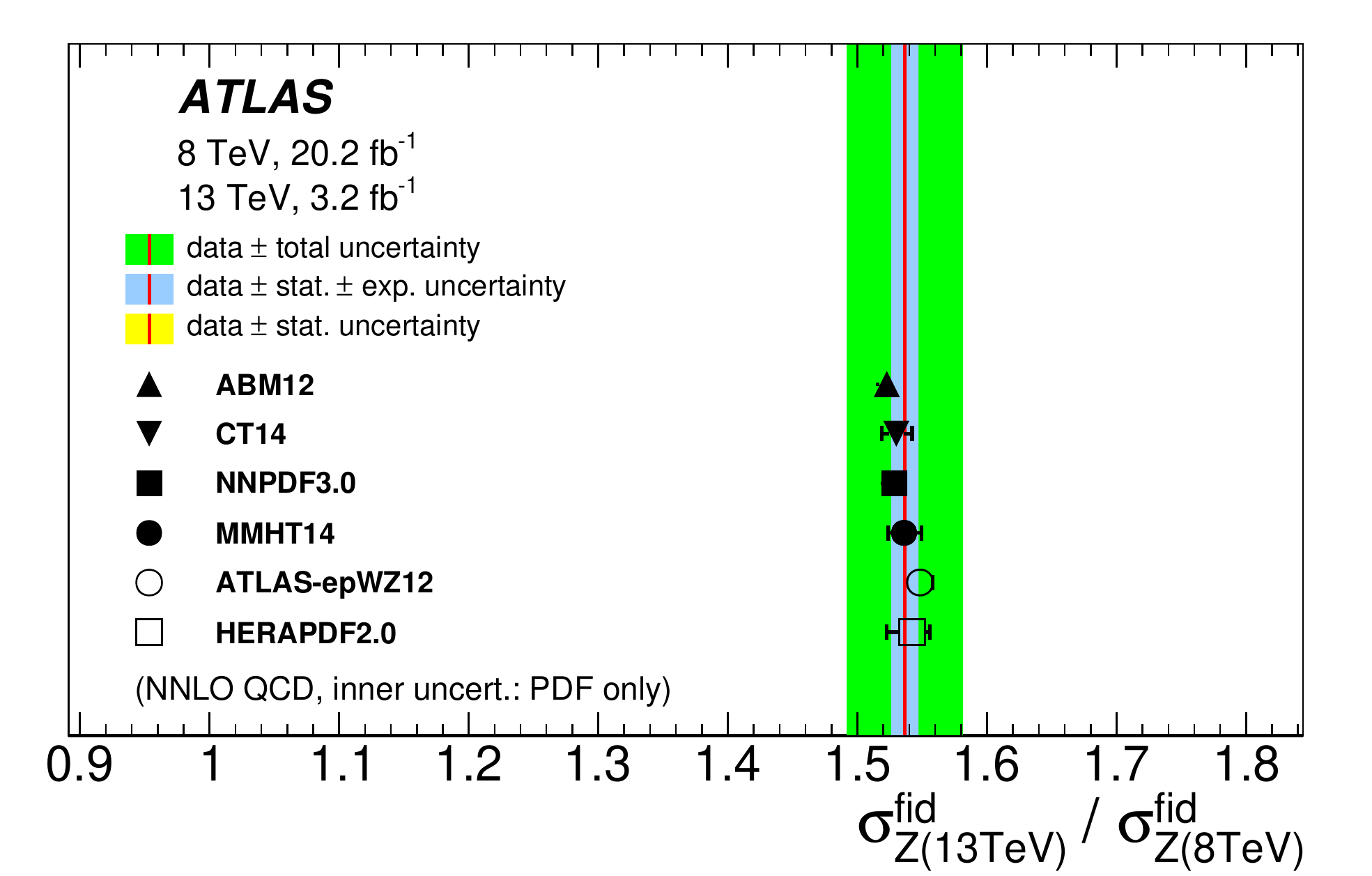}
    \includegraphics[width=0.49\textwidth]{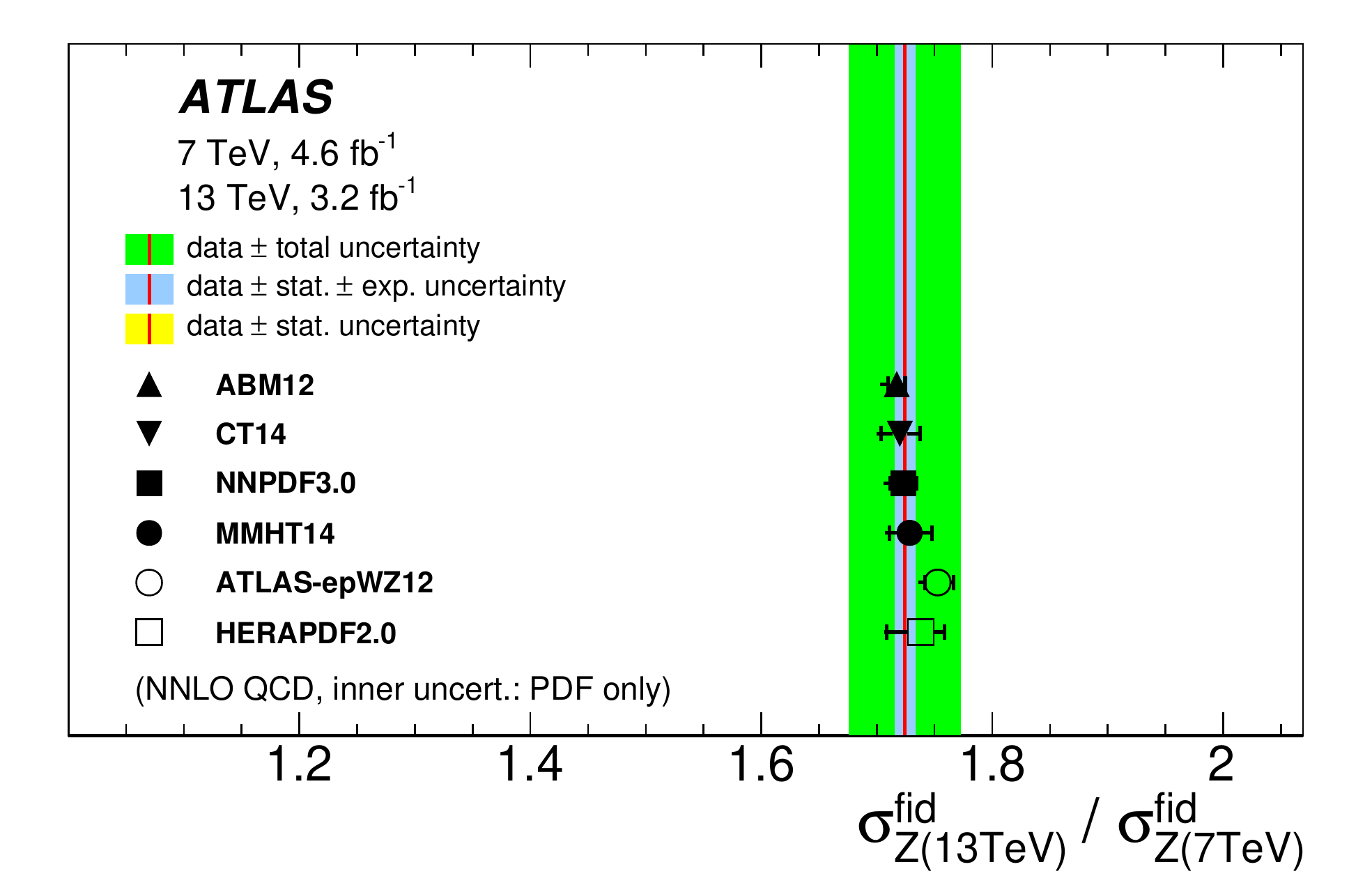}
    \includegraphics[width=0.49\textwidth]{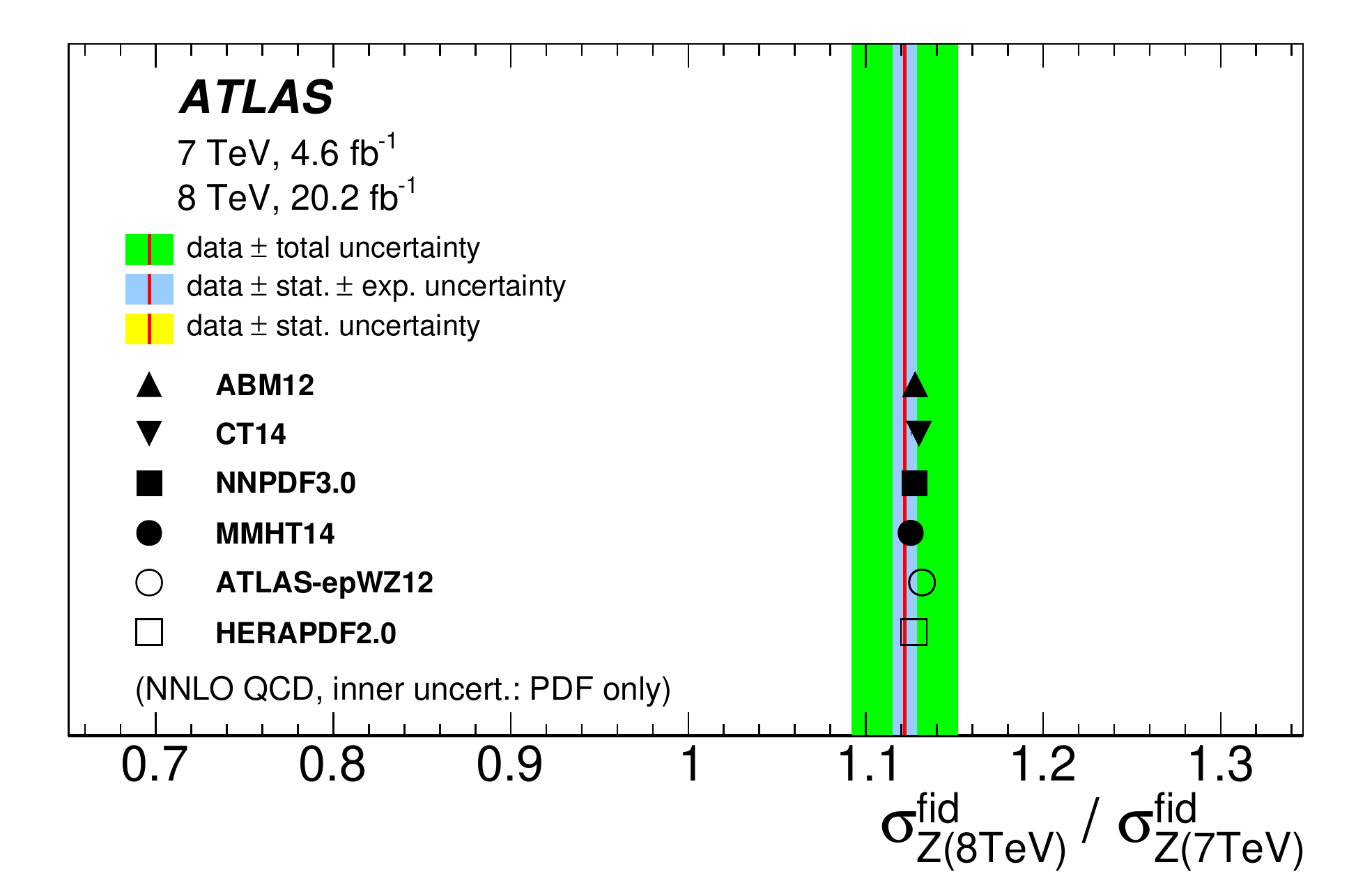}

  \caption{ 
The ratios \ratzf{i}{j}, for $i,j=13, 8, 7$  compared to predictions based on different PDF sets.
The inner shaded band (barely visible since it is small) corresponds to the statistical uncertainty, the middle band to the statistical  and experimental systematic uncertainties added in quadrature, while the outer band shows the total uncertainty, including the luminosity uncertainty. 
The theory predictions are given with the corresponding
PDF uncertainties shown as inner  bars while the outer bars include all other uncertainties added in quadrature.
  \label{fig:Z13ToZ8_fidfid}}
\end{figure}

The ratios of the fiducial $Z$-boson cross sections at various $\sqs$ values are compared in Figure~\ref{fig:Z13ToZ8_fidfid} to predictions employing different PDF sets.  The uncertainty in these ratios is dominated by the luminosity uncertainty. Even though the total luminosity uncertainties are of comparable magnitude at 7, 8 and 13~TeV, they are mostly uncorrelated and therefore do not cancel in the cross-section ratios.  

The measurements are consistent with the predictions for all PDF sets. Most of these predictions agree with the data within the experimental uncertainties, even omitting the luminosity uncertainty.  This observation may indicate that the luminosity-determination uncertainty in the measured ratio is conservative.  
The smallness of the PDF uncertainties for different predictions and the overall small spread among them suggest that 
the measured \Zboson-boson data could be used to cross-normalise the measurements at the different centre-of-mass energies, thereby avoiding the penalty associated with the combination of uncorrelated luminosity uncertainties. This aspect is explored in Section~\ref{sec:doublerat} by taking double ratios of $t\bar{t}$ to $Z$-boson  cross sections, but this approach can be used for other processes as well. 

The measured \ttbar ratios for different pairs of $\sqs$ are compared to the predictions in Figure~\ref{fig:ttbar13Tottbar8_tottot}.
These predictions follow a similar pattern for all ratios: the three predictions from PDF4LHC PDFs are the smallest, closely 
followed by \ATLASepWZ and \HERAPDF, and the \ABM prediction is the largest.
This pattern could be explained by the PDFs having different gluon distributions as a function of $x$.
At low $x$, all PDF sets have similar gluon content since the gluon PDF is primarily determined from a common source: scaling violations of the $F_2$ structure function measured at HERA.  At high $x$, the \ABM and HERA-based sets have a lower gluon density than other PDF sets. Thus, as the $\sqs$ increases, resulting in a decrease of the average value of $x$, the \ABM and HERA-based sets exhibit a stronger $\sqs$ dependence than the  PDF4LHC PDFs. 
Given the relative size of the experimental uncertainties and the spread of the theoretical predictions in these ratios, these measurements do not test the consistency of the luminosity calibrations at different centre-of-mass energies to the same precision as the \Zboson-boson cross-section ratios.

The ratio of 13~TeV to 8~TeV cross sections agrees with all predictions within experimental
uncertainties. The central value is closest to the \HERAPDF prediction.
For the ratios involving 7~TeV data, the measured ratios have central values lower than predicted by all the PDFs. This is especially so for the 8~TeV to 7~TeV ratio, which deviates from all predictions
by approximately two standard deviations. The deviation was observed previously by ATLAS~\cite{TOPADDENDUM2} and the results of this analysis 
are consistent with those published values.
 
\begin{figure}[t]
  \centering
    \includegraphics[width=0.49\textwidth]{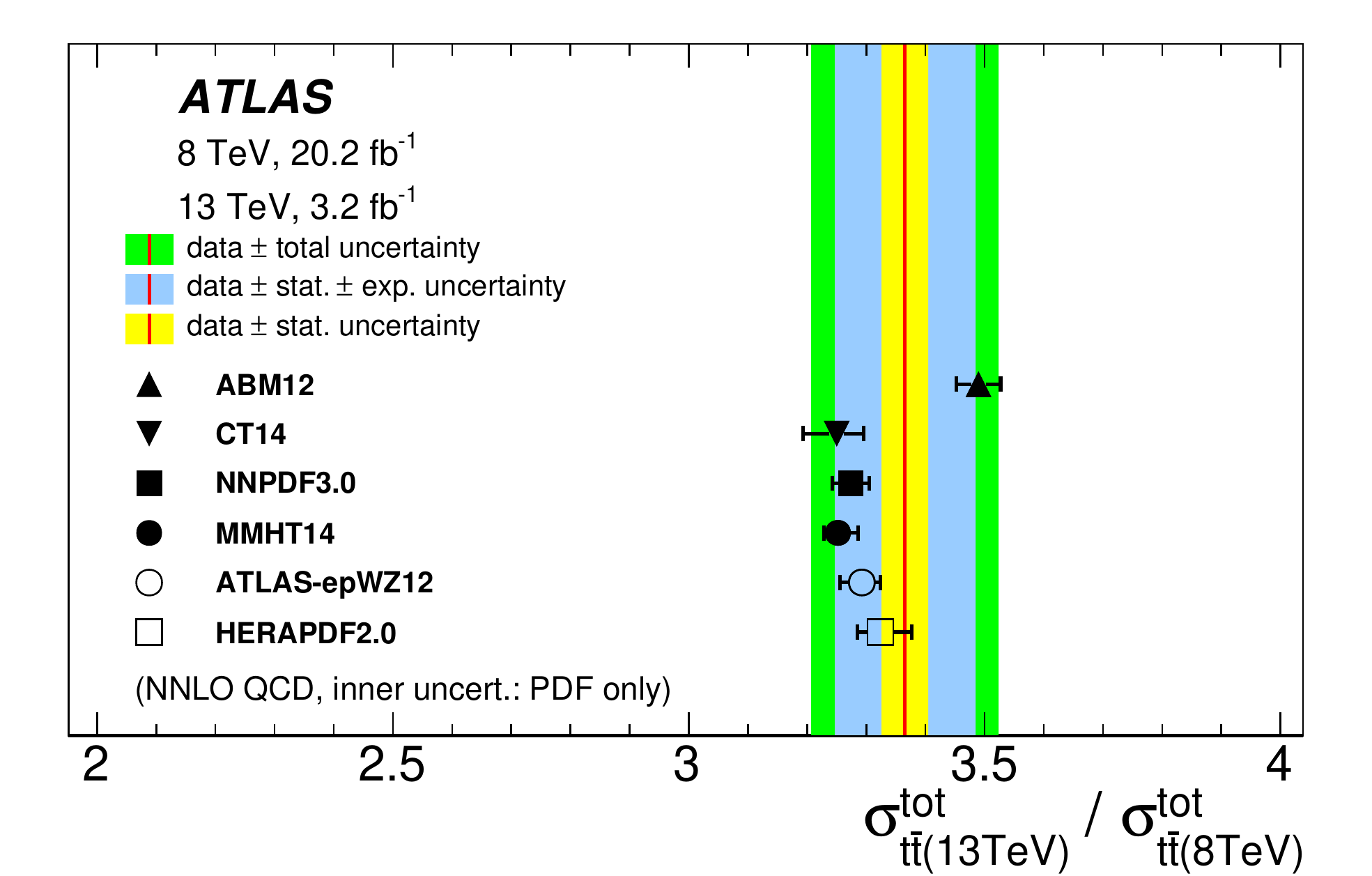}
    \includegraphics[width=0.49\textwidth]{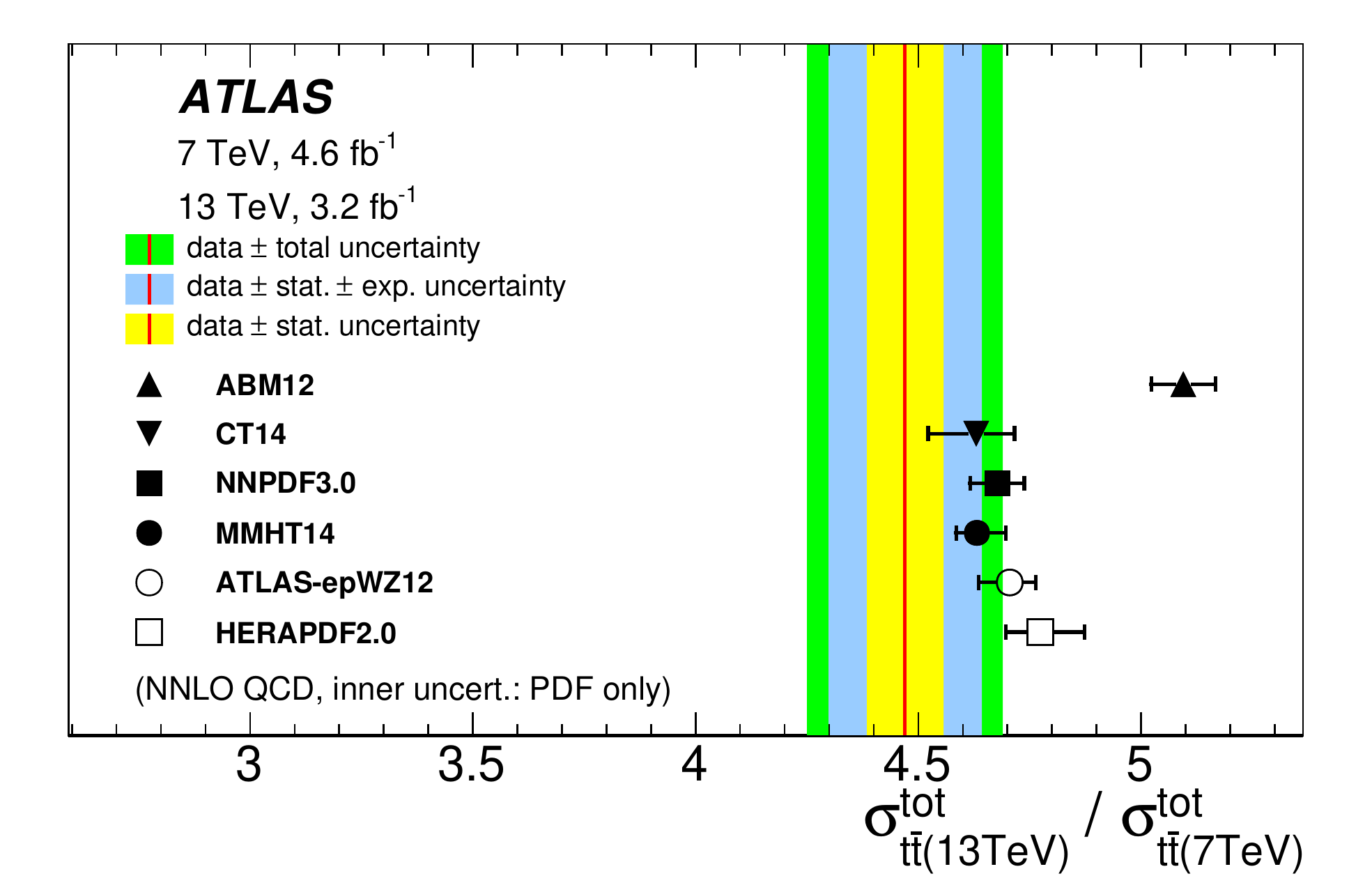}
    \includegraphics[width=0.49\textwidth]{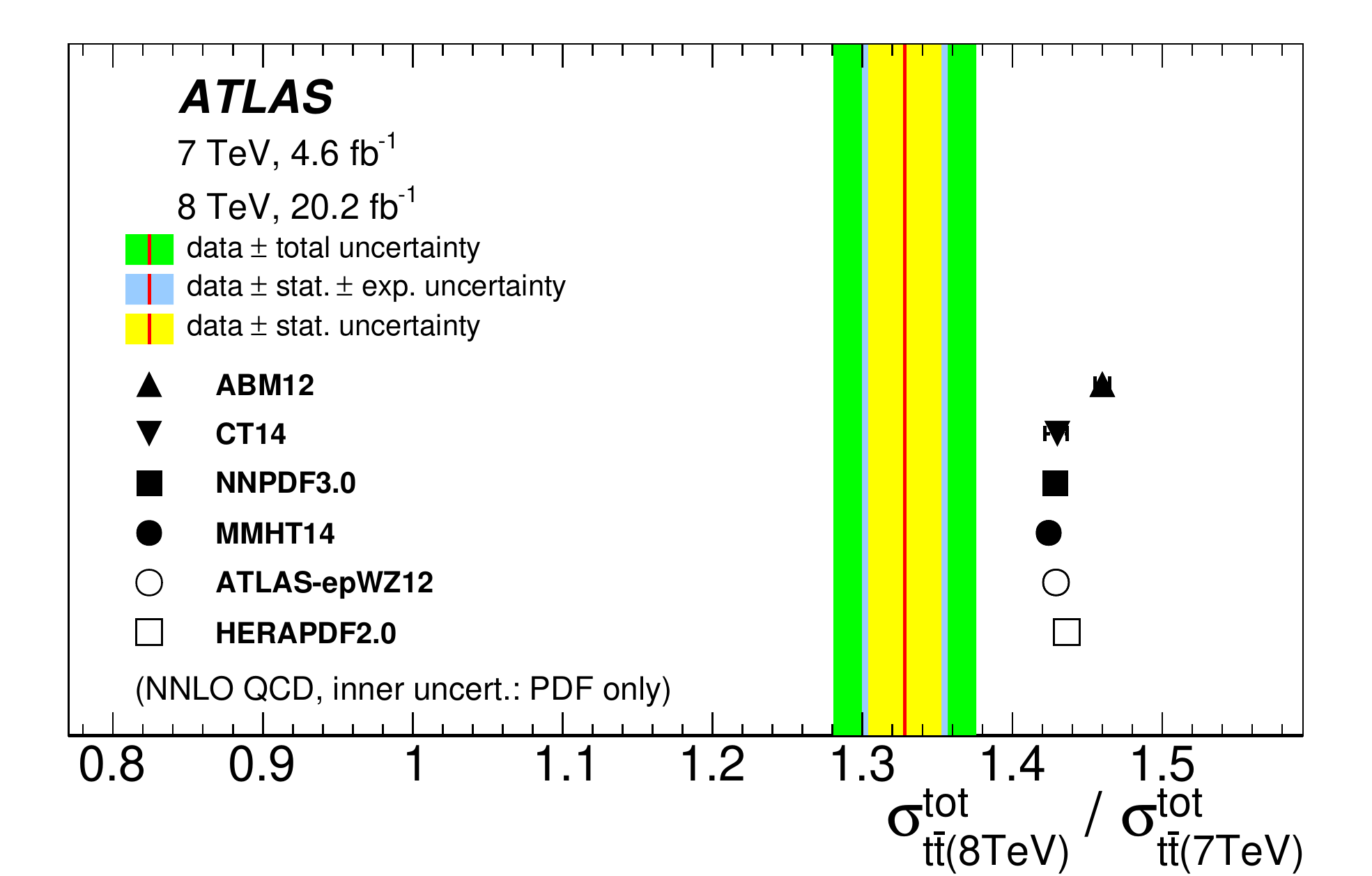}

  \caption{
The ratios \rattt{i}{j}, for $i,j=13, 8, 7$  compared to predictions based on different PDF sets.
The inner shaded band corresponds to the statistical uncertainty, the middle band to the statistical  and experimental systematic uncertainties added in quadrature, while the outer band shows the total uncertainty, including the luminosity uncertainty. For the 8-to-7~TeV ratio, the experimental systematic uncertainty band is too small to be clearly visible. The theory predictions are given with the corresponding
PDF uncertainties shown as inner  bars while the outer bars include all other uncertainties added in quadrature.
  }
  \label{fig:ttbar13Tottbar8_tottot}
\end{figure}

\subsubsection{Double ratios}
\label{sec:doublerat}

\begin{figure}[t]
  \centering
    \includegraphics[width=0.49\textwidth]{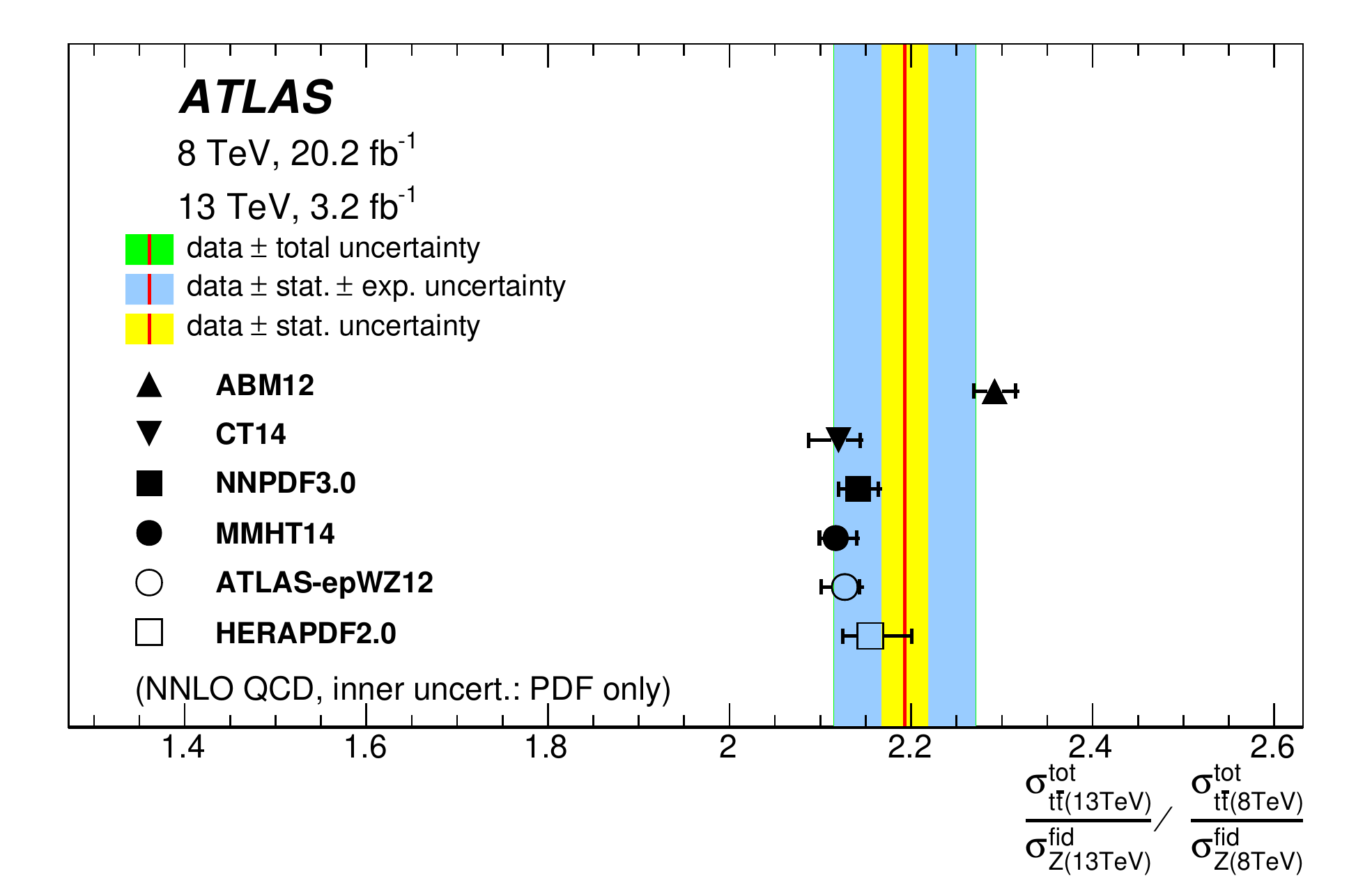}
    \includegraphics[width=0.49\textwidth]{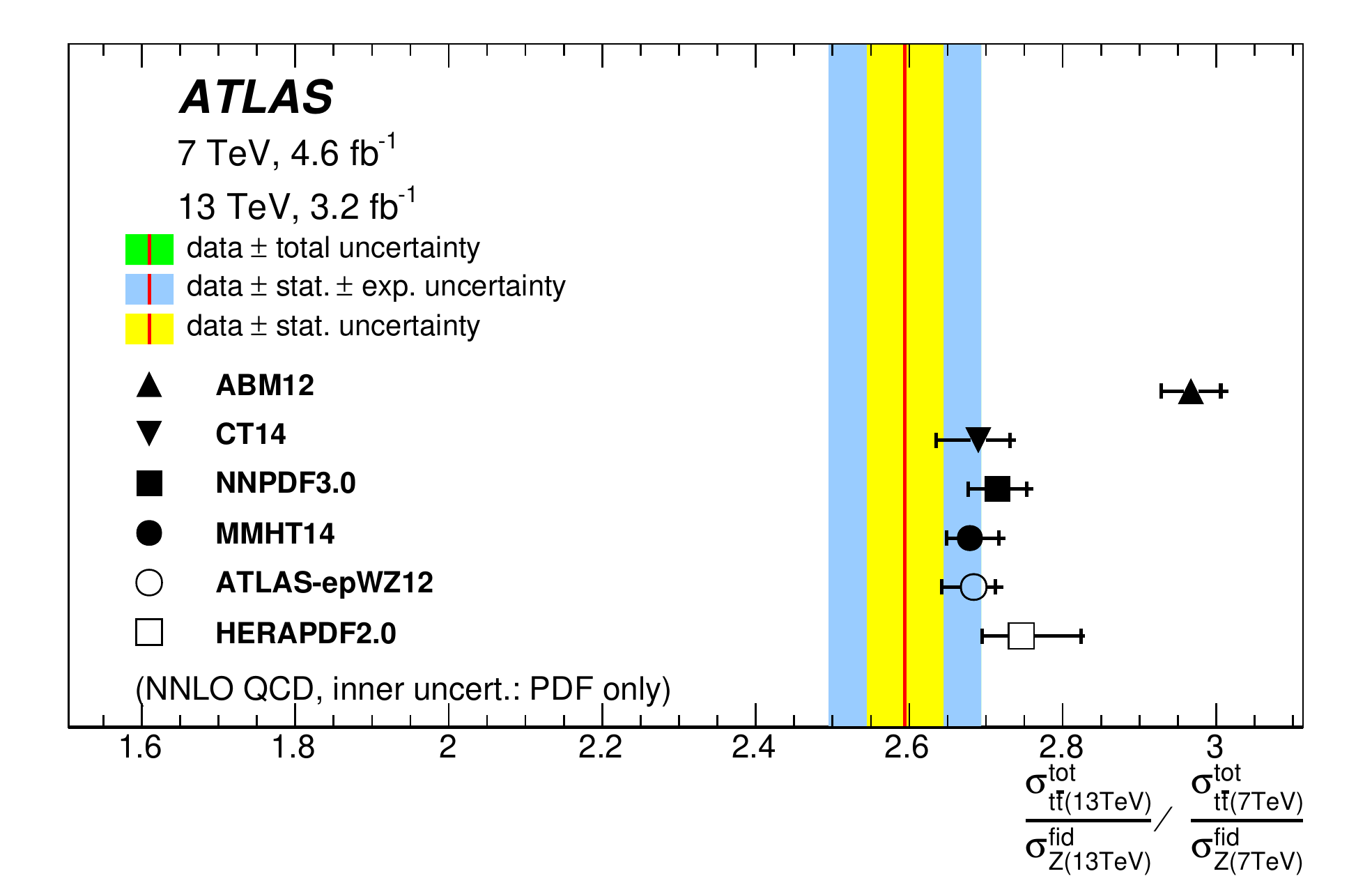}
    \includegraphics[width=0.49\textwidth]{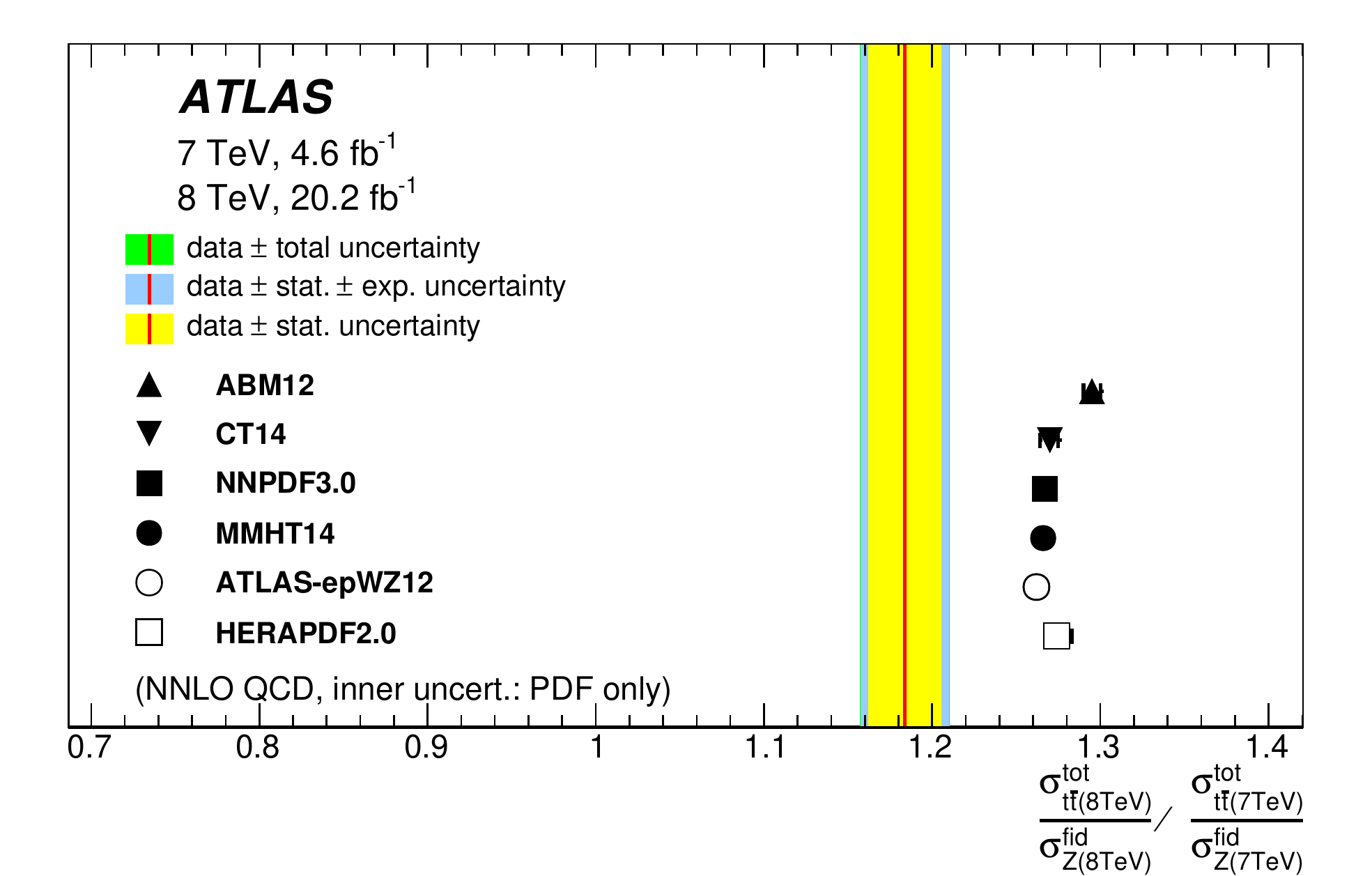}

  \caption{
The ratios $\ratttztf(i/j)$ where $i,j=13, 8, 7$  compared to predictions based on different PDF sets.
The inner shaded band corresponds to the statistical uncertainty, the middle band to the statistical  and experimental systematic uncertainties added in quadrature, while the outer band shows the total uncertainty, including the luminosity uncertainty. The latter is not visible since
the luminosity uncertainties almost entirely cancel in these ratios.
The theory predictions are given with the corresponding
PDF uncertainties shown as inner  bars while the outer bars include all other  uncertainties added in quadrature.
  }
  \label{fig:DoubleRat_totfid}
\end{figure}

The double ratios of total \ttbar to fiducial \Zboson-boson cross sections at different $\sqs$ are compared to predictions in
Figure~\ref{fig:DoubleRat_totfid}. The total uncertainties are smaller than those in the \ttbar cross-section ratios at different \sqs due to the almost complete cancellation of the luminosity uncertainty, which more than compensates for the uncertainties that the \Zboson-boson cross sections bring to these double ratios.
 
For the double ratios, the trends seen in comparisons between the  data and the predictions are similar to those observed for the single ratios of the \ttbar cross sections at different \sqs values. The double ratio of 13~TeV to 8~TeV results
is consistent with all predictions at the $1 \sigma$ level. The tension between the measured 8~TeV to 7~TeV ratio and the predictions is increased, due to the reduced uncertainty in the measurement that this double ratio brings.
This behaviour is difficult to ascribe to the $x$-dependence of the gluon distribution
since the change in the average $x$ is much larger for 13~TeV to 8~TeV than for 8~TeV to 7~TeV measurements.
The deviation from the \ABM PDFs is at the $4 \sigma$ level while for all other PDFs they are  at the $3 \sigma$ level.
The prediction closest to  the observed ratio is obtained from the \ATLASepWZ PDF set,
which predicts a stronger variation of the fiducial \Zboson-boson cross section as a function of $\sqrt{s}$. 

\subsubsection{Correlated cross-section measurements}
\label{sec:correlsigma}

As an alternative to taking ratios, the measured cross sections may be compared directly to theory, provided that the full correlation
information amongst the experimental results is evaluated. The electron and muon channel \sfidZ are combined,
accounting for the correlated systematic uncertainties, which as a result cause small shifts in all of the combined cross-sections values. 
The combination's $\chi^2$ per degree of freedom is $\chi^2/{\textrm{NDF}} = \AveChisqAllChanEnergZThirteenTevTTThirteenTevZEightTevTTEightTevZSevenTevTTSevenTevFidFidGlobalCorr $  for ${\textrm{NDF}}=3$, indicating excellent compatibility of the    
$Z\rightarrow  e^+e^-$ and $Z\rightarrow  \mu^+\mu^-$ measured cross sections.
The resulting \Zboson-boson fiducial and \ttbar total cross sections after combination
are given in Table~\ref{t:CombXsecZfidTTtot} with the correlation coefficients presented in Table~\ref{t:TabCorrelations}.
The correlations are large for the measurements at a given $\sqs$, due to the common luminosity uncertainty.
The corresponding table omitting both the luminosity and beam-energy uncertainties is given in Appendix~\ref{AppC}.
As expected from the ratio analysis, there is also a sizeable correlation between the \ttbar results at 7 and 8~TeV. 
It is verified that the uncertainties in the ratios are consistent with those of the direct evaluation of the combined cross section.

Figure~\ref{fig:contour} shows the results of this combination as two-dimensional 68\% CL contours of \sfidZ\ vs. \sigtttot\ at the three $\sqrt s$ values, overlayed with the theoretical cross-section predictions calculated from the error sets associated with each specific PDF. 
The correlations of the measured cross sections are opposite in sign to those of the predicted cross sections (with exception of \ABM set, which has a small positive correlation), providing discriminating input  to the determination of the PDFs.

\begingroup
\begin{table}[t]
\centering
\renewcommand*{\arraystretch}{1}
\begin{tabular}{|c|c@{$\,\pm\,$}c@{$\,\pm\,$}c@{$\,\pm\,$}c@{$\,\pm\,$}c|}
\hline
\sqs [TeV]  & \multicolumn{5}{c|}{Value $\pm$ stat $\pm$ syst $\pm$ beam $\pm$ lumi [pb]} \\
\hline \hline
  & \multicolumn{5}{c|}{\sfidZ} \T \B \\ \hline
$13$     &
\numRP{\AveCentralZThirteenTevabsvalttTotZFidGlobalCorr}{0}&
1
(\numRP{\AveStatZThirteenTevpercentsttTotZFidGlobalCorr}{1}\%) &                 
\numRP{\AveSystZThirteenTevabsvalttTotZFidGlobalCorrNoLumiNoBeam}{0} 
(\numRP{\AveSystZThirteenTevpercentsttTotZFidGlobalCorrNoLumiNoBeam}{1}\%) &                    
\numRP{\AveSystZThirteenTevabsvalttTotZFidGlobalCorrBeamOnly}{0} 
(\numRP{\AveSystZThirteenTevpercentsttTotZFidGlobalCorrBeamOnly}{1}\%) &
\numRP{\AveSystZThirteenTevabsvalttTotZFidGlobalCorrLumiOnly}{0} 
(\numRP{\AveSystZThirteenTevpercentsttTotZFidGlobalCorrLumiOnly}{1}\%) \\
$8$     &
\numRP{\AveCentralZEightTevabsvalttTotZFidGlobalCorr}{0} & 
$<1$
$(<0.1\%)$ &
\numRP{\AveSystZEightTevabsvalttTotZFidGlobalCorrNoLumiNoBeam}{0} 
(\numRP{\AveSystZEightTevpercentsttTotZFidGlobalCorrNoLumiNoBeam}{1}\%) &
\numRP{\AveSystZEightTevabsvalttTotZFidGlobalCorrBeamOnly}{0} 
(\numRP{\AveSystZEightTevpercentsttTotZFidGlobalCorrBeamOnly}{1}\%) &
\numRP{\AveSystZEightTevabsvalttTotZFidGlobalCorrLumiOnly}{0} 
(\numRP{\AveSystZEightTevpercentsttTotZFidGlobalCorrLumiOnly}{1}\%) \\
$7$     &
\numRP{\AveCentralZSevenTevabsvalttTotZFidGlobalCorr}{0} & 
$<1$
(\numRP{\AveStatZSevenTevpercentsttTotZFidGlobalCorr}{1}\%) &
\numRP{\AveSystZSevenTevabsvalttTotZFidGlobalCorrNoLumiNoBeam}{0} 
(\numRP{\AveSystZSevenTevpercentsttTotZFidGlobalCorrNoLumiNoBeam}{1}\%) &
\numRP{\AveSystZSevenTevabsvalttTotZFidGlobalCorrBeamOnly}{0} 
(\numRP{\AveSystZSevenTevpercentsttTotZFidGlobalCorrBeamOnly}{1}\%) &
\numRP{\AveSystZSevenTevabsvalttTotZFidGlobalCorrLumiOnly}{0} 
(\numRP{\AveSystZSevenTevpercentsttTotZFidGlobalCorrLumiOnly}{1}\%) \\
\hline \hline
 &  \multicolumn{5}{c|}{\stottt} \T \B \B \\ \hline
$13$  &
\numRP{\AveCentralTTThirteenTevabsvalttTotZFidGlobalCorr}{0} & 
\numRP{\AveStatTTThirteenTevabsvalttTotZFidGlobalCorr}{0} 
(\numRP{\AveStatTTThirteenTevpercentsttTotZFidGlobalCorr}{1}\%) &
\numRP{\AveSystTTThirteenTevabsvalttTotZFidGlobalCorrNoLumiNoBeam}{0} 
(\numRP{\AveSystTTThirteenTevpercentsttTotZFidGlobalCorrNoLumiNoBeam}{1}\%) &
\numRP{\AveSystTTThirteenTevabsvalttTotZFidGlobalCorrBeamOnly}{0} 
(\numRP{\AveSystTTThirteenTevpercentsttTotZFidGlobalCorrBeamOnly}{1}\%) &
\numRP{\AveSystTTThirteenTevabsvalttTotZFidGlobalCorrLumiOnly}{0}
(\numRP{\AveSystTTThirteenTevpercentsttTotZFidGlobalCorrLumiOnly}{1}\%) \\
$8$  &
\numRP{\AveCentralTTEightTevabsvalttTotZFidGlobalCorr}{0} & 
\numRP{\AveStatTTEightTevabsvalttTotZFidGlobalCorr}{0} 
(\numRP{\AveStatTTEightTevpercentsttTotZFidGlobalCorr}{1}\%) &
\numRP{\AveSystTTEightTevabsvalttTotZFidGlobalCorrNoLumiNoBeam}{0} 
(\numRP{\AveSystTTEightTevpercentsttTotZFidGlobalCorrNoLumiNoBeam}{1}\%) &
\numRP{\AveSystTTEightTevabsvalttTotZFidGlobalCorrBeamOnly}{0}
(\numRP{\AveSystTTEightTevpercentsttTotZFidGlobalCorrBeamOnly}{1}\%) &
\numRP{\AveSystTTEightTevabsvalttTotZFidGlobalCorrLumiOnly}{0} 
(\numRP{\AveSystTTEightTevpercentsttTotZFidGlobalCorrLumiOnly}{1}\%) \\
$7$  &
\numRP{\AveCentralTTSevenTevabsvalttTotZFidGlobalCorr}{0} & 
\numRP{\AveStatTTSevenTevabsvalttTotZFidGlobalCorr}{0} 
(\numRP{\AveStatTTSevenTevpercentsttTotZFidGlobalCorr}{1}\%) &
\numRP{\AveSystTTSevenTevabsvalttTotZFidGlobalCorrNoLumiNoBeam}{0} 
(\numRP{\AveSystTTSevenTevpercentsttTotZFidGlobalCorrNoLumiNoBeam}{1}\%) &
\numRP{\AveSystTTSevenTevabsvalttTotZFidGlobalCorrBeamOnly}{0} 
(\numRP{\AveSystTTSevenTevpercentsttTotZFidGlobalCorrBeamOnly}{1}\%) &
\numRP{\AveSystTTSevenTevabsvalttTotZFidGlobalCorrLumiOnly}{0} 
(\numRP{\AveSystTTSevenTevpercentsttTotZFidGlobalCorrLumiOnly}{1}\%) \\

\hline
\end{tabular}
\caption{\label{t:CombXsecZfidTTtot}
Combined fiducial \Zboson-boson and total \ttbar cross sections for $\sqs=13, 8, 7$~TeV. The uncertainties are listed as statistical, systematic, beam-energy, and luminosity. }
\end{table}
\endgroup

\begingroup
\renewcommand*{\arraystretch}{1}
\begin{table}[t]
\centering
\begin{tabular}{|c|cccccc|}
\hline
       &  $Z~13$~TeV  &   $t \bar{t}~13$~TeV  &  $Z~8$~TeV  &   $t \bar{t}~8$~TeV  &  $Z~7$~TeV  &   $t \bar{t}~7$~TeV  \\
\hline
 $Z~13$~TeV  & 
\numRP{\CorrCombZThirteenTevcorZThirteenTevttTotZFidGlobalCorr}{2} & 
\numRP{\CorrCombTTThirteenTevcorZThirteenTevttTotZFidGlobalCorr}{2} & 
\numRP{\CorrCombZEightTevcorZThirteenTevttTotZFidGlobalCorr}{2}  &  
\numRP{\CorrCombTTEightTevcorZThirteenTevttTotZFidGlobalCorr}{2}   & 
\numRP{\CorrCombZSevenTevcorZThirteenTevttTotZFidGlobalCorr}{2}     & 
\numRP{\CorrCombTTSevenTevcorZThirteenTevttTotZFidGlobalCorr}{2} \\
 $t \bar{t}~13$~TeV  &    
 - & 
\numRP{\CorrCombTTThirteenTevcorTTThirteenTevttTotZFidGlobalCorr}{2} & 
\numRP{\CorrCombZEightTevcorTTThirteenTevttTotZFidGlobalCorr}{2}  & 
\numRP{\CorrCombTTEightTevcorTTThirteenTevttTotZFidGlobalCorr}{2}   & 
\numRP{\CorrCombZSevenTevcorTTThirteenTevttTotZFidGlobalCorr}{2}     & 
\numRP{\CorrCombTTSevenTevcorTTThirteenTevttTotZFidGlobalCorr}{2} \\
 $Z~8$~TeV  & 
 - & - &
\numRP{\CorrCombZEightTevcorZEightTevttTotZFidGlobalCorr}{2}  &
\numRP{\CorrCombTTEightTevcorZEightTevttTotZFidGlobalCorr}{2}   & 
\numRP{\CorrCombZSevenTevcorZEightTevttTotZFidGlobalCorr}{2}     & 
\numRP{\CorrCombTTSevenTevcorZEightTevttTotZFidGlobalCorr}{2} \\
 $t \bar{t}~8$~TeV  & 
 - & - & - & 
\numRP{\CorrCombTTEightTevcorTTEightTevttTotZFidGlobalCorr}{2}   & 
\numRP{\CorrCombZSevenTevcorTTEightTevttTotZFidGlobalCorr}{2}     & 
\numRP{\CorrCombTTSevenTevcorTTEightTevttTotZFidGlobalCorr}{2} \\
 $Z~7$~TeV  & 
 - & - & - & - &
\numRP{\CorrCombZSevenTevcorZSevenTevttTotZFidGlobalCorr}{2}     & 
\numRP{\CorrCombTTSevenTevcorZSevenTevttTotZFidGlobalCorr}{2} \\
 $t \bar{t}~7$~TeV  & 
 - & - & - & - & - &
\numRP{\CorrCombTTSevenTevcorTTSevenTevttTotZFidGlobalCorr}{2} \\
\hline
\end{tabular}
\caption{\label{t:TabCorrelations} The correlation coefficients amongst the combined  \Zboson-boson fiducial and \ttbar total cross-section measurements at $\sqrt s = 13, 8, 7$~TeV.}
\end{table}
\endgroup

\begin{figure}[t]
  \centering
    \includegraphics[width=0.48\textwidth]{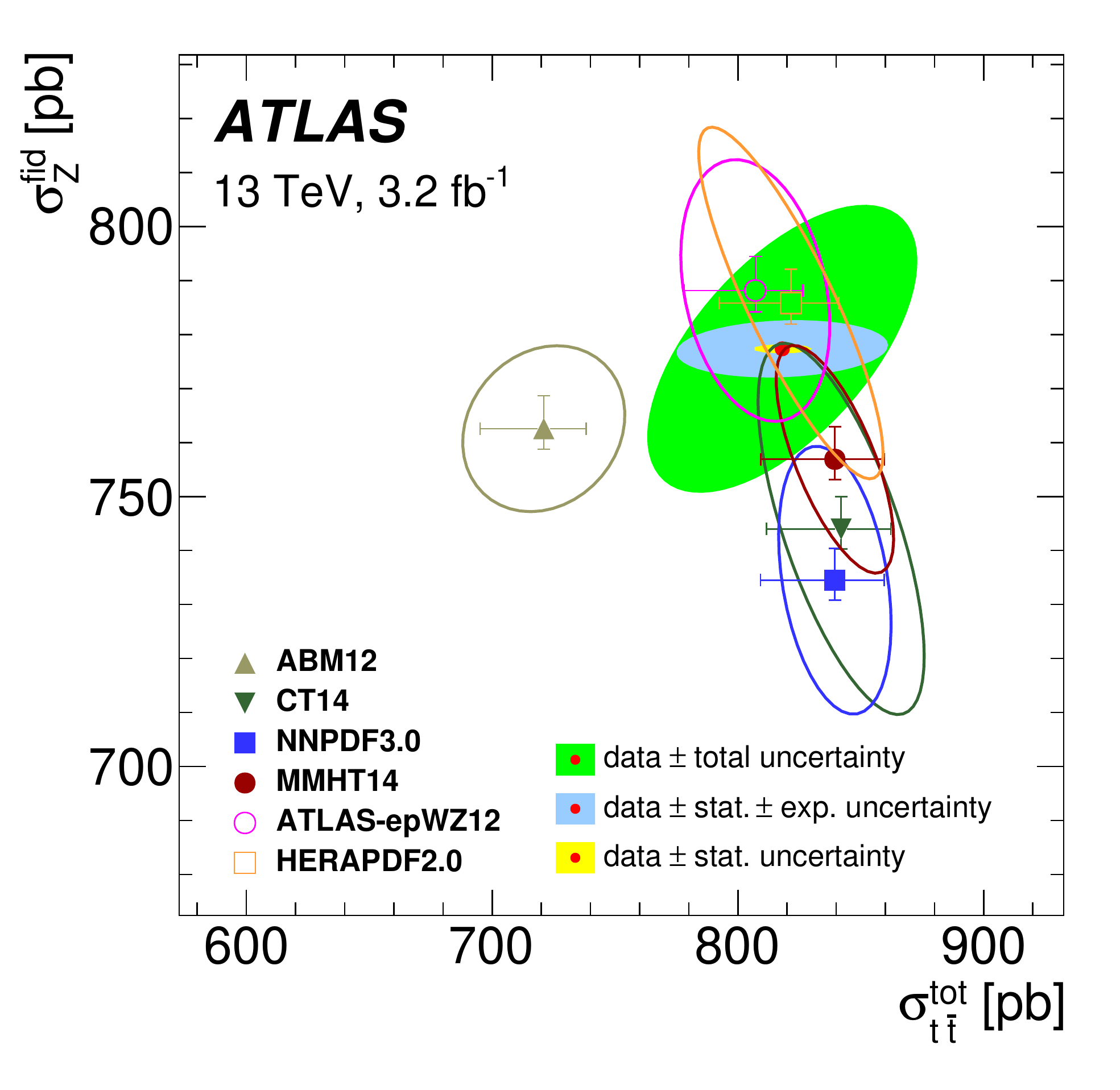}
    \includegraphics[width=0.48\textwidth]{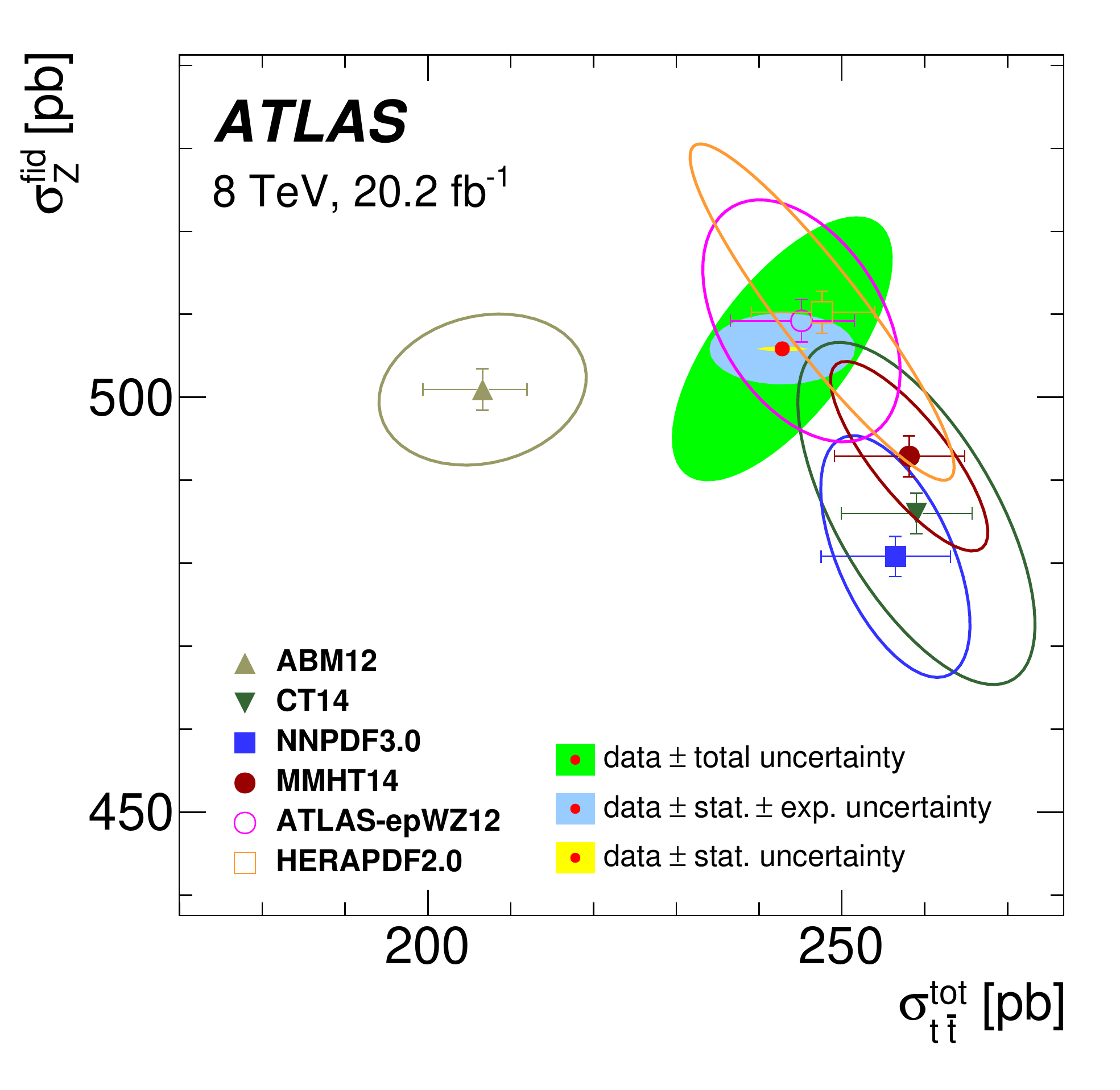}
    \includegraphics[width=0.48\textwidth]{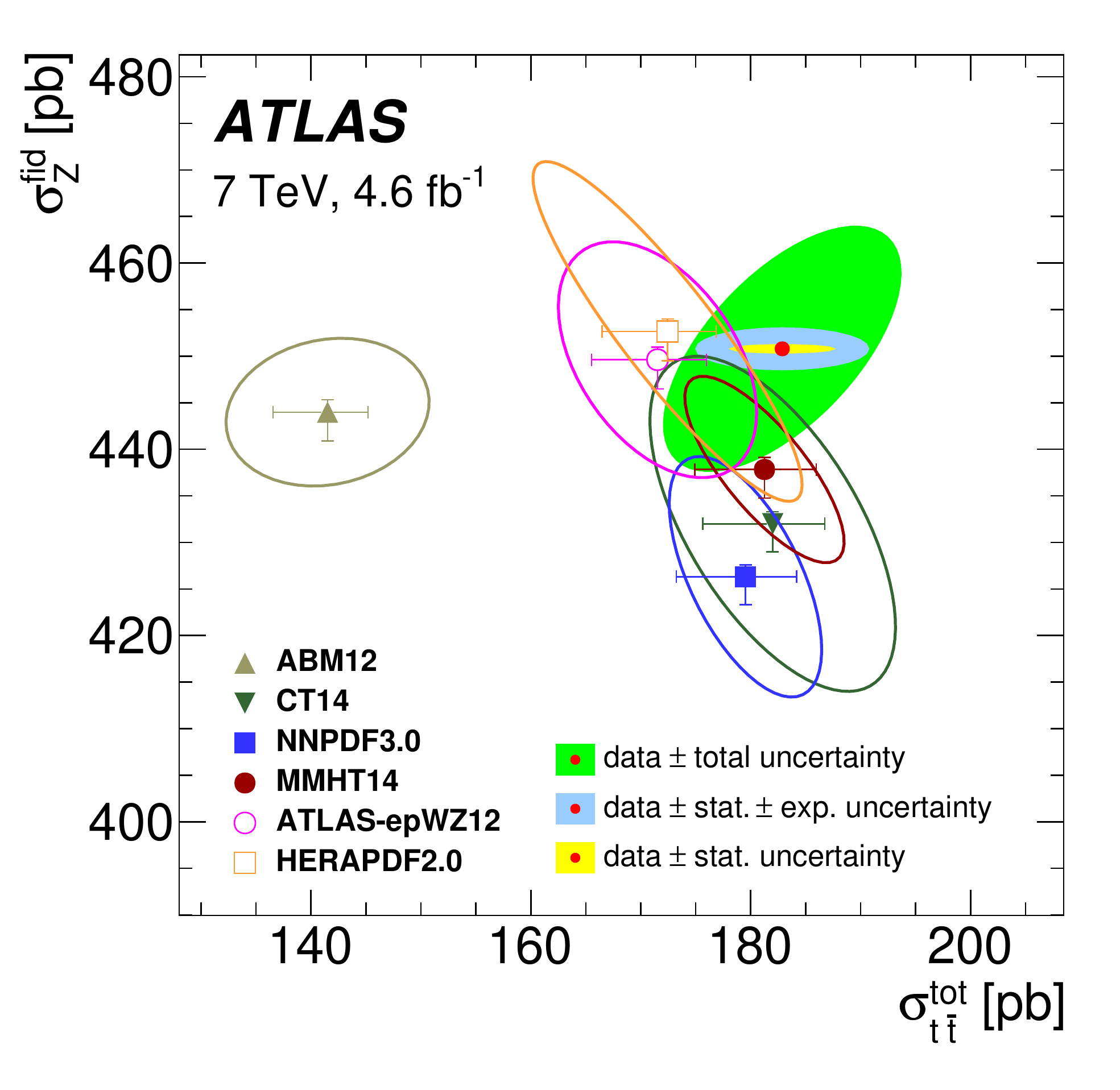}
\caption{\label{fig:contour}
Two-dimensional 68\% CL contours of \sfidZ\ vs. \sigtttot\ at 13~TeV (top, left), 8~TeV (top, right), and 7~TeV (bottom). The solid red circle shows the result of the combination, the yellow ellipse represents the statistical uncertainty, the blue ellipse adds the experimental uncertainty, while the green ellipse is the total uncertainty.  The results are overlayed with the theoretical cross-section predictions calculated from the error sets associated with each specific PDF, also plotted at 68\% CL. The ellipses  correspond to the PDF uncertainties, the asymmetric error bars inside the ellipses represent the scale uncertainties, and the coloured markers are the  central values. 
}
\end{figure}

\subsection{Quantitative comparison with predictions}
\label{sec:cor}

The measured cross sections along with the complete correlation information are compared in a quantitative way to the predictions based on different PDF sets. The comparison is performed using the xFitter package~\cite{Alekhin:2014irh}, 
which allows  PDF and other theoretical uncertainties to be included via asymmetric error propagation.
The comparison is performed for the total  $t\bar{t}$ and fiducial  $Z$-boson cross sections,
including their correlations, as reported in Section~\ref{sec:correlsigma}. 
The resulting $\chi^2$ values corresponding to the different PDFs are given in Table~\ref{tab:chi2pdf}.
All comparisons give an acceptable $\chi^2$ value except for the \ABM PDF set, which is disfavoured by the data.
The covariance matrix is decomposed so as to extract the uncorrelated component of the uncertainties.  Figure~\ref{fig:pdf} visually compares the measurements, with both the total and the uncorrelated components of the uncertainties, to the predictions.
From Figure~\ref{fig:pdf} and Table~\ref{tab:chi2pdf}, it can be observed that the \HERAPDF and \ATLASepWZ sets have good compatibility with the ATLAS data and agreement is improved  when the measurement of the \ttbar cross section at  7~TeV is excluded.

The impact of the ATLAS data  on the PDF uncertainties can be quantified by using the PDF profiling method~\cite{Paukkunen:2014zia,Camarda:2015zba}. 
It is preferable to quantify the impact of the ATLAS data by using PDFs that do not include the cross-section data used in this analysis. Both the \HERAPDF and  \ATLASepWZ sets satisfy these conditions. Given that the \ATLASepWZ set provides smaller uncertainties for the predicted cross sections compared to \HERAPDF, it is chosen for this purpose. 
The profiling of the \ATLASepWZ PDF set is performed only with the components related to the uncertainties of the HERA~\cite{Aaron:2009bp} and 2010 ATLAS~\cite{STDM-2011-06}  \Wboson, \Zboson-boson  data, to mimic the inclusion of the new ATLAS data in the PDF fit. The effect of additional uncertainties arising from model and PDF-parameterisation variations estimated in the  \ATLASepWZ PDF fit are not further investigated. 

Figure~\ref{fig:prof} shows the light-quark sea $\Sigma = \bar{u}+\bar{d}+\bar{s}$ and gluon $g$ distributions
before and after the profiling, including their uncertainties, at the scales $Q^2\approx m_Z^2$  and $Q^2\approx m_t^2$, respectively.
The upper plots show  the profiled distributions divided by
the central value of the \ATLASepWZ PDF set and demonstrate that the central values of the profiled distributions agree very well with the
original set. The lower plots show that the ATLAS \ttbar and $Z$-boson cross-section data impose visible constraints on the light-quark sea distribution at $x<0.02$ and on the gluon distribution at $x\sim 0.1$. These data constrain the least-well-understood component of the light-quark sea distribution, namely the strange-quark distribution while the other quark PDFs are not significantly constrained~\cite{STDM-2012-20}.  The lower plots also show the impact of the \ttbar\ data only, which contribute significantly to the constraint on the gluon distribution, while the \Zboson-boson data help to constrain both the light-quark-sea and gluon distributions.

\begin{table}[t]
\begin{small}
  \begin{center}
\small
    \begin{tabular}{|c|llllll|}
\hline
     & \ATLASepWZ   & \CT   & \MMHT   & \NNPDF   & \HERAPDF   & \ABM  \\ 
\hline
  $\chi^2/{\textrm{NDF}}$  & 8.3 / 6& 15 / 6& 13 / 6& 17 / 6& 10 / 6& 25 / 6  \T \B \\ 
\hline
 p-value  & 0.22 & 0.02 & 0.05 & 0.01 & 0.11 & $<0.001$  \T \B  \\ 
\hline
    \end{tabular}  
  \end{center}
\end{small}
\caption{\label{tab:chi2pdf}$\chi^2$ values for the comparisons of the ATLAS data to the  predictions based on
\ATLASepWZ, \CT, \MMHT, \NNPDF, \HERAPDF and \ABM PDF sets along with the probability of finding the observed value or larger.}
\end{table}
 
\begin{figure}[t]
\centerline{
\includegraphics[width=0.95\textwidth]{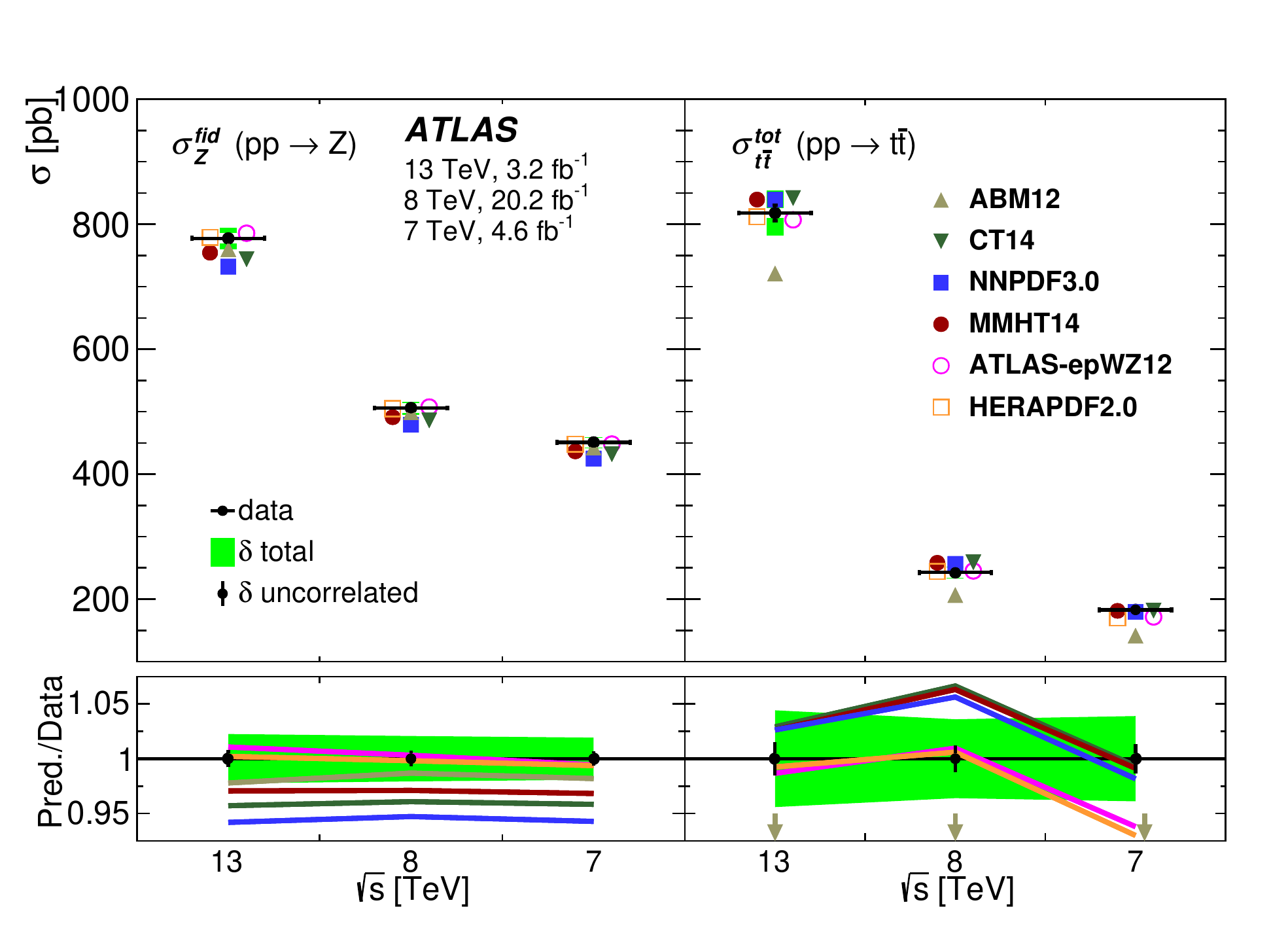}}
\caption{\label{fig:pdf}Comparison of the measured \sfidZ  (left) and \stottt (right) to predictions
based on different PDF sets.  The lower panel shows the total and uncorrelated uncertainties, $\delta$, associated with the ratios of the predictions to the data. In the lower-right panel, the \stottt \ABM predictions are outside of the plot, as indicated by the arrows. The uncertainties in the \Zboson-boson (\ttbar) predictions are typically 3\% (6\%). 
}
\end{figure}

\begin{figure}[t]
\begin{tabular}{cc}
\includegraphics[width=0.48\textwidth]{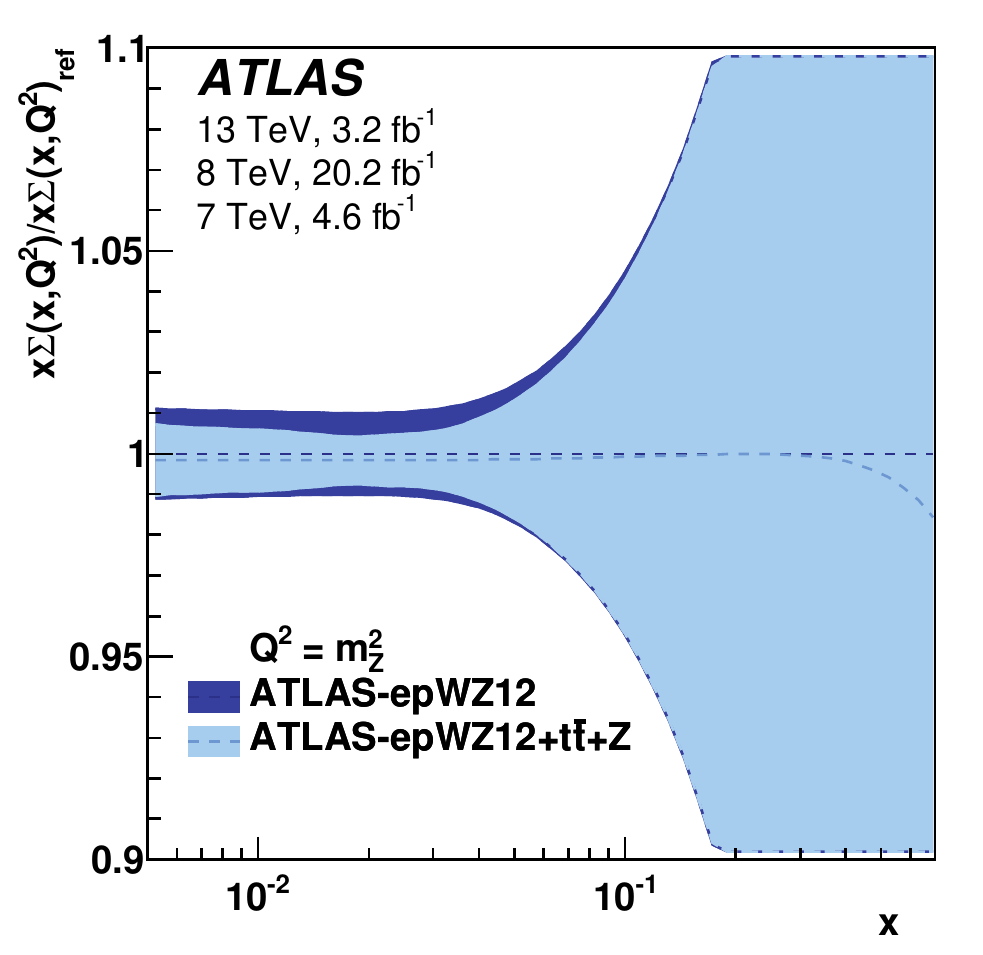}
&
\includegraphics[width=0.48\textwidth]{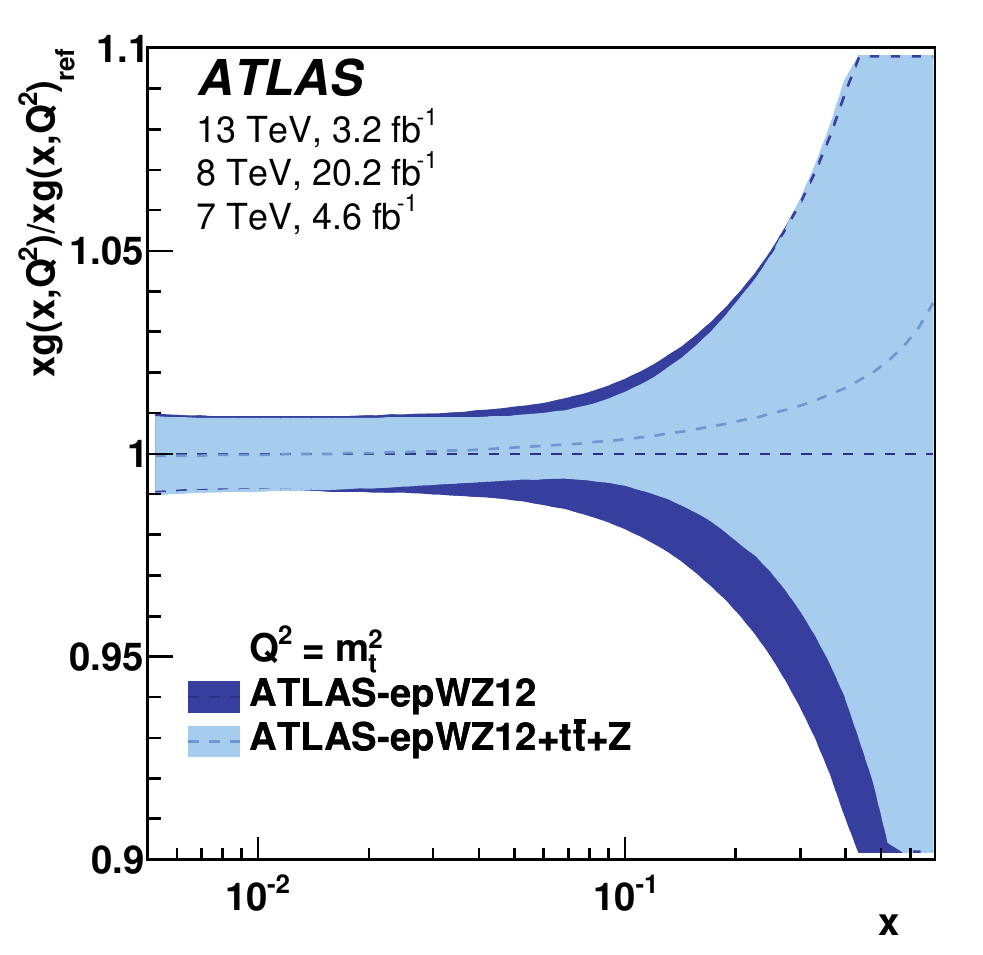}
\\
\includegraphics[width=0.48\textwidth]{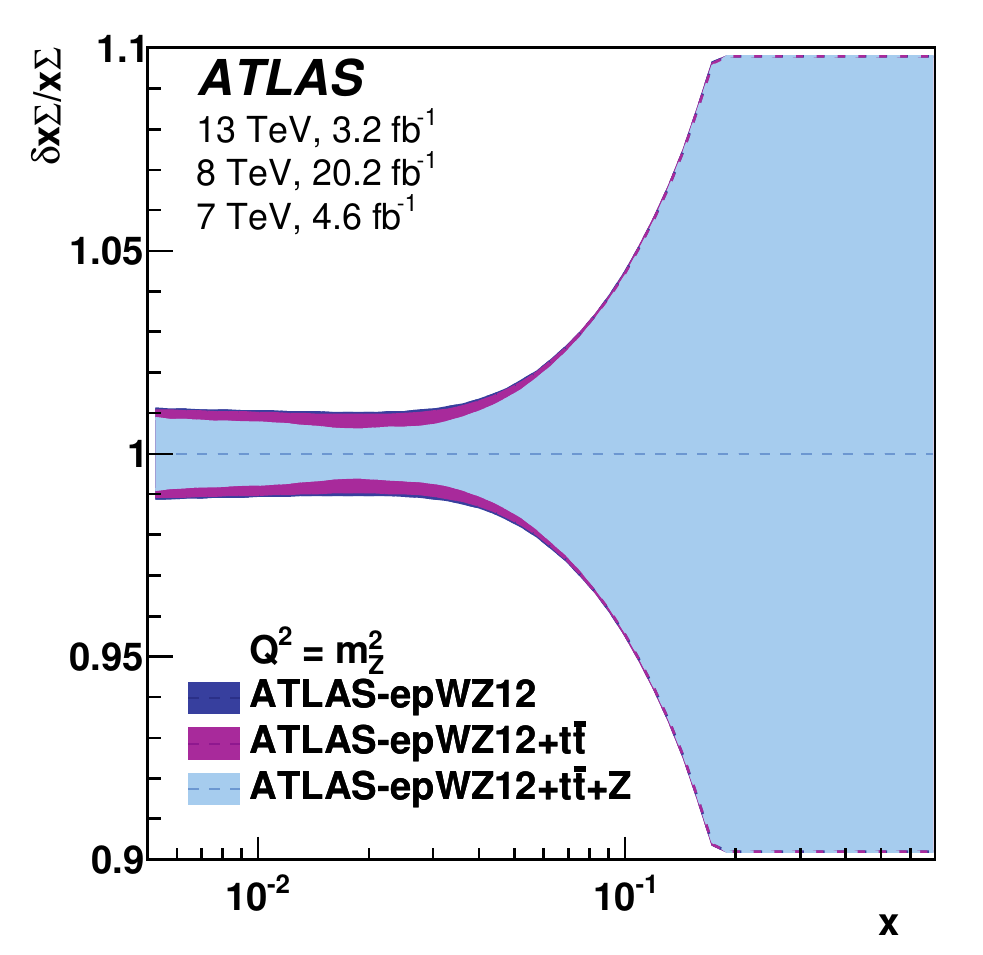}
&
\includegraphics[width=0.48\textwidth]{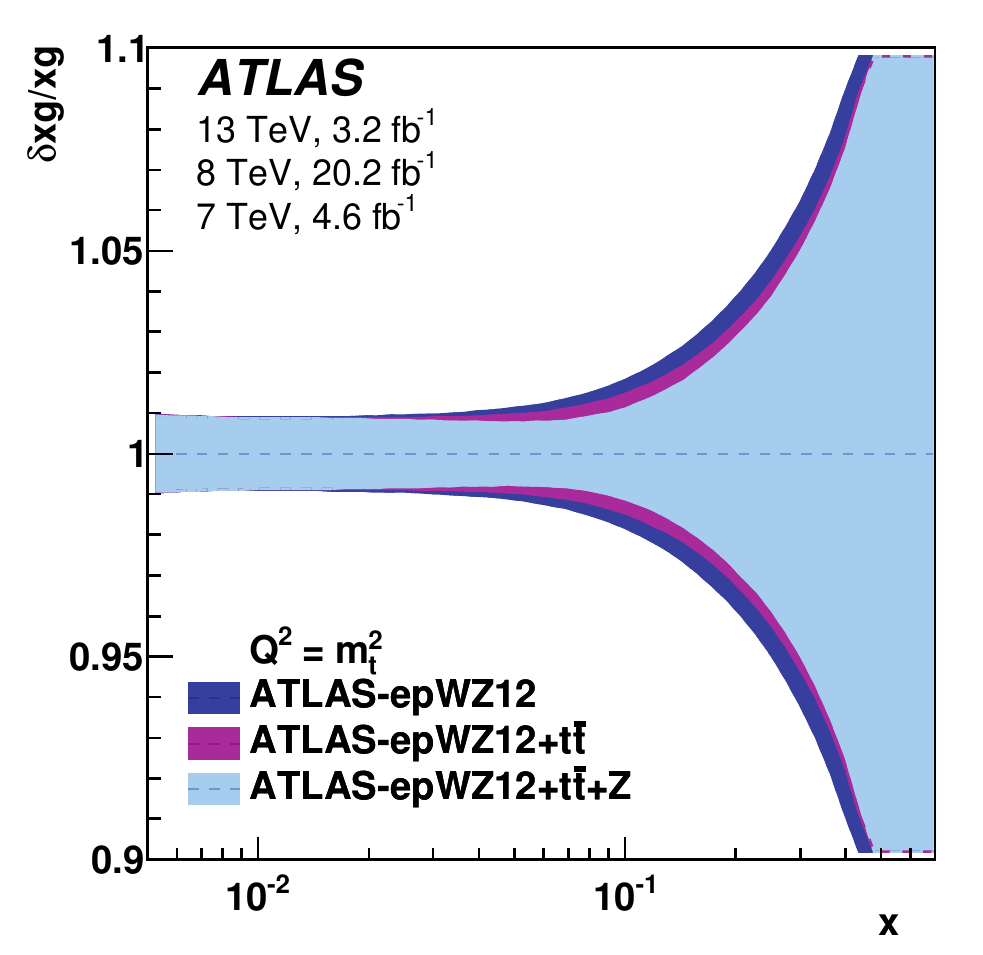}
\\
\end{tabular}
\caption{\label{fig:prof}
Impact of the ATLAS \Zboson-boson and \ttbar cross-section data on the determination of PDFs. The bands represent the uncertainty for the \ATLASepWZ PDF set and the uncertainty of the profiled  \ATLASepWZ PDF set using $t\overline{t}+Z$ data as a function of $x$ for the total light-quark-sea distribution, $x\Sigma$, at  $Q^2 \approx m_Z^2$  (left) and for the gluon density, $x g$, at $Q^2 \approx m_t^2$ (right). In the upper plots, the profiled PDF set is divided by the central value of  \ATLASepWZ PDF set, ``ref'',  while in the lower plots, the relative uncertainty, $\delta$, is given. The lower plots also show the impact  of only including the ATLAS \ttbar\ data set.
In the upper plots, the dashed blue curve represents the ratio of the central value of the profiled result to \ATLASepWZ PDF set. 
}
\end{figure}

\FloatBarrier

\section{Conclusion}
\label{sec:conclusion}
This paper reports a new measurement by the ATLAS Collaboration of the $Z$-boson production cross section at $\sqs=13$~TeV using $3.2$~\ifb of $pp$ collisions from the LHC, together with  the evaluations of single and double ratios
involving \Zboson-boson and \ttbar production cross sections, ($\ratzf{i}{j}, \rattt{i}{j}, \ratttztf(i~\mathrm{TeV})$, and $\ratttztf(i/j)$ where $i,j =  13, 8, 7$) using this new measurement and previously published cross-section measurements at $\sqrt s = 13, 8, 7$~TeV.  
The new measurement of \Zboson-boson production at $\sqs=13$~TeV is fully synchronised to the corresponding \ttbar analysis, to improve the cancellation
of the uncertainties in the ratios, while all other measurements also benefit significantly from the partial cancellation of uncertainties that evaluating ratios can bring.

The experimental results are compared to the state-of-the-art theoretical predictions, which are computed at NNLO (with NLO EW corrections) and NNLO+NNLL accuracy for \Zboson-boson and \ttbar production, respectively. 
Excellent agreement between data and predictions is observed in the \Zboson-boson cross-section ratios at the various centre-of-mass energies, even omitting the luminosity uncertainties. These results indicate that such measurements could be used to normalise cross-section measurements at different $\sqrt s$, as well as provide stringent cross-checks on the corresponding ratios of absolute integrated luminosity values.
The data are found to be in best agreement with the \ATLASepWZ  PDF set, closely followed by the \HERAPDF set, while the \CT, \NNPDF, and \MMHT PDF sets deviate from some of the ratio measurements at the 1--2$\sigma$ level. The \ABM PDF set is disfavoured by the data. A tension is  observed between data and predictions of the double ratio between 8~TeV and 7~TeV, which is difficult to ascribe entirely to the $\sqrt s$ dependence of the PDFs. 

The data presented here have significant power to constrain the gluon distribution function at Bjorken-$x\sim 0.1$ and the total light-quark sea at $x < 0.02$, as demonstrated from a profiling analysis involving the  \ATLASepWZ PDF set.


\section*{Acknowledgements}


We thank CERN for the very successful operation of the LHC, as well as the
support staff from our institutions without whom ATLAS could not be
operated efficiently.

We acknowledge the support of ANPCyT, Argentina; YerPhI, Armenia; ARC, Australia; BMWFW and FWF, Austria; ANAS, Azerbaijan; SSTC, Belarus; CNPq and FAPESP, Brazil; NSERC, NRC and CFI, Canada; CERN; CONICYT, Chile; CAS, MOST and NSFC, China; COLCIENCIAS, Colombia; MSMT CR, MPO CR and VSC CR, Czech Republic; DNRF and DNSRC, Denmark; IN2P3-CNRS, CEA-DSM/IRFU, France; GNSF, Georgia; BMBF, HGF, and MPG, Germany; GSRT, Greece; RGC, Hong Kong SAR, China; ISF, I-CORE and Benoziyo Center, Israel; INFN, Italy; MEXT and JSPS, Japan; CNRST, Morocco; FOM and NWO, Netherlands; RCN, Norway; MNiSW and NCN, Poland; FCT, Portugal; MNE/IFA, Romania; MES of Russia and NRC KI, Russian Federation; JINR; MESTD, Serbia; MSSR, Slovakia; ARRS and MIZ\v{S}, Slovenia; DST/NRF, South Africa; MINECO, Spain; SRC and Wallenberg Foundation, Sweden; SERI, SNSF and Cantons of Bern and Geneva, Switzerland; MOST, Taiwan; TAEK, Turkey; STFC, United Kingdom; DOE and NSF, United States of America. In addition, individual groups and members have received support from BCKDF, the Canada Council, CANARIE, CRC, Compute Canada, FQRNT, and the Ontario Innovation Trust, Canada; EPLANET, ERC, FP7, Horizon 2020 and Marie Sk{\l}odowska-Curie Actions, European Union; Investissements d'Avenir Labex and Idex, ANR, R{\'e}gion Auvergne and Fondation Partager le Savoir, France; DFG and AvH Foundation, Germany; Herakleitos, Thales and Aristeia programmes co-financed by EU-ESF and the Greek NSRF; BSF, GIF and Minerva, Israel; BRF, Norway; Generalitat de Catalunya, Generalitat Valenciana, Spain; the Royal Society and Leverhulme Trust, United Kingdom.

The crucial computing support from all WLCG partners is acknowledged gratefully, in particular from CERN, the ATLAS Tier-1 facilities at TRIUMF (Canada), NDGF (Denmark, Norway, Sweden), CC-IN2P3 (France), KIT/GridKA (Germany), INFN-CNAF (Italy), NL-T1 (Netherlands), PIC (Spain), ASGC (Taiwan), RAL (UK) and BNL (USA), the Tier-2 facilities worldwide and large non-WLCG resource providers. Major contributors of computing resources are listed in Ref.~\cite{ATL-GEN-PUB-2016-002}.

\clearpage
\appendix
\part*{Appendix}
\addcontentsline{toc}{part}{Appendix}
\section{Predictions involving \Zboson-boson total cross sections}
\label{AppA}

Tables~\ref{tab:xsec_predictions_ZtotZZtt3} and~\ref{tab:xsec_ratio_pred_ZZtt4} mirror the \Zboson-boson information in Tables~\ref{tab:xsec_predictions_Zfidtttot},~\ref{tab:xsec_ratio_pred_ZZtt} and~\ref{tab:xsec_ratio_pred_ZZtt2}, and use the same methodology as described in Section~\ref{sec:ztheory} except that the total \Zboson-boson production cross sections times the branching ratio into a lepton pair of flavour $\ell^{+}\ell^{-}$ are given rather than the fiducial cross sections.

\begin{table}[h]
\centering
\begin{tabular}{|l|l|l|l||l|l|l|} 
\hline  
					& \multicolumn{3}{c||}{$\stotZ(i~\mathrm{TeV})$} & \multicolumn{3}{c|}{\ratzt{i}{j}} \T \B \\
\hline
\multicolumn{1}{|c|}{$i$ or $i/j$}	& \multicolumn{1}{c|}{13} &  \multicolumn{1}{c|}{8} &  \multicolumn{1}{c||}{7} & \multicolumn{1}{c|}{13/7} &  \multicolumn{1}{c|}{13/8} &  \multicolumn{1}{c|}{8/7} \T	\\ \hline 
\multicolumn{1}{|l|}{Central value [pb]}     &\numRP{1886.2}{0}&\numRP{1110.0}{0} &\numRP{954.3}{0}  &\numRP{1.9766}{3}&\numRP{1.6994}{3} &\numRP{1.1631}{3}	\\ \hline
\multicolumn{1}{|l|}{Uncertainties [\%]}& & & & & & \\
\multicolumn{1}{|l|}{\hspace{0.2cm}PDF}               &$^{+\numRP{2.39}{1}}_{\numRP{-2.67}{1}}$&$^{+\numRP{2.20}{1}}_{\numRP{-2.52}{1}}$&$^{+\numRP{2.17}{1}}_{\numRP{-2.50}{1}}$ &$^{+\numRP{0.8651219}{1}}_{\numRP{-0.7993524}{1}}$&$^{+\numRP{0.6884783}{1}}_{\numRP{-0.6296340}{1}}$&$^{+\numRP{0.1805520}{2}}_{\numRP{-0.1719543}{2}}$\T \B \\
\multicolumn{1}{|l|}{\hspace{0.2cm}$\alpha_{\textrm{S}}$}   &$^{+\numRP{1.00}{1}}_{\numRP{-0.93}{1}}$&$^{+\numRP{0.75}{1}}_{\numRP{-0.81}{1}}$&$^{+\numRP{0.83}{1}}_{\numRP{-0.87}{1}}$ &$^{+\numRP{0.17}{1}}_{\numRP{-0.06}{1}}$&$^{+\numRP{0.249999}{1}}_{\numRP{-0.12}{1}}$&$^{-\numRP{0.08}{1}}_{+\numRP{0.06}{1}}$ \T \B \\
\multicolumn{1}{|l|}{\hspace{0.2cm}Scale}            &$^{+\numRP{0.68}{1}}_{\numRP{-1.12}{1}}$&$^{+\numRP{0.64}{1}}_{\numRP{-0.93}{1}}$&$^{+\numRP{0.47}{1}}_{\numRP{-0.88}{1}}$ &$^{+\numRP{0.21}{2}}_{-\numRP{0.30}{2}}$&$^{+\numRP{0.20}{2}}_{-\numRP{0.25}{2}}$&$^{+\numRP{0.19}{2}}_{-\numRP{0.05}{2}}$ \T \B \\  \cline{2-7}
\multicolumn{1}{|l|}{\hspace{0.2cm}Total} &$^{+\numRP{2.6785}{1}}_{-\numRP{3.0411}{1}}$&$^{+\numRP{2.4108}{1}}_{-\numRP{2.8056}{1}}$&$^{+\numRP{2.3704}{1}}_{-\numRP{2.7895}{1}}$&$^{+\numRP{0.906331}{1}}_{-\numRP{0.855900}{1}}$&$^{+\numRP{0.759278}{1}}_{-\numRP{0.687996}{1}}$&$^{+\numRP{0.268885}{2}}_{-\numRP{0.196133}{2}}$ \T \B \\ \hline

\end{tabular}  
\caption{Predictions of the total cross section \stotZ at $\sqrt s = 13, 8, 7$~TeV (left) and  of the cross-section ratio  \ratzt{i}{j} where $i/j=13/7, 13/8, $ and $8/7$ (right) using the \CT PDF.  The uncertainties, given in \%, correspond to variations of: \CT eigenvector set at 68\% CL, $\alpha_{\textrm{S}}$, and QCD scale, as described in the text.  The statistical uncertainties in the cross-section predictions are $< 1$~pb  and are $\le 0.002$ for the ratio predictions, and are not given in the table. } 
\label{tab:xsec_predictions_ZtotZZtt3}
\end{table}

\begin{table}[h]
\centering
\begin{tabular}{|l|l|l|l||l|l|l|} 
\hline 
					& \multicolumn{3}{c||}{$\ratttztt(i~\mathrm{TeV})$}	& \multicolumn{3}{c|}{$\ratttztt(i/j)$} \T \B \\
\hline
\multicolumn{1}{|c|}{$i$ or $i/j$}	& \multicolumn{1}{c|}{13} &  \multicolumn{1}{c|}{8} &  \multicolumn{1}{c||}{7}		&  \multicolumn{1}{c|}{13/7}	&  \multicolumn{1}{c|}{13/8}	&  \multicolumn{1}{c|}{8/7} \T	\\ \hline 
\multicolumn{1}{|l|}{Central value}     &\numRP{0.4463}{3}&\numRP{0.2333}{3} &\numRP{0.1904}{3}& \numRP{2.3442}{3}&\numRP{1.9132}{3}&\numRP{1.2253}{3}	\\ \hline
\multicolumn{1}{|l|}{Uncertainties [\%]}& & & & & & \\
\multicolumn{1}{|l|}{\hspace{0.2cm} PDF}               &$^{+\numRP{5.156951}{0}}_{\numRP{-4.708520}{0}}$&$^{+\numRP{6.008584}{0}}_{\numRP{-5.150215}{0}}$&$^{+\numRP{6.315789}{0}}_{\numRP{-5.263158}{0}}$&$^{+\numRP{1.83447}{1}}_{\numRP{-2.21843}{1}}$&$^{+\numRP{1.35912}{1}}_{\numRP{-1.67277}{1}}$&$^{+\numRP{0.408163}{1}}_{\numRP{-0.571429}{1}}$ \T \B \\
\multicolumn{1}{|l|}{\hspace{0.2cm}$\alpha_{\textrm{S}}$}     &$^{+\numRP{0.86}{1}}_{\numRP{-0.91}{1}}$&$^{+\numRP{1.36}{1}}_{\numRP{-1.25}{1}}$&$^{+\numRP{1.35}{1}}_{\numRP{-1.27}{1}}$&$^{-\numRP{0.49}{2}}_{+\numRP{0.36}{2}}$&$^{-\numRP{0.49}{2}}_{+\numRP{0.35}{2}}$&$^{-\numRP{0.00}{2}}_{+\numRP{0.01}{2}}$ \T \B \\
\multicolumn{1}{|l|}{\hspace{0.2cm}Scale}              &$^{+\numRP{2.69}{1}}_{-\numRP{3.67}{1}}$&$^{+\numRP{2.74}{1}}_{-\numRP{3.56}{1}}$&$^{+\numRP{2.77}{1}}_{-\numRP{3.50}{1}}$&$^{+\numRP{0.35}{2}}_{-\numRP{0.34}{2}}$&$^{+\numRP{0.38}{2}}_{-\numRP{0.28}{2}}$&$^{+\numRP{0.09}{2}}_{-\numRP{0.21}{2}}$ \T \B \\ 
\multicolumn{1}{|l|}{\hspace{0.2cm}\mtop}            &$^{+\numRP{2.79}{1}}_{\numRP{-2.70}{1}}$ & $^{+\numRP{3.00}{1}}_{\numRP{-2.90}{1}}$ & $^{+\numRP{3.07}{1}}_{\numRP{-2.97}{1}}$ & $^{+\numRP{0.29}{2}}_{\numRP{-0.29}{2}}$ & $^{+\numRP{0.22}{2}}_{\numRP{-0.22}{2}}$ & $^{+\numRP{0.07}{2}}_{\numRP{-0.07}{2}}$ \T \B \\  \cline{2-7}
\multicolumn{1}{|l|}{\hspace{0.2cm}Total} &$^{+\numRP{6.50799}{0}}_{-\numRP{6.61492}{0}}$&$^{+\numRP{7.37972}{0}}_{-\numRP{7.01219}{0}}$&$^{+\numRP{7.66873}{0}}_{-\numRP{7.09821}{0}}$&$^{+\numRP{1.924}{1}}_{-\numRP{2.314}{1}}$&$^{+\numRP{1.470}{1}}_{-\numRP{1.779}{1}}$&$^{+\numRP{0.423}{1}}_{-\numRP{0.611}{1}}$ \T \B \\ \hline
\end{tabular}  
\caption{
Predictions of the cross-section ratios $\ratttztt(i~\mathrm{TeV})$ and $\ratttztt(i/j)$ at the different $\sqrt s$ values where $i,j=13, 8, 7$ using the \CT PDF.  The uncertainties, given in \%, correspond to variations of: \CT eigenvector set at 68\% CL, $\alpha_{\textrm{S}}$, QCD scale,  intrinsic \Zboson-boson prediction, and top-quark mass, as described in the text.  The statistical uncertainties in the predictions are $< 0.001$ for $\ratttztt(i~\mathrm{TeV})$  and $\le 0.002$ for $\ratttztt(i/j)$  and are not given in the table. 
}
\label{tab:xsec_ratio_pred_ZZtt4} 
\end{table}

\section{Acceptance factors and results in the \Zboson-boson total phase space}
\label{AppB}

The combined \Zboson-boson fiducial cross sections in Section~\ref{sec:correlsigma} are extrapolated to the full phase space within the dilepton invariant mass range $ 66 < m_{\ell\ell} < 116{\GeV}$ in Table~\ref{t:CombXsecZtot} by use of acceptance factors $A$, as  described in Section~\ref{sec:theory}.  
The acceptance factors $A$ are expressed as the fraction of decays satisfying the fiducial acceptance at Monte Carlo generator level and are calculated using \textsc{DYNNLO}~1.5 with the \CT PDF for the central value and for the variations reflecting the PDF set's systematic uncertainty.
In addition, uncertainties due to parton showers and the hadronisation description are taken from a previous publication~\cite{STDM-2011-06}, after checking their validity for the 13~TeV result, and  were derived as the differences between the acceptances calculated with \POWHEG-Box~v1 but using different models for parton shower and hadronisation descriptions, namely the \HERWIG~\cite{Corcella:2000bw} or \PYTHIA~\cite{Sjostrand:2006za} programs. 
The acceptance factor used to extrapolate from fiducial to total cross sections, however, has a sizeable uncertainty which is treated as correlated in the ratio measurements for data at different $\sqs$ values.

\begingroup
\begin{table}[t]
\centering
\renewcommand*{\arraystretch}{1}
\begin{tabular}{|c|c@{$\,\pm\,$}c@{$\,\pm\,$}c@{$\,\pm\,$}c@{$\,\pm\,$}c|}
\hline
\sqs [TeV]  & \multicolumn{5}{c|}{$A$ $\pm$ total uncertainty} \\ \hline
$13$ & \multicolumn{5}{c|}{$0.395 \pm 0.007$} \\
$8$  & \multicolumn{5}{c|}{$0.466 \pm 0.008$} \\
$7$  & \multicolumn{5}{c|}{$0.505 \pm 0.009$} \\
\hline \hline
\sqs [TeV]  & \multicolumn{5}{c|}{\stotZ $\pm$ stat $\pm$ syst $\pm$ beam $\pm$ lumi [pb]} \T \B \\
\hline
$13$     &
\multicolumn{1}{r}{\numRP{\AveCentralZThirteenTevabsvalttTotZTotGlobalCorr}{0}}$\pm$ & 
\numRP{\AveStatZThirteenTevabsvalttTotZTotGlobalCorr}{0}
(\numRP{\AveStatZThirteenTevpercentsttTotZTotGlobalCorr}{1}\%) &
\numRP{\AveSystZThirteenTevabsvalttTotZTotGlobalCorrNoLumiNoBeam}{0}
(\numRP{\AveSystZThirteenTevpercentsttTotZTotGlobalCorrNoLumiNoBeam}{1}\%) &
\numRP{\AveSystZThirteenTevabsvalttTotZTotGlobalCorrBeamOnly}{0}
(\numRP{\AveSystZThirteenTevpercentsttTotZTotGlobalCorrBeamOnly}{1}\%) &
\numRP{\AveSystZThirteenTevabsvalttTotZTotGlobalCorrLumiOnly}{0}
(\numRP{\AveSystZThirteenTevpercentsttTotZTotGlobalCorrLumiOnly}{1}\%) \\
$8$     &
\multicolumn{1}{r}{\numRP{\AveCentralZEightTevabsvalttTotZTotGlobalCorr}{0}}$\pm$ &
$<1$
$(<0.1\%)$ & 
\numRP{\AveSystZEightTevabsvalttTotZTotGlobalCorrNoLumiNoBeam}{0}
(\numRP{\AveSystZEightTevpercentsttTotZTotGlobalCorrNoLumiNoBeam}{1}\%) &
\numRP{\AveSystZEightTevabsvalttTotZTotGlobalCorrBeamOnly}{0}
(\numRP{\AveSystZEightTevpercentsttTotZTotGlobalCorrBeamOnly}{1}\%) &
\numRP{\AveSystZEightTevabsvalttTotZTotGlobalCorrLumiOnly}{0}
(\numRP{\AveSystZEightTevpercentsttTotZTotGlobalCorrLumiOnly}{1}\%) \\
$7$     &
\multicolumn{1}{r}{\numRP{\AveCentralZSevenTevabsvalttTotZTotGlobalCorr}{0}}$\pm$ &
\numRP{\AveStatZSevenTevabsvalttTotZTotGlobalCorr}{0}
(\numRP{\AveStatZSevenTevpercentsttTotZTotGlobalCorr}{1}\%) & 
\numRP{\AveSystZSevenTevabsvalttTotZTotGlobalCorrNoLumiNoBeam}{0}
(\numRP{\AveSystZSevenTevpercentsttTotZTotGlobalCorrNoLumiNoBeam}{1}\%) & 
\numRP{\AveSystZSevenTevabsvalttTotZTotGlobalCorrBeamOnly}{0}
(\numRP{\AveSystZSevenTevpercentsttTotZTotGlobalCorrBeamOnly}{1}\%) & 
\numRP{\AveSystZSevenTevabsvalttTotZTotGlobalCorrLumiOnly}{0}
(\numRP{\AveSystZSevenTevpercentsttTotZTotGlobalCorrLumiOnly}{1}\%) \\

\hline
\end{tabular}
\caption{\label{t:CombXsecZtot} 
Acceptance factors $A$ and combined total \Zboson-boson cross sections times leptonic branching ratio within the invariant mass window $66 < m_{\ell\ell} < 116$~GeV for $\sqs=13, 8, 7$~TeV. The uncertainties in the cross sections are listed as statistical,  systematic, beam-energy, and luminosity while those for the $A$ factor include the total uncertainty. } 
\end{table}
\endgroup

\section{Tables of results}
\label{AppC}

Table~\ref{t:TabCorrelationsNoLumi} presents the correlation coefficients matrix as in Table~\ref{t:TabCorrelations} but omitting both the luminosity and beam-energy uncertainties.
Tables~\ref{t:singleratio} and~\ref{t:TabDoubRats} summarise all of the single and double-ratio results for the \ttbar and \Zboson-boson production cross sections at $\sqrt s = 13,8,7$~TeV.

\begingroup
\renewcommand*{\arraystretch}{1}
\begin{table}[t]
\centering
\begin{tabular}{|c|cccccc|}
\hline
       &  $Z~13$~TeV  &   $t \bar{t}~13$~TeV  &  $Z~8$~TeV  &   $t \bar{t}~8$~TeV  &  $Z~7$~TeV  &   $t \bar{t}~7$~TeV  \\
\hline
 $Z~13$~TeV  &
1. & 
\numRP{\CorrCombTTThirteenTevcorZThirteenTevttTotZFidGlobalCorrNoLumiUnc}{2} & 
\numRP{\CorrCombZEightTevcorZThirteenTevttTotZFidGlobalCorrNoLumiUnc}{2}  &  
\numRP{\CorrCombTTEightTevcorZThirteenTevttTotZFidGlobalCorrNoLumiUnc}{2} & 
\numRP{\CorrCombZSevenTevcorZThirteenTevttTotZFidGlobalCorrNoLumiUnc}{2} & 
\numRP{\CorrCombTTSevenTevcorZThirteenTevttTotZFidGlobalCorrNoLumiUnc}{2} \\
 $t \bar{t}~13$~TeV  &
- & 1. & 
\numRP{\CorrCombZEightTevcorTTThirteenTevttTotZFidGlobalCorrNoLumiUnc}{2}  & 
\numRP{\CorrCombTTEightTevcorTTThirteenTevttTotZFidGlobalCorrNoLumiUnc}{2} & 
\numRP{\CorrCombZSevenTevcorTTThirteenTevttTotZFidGlobalCorrNoLumiUnc}{2} & 
\numRP{\CorrCombTTSevenTevcorTTThirteenTevttTotZFidGlobalCorrNoLumiUnc}{2} \\
 $Z~8$~TeV  &
- & - & 1. & 
\numRP{\CorrCombTTEightTevcorZEightTevttTotZFidGlobalCorrNoLumiUnc}{2}   & 
\numRP{\CorrCombZSevenTevcorZEightTevttTotZFidGlobalCorrNoLumiUnc}{2}     & 
\numRP{\CorrCombTTSevenTevcorZEightTevttTotZFidGlobalCorrNoLumiUnc}{2} \\
 $t \bar{t}~8$~TeV  &
- & - & - & 1. & 
\numRP{\CorrCombZSevenTevcorTTEightTevttTotZFidGlobalCorrNoLumiUnc}{2}     & 
\numRP{\CorrCombTTSevenTevcorTTEightTevttTotZFidGlobalCorrNoLumiUnc}{2} \\
 $Z~7$~TeV  &
- & - & -  & - & 1. & 
\numRP{\CorrCombTTSevenTevcorZSevenTevttTotZFidGlobalCorrNoLumiUnc}{2} \\
 $t \bar{t}~7$~TeV  &
- & - & - & - & - & 1. \\
\hline
\end{tabular}
\caption{\label{t:TabCorrelationsNoLumi} The correlation coefficients amongst the combined  \Zboson-boson fiducial and \ttbar total cross-section measurements at $\sqrt s = 13, 8, 7$~TeV as in Table~\ref{t:TabCorrelations} but omitting the luminosity and beam-energy uncertainties.}
\end{table}
\endgroup

\label{sec:rattab}

\begingroup
\renewcommand*{\arraystretch}{2.4}
\begin{sidewaystable}
\begin{scriptsize}
\begin{tabular}{|l||c|c|c|}
\hline
             & $\sigma^{\textrm{tot}} / \sigma^{\textrm{tot}}$     &   $\sigma^{\textrm{tot}} / \sigma^{\textrm{fid}}$   &  $\sigma^{\textrm{fid}} / \sigma^{\textrm{fid}}$  \\
\hline
             &  Value $\pm$ stat $\pm$ syst $\pm$ lumi      &   Value $\pm$ stat $\pm$ syst $\pm$ lumi     &   Value $\pm$ stat $\pm$ syst $\pm$ lumi    \\
\hline
\hline
$\ttbar/Z(13)$    & $\numRP{\TTtotalThirtZtotalThirtRatioVal}{3}$~$\pm$
                         $\numRP{\TTtotalThirtZtotalThirtRatioErrSt}{3}$ ($\numRP{\TTtotalThirtZtotalThirtRatioErrStPercent}{1}$\%)~$\pm$
                         $\numRP{\TTtotalThirtZtotalThirtRatioErrSyNoLumiUnc}{3}$ ($\numRP{\TTtotalThirtZtotalThirtRatioErrSyPercentNoLumiUnc}{1}$\%)~$\pm$
                         $\numRP{\TTtotalThirtZtotalThirtRatioLumiErrAbs}{3}$ ($\numRP{\TTtotalThirtZtotalThirtRatioLumiErrPercent}{1}$\%) &
                         $\numRP{\TTtotalThirtZfiducialThirtRatioVal}{3}$~$\pm$
                         $\numRP{\TTtotalThirtZfiducialThirtRatioErrSt}{3}$ ($\numRP{\TTtotalThirtZfiducialThirtRatioErrStPercent}{1}$\%)~$\pm$
                         $\numRP{\TTtotalThirtZfiducialThirtRatioErrSyNoLumiUnc}{3}$ ($\numRP{\TTtotalThirtZfiducialThirtRatioErrSyPercentNoLumiUnc}{1}$\%)~$\pm$
                         $\numRP{\TTtotalThirtZfiducialThirtRatioLumiErrAbs}{3}$ ($\numRP{\TTtotalThirtZfiducialThirtRatioLumiErrPercent}{1}$\%) &
                         $\numRP{\TTfiducialThirtZfiducialThirtRatioVal}{5}$~$\pm$
                         $\numRP{\TTfiducialThirtZfiducialThirtRatioErrSt}{5}$ ($\numRP{\TTfiducialThirtZfiducialThirtRatioErrStPercent}{1}$\%)~$\pm$
                         $\numRP{\TTfiducialThirtZfiducialThirtRatioErrSyNoLumiUnc}{5}$ ($\numRP{\TTfiducialThirtZfiducialThirtRatioErrSyPercentNoLumiUnc}{1}$\%)~$\pm$
                         $\numRP{\TTfiducialThirtZfiducialThirtRatioLumiErrAbs}{5}$ ($\numRP{\TTfiducialThirtZfiducialThirtRatioLumiErrPercent}{1}$\%) \\
\hline
$\ttbar/Z(8)$&      $\numRP{\TTbarEighttotalToZeighttotalRatioVal}{3}$~$\pm$
                         $\numRP{\TTbarEighttotalToZeighttotalratioErrSt}{3}$ ($\numRP{\TTbarEighttotalToZeighttotalratioErrStPercent}{1}$\%)~$\pm$
                         $\numRP{\TTbarEighttotalToZeighttotalratioErrSyNoLumiUnc}{3}$ ($\numRP{\TTbarEighttotalToZeighttotalratioErrSyPercentNoLumiUnc}{1}$\%)~$\pm$
                         $\numRP{\TTbarEighttotalToZeighttotalratioLumiErrAbs}{3}$ ($\numRP{\TTbarEighttotalToZeighttotalratioLumiErrPercent}{1}$\%) &
                         $\numRP{\TTbarEighttotalToZeightfiducialRatioVal}{3}$~$\pm$
                         $\numRP{\TTbarEighttotalToZeightfiducialratioErrSt}{3}$ ($\numRP{\TTbarEighttotalToZeightfiducialratioErrStPercent}{1}$\%)~$\pm$
                         $\numRP{\TTbarEighttotalToZeightfiducialratioErrSyNoLumiUnc}{3}$ ($\numRP{\TTbarEighttotalToZeightfiducialratioErrSyPercentNoLumiUnc}{1}$\%)~$\pm$
                         $\numRP{\TTbarEighttotalToZeightfiducialratioLumiErrAbs}{3}$ ($\numRP{\TTbarEighttotalToZeightfiducialratioLumiErrPercent}{1}$\%) &
                         $\numRP{\TTbarEightfiducialToZeightfiducialRatioVal}{5}$~$\pm$
                         $\numRP{\TTbarEightfiducialToZeightfiducialratioErrSt}{5}$ ($\numRP{\TTbarEightfiducialToZeightfiducialratioErrStPercent}{1}$\%)~$\pm$
                         $\numRP{\TTbarEightfiducialToZeightfiducialratioErrSyNoLumiUnc}{5}$ ($\numRP{\TTbarEightfiducialToZeightfiducialratioErrSyPercentNoLumiUnc}{1}$\%)~$\pm$
                         $\numRP{\TTbarEightfiducialToZeightfiducialratioLumiErrAbs}{5}$ ($\numRP{\TTbarEightfiducialToZeightfiducialratioLumiErrPercent}{1}$\%) \\
\hline
$\ttbar/Z(7)$     & $\numRP{\TTbarSeventotalToZseventotalRatioVal}{3}$~$\pm$
                         $\numRP{\TTbarSeventotalToZseventotalratioErrSt}{3}$ ($\numRP{\TTbarSeventotalToZseventotalratioErrStPercent}{1}$\%)~$\pm$
                         $\numRP{\TTbarSeventotalToZseventotalratioErrSyNoLumiUnc}{3}$ ($\numRP{\TTbarSeventotalToZseventotalratioErrSyPercentNoLumiUnc}{1}$\%)~$\pm$
                         $\numRP{\TTbarSeventotalToZseventotalratioLumiErrAbs}{3}$ ($\numRP{\TTbarSeventotalToZseventotalratioLumiErrPercent}{1}$\%) &
                         $\numRP{\TTbarSeventotalToZsevenfiducialRatioVal}{3}$~$\pm$
                         $\numRP{\TTbarSeventotalToZsevenfiducialratioErrSt}{3}$ ($\numRP{\TTbarSeventotalToZsevenfiducialratioErrStPercent}{1}$\%)~$\pm$
                         $\numRP{\TTbarSeventotalToZsevenfiducialratioErrSyNoLumiUnc}{3}$ ($\numRP{\TTbarSeventotalToZsevenfiducialratioErrSyPercentNoLumiUnc}{1}$\%)$\pm$
                         $\numRP{\TTbarSeventotalToZsevenfiducialratioLumiErrAbs}{3}$ ($\numRP{\TTbarSeventotalToZsevenfiducialratioLumiErrPercent}{1}$\%) &
                         $\numRP{\TTbarSevenfiducialToZsevenfiducialRatioVal}{5}$~$\pm$
                         $\numRP{\TTbarSevenfiducialToZsevenfiducialratioErrSt}{5}$ ($\numRP{\TTbarSevenfiducialToZsevenfiducialratioErrStPercent}{1}$\%)~$\pm$
                         $\numRP{\TTbarSevenfiducialToZsevenfiducialratioErrSyNoLumiUnc}{5}$ ($\numRP{\TTbarSevenfiducialToZsevenfiducialratioErrSyPercentNoLumiUnc}{1}$\%)~$\pm$
                         $\numRP{\TTbarSevenfiducialToZsevenfiducialratioLumiErrAbs}{5}$ ($\numRP{\TTbarSevenfiducialToZsevenfiducialratioLumiErrPercent}{1}$\%) \\ 
\hline
$Z(13)/Z(8)$& $\numRP{\RatCombCentZthirteenTeVoverZeightTeVabsvalTot}{3}$~$\pm$
                         $\numRP{\RatCombStatZthirteenTeVoverZeightTeVabsvalTot}{3}$ ($\numRP{\RatCombStatZthirteenTeVoverZeightTeVpercentsTot}{1}$\%)~$\pm$ 
                         $\numRP{\RatCombSystNoLumiUncZthirteenTeVoverZeightTeVabsvalTot}{3}$ ($\numRP{\RatCombSystNoLumiUncZthirteenTeVoverZeightTeVpercentsTot}{1}$\%)~$\pm$
                         $\numRP{\RatCombLumiUncZthirteenTeVoverZeightTeVabsvalTot}{3}$ ($\numRP{\RatCombLumiUncZthirteenTeVoverZeightTeVpercentsTot}{1}$\%) &
                         -  &
                         $\numRP{\RatCombCentZthirteenTeVoverZeightTeVabsvalFid}{3}$~$\pm$
                         $\numRP{\RatCombStatZthirteenTeVoverZeightTeVabsvalFid}{3}$ ($\numRP{\RatCombStatZthirteenTeVoverZeightTeVpercentsFid}{1}$\%)~$\pm$
                          $\numRP{\RatCombSystNoLumiUncZthirteenTeVoverZeightTeVabsvalFid}{3}$ ($\numRP{\RatCombSystNoLumiUncZthirteenTeVoverZeightTeVpercentsFid}{1}$\%)~$\pm$
                         $\numRP{\RatCombLumiUncZthirteenTeVoverZeightTeVabsvalFid}{3}$ ($\numRP{\RatCombLumiUncZthirteenTeVoverZeightTeVpercentsFid}{1}$\%) \\
\hline
$Z(13)/Z(7)$& $\numRP{\RatCombCentZthirteenTeVoverZsevenTeVabsvalTot}{3}$~$\pm$
                         $\numRP{\RatCombStatZthirteenTeVoverZsevenTeVabsvalTot}{3}$ ($\numRP{\RatCombStatZthirteenTeVoverZsevenTeVpercentsTot}{1}$\%)~$\pm$
                         $\numRP{\RatCombSystNoLumiUncZthirteenTeVoverZsevenTeVabsvalTot}{3}$ ($\numRP{\RatCombSystNoLumiUncZthirteenTeVoverZsevenTeVpercentsTot}{1}$\%)~$\pm$
                         $\numRP{\RatCombLumiUncZthirteenTeVoverZsevenTeVabsvalTot}{3}$ ($\numRP{\RatCombLumiUncZthirteenTeVoverZsevenTeVpercentsTot}{1}$\%) &
                         -   &
                         $\numRP{\RatCombCentZthirteenTeVoverZsevenTeVabsvalFid}{3}$~$\pm$
                         $\numRP{\RatCombStatZthirteenTeVoverZsevenTeVabsvalFid}{3}$ ($\numRP{\RatCombStatZthirteenTeVoverZsevenTeVpercentsFid}{1}$\%)~$\pm$
                         $\numRP{\RatCombSystNoLumiUncZthirteenTeVoverZsevenTeVabsvalFid}{3}$ ($\numRP{\RatCombSystNoLumiUncZthirteenTeVoverZsevenTeVpercentsFid}{1}$\%)~$\pm$
                         $\numRP{\RatCombLumiUncZthirteenTeVoverZsevenTeVabsvalFid}{3}$ ($\numRP{\RatCombLumiUncZthirteenTeVoverZsevenTeVpercentsFid}{1}$\%) \\
\hline
$Z(8)/Z(7)$& $\numRP{\RatCombCentZeightTeVoverZsevenTeVabsvalTot}{3}$~$\pm$
                         $\numRP{\RatCombStatZeightTeVoverZsevenTeVabsvalTot}{3}$ ($\numRP{\RatCombStatZeightTeVoverZsevenTeVpercentsTot}{1}$\%)~$\pm$
                         $\numRP{\RatCombSystNoLumiUncZeightTeVoverZsevenTeVabsvalTot}{3}$ ($\numRP{\RatCombSystNoLumiUncZeightTeVoverZsevenTeVpercentsTot}{1}$\%)~$\pm$
                         $\numRP{\RatCombLumiUncZeightTeVoverZsevenTeVabsvalTot}{3}$ ($\numRP{\RatCombLumiUncZeightTeVoverZsevenTeVpercentsTot}{1}$\%) &
                         -  &
                         $\numRP{\RatCombCentZeightTeVoverZsevenTeVabsvalFid}{3}$~$\pm$
                         $\numRP{\RatCombStatZeightTeVoverZsevenTeVabsvalFid}{3}$ ($\numRP{\RatCombStatZeightTeVoverZsevenTeVpercentsFid}{1}$\%)~$\pm$
                         $\numRP{\RatCombSystNoLumiUncZeightTeVoverZsevenTeVabsvalFid}{3}$ ($\numRP{\RatCombSystNoLumiUncZeightTeVoverZsevenTeVpercentsFid}{1}$\%)~$\pm$
                         $\numRP{\RatCombLumiUncZeightTeVoverZsevenTeVabsvalFid}{3}$ ($\numRP{\RatCombLumiUncZeightTeVoverZsevenTeVpercentsFid}{1}$\%) \\
\hline
$\ttbar(13)/\ttbar(8)$& $\numRP{\RatCombCentTTthirteenTeVoverTTeightTeVabsvalTot}{3}$~$\pm$
                         $\numRP{\RatCombStatTTthirteenTeVoverTTeightTeVabsvalTot}{3}$ ($\numRP{\RatCombStatTTthirteenTeVoverTTeightTeVpercentsTot}{1}$\%)~$\pm$
                         $\numRP{\RatCombSystNoLumiUncTTthirteenTeVoverTTeightTeVabsvalTot}{3}$ ($\numRP{\RatCombSystNoLumiUncTTthirteenTeVoverTTeightTeVpercentsTot}{1}$\%)~$\pm$
                         $\numRP{\RatCombLumiUncTTthirteenTeVoverTTeightTeVabsvalTot}{3}$ ($\numRP{\RatCombLumiUncTTthirteenTeVoverTTeightTeVpercentsTot}{1}$\%)  &
                         -  &
                        $\numRP{\RatCombCentTTthirteenTeVoverTTeightTeVabsvalFid}{3}$~$\pm$
                        $\numRP{\RatCombStatTTthirteenTeVoverTTeightTeVabsvalFid}{3}$ ($\numRP{\RatCombStatTTthirteenTeVoverTTeightTeVpercentsFid}{1}$\%)~$\pm$
                        $\numRP{\RatCombSystNoLumiUncTTthirteenTeVoverTTeightTeVabsvalFid}{3}$ ($\numRP{\RatCombSystNoLumiUncTTthirteenTeVoverTTeightTeVpercentsFid}{1}$\%)~$\pm$
                        $\numRP{\RatCombLumiUncTTthirteenTeVoverTTeightTeVabsvalFid}{3}$ ($\numRP{\RatCombLumiUncTTthirteenTeVoverTTeightTeVpercentsFid}{1}$\%) \\
\hline
$\ttbar(13)/\ttbar(7)$& $\numRP{\RatCombCentTTthirteenTeVoverTTsevenTeVabsvalTot}{3}$~$\pm$
                         $\numRP{\RatCombStatTTthirteenTeVoverTTsevenTeVabsvalTot}{3}$ ($\numRP{\RatCombStatTTthirteenTeVoverTTsevenTeVpercentsTot}{1}$\%)~$\pm$
                         $\numRP{\RatCombSystNoLumiUncTTthirteenTeVoverTTsevenTeVabsvalTot}{3}$ ($\numRP{\RatCombSystNoLumiUncTTthirteenTeVoverTTsevenTeVpercentsTot}{1}$\%)~$\pm$
                         $\numRP{\RatCombLumiUncTTthirteenTeVoverTTsevenTeVabsvalTot}{3}$ ($\numRP{\RatCombLumiUncTTthirteenTeVoverTTsevenTeVpercentsTot}{1}$\%)  &
                         -  &
                         $\numRP{\RatCombCentTTthirteenTeVoverTTsevenTeVabsvalFid}{3}$~$\pm$
                         $\numRP{\RatCombStatTTthirteenTeVoverTTsevenTeVabsvalFid}{3}$ ($\numRP{\RatCombStatTTthirteenTeVoverTTsevenTeVpercentsFid}{1}$\%)~$\pm$
                         $\numRP{\RatCombSystNoLumiUncTTthirteenTeVoverTTsevenTeVabsvalFid}{3}$ ($\numRP{\RatCombSystNoLumiUncTTthirteenTeVoverTTsevenTeVpercentsFid}{1}$\%)~$\pm$
                         $\numRP{\RatCombLumiUncTTthirteenTeVoverTTsevenTeVabsvalFid}{3}$ ($\numRP{\RatCombLumiUncTTthirteenTeVoverTTsevenTeVpercentsFid}{1}$\%)  \\
\hline
$\ttbar(8)/\ttbar(7)$& $\numRP{\RatCombCentTTeightTeVoverTTsevenTeVabsvalTot}{3}$~$\pm$
                         $\numRP{\RatCombStatTTeightTeVoverTTsevenTeVabsvalTot}{3}$ ($\numRP{\RatCombStatTTeightTeVoverTTsevenTeVpercentsTot}{1}$\%)~$\pm$
                         $\numRP{\RatCombSystNoLumiUncTTeightTeVoverTTsevenTeVabsvalTot}{3}$ ($\numRP{\RatCombSystNoLumiUncTTeightTeVoverTTsevenTeVpercentsTot}{1}$\%)~$\pm$
                         $\numRP{\RatCombLumiUncTTeightTeVoverTTsevenTeVabsvalTot}{3}$ ($\numRP{\RatCombLumiUncTTeightTeVoverTTsevenTeVpercentsTot}{1}$\%)  &
                         -  &
                         $\numRP{\RatCombCentTTeightTeVoverTTsevenTeVabsvalFid}{3}$~$\pm$
                         $\numRP{\RatCombStatTTeightTeVoverTTsevenTeVabsvalFid}{3}$ ($\numRP{\RatCombStatTTeightTeVoverTTsevenTeVpercentsFid}{1}$\%)~$\pm$
                         $\numRP{\RatCombSystNoLumiUncTTeightTeVoverTTsevenTeVabsvalFid}{3}$ ($\numRP{\RatCombSystNoLumiUncTTeightTeVoverTTsevenTeVpercentsFid}{1}$\%)~$\pm$
                         $\numRP{\RatCombLumiUncTTeightTeVoverTTsevenTeVabsvalFid}{3}$ ($\numRP{\RatCombLumiUncTTeightTeVoverTTsevenTeVpercentsFid}{1}$\%)  \\
\hline
\end{tabular}
\end{scriptsize}
\caption{\label{t:singleratio} Summary of the $Z$-boson and $t \bar{t}$ production cross-section single ratios at $13, 8, 7$~TeV. The beam-energy uncertainty, which largely cancels in the ratios, is included in the systematic uncertainty.}
\end{sidewaystable}
\endgroup

\begingroup
\renewcommand*{\arraystretch}{3.7}
\begin{sidewaystable}
\centering
\begin{scriptsize}
\begin{tabular}{|l||c|c|c|c|}
\hline
             &  $\left[\sigma^{\textrm{tot}} / \sigma^{\textrm{tot}}\right]/\left[\sigma^{\textrm{tot}} / \sigma^{\textrm{tot}}\right]$       &  $\left[\sigma^{\textrm{tot}} / \sigma^{\textrm{tot}}\right]/\left[\sigma^{\textrm{fid}} / \sigma^{\textrm{fid}}\right]$    &   $\left[\sigma^{\textrm{fid}} / \sigma^{\textrm{fid}}\right]/\left[\sigma^{\textrm{fid}} / \sigma^{\textrm{fid}}\right]$         \\
\hline
             &  Value $\pm$ stat $\pm$ syst  $\pm$ lumi &   Value $\pm$ stat $\pm$ syst  $\pm$ lumi &   Value $\pm$ stat $\pm$ syst $\pm$ lumi   \\
\hline
\hline

 $\ttbar/Z(13/8)$                      & $\numRP{\RatCombCentDoubRTTthirtOverZthirtTOtteightOverZeightabsvalttTotZTot}{3}$~$\pm$
                         $\numRP{\RatCombStatDoubRTTthirtOverZthirtTOtteightOverZeightabsvalttTotZTot}{3}$ ($\numRP{\RatCombStatDoubRTTthirtOverZthirtTOtteightOverZeightpercentsttTotZTot}{1}$\%)~$\pm$ 
                         $\numRP{\RatCombSystNoLumiUncDoubRTTthirtOverZthirtTOtteightOverZeightabsvalttTotZTot}{3}$ ($\numRP{\RatCombSystNoLumiDoubRTTthirtOverZthirtTOtteightOverZeightpercentsttTotZTot}{1}$\%)~$\pm$
                         $\numRP{\RatCombLumiUncDoubRTTthirtOverZthirtTOtteightOverZeightabsvalttTotZTot}{3}$ ($\numRP{\RatCombLumiUncDoubRTTthirtOverZthirtTOtteightOverZeightpercentsttTotZTot}{1}$\%)  &
                         $\numRP{\RatCombCentDoubRTTthirtOverZthirtTOtteightOverZeightabsvalttTotZFid}{3}$~$\pm$
                         $\numRP{\RatCombStatDoubRTTthirtOverZthirtTOtteightOverZeightabsvalttTotZFid}{3}$ ($\numRP{\RatCombStatDoubRTTthirtOverZthirtTOtteightOverZeightpercentsttTotZFid}{1}$\%)~$\pm$
                         $\numRP{\RatCombSystNoLumiUncDoubRTTthirtOverZthirtTOtteightOverZeightabsvalttTotZFid}{3}$ ($\numRP{\RatCombSystNoLumiDoubRTTthirtOverZthirtTOtteightOverZeightpercentsttTotZFid}{1}$\%)~$\pm$
                         $\numRP{\RatCombLumiUncDoubRTTthirtOverZthirtTOtteightOverZeightabsvalttTotZFid}{3}$ ($\numRP{\RatCombLumiUncDoubRTTthirtOverZthirtTOtteightOverZeightpercentsttTotZFid}{1}$\%) &
                         $\numRP{\RatCombCentDoubRTTthirtOverZthirtTOtteightOverZeightabsvalttFidZFid}{3}$~$\pm$
                         $\numRP{\RatCombStatDoubRTTthirtOverZthirtTOtteightOverZeightabsvalttFidZFid}{3}$ ($\numRP{\RatCombStatDoubRTTthirtOverZthirtTOtteightOverZeightpercentsttFidZFid}{1}$\%)~$\pm$
                         $\numRP{\RatCombSystNoLumiUncDoubRTTthirtOverZthirtTOtteightOverZeightabsvalttFidZFid}{3}$ ($\numRP{\RatCombSystNoLumiDoubRTTthirtOverZthirtTOtteightOverZeightpercentsttFidZFid}{1}$\%)~$\pm$
                         $\numRP{\RatCombLumiUncDoubRTTthirtOverZthirtTOtteightOverZeightabsvalttFidZFid}{3}$ ($\numRP{\RatCombLumiUncDoubRTTthirtOverZthirtTOtteightOverZeightpercentsttFidZFid}{1}$\%) \\
\hline
$\ttbar/Z(13/7)$                       & $\numRP{\RatCombCentDoubRTTthirtOverZthirtTOttsevenOverZsevenabsvalttTotZTot}{3}$~$\pm$
                         $\numRP{\RatCombStatDoubRTTthirtOverZthirtTOttsevenOverZsevenabsvalttTotZTot}{3}$ ($\numRP{\RatCombStatDoubRTTthirtOverZthirtTOttsevenOverZsevenpercentsttTotZTot}{1}$\%)~$\pm$
                         $\numRP{\RatCombSystNoLumiUncDoubRTTthirtOverZthirtTOttsevenOverZsevenabsvalttTotZTot}{3}$ ($\numRP{\RatCombSystNoLumiDoubRTTthirtOverZthirtTOttsevenOverZsevenpercentsttTotZTot}{1}$\%)~$\pm$
                         $\numRP{\RatCombLumiUncDoubRTTthirtOverZthirtTOttsevenOverZsevenabsvalttTotZTot}{3}$ ($\numRP{\RatCombLumiUncDoubRTTthirtOverZthirtTOttsevenOverZsevenpercentsttTotZTot}{1}$\%) &
                         $\numRP{\RatCombCentDoubRTTthirtOverZthirtTOttsevenOverZsevenabsvalttTotZFid}{3}$~$\pm$
                         $\numRP{\RatCombStatDoubRTTthirtOverZthirtTOttsevenOverZsevenabsvalttTotZFid}{3}$ ($\numRP{\RatCombStatDoubRTTthirtOverZthirtTOttsevenOverZsevenpercentsttTotZFid}{1}$\%)~$\pm$
                         $\numRP{\RatCombSystNoLumiUncDoubRTTthirtOverZthirtTOttsevenOverZsevenabsvalttTotZFid}{3}$ ($\numRP{\RatCombSystNoLumiDoubRTTthirtOverZthirtTOttsevenOverZsevenpercentsttTotZFid}{1}$\%)~$\pm$
                         $\numRP{\RatCombLumiUncDoubRTTthirtOverZthirtTOttsevenOverZsevenabsvalttTotZFid}{3}$ ($\numRP{\RatCombLumiUncDoubRTTthirtOverZthirtTOttsevenOverZsevenpercentsttTotZFid}{1}$\%)&
                         $\numRP{\RatCombCentDoubRTTthirtOverZthirtTOttsevenOverZsevenabsvalttFidZFid}{3}$~$\pm$
                         $\numRP{\RatCombStatDoubRTTthirtOverZthirtTOttsevenOverZsevenabsvalttFidZFid}{3}$ ($\numRP{\RatCombStatDoubRTTthirtOverZthirtTOttsevenOverZsevenpercentsttFidZFid}{1}$\%)~$\pm$
                         $\numRP{\RatCombSystNoLumiUncDoubRTTthirtOverZthirtTOttsevenOverZsevenabsvalttFidZFid}{3}$ ($\numRP{\RatCombSystNoLumiDoubRTTthirtOverZthirtTOttsevenOverZsevenpercentsttFidZFid}{1}$\%)~$\pm$
                         $\numRP{\RatCombLumiUncDoubRTTthirtOverZthirtTOttsevenOverZsevenabsvalttFidZFid}{3}$ ($\numRP{\RatCombLumiUncDoubRTTthirtOverZthirtTOttsevenOverZsevenpercentsttFidZFid}{1}$\%)  \\
\hline
$\ttbar/Z(8/7)$                       & $\numRP{\RatCombCentDoubRTTeightOverZeightTOttsevenOverZsevenabsvalttTotZTot}{3}$~$\pm$
                         $\numRP{\RatCombStatDoubRTTeightOverZeightTOttsevenOverZsevenabsvalttTotZTot}{3}$ ($\numRP{\RatCombStatDoubRTTeightOverZeightTOttsevenOverZsevenpercentsttTotZTot}{1}$\%)~$\pm$
                         $\numRP{\RatCombSystNoLumiUncDoubRTTeightOverZeightTOttsevenOverZsevenabsvalttTotZTot}{3}$ ($\numRP{\RatCombSystNoLumiDoubRTTeightOverZeightTOttsevenOverZsevenpercentsttTotZTot}{1}$\%)~$\pm$
                         $\numRP{\RatCombLumiUncDoubRTTeightOverZeightTOttsevenOverZsevenabsvalttTotZTot}{3}$ ($\numRP{\RatCombLumiUncDoubRTTeightOverZeightTOttsevenOverZsevenpercentsttTotZTot}{1}$\%)  &
                         $\numRP{\RatCombCentDoubRTTeightOverZeightTOttsevenOverZsevenabsvalttTotZFid}{3}$~$\pm$
                         $\numRP{\RatCombStatDoubRTTeightOverZeightTOttsevenOverZsevenabsvalttTotZFid}{3}$ ($\numRP{\RatCombStatDoubRTTeightOverZeightTOttsevenOverZsevenpercentsttTotZFid}{1}$\%)~$\pm$
                         $\numRP{\RatCombSystNoLumiUncDoubRTTeightOverZeightTOttsevenOverZsevenabsvalttTotZFid}{3}$ ($\numRP{\RatCombSystNoLumiDoubRTTeightOverZeightTOttsevenOverZsevenpercentsttTotZFid}{1}$\%)~$\pm$
                         $\numRP{\RatCombLumiUncDoubRTTeightOverZeightTOttsevenOverZsevenabsvalttTotZFid}{3}$ ($\numRP{\RatCombLumiUncDoubRTTeightOverZeightTOttsevenOverZsevenpercentsttTotZFid}{1}$\%)  &
                         $\numRP{\RatCombCentDoubRTTeightOverZeightTOttsevenOverZsevenabsvalttFidZFid}{3}$~$\pm$
                         $\numRP{\RatCombStatDoubRTTeightOverZeightTOttsevenOverZsevenabsvalttFidZFid}{3}$ ($\numRP{\RatCombStatDoubRTTeightOverZeightTOttsevenOverZsevenpercentsttFidZFid}{1}$\%)~$\pm$
                         $\numRP{\RatCombSystNoLumiUncDoubRTTeightOverZeightTOttsevenOverZsevenabsvalttFidZFid}{3}$ ($\numRP{\RatCombSystNoLumiDoubRTTeightOverZeightTOttsevenOverZsevenpercentsttFidZFid}{1}$\%)~$\pm$
                         $\numRP{\RatCombLumiUncDoubRTTeightOverZeightTOttsevenOverZsevenabsvalttFidZFid}{3}$ ($\numRP{\RatCombLumiUncDoubRTTeightOverZeightTOttsevenOverZsevenpercentsttFidZFid}{1}$\%)  \\
\hline
\end{tabular}
\end{scriptsize}
\caption{\label{t:TabDoubRats} Summary of the $Z$-boson and $t \bar{t}$ production cross-section double ratios at $13, 8, 7$~TeV. The beam-energy uncertainty, which largely cancels in the ratios, is included in the systematic uncertainty. The luminosity uncertainty mostly cancels in this ratio. }
\end{sidewaystable}
\endgroup

\label{AppendixA}

\clearpage

\printbibliography

\newpage

\newpage 
\begin{flushleft}
{\Large The ATLAS Collaboration}

\bigskip

M.~Aaboud$^\textrm{\scriptsize 137d}$,
G.~Aad$^\textrm{\scriptsize 88}$,
B.~Abbott$^\textrm{\scriptsize 115}$,
J.~Abdallah$^\textrm{\scriptsize 8}$,
O.~Abdinov$^\textrm{\scriptsize 12}$,
B.~Abeloos$^\textrm{\scriptsize 119}$,
S.H.~Abidi$^\textrm{\scriptsize 161}$,
O.S.~AbouZeid$^\textrm{\scriptsize 139}$,
N.L.~Abraham$^\textrm{\scriptsize 151}$,
H.~Abramowicz$^\textrm{\scriptsize 155}$
% ...
\bigskip
\\
% ...
$^{8}$ Department of Physics, The University of Texas at Arlington, Arlington TX, United States of America\\
% ...
$^{12}$ Institute of Physics, Azerbaijan Academy of Sciences, Baku, Azerbaijan\\
% ...
$^{88}$ CPPM, Aix-Marseille Universit{\'e} and CNRS/IN2P3, Marseille, France\\
% ...
$^{115}$ Homer L. Dodge Department of Physics and Astronomy, University of Oklahoma, Norman OK, United States of America\\
% ...
$^{119}$ LAL, Univ. Paris-Sud, CNRS/IN2P3, Universit{\'e} Paris-Saclay, Orsay, France\\
% ...
$^{137}$ $^{(a)}$ Facult{\'e} des Sciences Ain Chock, R{\'e}seau Universitaire de Physique des Hautes Energies - Universit{\'e} Hassan II, Casablanca; $^{(b)}$ Centre National de l'Energie des Sciences Techniques Nucleaires, Rabat; $^{(c)}$ Facult{\'e} des Sciences Semlalia, Universit{\'e} Cadi Ayyad, LPHEA-Marrakech; $^{(d)}$ Facult{\'e} des Sciences, Universit{\'e} Mohamed Premier and LPTPM, Oujda; $^{(e)}$ Facult{\'e} des sciences, Universit{\'e} Mohammed V, Rabat, Morocco\\
% ...
$^{139}$ Santa Cruz Institute for Particle Physics, University of California Santa Cruz, Santa Cruz CA, United States of America\\
% ...
$^{151}$ Department of Physics and Astronomy, University of Sussex, Brighton, United Kingdom\\
% ...
$^{155}$ Raymond and Beverly Sackler School of Physics and Astronomy, Tel Aviv University, Tel Aviv, Israel\\
% ...
$^{161}$ Department of Physics, University of Toronto, Toronto ON, Canada\\
% ...
$^{a}$ Also at Department of Physics, King's College London, London, United Kingdom\\
$^{b}$ Also at Institute of Physics, Azerbaijan Academy of Sciences, Baku, Azerbaijan\\
$^{c}$ Also at Novosibirsk State University, Novosibirsk, Russia\\
$^{d}$ Also at TRIUMF, Vancouver BC, Canada\\
$^{e}$ Also at Department of Physics {\&} Astronomy, University of Louisville, Louisville, KY, United States of America\\
% ...
\end{flushleft}

\end{document}